\documentclass{phrauth} 

\usepackage{amsmath,amssymb}
\usepackage{epsfig}

\newcommand{\pabl}[2]{\frac{\partial #1}{\partial #2}}

\newcommand{\imai}{\mathrm{i}}
\renewcommand{\Re}{\mathrm{Re}}
\renewcommand{\Im}{\mathrm{Im}}

\begin{document}

\begin{frontmatter}        
\title{Semiconductor Superlattices: A model system 
for nonlinear transport}

\author{Andreas Wacker}

\address{Institut f{\"u}r Theoretische Physik,
Technische Universit{\"a}t Berlin,
Hardenbergstra{\ss}e 36, 10623 Berlin, Germany}
\ead{wacker@physik.tu-berlin.de}

\date{\today}
\maketitle

\begin{abstract}
Electric transport in semiconductor superlattices
is dominated by pronounced negative differential
conductivity. In this report the standard 
transport theories for superlattices, i.e.
miniband conduction, Wannier-Stark-hopping, and sequential
tunneling, are reviewed in detail. Their relation to each other is
clarified by a comparison with a quantum transport model based
on nonequilibrium Green functions. 
It is demonstrated how the occurrence of negative differential
conductivity causes 
inhomogeneous electric field distributions, yielding either a 
characteristic sawtooth shape of the current-voltage characteristic
or self-sustained current oscillations.
An additional ac-voltage in the THz range is included in the
theory as well. The results display absolute negative
conductance, photon-assisted tunneling, the possibility of gain,
and a negative tunneling capacitance.

\end{abstract}

 
\end{frontmatter}

\newpage
\tableofcontents

\newpage
\section*{Notation and list of symbols}
\addcontentsline{toc}{section}{{\rm Notation and list of symbols}}

Throughout this work we consider a superlattice, which is
grown in the $z$  direction. Vectors within the $(x,y)$ plane
parallel to the interfaces are denoted by bold face letters
${\bf k},{\bf r}$, while vectors in 3 dimensional space are 
$\pol{r},\pol{k},\ldots$.
All sums and integrals extend from $-\infty$ to  $\infty$ if not stated
otherwise.

The following relations are frequently used in this work and are given here
for easy reference:
\begin{equation*}\begin{split}
J_{-n}(\alpha)&=(-1)^nJ_{n}(\alpha)\\
\sum_n J_n(\alpha)J_{n+h}(\alpha)&=\delta_{h,0}\\
\e^{\imai \alpha\sin(x)}&=\sum_n J_n(\alpha)\e^{\imai nx}\\
J_{n+1}(\alpha)+J_{n-1}(\alpha)&=\frac{2n}{\alpha}J_{n}(\alpha)\\
\frac{1}{x-x_0\pm \imai 0^+}&=\mathcal{P}\left\{\frac{1}{x-x_0}\right\}\mp
\imai\pi \delta(x-x_0)
\end{split}\end{equation*}

\newpage
{\def\baselinestretch{1}\large\normalsize 
\noindent
\begin{tabular}{|c|l|}
\hline
$A$ & cross section\\
$A({\bf k},E)$ & spectral function\\
$a,a^{\dag}$ & electron annihilation and creation operators\\
$b,b^{\dag}$ & phonon annihilation and creation operators\\
$d$ & period of the superlattice structure \\
d & integration and differentiation symbol\\
$E$ & energy\\
$E^{\nu}$ & center of energy for miniband  $\nu$\\
$E_k$ & $=\hbar^2k^2/2m_c$ kinetic energy in the direction
parallel to the layers\\
$\e$ & $=2.718\ldots$ base of natural logarithm\\
$e$ & charge of the electron ($e<0$)\\
$F$ & electric field in the superlattice direction\\
$f({\bf k})$ & semiclassical distribution function\\
$\hat{H}$ & Hamilton operator\\
$\imai$ & imaginary unit\\
$I$ & $=AJ$ electric current. In section \ref{ChapDomains} 
there is an additional prefactor
sgn$(e)$\\
& so that the direction is
identical with the electron flow.\\
$J$ & current density in the superlattice direction\\
$J_l(x)$ & Bessel function of first kind and order $l$\\
${\bf k}$ & wavevector in $(x,y)$-plane [{\it i.e.}, plane $\|$ to
superlattice interfaces]\\
$k_B$ & Boltzmann constant\\
$L$ & length in superlattice direction\\
$m,n$ & well indices\\
$m_c$ & effective mass of conduction band\\
$m_0$ & electron mass $9.11\times 10^{-31}$ kg.\\
$N$ & number of wells \\
$N_D$ & doping density per period and area (unit [cm$^{-2}$]) \\
$n_B(E)$ & $=(e^{E/k_BT}-1)^{-1}$ Bose distribution function\\
$n_F(E)$ & $=(e^{E/k_BT}+1)^{-1}$ Fermi distribution function\\
\hline
\end{tabular}
\newpage
\noindent
\begin{tabular}{|c|l|}
\hline
$n_m$ & electron density per period and area (unit [cm$^{-2}$]) in well $m$\\
$q$ & Bloch vector in superlattice direction\\
$T_h^{\nu}$ & coupling between Wannier-states of miniband $\nu$ separated by
$h$ barriers\\
$T$ & temperature \\
$U$ & bias applied to the superlattice\\
$\alpha$ & $=eF_{\rm ac}d/\hbar\Omega$ argument of Bessel function for irradiation\\
$\beta$ & $=2T_1/eFd$ argument of Bessel functions for Wannier-Stark states\\
$\rho_0$ & $=m_c/\pi\hbar^2$ free-particle density of states
for the 2D electron gas\\
$\hat{\rho}$ & density operator\\
$\rho_{\alpha\beta}$ & one-particle density matrix\\
$\mu,\nu$ & indices of energy bands/levels\\
$\mu_m$ & chemical potential in well $m$, measured with respect
to the bottom of the well\\
$\Omega$ & frequency of the radiation field\\
$\phi$ & electrical potential\\
$\varphi_q^{\nu}(z)$ & Bloch function of band $\nu$\\
$\Psi_m^{\nu}(z)$ & Wannier function of band $\nu$ localized in well $m$\\
$\Phi_j^{\nu}(z)$ & Wannier-Stark function of band $\nu$ centered around
well $j$\\
$\tau$ & scattering time\\
$\Theta(x)$ & Heavyside function $\Theta(x)=0$ for $x<0$ and
$\Theta(x)=1$ for $x\ge 0$\\
$\Im\{\}$ & imaginary part\\
$\Re\{\}$ & real part\\
$\mathcal{P}\{\}$ & principal value\\
$\mathcal{O}(x^n)$ & order of $x^n$\\
$[a,b]$ & $=ab-ba$ commutator\\
$\{a,b\}$ & $=ab+ba$ anticommutator\\
\hline
\end{tabular}}

\section{Introduction} 
In this review, the transport properties of semiconductor 
superlattices are studied. These nanostructures consist 
of two different semiconductor materials 
(exhibiting similar lattice constants, e.g., GaAs and AlAs), 
which are deposited alternately on each other to 
form a periodic structure in the growth direction. 
The technical development of growth techniques allows one to 
control the thicknesses of these layers with a high precision, so that 
the interfaces are well defined within one 
atomic monolayer. In this way it is possible to tailor 
artificial periodic structures which show 
similar features to conventional crystals. 
 
Crystal structures exhibit a periodic arrangement 
of the atoms with a lattice period $a$. 
This has strong implications for the energy spectrum of the electronic 
states:  Energy bands \cite{BLO28} appear instead of discrete levels, 
which are characteristic for atoms and molecules. 
The corresponding extended states are called {\em Bloch states} and 
are characterized by  the band index $\nu$ and the Bloch vector $\pol{k}$. 
Their energy is given by the dispersion relation 
$E^{\nu}(\pol{k})$. 
If an electric field $\pol{F}$ is applied, the Bloch states 
are no longer eigenstates of the Hamiltonian, but the Bloch vector 
$\pol{k}$ becomes time dependent according to the acceleration theorem 
\begin{equation} 
\hbar\frac{\d \pol{k}}{\d t}=e\pol{F}\, , 
\end{equation} 
where $e<0$ is the 
charge of the electron. 
Since the Bloch vectors are restricted to the 
Brillouin zone, which has a size $\sim 2\pi/a$, a special feature arises 
when the acceleration of the state lasts for a time of 
$\tau_{\rm Bloch}\approx 
2\pi\hbar/(eFa)$: If interband transitions are neglected, 
the initial state $\pol{k}$ is then reached again, 
and the electron performs a periodic motion both in 
the Brillouin zone and in real space \cite{ZEN34}, which is conventionally 
referred to as a {\em Bloch oscillation}. 
For typical materials and electric fields, 
$\tau_{\rm Bloch}$ is much larger than the scattering time, and thus this 
surprising effect has not been observed yet in standard crystals. 

\begin{figure} 
\noindent\epsfig{file=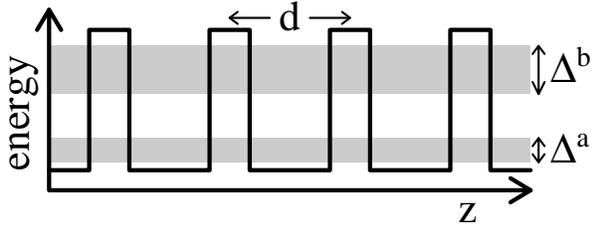,width=8cm} 
\caption[a]{Sketch of the spatial variation for the 
conduction band edge $E_c(z)$, together 
with minibands $\nu=a,b$ (shaded areas) for a semiconductor superlattice.} 
\label{Fig1uebergitter} 
\end{figure}

In 1970, Esaki and Tsu suggested that {\em superlattice structures} 
with an artificial period $d$ can be realized by the 
periodically repeated deposition of alternate  
layers from different materials \cite{ESA70}. 
This leads to spatial variations 
in the conduction and valence band of the material 
with period $d$ implying the formation 
of energy bands as sketched in Fig.~\ref{Fig1uebergitter}. 
Both the energy width $\Delta$ of these bands, as well as 
the extension $2\pi/d$ of the Brillouin zone, are much smaller than 
the corresponding values for  conventional conduction bands. 
Thus, the energy bands originating from the superlattice structure 
are called {\em minibands}. 
As $d$ can be significantly larger than the period $a$ of 
the crystal, $\tau_{\rm Bloch}$ can become smaller than 
the scattering time for available structures and applicable electric 
fields. 

It is crucial to note that the picture of Bloch-oscillations
is not the only possibility to understand the behavior
of semiconductor superlattices in an electric field.
The combination of a constant electric field 
and a periodic structure causes the formation of 
a {\em Wannier-Stark ladder} \cite{WAN60}, a periodic sequence
of energy levels separated by $eFd$ in energy space. 
This concept is complementary to the Bloch-oscillation picture,
where the frequency $\omega_{\rm Bloch}=eFd/\hbar$ corresponds to
the energy difference between the Wannier-Stark levels.

Both the occurrence of Bloch oscillations and the nature of the
Wannier-Stark states predict an increasing localization of the 
electrons with increasing electric field. 
This causes a significant drop of the conductivity at moderate fields, 
associated with the occurrence of {\em negative differential conductivity} 
\cite{ESA70}. 
Similar to the Gunn diode, this effect is likely to
cause the formation of {\em inhomogeneous field distributions}. 
These provide various kinds of interesting nonlinear behavior,
but make it difficult to observe the  Bloch oscillations.

The presence of a  strong alternating electric field (with frequency
$\Omega$ in the THz range) 
along  the superlattice structure provides further interesting features.
Both photon-assisted resonances (shifted by $\hbar \Omega$ from the original
resonance) and negative 
dynamical conductance have been predicted on the basis 
of a simple analysis \cite{KTI72}.
For specific ratios between the 
field strength and its frequency,
{\em dynamical localization} \cite{DUN86} occurs,
i.e. the dc-conductance becomes zero, which can be 
attributed to the collapse of the miniband \cite{HOL92}. 

Some fundamental aspects of superlattice physics 
have already been reviewed in Ref.~\cite{SHI75}. 
Refs.~\cite{SMI90,IVC95} consider the electronic structure  
in detail and
the review article \cite{HEL95} focuses on infrared 
spectroscopy.  Much information regarding the 
growth processes as well 
as transport measurements can be found in Ref.~\cite{GRA95d}.
The relation between Bloch oscillations and 
Wannier-Stark states has been analyzed in \cite{ROS98}. 
Ref.~\cite{BAS86} provides an early review on high-frequency phenomena. 
In addition  to superlattices consisting of different 
semiconductor materials, it is possible to achieve 
similar properties by a periodic sequence of n- and p-type doped
layers \cite{DOE84}.

\subsection{Experimental summary}

A large variety of superlattice structures has been studied 
since the original proposal of  Esaki and Tsu in 1970.
These investigations can be divided  into 
four different areas:
the nonlinear current-field relation and its implications,
the Wannier-Stark ladder,
the search for Bloch oscillations,
and the interaction with THz-fields.

The simple model by Esaki and Tsu \cite{ESA70} predicts a 
{\em nonlinear current-field relation}
exhibiting a maximum for field strengths of 
$eFd=\hbar/\tau$, where $\tau$ denotes the scattering time.
For higher fields, the current drops with increasing field,
yielding a region of {\em negative differential conductivity}.
Such behavior was first observed in the experiment
by Esaki and Chang \cite{ESA74} in 1974, where the conductance exhibited
a sequence of dips, reaching negative values,
as a function of bias voltage.
This complicated behavior was attributed to the formation
of domains with two different field values in the superlattice.
With improving sample quality the sawtooth structure
of the current-voltage characteristic due to domain formation
could be resolved \cite{KAW86}
more than a decade later. Traveling field domains
(already proposed in 1977 \cite{BUE77}) were observed as
well \cite{KAS95}. They cause self-generated current oscillations
with frequencies up to 150 GHz \cite{SCH99h}.
Domain formation effects typically hinder 
the direct observation of negative differential conductance,
as there is no simple proportionality between the measured bias and
local electric field in the sample.
In Ref.~\cite{SIB89}, the local relation between
current and field could be extracted from an analysis of the 
global current-voltage characteristic. 
A direct observation of the Esaki-Tsu
shape was possible from time-of-flight-measurements \cite{GRA91a}, and
the analysis of the frequency response \cite{SCH96e}.

The concept of the  {\em Wannier-Stark ladder} 
could be corroborated by the observation of the 
typical spacing $eFd$ in the optical excitation spectrum
of  superlattice structures
\cite{MEN88,VOI88}. More recent studies
refer to the transition between the Franz-Keldysh oscillations
and the Wannier-Stark ladder \cite{SCH94f}, and the influence of
higher valleys in the band structure \cite{OHT00}.

The dynamical nature of
{\em Bloch oscillations} with period $\tau_{\rm Bloch}$  
was observed by transient 
four-wave mixing \cite{FEL92} and by a direct
observation of the THz emission at $\omega_{\rm Bloch}$ \cite{WAS93}. 
Under stationary conditions, the phases 
for the oscillation cycles of individual electrons are 
randomized by scattering processes, and the global signal 
averages out. Therefore, decaying signals 
have been observed in these experiments after a short pulse 
excitation, which synchronizes the dynamics in the very 
beginning. More  recently, the spatial extension of the Bloch-oscillation
was resolved by measuring its dipole field \cite{LYS97}.

With the development of strong THz sources,
the {\em interaction of THz fields} with transport through superlattices
have been  studied in the last few years.
In particular, dynamical localization, photon-assisted tunneling,
and absolute negative conductance were observed
under irradiation by a free-electron laser \cite{KEA95b}.
Recent work aims at applying these effects to the detection of 
THz signals \cite{WIN98}.

Further experiments will be discussed in the subsequent sections
in direct comparison with the theory.

\subsection{Outline of this work}
 
In this work the theory of electrical transport in 
semiconductor superlattices is reviewed with a strong emphasis 
towards {\em nonlinear electric transport}. 
Here two different issues arise, which will 
be treated thoroughly: 
 
{\em How can the electric transport in semiconductor structures 
be described quantitatively?} This is not a straightforward issue 
as different energy scales like the miniband width, the scattering 
induced broadening, and the potential drop per period are typically 
of the same order of magnitude in semiconductor superlattices. 
For this reason, standard concepts from bulk transport (like 
the semiclassical Boltzmann equation), which rely on the 
large band width, become questionable. Therefore, different 
approaches, such as miniband transport \cite{ESA70}, 
Wannier-Stark hopping \cite{TSU75}, 
or sequential tunneling \cite{MIL94,WAC98}, have been suggested to study the 
transport properties of semiconductor superlattices. 
These standard approaches imply different approximations, 
and their relation to each other was only recently resolved within 
a quantum transport theory \cite{WAC98a}. 
As a result, one finds that all standard approaches are 
likely to fail if the  miniband width, the scattering 
induced broadening, and the potential drop per period take similar values. 
In this case, one has to apply a full quantum transport calculation. 
For the linear response, the quantum aspects of the problem can be 
treated within the Kubo formula \cite{KUB57}, which is evaluated in 
thermal equilibrium. 
For the nonlinear transport discussed here, this is not sufficient and 
a more involved treatment of  nonequilibrium quantum transport 
is necessary. An overview regarding 
different aspects of quantum transport 
in mesoscopic systems can be found in recent 
textbooks \cite{DAT95,HAU96,FER97,SCH98,DIT98}. 
 
{\em What is the implication of a strongly nonlinear 
relation between the current-density and the local field?} 
As long as the field and current distribution 
remain (approximately) homogeneous, the ratio between this local 
relation and  the global current-voltage 
characteristic is given by geometrical factors. 
In the region of negative differential conductivity, 
the stationary 
homogeneous field distribution becomes unstable, which may lead to 
complex spatio-temporal behavior. Typically, one observes complex scenarios, 
where current filaments or electric field domains form, which may 
yield both stationary or oscillating behavior. 
In some cases, chaotic behavior is observed as well. 
Such effects can be treated within standard concepts 
of nonlinear dynamics 
for a variety of different semiconductor systems
\cite{SCH87,ABE89,SHA92,NIE95,SCH01}. 
 
The nature of quantum transport as well as pronounced 
nonlinearities are characteristic problems of high-field 
transport in semiconductor nano\-struc\-tures. In such structures 
the electric transport is determined by various 
quantum phenomena such as  resonant tunneling 
(e.g. the resonant tunneling diode \cite{CHA74}), 
or transmission through funnel injectors (e.g. in the quantum cascade laser 
\cite{FAI94a}). In such cases, neither standard semiclassical 
bulk transport models nor linear-response theories apply,
and more advanced simulation techniques are required.
The excellent possibilities for tailoring different structures 
with specific superlattice periods, miniband widths, or doping densities
make  semiconductor superlattices an ideal  testing ground
for nonlinear quantum transport.
 
This review is organized as follows: 
Section \ref{ChapStates} introduces the basis state functions, 
such as Bloch, Wannier, or Wannier-Stark states, which will be
used in the subsequent sections. In particular, it is shown
how the miniband widths and coupling parameters can be calculated from the 
material parameters on the basis of the envelope function theory. 
Section \ref{ChapStandard} reviews the 
three standard approaches of superlattice transport: miniband transport, 
Wannier-Stark hopping, and sequential tunneling. Each of these 
approaches is valid in a certain parameter range and allows for 
a quantitative determination of the current density. 
It is a common feature of these approaches that they 
display negative differential conductivity in qualitative agreement 
with the simple Esaki-Tsu result. 
These approaches can be viewed as limiting cases of a quantum 
transport theory, which is derived in section \ref{ChapNGFT} 
on the basis of nonequilibrium Green functions. 
The occurrence of stationary and traveling field domains 
in long superlattices structures is discussed in 
section \ref{ChapDomains}. Here, specific criteria are presented, 
which allow the prediction of the global behavior on the basis 
of the current-field relation and the contact conditions. 
Finally, transport under irradiation by a THz field is addressed 
in section \ref{ChapIrr}. Some technical matters are presented in 
the appendices.

\section{States in superlattices\label{ChapStates}}
In order to perform any quantum calculation one
has to define a basis set of states to be used. While in principle
the exact result of any calculation must not depend on the choice
of basis states, this does not hold if approximations are made,
which is necessary for almost any realistic problem.
Now different sets of basis states suggest different kinds of approximations
and therefore a good choice of basis states is a crucial question.
For practical purposes the basis set is usually chosen as
the set of eigenstates of a soluble part $H_0$ of the total Hamiltonian.
If the remaining part $H-H_0$ is small, it can be treated
in lowest order of perturbation theory (e.g., Fermi's golden rule for
transition rates) which allows for a significant simplification.
This provides an indication for the practicability of a set of states.

In the following discussion we restrict ourselves to the states
arising from the conduction band of the superlattice, which is
assumed to be a single band with spin degeneracy. All wave functions 
employed in the following have to be considered as envelope functions
$f_c(\pol{r})$ with respect to this conduction band, which are
determined by the Schr{\"o}dinger-like equation
\begin{equation}
\left[E_c(\pol{r})-\nabla\frac{\hbar^2}{2m_c}\nabla+e\phi(\pol{r})\right]
f_{c}(\pol{r})=Ef_{c}(\pol{r})\label{Eq2Henvelope}
\end{equation}
where $m_c$ denotes the effective mass in the conduction band.
Assuming ideal interfaces, the
structure is translational invariant within
the $x$ and $y$ direction perpendicular to the growth direction.
Therefore the $(x,y)$ dependence can be taken in the form
of plane waves $\e^{\imai {\bf k}\cdot {\bf r}}$ where
${\bf k}$ and ${\bf r}$ are vectors within the two-dimensional
$(x,y)$ plane.
The crucial point in this section
is the choice of the $z$ dependence of the basis states. This reflects
the current direction considered here and has therefore strong
implications on the description of transport.

Semiconductor superlattices are designed as periodic structures
with period $d$ in the growth direction. Thus their eigenstates can be
chosen as Bloch-states $\varphi^{\nu}_q(z)$
(where $q\in [-\pi/d,\pi/d]$ denotes
the Bloch-vector and $\nu$ is the band index) which extend over
the whole structure.
The corresponding eigenvalues $E^{\nu}(q)$ of the Hamiltonian
form the miniband (subsection \ref{SecMiniband}).
This provides the exact solution for a
perfect superlattice without applied electric field.
An alternative set of basis functions can be constructed by
employing localized wave functions which resemble eigenstates of single
quantum wells labeled by the index $n$.  Here we use the
Wannier-states $\Psi^{\nu}(z-nd)$, which can be constructed
separately for each miniband $\nu$ (subsection \ref{SecWannier}).
At third we may consider the Hamiltonian of the superlattice in the
presence of a finite electric field $F$. Then the energy levels
take the form $E_n^{\nu}=E^{\nu}_{0}-neFd$ and one obtains the
Wannier-Stark states
$\Phi^{\nu}(z-nd)$, where we neglect the field-dependent coupling
between the subbands (subsection \ref{SecWS}).
The spatial extension of these states is inversely proportional to the
electric field.
In the subsequent subsections the different basis sets will be derived and
their properties will be studied in detail.

\subsection{Minibands \label{SecMiniband}}
\begin{figure}
\noindent\epsfig{file=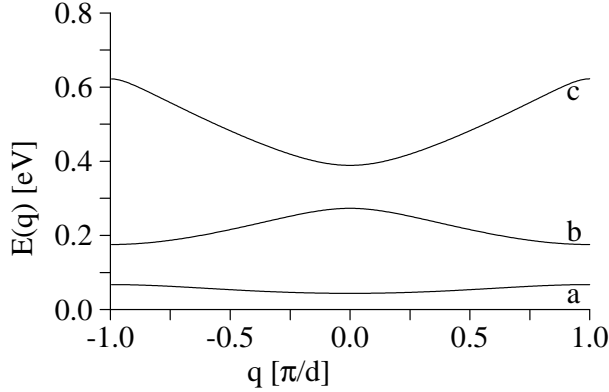,width=8cm}
\caption[a]{Calculated miniband structure for the GaAs-Al$_{0.3}$Ga$_{0.7}$As
superlattice used in  \cite{RAU98} with well width 6.5 nm and barrier width
2.5 nm.
\label{Fig2MBrauch}}
\end{figure}

The periodicity of the superlattice structure within the $z$ direction
implies that the eigenstates of the Hamiltonian can be written
as Bloch states $\varphi^{\nu}_q(z)$, where $q\in [-\pi/d,\pi/d]$ denotes
the Bloch vector.
The construction of these eigenstates can be performed straightforward
within the transfer matrix formulation, see e.g.
 \cite{AND87,YU99}. Within a region $z_j<z<z_{j+1}$
of constant potential and constant material composition the envelope function
can be written as $f(z)=A_j\e^{\imai k_j(E)(z-z_j)}+B_j\e^{-\imai k_j(E)(z-z_j)}$.
Then the connection rules \cite{BEN66} 
(see also Refs.~\cite{IVC95,YU99,BAS88} for a detailed discussion)
\begin{eqnarray}
f_c(\pol{r})_{z\to z_{j+1}+0^-}&=&f_c(\pol{r})_{z\to z_{j+1}+0^+}\label{Eq2Connection1}\\
\frac{1}{m_{c,j}}\pabl{f_c(\pol{r})}{z}_{z\to z_{j+1}+0^-}&=&
\frac{1}{m_{c,j+1}}\pabl{f_c(\pol{r})}{z}_{z\to z_{j+1}+0^+}\label{Eq2Connection2}
\end{eqnarray}
apply\footnote{Throughout this work we use the energy dispersion
$E(\pol{k})= E_c+\frac{\hbar^2k^2}{2m_c}$
with the band edge $E_c=0.8x$ meV and the
effective mass $m_c=(0.067+0.083x)m_0$ for the conduction band
of Al$_x$Ga$_{1-x}$As with $x<0.45$ \cite{ADA93}.
For GaAs/AlAs structures, nonparabolicity effects are included
using the energy-dependent effective mass
$m_c(E)=m_c(E-E_{v})/(E_c-E_v)$ with
the parameters $m_c=0.067 m_0$, $E_c=0$, $E_v=-1.52$ meV
for GaAs and $m_c=0.152 m_0$, $E_c=1.06$ meV, $E_v=-2.07$ meV for AlAs
\cite{WHI81,BRO90}, where $E_v$ denotes the edge of the valence 
band. X,L-related effects are neglected for simplicity. They
become relevant in some transport studies \cite{HOS98}.
Some approaches to the theoretical study of tunneling via these minima 
can be found in  \cite{STO94,TIN95,OGA99}.}
where $m_{c,j}$ is the effective mass in region $z_j<z<z_{j+1}$.
\begin{equation}
\left(\begin{array}{c}
A_{j+1}\\B_{j+1}
\end{array}\right)
=\mathcal{M}_{j}(E)
\left(\begin{array}{c}
A_{j}\\B_{j}
\end{array}\right)\end{equation}
with
\begin{equation}
\mathcal{M}_j(E)=\frac{1}{2}\left(\begin{array}{cc}
\left(1+\frac{k_jm_{c,j+1}}{k_{j+1}m_{c,j}}\right)\e^{\imai k_j(z_{j+1}-z_j)}
& \left(1-\frac{k_jm_{c,j+1}}{k_{j+1}m_{c,j}}\right)\e^{-\imai k_j(z_{j+1}-z_j)}
\\[0.2cm]
\left(1-\frac{k_jm_{c,j+1}}{k_{j+1}m_{c,j}}\right)\e^{\imai k_j(z_{j+1}-z_j)}
& \left(1+\frac{k_jm_{c,j+1}}{k_{j+1}m_{c,j}}\right)\e^{-\imai k_j(z_{j+1}-z_j)}
\end{array}\right)\, .
\end{equation}
If a single period of the superlattice consists of $M$ regions
with constant material composition,
the Bloch-condition $\varphi_q(z+d)=\e^{\imai qd}\varphi_q(z)$ implies
\begin{equation}
\left(\begin{array}{c}
A_{M+1}\\ B_{M+1}
\end{array}\right)
=\prod_{j=1}^M \mathcal{M}_j(E)
\left(\begin{array}{c}
A_{1}\\ B_{1}
\end{array}\right)
\stackrel{!}{=}
\e^{\imai qd}
\left(\begin{array}{c}
A_{1}\\ B_{1}
\end{array}\right) \label{EqBlochbed} \, .
\end{equation}
For standard superlattices ($M=2$) the solutions resemble those of
the Kronig-Penney model \cite{KRO31}, except for the use of effective masses
associated with the connection rule (\ref{Eq2Connection2}). Within
the transfer formalism the extension to superlattices with a
basis \cite{KRI98} (larger $M$) is straightforward.
For given $q$, Eq.~(\ref{EqBlochbed}) is only solvable for selected values
of $E$ which define the miniband-structure $E^{\nu}(q)$.
An example is shown in Fig.~\ref{Fig2MBrauch}.
Next to the envelope function approximation discussed here,
different approaches can be used to calculate the superlattice
band structure \cite{SMI90,WAN99d}.

For a given miniband $\nu$  the following quantities may be defined:
\begin{alignat}{2}
&\mbox{center of miniband: }&
E^{\nu}&=d/(2\pi)\int_{-\pi/d}^{\pi/d}\d q\, E^{\nu}(q)\\
&\mbox{miniband width: }&
\Delta^{\nu}&={\rm Max}_q\{E^{\nu}(q)\}-{\rm Min}_q\{E^{\nu}(q)\}
\end{alignat}
which characterize the miniband structure.
For the miniband structure shown in Fig.~\ref{Fig2MBrauch}
the values $E^a=54.5$ meV, $\Delta^a=23.6$ meV, $E^b=220$ meV,
$\Delta^b=98$ meV,
$E^c=491$ meV, and $\Delta^c=233$ meV are found, where the band indices
are labeled by $\nu=a,b,c\ldots $.
The increase of $\Delta$ with $\nu$ can be easily understood in terms of
the increasing transparency of the barrier with the electron energy.
As $E^{\nu}(q)$ is restricted to $-\pi/d<q<\pi/d$ and
$E^{\nu}(q)=E^{\nu}(-q)$ by Kramers degeneracy, the function $E^{\nu}(q)$
can be expanded as follows:
\begin{equation}
E^{\nu}(q)=E^{\nu}
+\sum_{h=1}^{\infty} 2 T^{\nu}_h  \cos(hdq)\label{Eq2fourier}
\end{equation}
Typically, the terms $T^{\nu}_h$ for $h\ge 2$ are much smaller than
the $T^{\nu}_1$ term.
E.g., for the bandstructure of Fig.~\ref{Fig2MBrauch} one obtains
$T_1^a=-5.84$ meV, $T_2^a=0.48$ meV for the lowest band,
$T_1^b=23.7$ meV, $T_2^b=2.1$ meV for the second band, and
$T_1^c= -53.3$ meV, $T_2^c=5.3$ meV for the third band.
This demonstrates that the band structure is essentially
of cosine-shape and thus $\Delta^{\nu}\approx 4|T_1^{\nu}|$.
The dispersion $E^{\nu}(q)=E^{\nu}
+ 2 T^{\nu}_1  \cos(dq)$ can be viewed as the result of a standard
tight-binding calculation with next-neighbor coupling $T_1^{\nu}$.

In order to perform many-particle calculations
the formalism of second quantization, see, e.g.,  \cite{MAH90}, is
appropriate.  Let  $a^{\nu\dag}_q$ and
$a^{\nu}_q$ be the creation and annihilation operator for electrons in
the Bloch-state of band $\nu$ with Bloch-vector $q$. Then the Hamiltonian
reads
\begin{equation}
\hat{H}_{\rm SL}=\sum_{\nu}\int_{-\pi/d}^{\pi/d}\d q\,
E^{\nu}(q) a^{\nu\dag}_q a^{\nu}_q\label{Eq2HamMB}
\end{equation}
which is diagonal in the Bloch-states, as the Bloch-states are eigenstates
of the unperturbed superlattice.

\subsection{Bloch-states of the three-dimensional superlattice}
In the preceding subsection only the
$z$ direction of the superlattice was taken into account.
For an ideal superlattice Eq.~(\ref{Eq2Henvelope}) does not
exhibit an $(x,y)$-dependence and thus a complete set of
eigenstates states can be constructed by products of plane waves
$\e^{\imai {\bf k}\cdot {\bf r}}/(2\pi)$ and a $z$-dependent function
$f_k(z)$ which satisfies the eigenvalue equation
\begin{equation}
\left(E_c(z)-\pabl{}{z}\frac{\hbar^2}{2m_c(z)}\pabl{}{z}
+\frac{\hbar^2 {\bf k}^2}{2m_c(z)}\right)f_k(z) =E f_k(z)
\label{Eq2Henvelope3D}
\end{equation}
As $E_c(z)$ and $m_c(z)$ are periodic functions with the
superlattice period $d$, the eigenstates
are Bloch state $f_k(z)=\varphi^{\nu}_{q,{\bf k}}(z)$ with energy
$E^{\nu}(q,{\bf k})$.\footnote{Eq.~(\ref{Eq2Henvelope3D})
shows that the effective Hamiltonian is {\em not exactly}
separable in a $z$ and {\bf r}-dependent part, as the $z$-dependent
effective mass affects the {\bf k}-dependence, describing the
behavior in the $(x,y)$ plane. Nevertheless, this subtlety is not
taken into account here, as discussed below.}
Within first order perturbation theory
in ${\bf k}^2$ one obtains the energy
\begin{equation}
E^{\nu}(q,{\bf k})\approx E^{\nu}(q,{\bf 0})+
\langle \varphi^{\nu}_{q,{\bf 0}}|
\frac{\hbar^2 {\bf k}^2}{2m_c(z)}
|\varphi^{\nu}_{q,{\bf 0}}\rangle
\end{equation}
Now $\varphi^{\nu}_{q,{\bf 0}}(z)$ will exhibit a larger probability
in the well, so that it seems reasonable to replace the
second term by
\begin{equation}
E_k=\frac{\hbar^2 {\bf k}^2}{2m_w}
\end{equation}
where $m_w$ is the effective mass of the quantum well.
In analogy to Eq.~(\ref{Eq2HamMB}) the full Hamiltonian
reads
\begin{equation}
\hat{H}_{\rm SL}=\sum_{\nu}\int_{-\pi/d}^{\pi/d}\d q\,\int \d ^2k
\left[E^{\nu}(q)+E_k\right]
a^{\nu\dag}_q({\bf k}) a^{\nu}_q({\bf k})\label{Eq2HamMBvoll}
\end{equation}
where $a^{\nu\dag}_q({\bf k})$ and
$a^{\nu}_q({\bf k})$ are the creation and annihilation operator
for electrons in
the Bloch-state of band $\nu$ with Bloch-vector $q$ and wave vector ${\bf k}$
in $(x,y)$ plane.
In order to evaluate matrix elements for scattering processes
the zeroth order envelope wave-functions
$\varphi^{\nu}_{q,{\bf 0}}(z)
\e^{\imai {\bf k}\cdot {\bf r}}/(2\pi)$ are applied in subsequent sections.
The treatment is completely analogous for  Wannier and  Wannier-Stark states
discussed in the subsequent subsections.

\subsection{Wannier functions\label{SecWannier}}
By definition the Bloch-functions are delocalized over the
whole superlattice structure. This may provide difficulties
if electric fields are applied or effects due to
the finite length of the superlattice are considered.
Therefore it is often helpful to use different
sets of basis states which are better localized.
A tempting choice would be the use of eigenstates of
single quantum wells, see, e.g., \cite{PRE94,AGU97}.
Nevertheless such a choice has a severe
shortcoming:
The corresponding states are solutions
of two different Hamiltonians, each neglecting the presence
of the other well. Thus these states  are not orthogonal
which provides  complications. Typically, the coupling
is estimated by the transfer Hamiltonian \cite{BAR61}
within this approach.

\begin{figure}
\noindent\epsfig{file=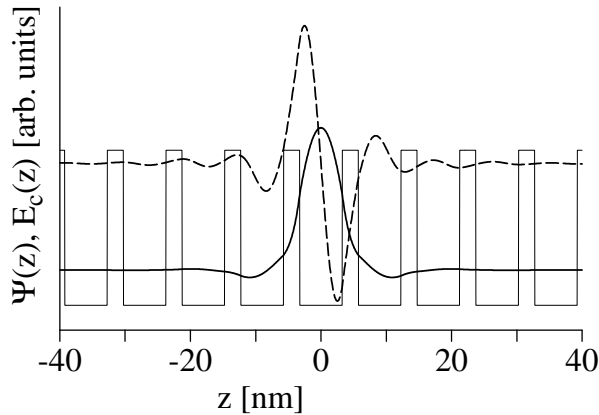,width=8cm}
\caption[a]{Wannier functions for the superlattice
from Fig.~\ref{Fig2MBrauch}. Full line: $\Psi^{a}(z)$, dashed line
$\Psi^{b}(z)$, The thin line indicates the conduction band edge profile.
\label{Fig2Wanstates}}
\end{figure}

For these reasons it is more convenient to use the set of
Wannier functions \cite{WAN37}
\begin{eqnarray}
\Psi^{\nu}(z-nd)=\sqrt{\frac{d}{2\pi}}\int_{-\pi/d}^{\pi/d} \d q \,
\e^{-\imai nqd} \varphi_q^{\nu}(z)\label{Eqwannier}
\end{eqnarray}
which are constructed from the Bloch functions with the normalization
$\int \d z\, [\varphi_{q'}^{\nu'}(z)]^*\varphi_q^{\nu}(z)=
\delta_{\nu\nu'}\delta(q-q')$.
Here some care has to be
taken: The Bloch functions are only defined up to a complex phase
which can be chosen arbitrarily for each value $q$. The functions
$\Psi(z)$ depend strongly on the choice of these phases.
In  \cite{KOH59} it has been shown that
the Wannier functions are maximally localized if the phase
is chosen in the following way at a symmetry point $z_{\rm sym}$ of the
superlattice:
If $\varphi^{\nu}_0(z_{\rm sym})\neq 0$ choose
$\varphi^{\nu}_{q}(z_{\rm sym})\in \mathbb{R}$ (i).
Otherwise choose
$\imai\varphi^{\nu}_{q}(z_{\rm sym})\in \mathbb{R}$ (ii).
In both cases the phase is chosen such, that
$\varphi^{\nu}_{q}(z)$ is an analytic function in $q$.
(The latter requirement defines the phase when
$\varphi^{\nu}_{q}(z_{\rm sym})$ becomes zero and prevents
from arbitrary sign changes of $\varphi_{q}(z_{\rm sym})$.)
For such a choice the Wannier functions $\Psi^{\nu}(z)$
are real and  symmetric (i) or
antisymmetric (ii) around $z_{\rm sym}$. Now there are two
symmetry points, one in the center of the well ($z_{\rm sym}^{\rm well}$)
and one in the barrier ($z_{\rm sym}^{\rm barrier}$), for
a typical superlattice.  If the energy of the miniband is below the
conduction band of the barrier, the Wannier functions seem to be strongly
localized for $z_{\rm sym}^{\rm well}$, while $z_{\rm sym}^{\rm barrier}$
may be suited as well for larger energies, where the minibands
are above the barrier. This point has also been addressed in
 \cite{PED91}.
In Fig.~\ref{Fig2Wanstates} the Wannier functions for the
first two bands (using $z_{\rm sym}^{\rm well}$) are displayed.
One finds that both functions
are essentially localized to the central quantum well where they
resemble the bound states.
Outside the well they exhibit a decaying oscillatory behavior which ensures
the orthonormality relation\footnote{Note that the orthonormality is not strictly fulfilled for different
bands $\nu,\mu$ if an energy dependent effective mass is used.
In this case, energy-dependent  Hamiltonians (\ref{Eq2Henvelope})
are used for the envelope functions, and therefore the orthonormality
of eigenfunctions belonging to different energies is not guaranteed.
In principle this problem could be cured by reconstructing the
full wave functions from the envelope functions under consideration of
the admixtures from different bands.}
\begin{eqnarray}
\int \d z\,  \Psi^{\nu}(z-nd)\Psi^{\mu}(z-md)=\delta_{n,m}\delta_{\nu,\mu}
\end{eqnarray}

Within second quantization the creation  $a^{\nu\dag}_n$ and
annihilation $a^{\nu}_n$ operators
of the states associated with the Wannier functions $\Psi^{\nu}(z-nd)$
are defined via
\begin{equation}
a_{q}^{\nu}=\sqrt{\frac{d}{2\pi}}
\sum_{n}\e^{-\imai qnd}a_{n}^{\nu} \, .
\end{equation}
Inserting into Eq.~(\ref{Eq2HamMB}) and using Eq.~(\ref{Eq2fourier})
one obtains the Hamiltonian within the Wannier basis
\begin{equation}
\hat{H}_{\rm SL}=\sum_{n,\nu} \left[
E^{\nu}a^{\nu\dag}_na^{\nu}_n
+\sum_{h=1}^{\infty}T^{\nu}_h \left(a^{\nu\dag}_{n+h}a^{\nu}_n
+a^{\nu\dag}_{n-h}a^{\nu}_n \right)\right]\, .
\label{Eq2hamWS}
\end{equation}
As the Wannier functions are linear combinations of Bloch functions
with different energies, they do not represent stationary states.
Neglecting terms with $h>1$ the time evolution of the annihilation operators
in the Heisenberg representation
is given by
\begin{equation}
\imai\hbar \frac{\d }{\d t}a^{\nu}_n(t)=E^{\nu}a^{\nu}_n(t)
+T_1^{\nu}\left[a^{\nu}_{n+1}(t)+ a^{\nu}_{n-1}(t)\right]\label{Eqcoherent}\, .
\end{equation}
For the initial condition $a^{\nu}_n(t=0)=\delta_{n,0}a^{\nu}_0$ this
set of equations has the solution
\begin{equation}
a^{\nu}_n(t)=
\imai^{-n}J_n\left(\frac{2T^{\nu}_1}{\hbar}t\right)\e^{-\imai E^{\nu}t/\hbar} 
a^{\nu}_0
\end{equation}
where $J_{n}(x)$ is the Bessel function of first kind \cite{ABR66}.
This shows that the initially occupied Wannier state
decays on a time scale of
\begin{equation}
\tau_{\rm Wannier}\sim \frac{\hbar}{2T_1^{\nu}}\, .\label{Eq2Wtime}
\end{equation}
At this time
$\langle a^{\nu\, \dag}_0(t)a^{\nu}_0(t)\rangle=J_0(1)^2\approx 0.586$, thus
$\tau_{\rm Wannier}$ may be viewed as a kind of half-life period,
although there is no exponential decay.

If an electric field $F$ is applied to the superlattice, the additional
potential $\phi(z)=-Fz$ has to be taken into account. Within the Wannier basis
the corresponding terms of the Hamiltonian can be evaluated directly
by the corresponding matrix elements
$\int dz\,  \Psi^{\mu}(z-md)eFz\Psi^{\nu}(z-nd)$.
Including the parallel degrees of freedom ${\bf k}$,
the total Hamiltonian $\hat{H}=\hat{H}_0+\hat{H}_1+\hat{H}_2$ reads:
\begin{eqnarray}
\hat{H}_0&=&\sum_{n,\nu}\int \d^2k
(E^{\nu}+E_k-eFR^{\nu\nu}_0-eFdn)a_n^{\nu\dag}({\bf k})
a_n^{\nu}({\bf k})\label{Eq2hamW0}\\
\hat{H}_1&=&\sum_{n,\nu,\mu}\sum_{h=1}^{\infty}
\int \d^2k \big\{
T_h^{\nu} \left[a_{n+h}^{\nu\dag}({\bf k})
a_n^{\nu}({\bf k})+a_{n}^{\nu\dag}({\bf k})
a_{n+h}^{\nu}({\bf k})\right]\delta_{\mu,\nu}
\nonumber \\
&&\phantom{\sum_{n,\nu,\mu}\int \d^2k \Big\{}
-eFR^{\mu\nu}_h\left[a_{n+h}^{\mu\dag}({\bf k})a_n^{\nu}({\bf k})
+a_{n}^{\nu\dag}({\bf k})a_{n+h}^{\mu}({\bf k})\right]\Big\}
\label{Eq2hamW1}\\
\hat{H}_2&=&\sum_{\stackrel{n,\nu,\mu}{\nu\neq \mu}}\int \d^2k
(-eFR_0^{\mu\nu})a_{n}^{\mu\dag}({\bf k})a_n^{\nu}({\bf k})
\label{Eq2hamW2}
\end{eqnarray}
with the couplings
$R_h^{\mu\nu}=\int \d z\,  \Psi^{\mu}(z-hd)z\Psi^{\nu}(z)$.
If the superlattice exhibits inversion symmetry the coefficients
$R_h^{\nu\nu}$ vanish for $h>0$.
Finally note, that the expression of $\hat{H}$ is still exact, except
for the separation of $z$ and $(x,y)$ direction.

The term $\hat{H}_0$ describes the energy of the states
in the superlattice neglecting any couplings
to different bands or different wells.
$\hat{H}_1$ gives the coupling between different wells.
Finally $\hat{H}_2$ describes the field-dependent mixing
of the levels inside a given well. In particular it is responsible
for the Stark shift.

The term $\hat{H}_0+\hat{H}_2$ for $n=0$ can be diagonalized \cite{KAZ72} by
constructing the new basis
\begin{equation}
|\tilde{\mu}\rangle=\sum_{\nu} U^{\nu\tilde{\mu}}|\nu \rangle
\end{equation}
satisfying $(\hat{H}_0+\hat{H}_2)|\tilde{\mu}\rangle=
\tilde{E}^{\tilde{\mu}}|\tilde{\mu}\rangle$ where
the columns of $U^{\nu\tilde{\mu}}$ are the components of
$|\tilde{\mu}\rangle$ with respect to the basis $|\nu \rangle$.
This shows, that the level separation becomes
field dependent, which has been recently observed in superlattice transport
\cite{VIE98} under irradiation.
In the new basis the Hamiltonian is given by
$\hat{H}=\hat{H}_0^{\rm ren}+\hat{H}_1^{\rm ren}$ with
\begin{eqnarray}
\hat{H}_0^{\rm ren}&=& \sum_{n,\nu} \int \d^2k\,
(\tilde{E}^{\nu}+E_k-eFdn)a_n^{\nu\dag}({\bf k})a_n^{\nu}({\bf k})
\label{Eq2hamW0ren}\\
\hat{H}_1^{\rm ren}&=&
\sum_{n,\nu,\mu}\sum_{h=1}^{\infty}
\int \d^2k\, \tilde{H}^{\mu\nu}_h\left[
a_{n+h}^{\mu\dag}({\bf k})a_n^{\nu}({\bf k})+a_{n}^{\nu\dag}({\bf k})
a_{n+h}^{\mu}({\bf k})\right]\label{Eq2hamW1ren}
\end{eqnarray}
and the matrix elements
\begin{equation}
\tilde{H}^{\tilde{\mu}\tilde{\nu}}_h=\sum_{\mu,\nu}
\left(U^{\mu\tilde{\mu}}\right)^*(T_h^{\nu}\delta_{\nu,\mu}-eFR^{\mu\nu}_h)
U^{\nu\tilde{\nu}}\, .
\label{Eq2Matren}\end{equation}
It will turn out later that in the limit of sequential tunneling
it is more appropriate to use  $\hat{H}_1^{\rm ren}$  as a
perturbation instead of $\hat{H}_1$.

\subsection{Wannier-Stark ladder\label{SecWS}}
If an electric field $F$ is applied to the superlattice structure
the Hamiltonian exhibits an additional scalar potential
$e\phi(z)=-eFz$ which destroys the  translational invariance. In this case
we can easily see:
If there exists an eigenstate with wavefunction
$\Phi_0(z)$ and energy $E_0$, then the set of states
corresponding to
wavefunctions $\Phi_j(z)=\Phi_0(z-jd)$ are eigenstates
of the Hamiltonian with energies $E_j=E_0-jeFd$ as well.
These states are equally spaced both in energy and real space
and form the so-called {\em Wannier-Stark ladder} \cite{WAN60}.
This feature has to be considered with some care,
as the potential $e\phi(z)$ is not bounded
for the infinite crystal, which implies a continuous energy spectrum
\cite{ZAK68}. Nevertheless, the characteristic energy spectrum
of these Wannier-Stark ladders could be resolved experimentally
\cite{MEN88,VOI88} in semiconductor superlattices.
For a more detailed discussion of this subject see
 \cite{ROS98,NEN91,AGU95}.

\begin{figure}
\noindent\epsfig{file=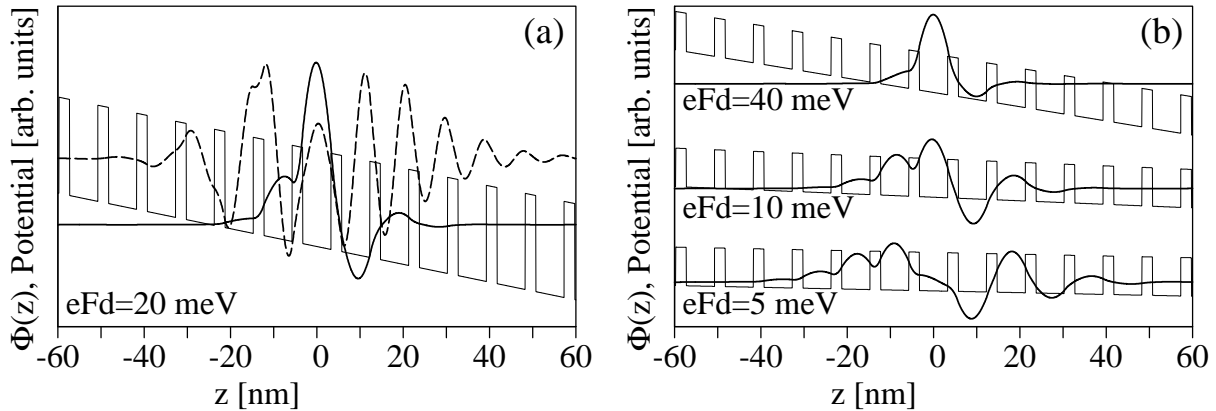,width=16cm}
\caption[a]{Wannier-Stark states calculated from Eq.~(\ref{Eq2WS})
for the superlattice from Fig.~\ref{Fig2MBrauch}.
The thin line indicates the conduction band edge profile.
(a) Full line: $\Phi^{a}(z)$, dashed line $\Phi^{b}(z)$,
(b) $\Phi^{a}(z)$ for different fields.
\label{Fig2WSstates}}
\end{figure}

If one restricts the Hamiltonian $\hat{H}_{\rm SL}$ in
Eq.~(\ref{Eq2HamMB})
to a given miniband $\nu$, an analytical solution for the
eigenstates of $\hat{H}_{\rm SL}-eFz$ exists \cite{KAN59}:
\begin{equation}
|\Phi^{\nu}_j\rangle=\sqrt{\frac{d}{2\pi}}\int_{-\pi/d}^{\pi/d}\d q\,
\exp\left\{\frac{\imai}{eF}\int_0^q \d q'
\left[E_j^{\nu}-E^{\nu}(q')\right]\right\}|\varphi_q^{\nu}\rangle
\label{Eq2WS}
\end{equation}
where $E_j^{\nu}=E^{\nu}-jeFd$ is the ladder of energies
corresponding to the $\nu^{\rm th}$ miniband with average energy $E^{\nu}$
similar to the discussion of the Wannier states\footnote{This
representation depends crucially on the relative
choice of phases in the Bloch functions $\varphi_q^{\nu}(z)$. The
situation resembles the construction of Wannier states (\ref{Eqwannier})
and it is suggestive to use the same choice of phase although I am not
aware of a proof. For consistency the origin of $z$ has to be chosen such
that $z_{\rm sym}=0$ holds for the symmetry point of the superlattice.}.
The field-induced coupling to different bands induces a finite lifetime
of these single band Wannier-Stark states due to Zener tunneling
\cite{KAN59}. Thus these Wannier-Stark states can be viewed
as resonant states (an explicit calculation of these resonances has been
performed in  \cite{CHA93}). For a cosine-shaped band
$E^{\nu}(q)=E^{\nu}+2T_1^{\nu}\cos(qd)$ the Wannier-Stark states
from Eq.~(\ref{Eq2WS}) can be expanded in Wannier states
$|\Psi^{\nu}_{n}\rangle$ \cite{FUK73}
\begin{equation}
|\Phi^{\nu}_j\rangle=\sum_{n=-\infty}^{\infty}
J_{n-j}\left(\frac{2T_1^{\nu}}{eFd}\right)|\Psi^{\nu}_{n}\rangle
\quad \mbox{for next neighbor coupling}\label{Eq2WSsimp}
\end{equation}
where the definition  (\ref{Eqwannier}) has been used.
This relation can be obtained directly by diagonalizing
Eqs.~(\ref{Eq2hamW0},\ref{Eq2hamW1}) within
the restriction to a single band, nearest-neighbor coupling $T_1$,
and $R^{\nu\nu}_0=0$, i.e. $z_{\rm sym}=0$.
In Fig.~\ref{Fig2WSstates} examples for the Wannier-Stark states
are shown.
It can be clearly seen that the localization of these states
increases with the electric field. They exhibit an oscillatory structure
within a region of approximately $\Delta^{\nu}/eFd$ periods and a
strong decay outside this region.
This magnitude can be  estimated via Eq.~(\ref{Eq2WSsimp}).
As $\sum_{n=-\infty}^{\infty}n^2J_n^2(x)=x^2/2$ [Eq.~(8.536(2) of
 \cite{GRA80}]
we can conclude that $J_n\left(\frac{2T_1^{\nu}}{eFd}\right)$ deviates
from zero essentially in the range
$-2|T_1^{\nu}|/eFd\lesssim n\lesssim 2|T_1^{\nu}|/eFd$ which,
together with $\Delta^{\nu}\approx 4|T_1^{\nu}|$, provides
the result given above.

\section[Standard approaches]{The standard approaches for superlattice 
transport\label{ChapStandard}} 
If an external bias is applied to a conductor, such as a metal 
or a semiconductor, typically an electrical current is generated. 
The magnitude of this current is determined by the 
band structure of the material, scattering processes, 
the applied field strength, as well as the equilibrium 
carrier distribution of the conductor. 
In this section the question is addressed, how the special 
design of a semiconductor superlattice, which allows 
to vary the band-structure in a wide range, 
influences the transport behavior. 
Throughout this section we assume that a homogeneous electrical field 
$F$ is applied in the direction of the superlattice (the $z$ direction) 
and consider the current parallel to this field. 
Due to symmetry reasons the transverse current parallel to the layers 
should vanish. 
 
\begin{figure} 
\noindent\epsfig{file=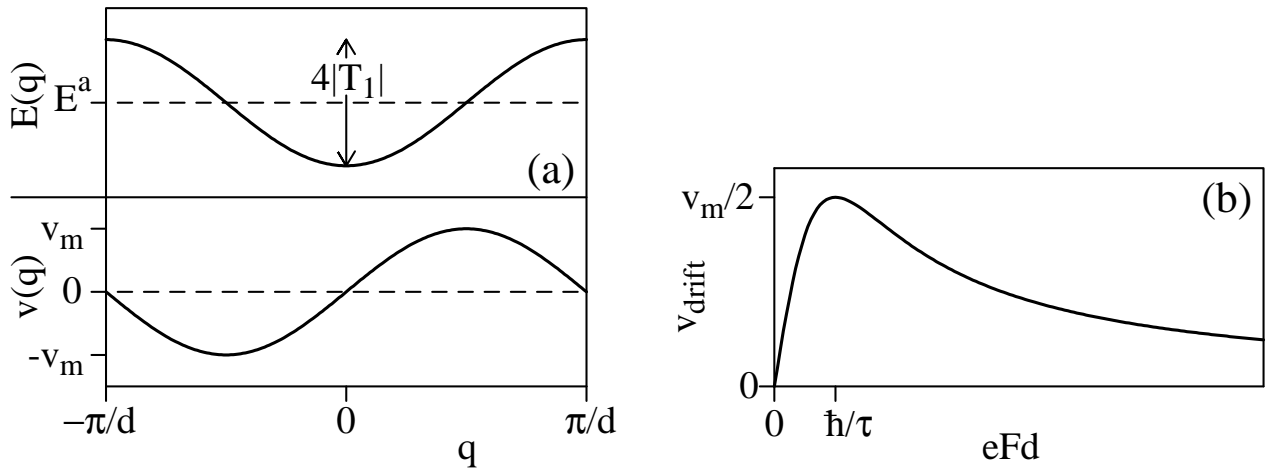,width=16cm} 
\caption[a]{(a) Dispersion $E^a(q)$ and velocity $v^a(q)$ for the lowest 
miniband. (b) Velocity-field relation according to Esaki-Tsu \cite{ESA70}. 
\label{Fig3esakitsu}} 
\end{figure}

A very elementary solution to the problem 
has been provided by Esaki and Tsu in their pioneering paper \cite{ESA70}. 
Consider the lowest miniband of the superlattice labeled by the 
superscript $a$. The eigenstates are the Bloch-states $\varphi^{a}_q(z)$ 
with the Bloch-vector $q$ and the dispersion is approximately given by 
$E^{a}(q)\approx E^{a}-2|T_1^a|\cos(qd)$ 
(see Sec.~\ref{SecMiniband} for details) as depicted 
in Fig.~\ref{Fig3esakitsu}a. At low temperatures 
the states close to the minimum at $q\approx0$ are occupied 
in thermal equilibrium. 
If an electric field is applied (in $z$-direction) the Bloch-states 
are no longer eigenstates of the full Hamiltonian but change in time. 
According to the acceleration theorem \cite{BLO28} 
\begin{equation} 
\frac{\d  q}{\d t}=\frac{eF}{\hbar} 
\label{Eq3acceltheorem} 
\end{equation} 
the states remain Bloch states in time, but the Bloch-vector becomes 
time dependent and we find $q(t)=eFt/\hbar$ if 
the electron starts in the minimum of the band at $t=0$. 
For $t=\pi\hbar/(eFd)$ the boundary of the Brillouin zone ($q=\pi/d$) is 
reached. This point is equivalent with the point at $q=-\pi/d$, so that the 
trajectory continues there which is often called Bragg-reflection. 
Finally, at $t=\tau_{\rm Bloch}=2\pi\hbar/(eFd)$ the origin is reached 
again. 
Neglecting transitions to different bands (Zener transitions, 
whose probability is extremely small for low fields) 
the state remains in the given band and thus the same state is reached after 
$\tau_{\rm Bloch}$ resulting in a periodical motion of the state through 
the Brillouin zone \cite{ZEN34}. This oscillation is called 
{\em Bloch-oscillation} 
and is quite general for arbitrary bandstructures. 
It could be observed in superlattices \cite{FEL92,WAS93}. 
The Bloch-states $q$ travel with the velocity 
\begin{equation} 
v^a(q)=\frac{1}{\hbar}\pabl{E^a(q)}{q}\approx \frac{2d|T_1^a|}{\hbar}\sin (qd) 
\end{equation} 
Thus, we find $v(t)=v_m\sin(eFdt/\hbar)$ and 
the position of a wave packet 
$z(t)=z_0+z_m\{1-\cos(eFdt/\hbar)\}$ 
with $v_m=2d|T_1^a|/\hbar$ and $z_m=2|T_1^a|/eF$. 
In \cite{BOU95} this behavior has been nicely demonstrated by an 
explicit solution of the Schr{\"o}dinger equation. 
The spatial amplitude of this oscillation has been resolved 
recently \cite{LYS97}. 
 
Scattering processes will interrupt this oscillatory behavior. 
As scattering processes are likely to restore thermal equilibrium 
it makes sense to assume that the scattered electron will be found 
close to $q=0$, the initial point used before. As long as the average 
scattering time $\tau$ is much smaller than $\pi\hbar/(2eFd)$ the 
electrons will remain in the range $0\lesssim q< \pi/(2d)$ where 
the velocity increases with $q$ and thus an increase of $F$ will generate 
larger average drift velocities. Thus for $eFd\ll \pi\hbar/(2\tau)$ 
a linear increase of $v_{\rm drift}(F)$ is expected. 
In contrast, if  $\tau\gtrsim \pi\hbar/(eFd)$ the electrons reach 
the region $-\pi/d< q< 0$ with negative velocities and thus the average 
drift velocity can be expected to drop with the field for 
$eFd\gtrsim \pi\hbar/\tau$. For high fields, $eFd\gg 2\pi\hbar/\tau$, 
the electrons perform many periods of the Bloch-oscillation before they 
are scattered and thus the average drift velocity tends to zero 
for $F\to \infty$. A detailed analysis for a constant (momentum-independent) 
scattering time $\tau$ gives the Esaki-Tsu relation \cite{ESA70}: 
\begin{equation} 
v_{\rm drift}(F)=v_{\rm ET}(F)= 
v_m\frac{eFd \hbar/\tau}{(eFd)^2+(\hbar/\tau)^2} 
\label{Eq3EsakiTsu} 
\end{equation} 
This result will be derived 
in Sec.~\ref{SecMBTconstscat} as well. 
The drift velocity exhibits a linear increase with $F$ for low fields, 
a maximum at $eFd=\hbar/\tau$ and negative differential conductivity for 
$eFd>\hbar/\tau$, see Fig.~\ref{Fig3esakitsu}b. 
This general behavior could be observed experimentally \cite{SIB90,GRA91a}. 
 
The rather simple argument given above neglects 
the plane wave states in $(x,y)$-direction, the thermal distribution 
of carriers, and treats scattering processes in an extremely 
simplified manner. In Section \ref{SecMBT} a more realistic treatment 
is given within the {\em miniband transport} model 
where the electrons occupy Bloch-states and the dynamical 
evolution of the single states is described by the acceleration theorem 
(\ref{Eq3acceltheorem}). 
 
A complementary approach to miniband transport is the use 
of Wannier-Stark states, which are the 'real' eigenstates of the 
superlattice in an electric field [see Sec.~\ref{SecWS} for a discussion 
of the problems involved with these states]. Scattering processes cause 
transitions between these states yielding a net current in the 
direction of the electric field \cite{TSU75}. This approach is called 
{\em Wannier-Stark hopping} and will be described in detail in 
Sec.~\ref{SecWSH}. 
 
For superlattices with thick barriers (i.e. narrow minibands) 
it seems more appropriate to view the structure as a series of 
weakly-coupled quantum wells with localized eigenstates. 
Due to the residual coupling between the wells  tunneling processes 
through the barriers are possible and the electrical transport 
results from  {\em sequential tunneling} from well to well, which will be 
discussed in  Sec.~\ref{SecST}. 
Generally, the lowest states in adjacent 
wells are energetically aligned for zero potential drop. Thus, the 
energy conservation in Fermi's 
golden rule would forbid tunneling transitions for finite fields. 
Here it is essential to include the scattering induced 
broadening of the states which allows for such transitions. 
 
\begin{figure} 
\begin{tabular}{|c|c|c|c|} 
\hline 
 & coupling $ T_1$ & voltage drop $eFd$ & 
scattering $\Gamma=\hbar/\tau$\\ \hline
\begin{minipage}{5.2cm} 
Miniband conduction\\ 
\epsfig{file=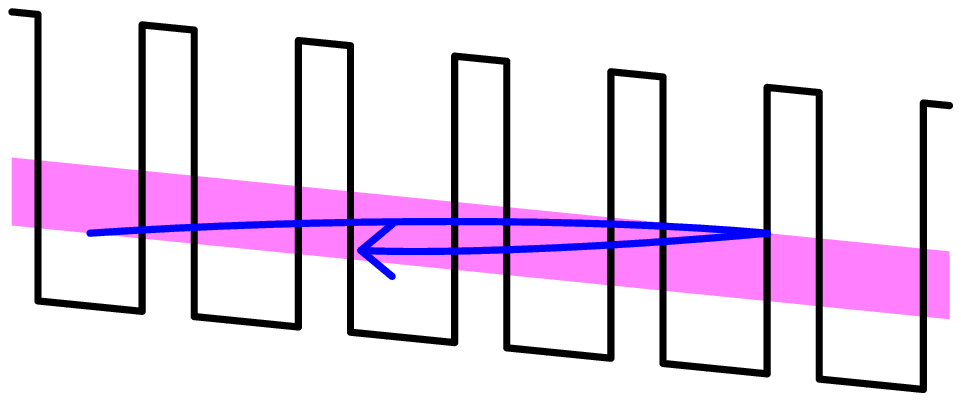,width=5cm} 
\end{minipage} 
&\begin{minipage}{2.5cm} exact:\\ 
miniband\end{minipage} & 
acceleration & 
golden rule \\ \hline
\begin{minipage}{5.2cm} 
Wannier-Stark hopping\\ 
\epsfig{file=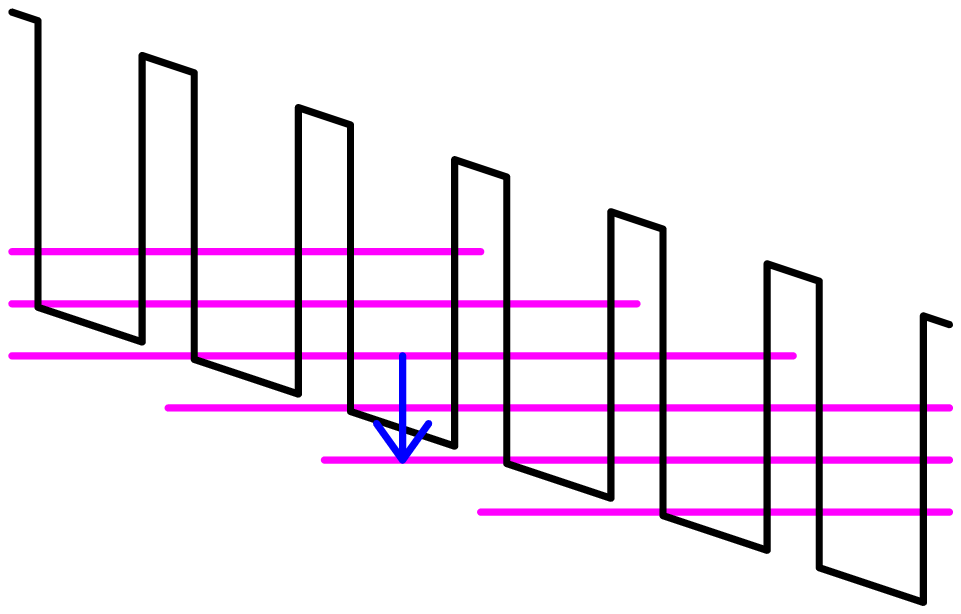,width=5cm}
\end{minipage} & 
\multicolumn{2}{|c|}  {exact: Wannier Stark states} & 
golden rule \\ \hline
\begin{minipage}{5.2cm}Sequential tunneling\\ 
\epsfig{file=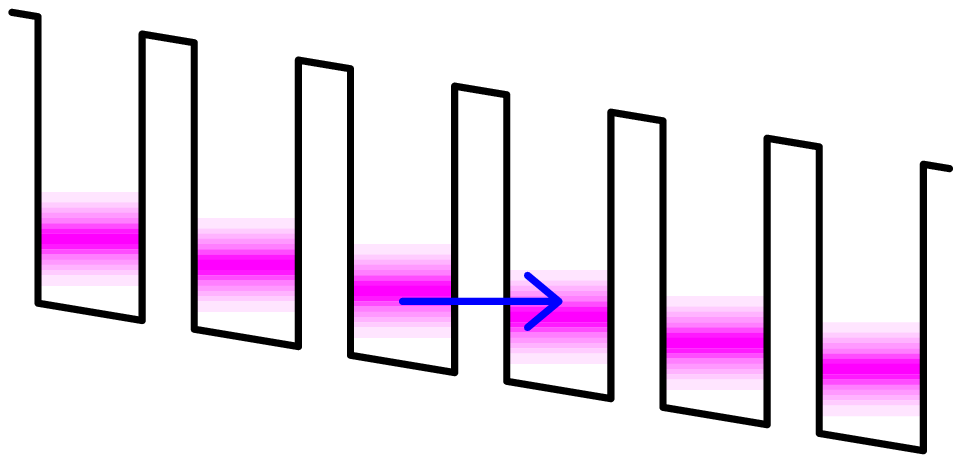,width=5cm}
\end{minipage} & 
 lowest order& \begin{minipage}{3.1cm} energy mismatch\end{minipage} & 
\begin{minipage}{3.0cm} "exact"\\ spectral function 
\end{minipage} \\ \hline 
\end{tabular} 
\caption[a]{Overview of the  different standard approaches for 
superlattice transport. 
\label{Fig3skizzesimp}} 
\end{figure} 
 
\begin{figure} 
\noindent\epsfig{file=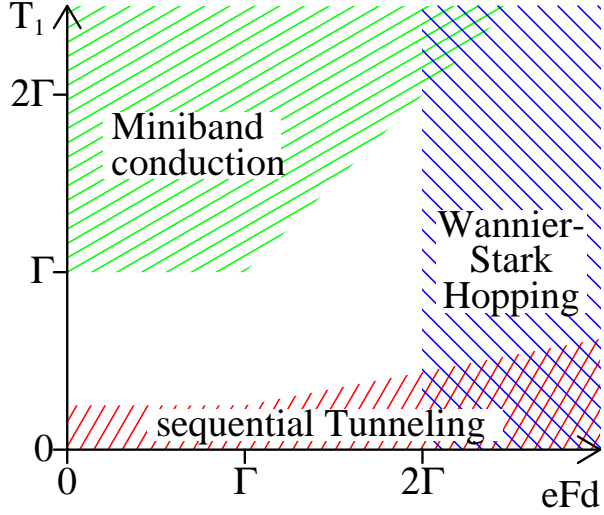,width=8cm} 
\caption[a]{Ranges of validity for the different standard approaches for 
superlattice transport. Miniband transport holds for $eFd\ll 2|T_1|$ and 
$\Gamma\ll 2|T_1|$; Wannier-Stark hopping holds for 
$\Gamma\ll eFd$; Sequential tunneling holds for $2|T_1|\ll eFd$ or 
$2|T_1|\ll \Gamma$, from   \cite{WAC98a}. 
\label{Fig3regimes}} 
\end{figure}

These three complementary approaches are  schematically depicted 
in Fig.~\ref{Fig3skizzesimp}. 
They treat the basic 
ingredients to transport, band structure (coupling $T_1$), 
field strength ($eFd$), and scattering ($\Gamma=\hbar/\tau$), 
in completely different ways. In section \ref{ChapNGFT} and 
appendix \ref{AppDerivation} these approaches 
will be compared with a quantum transport model, which 
will determine their respective range of validity as sketched in 
Fig.~\ref{Fig3regimes}. 
 
It is an intriguing feature that all three approaches provide a 
velocity-field relation in 
qualitative agreement with Fig.~\ref{Fig3esakitsu}b, except 
that the linear increase is missing in the Wannier-Stark hopping 
model. Therefore the qualitative features from the Esaki-Tsu 
model persist but details as well as the magnitude of the current may 
be strongly altered. These points will be discussed in detail 
in the subsequent subsections.

\subsection{Miniband transport\label{SecMBT}} 
Conventionally, the electrical transport in semiconductors or metals 
is described within a semiclassical approach. 
Due to the periodicity of the crystal, a basis of eigenstates 
from Bloch-functions can be constructed. 
For the superlattice structure considered here it is convenient 
to treat the Bloch-vector $q$ in superlattice direction 
and the Bloch-vector ${\bf k}$ in the direction parallel to the layers 
separately. The eigenfunctions are 
$\varphi_q^{\nu}(z)\e^{\imai {\bf k}\cdot {\bf r}}/2\pi$ 
and the corresponding energies are $E(q,{\bf k},\nu)=E^{\nu}(q)+E_{k}$ 
with the superlattice dispersion $E^{\nu}(q)$ and the in-plane energy 
$E_{k}=\hbar^2{\bf k}^2/2m_c$, 
for details see Sec.~\ref{SecMiniband}. 
The occupation of these states is given by the distribution 
function $f(q,{\bf k},\nu,t)$ describing 
the probability that the state $(q,{\bf k},\nu)$ 
is occupied and  $f(q,{\bf k},\nu,t) \d q \d^2k/(2\pi)^3$ is the 
particle density within the the volume element $\d q \d^2k$ 
in momentum space. 
 
An electric field breaks the translational invariance 
of the system  and the Bloch states are no 
longer eigenstates. Within the semiclassical theory 
the temporal evolution of the distribution function is given by the 
Boltzmann equation\footnote{If spatially inhomogeneous distributions 
$f(\vec{r},q,{\bf k},\nu,t)$ are considered, the convection term 
$v^{\nu}(q)\pabl{f}{z}+\hbar{\bf k}/m_c \cdot \pabl{f}{{\bf r}}$ 
has to be added on the left hand side.} 
\begin{equation} 
\pabl{f(q,{\bf k},\nu,t)}{t}+\frac{eF}{\hbar} 
\pabl{f}{q} = \left(\pabl{f}{t}\right)_{\rm scatt} 
\label{Eq3Boltzmann} 
\end{equation} 
which is derived in most textbooks on solid state physics. 
(The inclusion of magnetic fields 
is straightforward and results for superlattice structures 
are given in  \cite{PAL92,MIL95}; see also \cite{HUT94} for corresponding 
experimental results. Recently, a mechanism for spontaneous current 
generation due to the presence of hot electron in a magnetic field
perpendicular to the superlattice direction was proposed \cite{CAN00}.) 
The right side  describes the change of the distribution 
function due to  scattering. 
For impurity or phonon scattering 
the scattering term reads 
\begin{equation}\begin{split} 
&\left(\pabl{f(q,{\bf k},\nu,t)}{t}\right)_{\rm scatt}=\\
&\sum_{\nu'} 
\int_{-\pi/d}^{\pi/d} \d q'\int\d^2 k'\, 
\left\{P(q',{\bf k}',\nu'\to q,{\bf k},\nu) 
f(q',{\bf k}',\nu',t)[1-f(q,{\bf k},\nu,t)] \right.\\ 
&\phantom{\sum_{\nu'}\int_{-\pi/d}^{\pi/d} \d q'\int\d^2 k'\,}
-\left. P(q,{\bf k},\nu\to q',{\bf k}',\nu') 
f(q,{\bf k},\nu,t)[1-f(q',{\bf k}',\nu',t)]\right\} \, . 
\end{split}\end{equation} 
$P(q,{\bf k},\nu\to q',{\bf k}',\nu')$ denotes the scattering probability 
from state $(q,{\bf k},\nu)$ to state $(q',{\bf k}',\nu')$ 
which can be calculated by 
Fermi's golden rule using appropriate matrix elements for the different 
scattering processes. Details regarding these scattering processes 
(for bulk systems) can be found textbooks, such as   \cite{FER91,SEE96}. 
Specific calculations for superlattice structures can be found in 
\cite{DHA90,ETE93,ROT99a}. 
Once the distribution function is known, the current density $J_{\rm MBT}$ 
for miniband transport 
in $z$ direction can be 
evaluated directly by 
\begin{equation} 
J_{\rm MBT}=\frac{2\mbox{(for Spin)}e}{(2\pi)^3}\sum_{\nu} 
\int_{-\pi/d}^{\pi/d} \d q \int\d^2 k\, f(q,{\bf k},\nu,t)v^{\nu}(q) 
\label{Eq3Boltzmann-J} 
\end{equation} 
and the electron density per superlattice 
period (in units [1/cm$^2$]) is given by 
\begin{equation} 
n=\frac{2\mbox{(for Spin)}d} 
{(2\pi)^3}\sum_{\nu} 
\int_{-\pi/d}^{\pi/d} \d q\int\d^2 k\, f(q,{\bf k},\nu,t)
\label{Eq3Boltzmann-n} \, . 
\end{equation} 
This approach for the electric transport is called 
miniband transport. An earlier review has been given in   \cite{SIB95} 
where several experimental details are provided. 
 
One has to be aware that Boltzmann's equation holds for 
classical particles under the assumption of independent 
scattering events. The only quantum mechanical ingredient is 
the use of the dispersion $E^q(q,{\bf k})$, thus the term 
semiclassical approach is often used. Therefore deviations 
may result from various quantum features, such as 
scattering induced broadening of the states, the intracollisional 
field effect, or correlations between scattering effects 
leading, e.g., to weak localization. While these features are notoriously 
difficult to describe, operational solution methods like Monte-Carlo methods 
\cite{JAC83,JAC98} exist for the 
Boltzmann equation explaining the popularity of the 
semiclassical approach.

\subsubsection{Relaxation time approximation\label{SecMBTconstscat}} 
Boltzmann's equation can be solved easily if the scattering term 
is approximated by 
\begin{equation} 
\left(\pabl{f(q,{\bf k},t)}{t}\right)_{\rm scatt}= 
\frac{n_F(E(q,{\bf k})-\mu)-f(q,{\bf k},t)}{\tau_{\rm scatt}(q,{\bf k})} 
\end{equation} 
with the relaxation time $\tau_{\rm scatt}(q,{\bf k})$. 
The relaxation time approximation 
is correct in the linear response regime 
for a variety of scattering processes. 
Here it is applied to nonlinear transport, in order to obtain 
some insight into the general features. 
The underlying assumption is that any scattering process restores the thermal 
equilibrium described by the Fermi function $n_F(E)=[\exp(E/k_BT)+1]^{-1}$ 
and the chemical potential $\mu$. 
(The discussion is restricted  to the lowest miniband here and thus 
the miniband index $\nu$ is neglected.) 
Then the stationary Boltzmann equation reads: 
\begin{equation} 
\frac{eF\tau_{\rm scatt}(q,{\bf k})}{\hbar} 
\pabl{f(q,{\bf k})}{q}+f(q,{\bf k})=n_F(E(q,{\bf k})-\mu)\, . 
\end{equation} 
This is an inhomogeneous linear partial differential equation 
which, together with the boundary condition 
$f(-\pi/d,{\bf k},t)=f(\pi/d,{\bf k},t)$, 
can be integrated directly and one finds: 
\begin{equation}\begin{split} 
f(q,{\bf k})= 
\int_{-\pi/d}^{\pi/d}\d q_0& 
\frac{\hbar n_F(E(q_0,{\bf k})-\mu)}{eF\tau_{\rm scatt}(q_0,{\bf k})} 
\exp\left[-\int_{q_0}^{q}\d q' 
\frac{\hbar}{eF\tau_{\rm scatt}(q',{\bf k})}\right]\\ 
&\times\left\{ \Theta(q-q_0)+\frac{1} 
{\exp\left[\int_{-\pi/d}^{\pi/d}\d q' 
\frac{\hbar}{eF\tau_{\rm scatt}(q',{\bf k})}\right]-1} 
\right\}\end{split} 
\end{equation} 
Assuming a constant scattering rate $\tau_{\rm scatt}(q,{\bf k})=\tau$, 
the $q'$ integrals become trivial. For 
the simplified miniband structure $E(q)=E^a-2|T_1^a|\cos(qd)$ one obtains 
the electron density from Eq.~(\ref{Eq3Boltzmann-n}) 
\begin{equation} 
n=n_{\rm eq}(\mu,T)=\frac{2d}{(2\pi)^3}\int_{-\pi/d}^{\pi/d}\d q_0
\int\d^2k\, 
n_F(E(q_0,{\bf k})-\mu)\label{Eq3n0} 
\end{equation} 
and the current density from Eq.~(\ref{Eq3Boltzmann-J}) 
\begin{equation} 
J_{\rm MBT}=e\frac{2|T_1^a|}{\hbar} 
\frac{eFd\hbar/\tau}{(\hbar/\tau)^2+(eFd)^2} c_{\rm eq}(\mu,T)\label{Eq3relaxJ} 
\end{equation} 
with 
\begin{equation} 
c_{\rm eq}(\mu,T)=\frac{2d}{(2\pi)^3} 
\int_{-\pi/d}^{\pi/d}\d q_0\int\d^2k\, 
\cos(q_0d)n_F(E(q_0,{\bf k})-\mu)\, .\label{Eq3c0} 
\end{equation} 
Note that the field dependence of the current density 
is identical with the simple Esaki-Tsu result (\ref{Eq3EsakiTsu}), 
but the prefactor has a complicated form, which will be analyzed in the 
following. 
The $k$-integration can be performed analytically 
in $n_{\rm eq}(\mu,T)$ and $c_{\rm eq}(\mu,T)$. 
\begin{equation} 
\frac{2}{(2\pi)^2}\int\d^2k\, n_F(E(q,{\bf k})-\mu)= 
\rho_0 \int_0^{\infty} \d E_k \frac{1}{\e^{\frac{E(q)+E_k-\mu}{k_BT}}+1} 
=\rho_0k_BT\log 
\left[\e^{\left(\frac{\mu-E(q)}{k_BT}\right)}+1\right] 
\end{equation} 
Here $\rho_0=m_c/\pi\hbar^2$ is the density of states of the 
two-dimensional electron gas parallel to the layers including spin 
degeneracy. 
 
At first consider the {\em degenerate} case $T=0$ which holds 
for low temperatures $k_BT\ll 2|T_1^a|$.
If $\mu>E^a+2|T_1^a|$ one obtains \cite{LEB70} $n=\rho_0(\mu-E^a)$ 
and 
\begin{equation} 
J_{\rm MBT}=e\frac{2\rho_0|T_1^a|^2}{\hbar} 
\frac{eFd\hbar/\tau}{(\hbar/\tau)^2+(eFd)^2} 
\quad \mbox{for}\quad 
\left\{\begin{array}{rl} 
k_BT&\ll 2|T_1^a|\\ 
n&>2|T_1^a|\rho_0 
\end{array}\right.\label{Eq3MBT-deg} 
\end{equation} 
This expression is independent from the carrier density.
(For superlattices with very thin barriers a different behavior
has been reported \cite{PUS97} which was explained by 
the one-dimensional character of the tranport due to  inhomogeneities.) 
A second instructive result is obtained for very low densities 
$0<(\mu-E^a)/2|T_1^a|+1\ll 1$ when 
\begin{equation} 
n_{\rm eq}\approx c_{\rm eq}\approx \rho_0|T_1^a|
\frac{2}{3\pi} 
\left[\arccos\left(\frac{E_a-\mu}{2|T_1^a|}\right)\right]^3 
\end{equation} 
holds and thus 
\begin{equation} 
J_{\rm MBT}=en\frac{2|T_1^a|}{\hbar}\frac{eFd\hbar/\tau}{(\hbar/\tau)^2+(eFd)^2} 
\quad \mbox{for}\quad 
\left\{\begin{array}{rl} 
k_BT&\ll 2|T_1^a|\\ 
n&\ll 2|T_1^a|\rho_0 
\end{array}\right. \label{Eq3MBT-ET} 
\end{equation} 
which gives $J=e(n/d)v_{\rm ET}(F)$ providing the 
Esaki-Tsu result (\ref{Eq3EsakiTsu}) mentioned above. 
 
In the {\em non-degenerate} case $\mu\ll E^a-2|T_1^a|$ one obtains \cite{SHI75} 
\begin{eqnarray} 
n&=&\rho_0k_BT\e^{\left(\frac{\mu-E^{a}}{k_BT}\right)} 
I_0\left(\frac{2|T_1^a|}{k_BT}\right)\label{Eq3nBessel}\\ 
J&=&e\rho_0k_BT\e^{\left(\frac{\mu-E^{a}}{k_BT}\right)} 
I_1\left(\frac{2|T_1^a|}{k_BT}\right) 
\frac{2|T_1^a|}{\hbar}\frac{eFd\hbar/\tau}{(\hbar/\tau)^2+(eFd)^2} 
\label{Eq3JBessel}\end{eqnarray} 
with the modified Bessel functions $I_0,I_1$, see Eq.~(9.6.19) of 
  \cite{ABR66}. 
For low temperatures $k_BT\ll 2|T_1^a|$ the argument $x$ of the 
Bessel functions becomes large and 
$I_0(x)\sim I_1(x)$. Then Eq.~(\ref{Eq3MBT-ET}) 
is recovered again. For high temperatures $k_BT\gg 2|T_1^a|$ 
the Bessel functions behave as $I_0(x)\sim 1$ and $I_1(x)\sim x/2$, and 
\begin{equation} 
J_{\rm MBT}=en \frac{2|T_1^a|^2}{k_BT\hbar} 
\frac{eFd\hbar/\tau}{(\hbar/\tau)^2+(eFd)^2} 
\quad \mbox{for}\quad 
\left\{\begin{array}{rl} 
k_BT&\gg 2|T_1^a|\\ 
n&\ll \rho_0k_BT 
\end{array}\right. \, . \label{Eq3MBT-T} 
\end{equation} 
Such a $1/T$ dependence of the current density has 
been observed experimentally in   \cite{BRO90a,SIB93a} albeit the 
superlattices considered there exhibit a rather small miniband width 
and the justification of the miniband transport approach is not 
straightforward. 
 
\subsubsection{Two scattering times} 
A severe problem of the relaxation-time model is the fact that 
all scattering processes restore thermal equilibrium. 
While this may be correct for phonon scattering, where energy 
can be transferred to the phonon systems, this assumption is clearly 
wrong for impurity scattering, which does not change the energy 
of the particle. The significance of this distinction can be studied 
by applying the following scattering term \cite{KTI72,IGN91} 
\begin{equation} 
\left(\pabl{f(q,{\bf k},t)}{t}\right)_{\rm scatt}= 
\frac{n_F(E(q,{\bf k})-\mu)-f(q,{\bf k},t)}{\tau_e}+ 
\frac{f(-q,{\bf k},t)-f(q,{\bf k},t)}{2\tau_{\rm elast}}\, . 
\label{Eq3Scatt2relax} 
\end{equation} 
Here scattering processes which change both momentum and energy 
are contained in the energy scattering time $\tau_e$ and 
elastic scattering events, changing only the momentum, are taken into 
account by $\tau_{\rm elast}$. While the Boltzmann equation 
was solved explicitly in Sec.~\ref{SecMBTconstscat}, 
dynamical equations for the physical quantities of interest 
are derived here by taking the appropriate averages with the 
distribution function. At first consider the electron 
density (\ref{Eq3Boltzmann-n}). Performing the 
integral $\frac{2d}{(2\pi)^3}\int \d q\d^2k$ of 
Eq.~(\ref{Eq3Boltzmann}) one obtains 
\begin{equation} 
\frac{\d n(t)}{\d t}=\frac{n_{\rm eq}-n(t)}{\tau_e} 
\end{equation} 
where the periodicity of $f(q)$ has been used to eliminate the term 
$\propto F$. Therefore the electron density is again 
given by $n_{\rm eq}(\mu,T)$, see Eq.~(\ref{Eq3n0}). In the same manner 
one obtains (using integration by parts in the term 
$\propto F$) 
\begin{equation} 
\frac{\d J(t)}{\d t}-\frac{\e^2Fd}{\hbar^2} 2|T_1^a|c(t) 
=-\frac{J(t)}{\tau_m} \label{Eq3dynJ} 
\end{equation} 
with the momentum relaxation time $1/\tau_m=1/\tau_e+1/\tau_{\rm elast}$ 
and the average 
\begin{equation} 
c(t)=\frac{2d}{(2\pi)^3}\int_{-\pi/d}^{\pi/d}\d q\int\d^2k\, 
\cos(qd)f(q,{\bf k},t)\, . 
\end{equation} 
Finally, the dynamical evolution of $c(t)$ is given by 
\begin{equation} 
\frac{\d c(t)}{\d t}+\frac{Fd}{2|T_1^a|} J(t) 
=\frac{c_{\rm eq}(\mu,T)-c(t)}{\tau_e}\, .\label{Eq3dync} 
\end{equation} 
This equation can be considered as a balance equation 
for the kinetic energy in the superlattice direction 
as $\frac{2d}{(2\pi)^3}\int\d q\d^2k\, 
E(q)f(q,{\bf k},t)=E^an(t)-2|T_1^a|c(t)$. 
The stationary solution of Eqs.~(\ref{Eq3dynJ},\ref{Eq3dync}) gives 
\begin{equation} 
J_{\rm MBT}=e\delta \frac{2|T_1^a|}{\hbar} 
\frac{eFd\hbar/\tau_{\rm eff}}
{(\hbar/\tau_{\rm eff})^2+(eFd)^2}c_{\rm eq}(\mu,T)\label{Eq3J2relax} 
\end{equation} 
with the effective scattering time $\tau_{\rm eff}=\sqrt{\tau_e\tau_m}$ 
and $\delta=\sqrt{\tau_m/\tau_e}$. 
This is just the result  (\ref{Eq3relaxJ}) with the 
additional factor $\delta$ reducing the magnitude of the current, 
as $\tau_m\le \tau_e$. The prefactors $n_{\rm eq}(\mu,T)$ and 
$c_{\rm eq}(\mu,T)$ are identical with those introduced 
in the last subsection. 
 
The relaxation time approximation has proven to be useful for 
the analysis of experimental data by fitting the phenomenological 
scattering times $\tau_e,\tau_{\rm elast}$. In   \cite{SCH98h} 
the times $1/\tau_e=9\times 10^{12}/$s and $1/\tau_{\rm elast}=2\times 10^{13}/$s 
has been obtained for a variety of highly-doped and 
strongly-coupled superlattices at $T=300$ K. 
 
It should be noted that there is an instructive interpretation \cite{ROT99} 
of Eq.~(\ref{Eq3J2relax}) which may be rewritten as 
\begin{equation} 
\frac{1}{J}=\frac{d}{ec_{\rm eq}(\mu,T)}\left(\frac{1}{v_{\rm lf}(F)} 
+\frac{1}{v_{\rm hf}(F)}\right)\label{Eq3Jmatthiessen} 
\end{equation} 
with the low-field velocity 
\begin{equation} 
v_{\rm lf}(F)=\frac{e\tau_m}{m_{\rm miniband}}F 
\end{equation} 
where 
\begin{equation} 
\frac{1}{m_{\rm miniband}}=\frac{1}{\hbar^2} 
\frac{\partial^2 E^a(q)}{\partial q^2}_{|q=0}=\frac{2|T_1^a|d^2}{\hbar^2}\, . 
\end{equation} 
This is just the standard expression for the linear conductivity 
which is dominated by momentum relaxation. 
The high-field velocity 
\begin{equation} 
v_{\rm hf}(F)=\frac{P_{\rm loss}}{eF}\label{Eq3Venergyloss} 
\end{equation} 
is determined by the maximal energy loss per particle given by 
$P_{\rm loss}=2|T_1^a|/\tau_e$ for the particular scattering 
term (\ref{Eq3Scatt2relax}) 
where {\bf k} is conserved. This relation results from the 
energy balance providing negative differential conductivity as already 
pointed out in   \cite{KRO53}. While both expressions for the 
low-field velocity and the  high-field velocity are quite 
general, it is not clear if the interpolation (\ref{Eq3Jmatthiessen}) 
in the form of a generalized Matthiessen's rule \cite{ROT99} holds 
beyond the relaxation time model.

\subsubsection{Results for real scattering processes\label{SecMBTsophist}} 
The relaxation time approximations discussed above contain several problems: 
\begin{itemize} 
\item The scattering processes conserve {\bf k}, which is artificial. 
An adequate improvement to this point has been suggested in   \cite{GER93}. 
\item The magnitude of the scattering times is not directly 
related to physical scattering processes. 
\item Energy relaxation is treated in a very crude way by assuming 
that in-scattering occurs from a thermal distribution. 
\end{itemize} 
 
In   \cite{LEI91} balance equations have been derived 
for the condition of stationary drift velocity and stationary mean energy. 
Here the distribution function was parameterized by a drifted 
Fermi-function similar to the concepts of the 
hydrodynamic model for semiconductor transport (see \cite{RUD98} and 
references cited therein). This approach allows for taking into account 
the microscopic scattering matrix elements for impurity and 
electron--phonon scattering and good results were obtained for the 
peak position and peak velocity observed in   \cite{SIB90}. 
 
\begin{figure} 
\epsfig{file=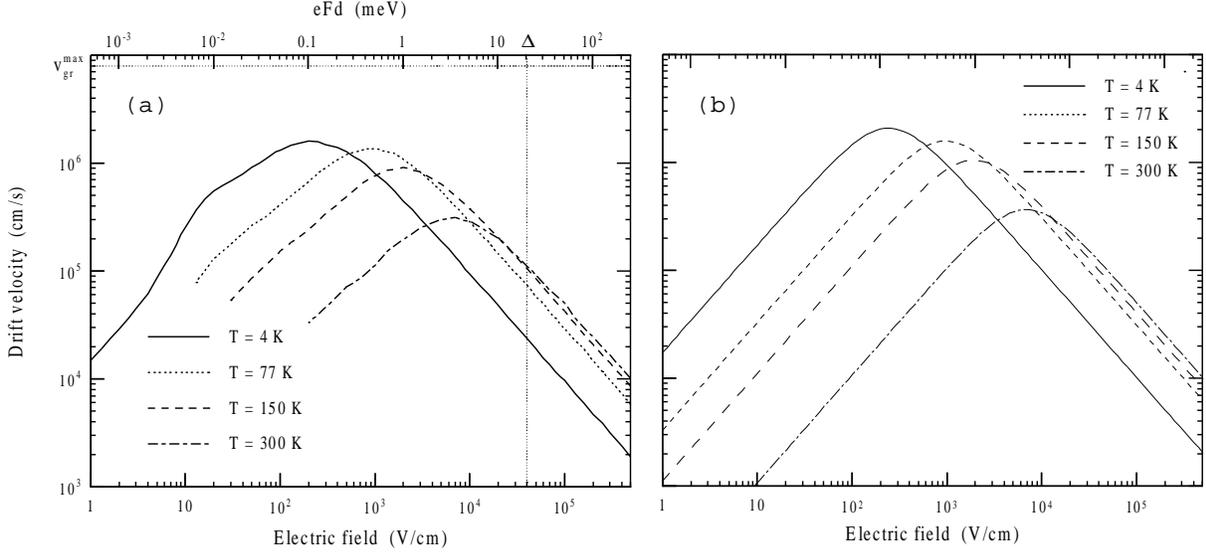,angle=270,width=16cm} 
\caption[a]{(a)Drift-velocity versus electric field for a 
superlattice with miniband width 20.3 meV and period $d=5.1$ nm 
with homogeneous doping $N_{3D}=10^{16}/{\rm cm}^3$. The calculation has 
been performed within the miniband transport model 
employing the Monte-Carlo method (Fig. 3.8 from   \cite{ROT99a}).
(b) Fit by the two-time model (\ref{Eq3J2relax}) using scattering times 
$\tau_{\rm elast}=4 {\rm ps, }
1.6{\rm ps, }1.1{\rm ps, }0.12$ps and 
$\tau_e=10{\rm ps, }2{\rm ps, }0.9{\rm ps, }0.4$ps 
for $T=4{\rm K, }77{\rm K, }150{\rm K, }300$K, respectively. 
\label{Fig3MCrott}} 
\end{figure}

Self-consistent solutions of the Boltzmann equation 
have been performed by various groups. 
In   \cite{ETE93} results for optical phonon and interface roughness 
have been presented where Boltzmann's equation was solved 
using a conjugate gradient algorithm. 
Using Monte-Carlo methods \cite{JAC83} the Boltzmann equation can be 
solved to a desired degree of numerical accuracy in a rather straightforward 
way (at least in the non-degenerate case and without electron-electron 
scattering). Results have been given in   \cite{AND73a} for acoustic 
phonon scattering and in   \cite{PRI73} for optical phonon and impurity 
scattering (using constant matrix elements). 
Modified scattering rates due to collisional broadening have been 
applied in   \cite{ART85} without significant changes in the result. 
Recently, extensive Monte-Carlo 
simulations \cite{ROT98,MOR98a,ROT99a} have been performed 
where both optical and acoustic phonon scattering as well as 
impurity  scattering has been considered using the microscopic matrix elements. 
Results of these calculations are presented in Fig.~\ref{Fig3MCrott}a. 
The general shape of the velocity-field relations 
resembles the Esaki-Tsu result shown in Fig.~\ref{Fig3esakitsu}b
both here and in all other calculations mentioned above. 
This is demonstrated by a comparison with the two-time model (\ref{Eq3J2relax}),
where the scattering times have been chosen to give good agreement with the
Monte-Carlo simulations, see Fig.~\ref{Fig3MCrott}b.
The increase of the scattering rate with lattice temperature 
can be attributed to the enhanced phonon occupation. In contrast, the 
high-field behavior does not strongly depend on lattice temperature. 
Here the drift velocity is limited by energy relaxation (\ref{Eq3Venergyloss}) 
which is dominated by spontaneous 
emission of phonons and thus does not depend on the thermal 
occupation of phonon modes.

\subsection{Wannier-Stark hopping\label{SecWSH}} 
 
\begin{figure} 
\epsfig{file=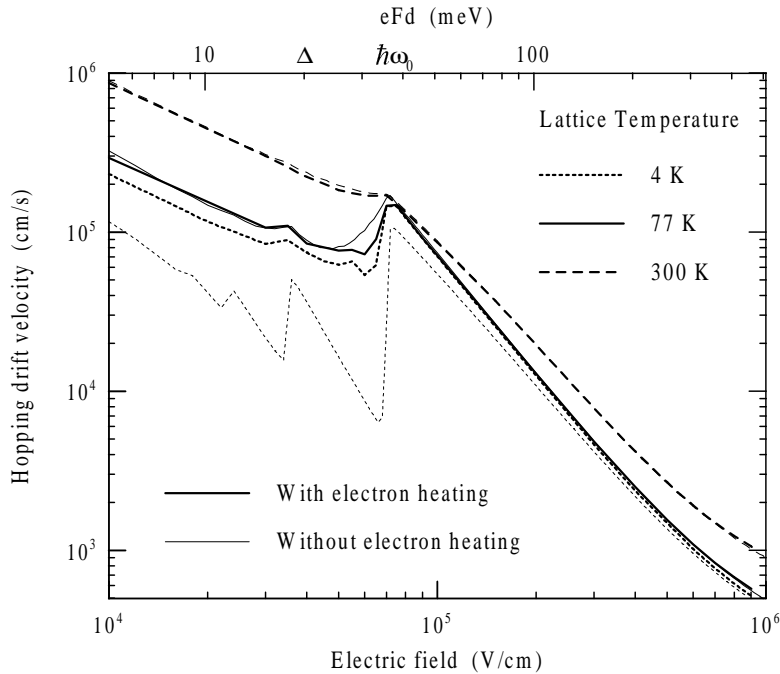,width=11cm} 
\caption[a]{Drift-velocity versus electric field for a 
superlattice with miniband width 20.3 meV and period $d=5.1$ nm 
with homogeneous doping $N_{3D}=10^{16}/{\rm cm}^3$. 
The transport model of Wannier-Stark-Hopping has been applied. 
Both self-consistent (thick lines) and thermal (thin lines) 
distributions of electrons 
in the {\bf k} direction have been used (Fig. 4.20 from   \cite{ROT99a}). 
\label{Fig3WSHrott}} 
\end{figure}

If a finite electric field is applied to a semiconductor superlattice, 
the Bloch states are no longer eigenstates. Within the restriction to 
a given miniband $\nu$,  Wannier-Stark states 
$\Phi^{\nu}_{j,{\bf k}}({\bf r},z)= 
\Phi^{\nu}_j(z)\e^{\imai {\bf k}\cdot{\bf r}}/\sqrt{A}$ 
with energy $E^{\nu}_{j,{\bf k}}=E^{\nu}_{\rm WS}-j eFd+E_k$ 
diagonalize the Hamiltonian as discussed in Sec.~\ref{SecWS}. 
(We apply a normalization area $A$ in the $(x,y)$ direction here 
yielding discrete values of ${\bf k}$ and normalizable states. 
For practical calculations the continuum limit 
$\sum_{\bf k}\to A/(2\pi)^2\int \d^2k$ is applied.) 
These states are approximately centered around well $n=j$. 
In the following we restrict ourselves to the lowest band $\nu=a$ 
and omit the index $\nu$. 
In a semiclassical approach the occupation of the states is given by the 
distribution function $f_j({\bf k})$. 
Scattering causes hopping between these states \cite{TSU75,CAL84}. 
Thus, this approach is called {\em Wannier-Stark hopping}. 
Within Fermi's golden rule the hopping rate is given by 
\begin{equation} 
R_{i,{\bf k}\to j,{\bf k}'}=\frac{2\pi}{\hbar} 
\left|\langle \Phi^{a}_{j,{\bf k}'}|\hat{H}^{\rm scatt} 
|\Phi^{a}_{i,{\bf k}} 
\rangle\right|^2 \delta\left(E_{k'}-j eFd-E_{k}+i eFd\, 
[\pm \hbar \omega_{\rm phonon}]\right) 
\left[1-f_j({\bf k}')\right]\label{Eq3WShopp} 
\end{equation} 
where the term $[\pm\hbar \omega_{\rm phonon}]$ has to be included if 
emission or absorption of phonons is considered. For details 
regarding the evaluation of scattering matrix elements see 
  \cite{TSU75,ROT99a,ROT97}. 
The current through the barrier between the wells $m=0$ and $m=1$ 
is then obtained by the sum of all transitions between states 
centered around wells $m\le 0$ and those centered around $m\ge 1$, i.e. 
\begin{equation} 
J_{\rm WSH}=\frac{2\mbox{(for Spin)}e}{A}\sum_{i\le 0}\sum_{j\ge 1} 
\sum_{{\bf k},{\bf k}'} 
\left[R_{i,{\bf k}\to j,{\bf k}'}f_i({\bf k})- 
R_{j,{\bf k'}\to i,{\bf k}}f_j({\bf k}')\right] 
\end{equation} 
If the occupation $f_i({\bf k})=f({\bf k})$ is independent of the index $i$, 
i.e., the electron distribution is homogeneous in superlattice direction, 
one finds 
\begin{equation} 
J_{\rm WSH}=\frac{2e}{A}\sum_{h \ge 1} 
\sum_{{\bf k},{\bf k}'} 
h\left[ 
R_{0,{\bf k}\to h,{\bf k}'}f({\bf k})- 
R_{h,{\bf k}'\to 0,{\bf k}}f({\bf k}')\right] 
\label{Eq3WSHcurrent} 
\end{equation} 
where $R_{i,{\bf k}\to j,{\bf k}'}=R_{0,{\bf k}\to (j-i),{\bf k}'}$ 
has been used. 
Typically, in the evaluation of Eq.~(\ref{Eq3WSHcurrent}) thermal distribution 
functions $f_i({\bf k})=n_F(E^a_{\rm WS}+E_k-\mu_i)$ are employed. 
The underlying idea is the assumption that the scattering 
rates $R_{i,{\bf k}\to i,{\bf k}'}$ inside each Wannier-Stark state are 
sufficiently fast to restore thermal equilibrium. In this case 
Eq.~(\ref{Eq3WSHcurrent}) can be further simplified to 
\begin{equation} 
J_{\rm WSH}=\frac{2e}{A}\sum_{h \ge 1} 
\sum_{{\bf k},{\bf k}'} 
hR_{0,{\bf k}\to h,{\bf k}'}n_F(E^a_{\rm WS}+E_k-\mu) 
\left[1-\exp\left(-\frac{heFd}{k_BT}\right)\right]\, . 
\label{Eq3WSHthermcurrent}\end{equation} 
Evaluating this expression for various types 
of scattering processes one obtains a drop of the 
current density with electrical field as shown in Fig.~\ref{Fig3WSHrott}. 
This is caused by the increasing 
localization of the Wannier-Stark functions (see Fig.~\ref{Fig2WSstates}) 
which reduces the matrix elements 
$\langle \Phi_{j,{\bf k}'}|\hat{H}^{\rm scatt}|\Phi_{i,{\bf k}} 
\rangle$ with increasing field. 
This behavior can be analyzed by 
expanding the scattering matrix elements in terms of Wannier states 
$|\Psi^a_{n,{\bf k}}\rangle$ from Eq.~(\ref{Eq2WSsimp}): 
\begin{equation} 
\langle \Phi_{i,{\bf k}'}|\hat{H}^{\rm scatt}|\Phi_{j,{\bf k}} 
\rangle= 
\sum_{n,m} J_{m-i}\left(\frac{2T_1^a}{eFd}\right) 
J_{n-j}\left(\frac{2T_1^a}{eFd}\right) 
\langle \Psi^a_{m,{\bf k}'}|\hat{H}^{\rm scatt}|\Psi^a_{n,{\bf k}}\rangle\, . 
\end{equation} 
As the Wannier states are essentially localized to single quantum wells 
the diagonal parts $m=n$  dominate. Neglecting correlations between 
the  matrix elements 
$\langle \Psi^a_{m,{\bf k}'}|\hat{H}^{\rm scatt}|\Psi^a_{m,{\bf k}}\rangle$ 
for different wells $m$ one obtains: 
\begin{equation}\begin{split} 
R_{0,{\bf k}\to h,{\bf k}'}= 
\frac{2\pi}{\hbar}\sum_m & \left[ 
J_{m}\left(\frac{2T_1^a}{eFd}\right) 
J_{m-h}\left(\frac{2T_1^a}{eFd}\right)\right]^2 
\left|\langle \Psi^a_{m,{\bf k}'}|\hat{H}^{\rm scatt}|\Psi^a_{m,{\bf k}} 
\rangle\right|^2\\ 
&\times\delta\left(E_{k'}-h eFd-E_{k}\, [\pm\hbar \omega_{\rm phonon}]\right) 
\left[1-f({\bf k'})\right]\, \label{Eq3R-WSH} 
\end{split}\end{equation} 
For $eFd\gg 2T^a_1$ the Bessel functions behave as 
$J_{n}(2x)\sim x^n/n!$ giving a field dependence 
$\propto (T_1^a/eFd)^{2|h|}$. Therefore the transitions $0\to 1$ dominate 
and 
\begin{equation}\begin{split} 
R_{0,{\bf k}\to 1,{\bf k}'}\sim & 
\frac{2\pi}{\hbar}\left(\frac{T_1^a}{eFd}\right)^2 
\left( 
\left|\langle \Psi^a_{0,{\bf k}'}|\hat{H}^{\rm scatt}|\Psi^a_{0,{\bf k}} 
\rangle\right|^2 
+\left|\langle \Psi^a_{1,{\bf k}'}|\hat{H}^{\rm scatt}|\Psi^a_{1,{\bf k}} 
\rangle\right|^2\right)\\ 
&\times 
\delta\left(E_{k'}-E_{k}-eFd\, [\pm\hbar \omega_{\rm phonon}]\right) 
\left[1-f({\bf k'})\right]\, . 
\label{Eq3WSHhf} 
\end{split}\end{equation} 
For high electric fields the wave vector ${\bf k}'$ must be large in order 
to satisfy energy conservation and thus the scattering process 
transfers a large momentum. If the scattering matrix element 
does not strongly depend on momentum (such as deformation potential 
scattering at acoustic phonons) 
$J_{\rm WSH}\propto 1/F^2$ is found, while different power laws occur for 
momentum dependent  matrix elements (such as $J_{\rm WSH}\propto 1/F^{3.5}$ 
for impurity scattering \cite{ROT97}). For optical phonon scattering, 
resonances can be found at $h eFd=\hbar\omega_{\rm opt}$ when hopping to 
states in distance $h$ becomes possible under the emission of 
one optical phonon \cite{ROT98}, see also Fig.~\ref{Fig3WSHrott}. 
 
If the sum in Eq.~(\ref{Eq3WSHcurrent}) is restricted to $h\le h_{\rm max}$ 
one obtains a linear increase of the current-field relation 
for low fields \cite{TSU75,CAL84} and a maximum at intermediate fields 
before the current drops with higher fields as discussed above. 
In   \cite{ROT97} it has been shown that this is an artifact and 
the correct $J(F)$ relation is proportional to $1/F$ for low fields. 
Thus the linear response region for low fields 
cannot be recovered by the Wannier-Stark hopping approach. 
 
While most calculations have been performed assuming thermal 
distribution functions 
$f({\bf k})$ recently self-consistent calculations of the 
distribution functions have been obtained by solving 
the semiclassical Boltzmann equation for the Wannier-Stark states: 
\begin{equation} 
\pabl{f_i({\bf k},t)}{t} 
= \sum_{j,{\bf k}'} R_{j,{\bf k}'\to i,{\bf k}}f_j({\bf k}',t) 
-R_{i,{\bf k}\to j,{\bf k}'}f_i({\bf k},t) 
\label{Eq3WSHselfconsist} 
\end{equation} 
The self-consistent stationary solution 
of this equation can be used for the evaluation of Eq.~(\ref{Eq3WSHcurrent}). 
As can be seen in Fig.~\ref{Fig3WSHrott} significant deviations 
between both approaches occur for low lattice temperatures, when electron 
heating effects become important.

\subsection{Sequential tunneling\label{SecST}} 
 
\begin{figure} 
\epsfig{file=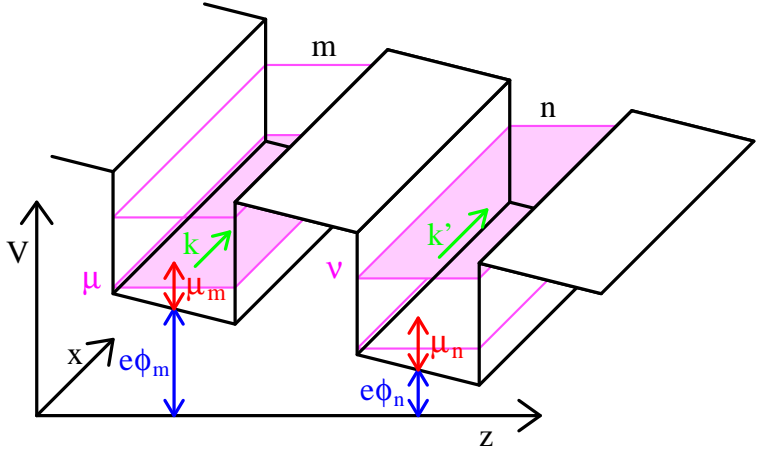,width=10cm} 
\caption[a]{Sketch of the levels in neighboring wells 
$n=m+1$, where  $\mu=a$ is the lowest  and $\nu=b$ is the first excited level 
in the respective quantum wells. The chemical potentials $\mu_m$ 
as well as the electrical potentials $e\phi_m$ are marked as well. 
For a constant electric field $F$ one finds $\phi_{m+1}=\phi_{m}-Fd$. 
\label{Fig3tunnelformel}} 
\end{figure}

If the barrier width of a superlattice is large, the structure 
essentially consists of several decoupled quantum wells. 
In each quantum well $n$ we have a basis set of wave functions 
$\Psi_n^{\nu}(z)\e^{\imai {\bf k}\cdot{\bf r}}/\sqrt{A}$, where 
$\Psi_n^{\nu}(z)$ is the $\nu^{\rm th}$ eigenfunction of the quantum well 
potential. 
The states have the energy $E^{\nu}+E_k+e\phi_n$, where 
the potential energy due to an electrical potential $\phi_n$ has 
been considered separately. The notation is clarified in 
Fig.~\ref{Fig3tunnelformel}. If the wells are not coupled 
to each other  (infinite barrier width or height)  no current 
flows in the superlattice direction. For finite 
barrier width the states from different wells 
become coupled to each other  which can be described by a tunnel matrix element 
$H_{n,m}^{\nu,\mu}$ in the spirit of   \cite{BAR61} inducing 
transitions between the wells. 
In lowest order perturbation theory the transition rate is given by 
Fermi's golden rule 
\begin{equation} 
R(m,\mu,{\bf k} \to n,\nu,{\bf k}')= 
\frac{2\pi}{\hbar}|H_{m,n}^{\mu,\nu}({\bf k},{\bf k}')|^2\delta 
(E^{\nu}+E_{k'}+e\phi_n-E^{\mu}-E_k-e\phi_m)\, .\label{Eq3goldenrule} 
\end{equation} 
As the superlattice is assumed to be translational 
invariant in the $(x,y)$-plane, the matrix element is 
diagonal in ${\bf k}$. Thus, transitions are only 
possible if $E^{\nu}-E^{\mu}=e\phi_m-e\phi_n$, suggesting 
sharp resonances when the potential drop between different wells 
$\phi_m-\phi_n$ equals the energy spacing of the bound states. 
This could be nicely demonstrated experimentally in  Ref.~\cite{CAP86} 
for simple superlattices and in Ref.~\cite{KRI98} for superlattices 
with a basis where even tunneling over 5 barriers was observed. 
In the presence of a strong magnetic field along the superlattice direction
further peaks due to transitions between different Landau levels 
are observed \cite{CAN96}.

The resonance condition from Eq.~(\ref{Eq3goldenrule}) 
implies vanishing electric field for transitions between 
equivalent levels (in particular the lowest level). 
As for zero field the current vanishes (provided the 
electron density is equal in both wells), it was concluded that 
only phonon-assisted tunneling processes are possible. 
Thus neither a linear increase of the current for low fields 
nor a peak  at low fields was expected for weakly coupled 
superlattices \cite{SHI75}. This conclusion is in contradiction 
with experimental findings \cite{GRA91a}, where a drift-velocity 
in qualitative agreement with the Esaki-Tsu result 
(Fig.~\ref{Fig3esakitsu}) has been obtained for weakly coupled superlattices 
as well. This discrepancy is due to the neglect 
of broadening in the argument given above.\footnote{Broadening had been 
included in earlier theories \cite{KAZ72} but there 
a term was missing which is essential for the 
transition between equivalent levels. This point is discussed in 
Appendix \ref{AppDMT}.} 
 
\subsubsection{General theory\label{SecSTgeneral}} 
In a real quantum well the states 
$\Psi_n^{\nu}(z)\e^{\imai {\bf k}\cdot{\bf r}}/\sqrt{A}$ 
with energy $E^{\nu}+E_k+e\phi_n$ are not exact eigenstates of the full 
Hamiltonian due to the presence of phonons and nonperiodic 
impurity potentials. The respective scattering processes 
lead to an energy shift $\Delta E$ and 
a finite lifetime $\tau$ of the states. 
These features can be treated within the theory of Green functions
(see, e.g.,  Ref.~\cite{MAH90}). While a general treatment is postponed
to Chapter \ref{ChapNGFT}, a motivation of the concept and a heuristic
derivation of the current formula (\ref{Eq3JST}) is given in the following.

For a stationary fluctuating 
potential $V(x,y)$ due to impurities or interface fluctuations 
one finds 
\begin{eqnarray} 
\Delta E({\bf k})&\approx &\sum_{\bf k'} 
\frac{\left|\langle \Psi_{\bf k}|V|\Psi_{\bf k'}\rangle\right|^2}{E_k-E_{k'}} 
\label{Eq3ReBorn}\\ 
\frac{1}{\tau({\bf k})}&\approx &\frac{2\pi}{\hbar}\sum_{\bf k'} 
\left|\langle \Psi_{\bf k}|V|\Psi_{\bf k'}\rangle\right|^2\delta(E_k-E_{k'}) 
\label{Eq3ImBorn} 
\end{eqnarray} 
where the second order of stationary perturbation theory 
as well as Fermi's golden rule was applied for 
a stationary fluctuating 
potential $V(x,y)$ due to impurities or interface fluctuations. 
For a particle, which is injected at 
$t=0$, the time dependence of the wave function is then given by 
\begin{equation} 
\Psi_{n,{\bf k}}^{\nu}(t)= 
G^{\nu\,{\rm ret}}_n({\bf k};t,0)=-\imai \Theta(t) 
\e^{-\imai (E^{\nu}+E_k+e\phi_n+\Sigma^{\rm ret})t/\hbar} 
\end{equation} 
with 
$\Sigma^{\rm ret}=\Delta E({\bf k})-\imai\hbar/2\tau({\bf k})$, so that 
$|\Psi_{n,{\bf k}}^{\nu}(t)|^2=\e^{-t/\tau}$. 
This motivates the meaning of the (retarded) Green function $G^{\rm ret}$
and the (retarded) self energy $\Sigma^{\rm ret}$, which are key quantities
in the theory of Green functions. (A nice introduction can be found
in Ref.~\cite{MAT92}.)
The Fourier transformation is given by 
\begin{equation} 
G^{\nu\,{\rm ret}}_n({\bf k},E)=\frac{1}{\hbar} 
\int \d t \e^{\imai Et/\hbar}G^{\rm ret}(t,0)= 
\frac{1}{E-E^{\nu}-E_k-e\phi_n-\Sigma^{\rm ret}}\label{Eq3Gret} 
\end{equation} 
and the spectral function is  defined by 
\begin{equation} 
A^{\nu}_n({\bf k},E)=-2\Im\left\{G^{\nu\,{\rm ret}}_n({\bf k},E)\right\}\, . 
\label{Eq3Spekt}\end{equation} 
For infinite lifetime $\tau\to \infty$ one finds 
$A_n^{\nu}({\bf k},E)\to 2\pi \delta(E-[E^{\nu}+E_k+e\phi_n+\Delta E({\bf k})])$ 
which is (except for the factor $2\pi$) just the contribution of 
the state $(n,{\nu},{\bf k})$ to the total density of states. 
This relation is more general and $A_n^{\nu}({\bf k},E)/(2\pi)$ can be 
viewed as the contribution 
of the  state $(n,{\nu},{\bf k})$ if the system is probed 
with an energy $E$. 
Here the expressions (\ref{Eq3ReBorn},\ref{Eq3ImBorn}) approximately 
correspond to the Born approximation for $\Sigma^{\rm ret}$. 
Results for higher order approximations  as well as for 
phonon or electron-electron scattering can be obtained in a systematic 
manner by the theory of Green functions \cite{MAH90}. Typically 
$\Sigma^{\rm ret}({\bf k},E)$ becomes a function of energy and momentum, 
but the structure of Eqs.~(\ref{Eq3Gret},\ref{Eq3Spekt}) persists. 
 
As the wells are considered to be almost uncoupled, the electronic 
distribution is described by $n_F(E-\mu_n-e\phi_n)$ in each well 
with the local chemical potential $\mu_n$ measured with respect 
to the energy of the lowest level (see Fig.~\ref{Fig3tunnelformel}). 
This provides an 
electron density $n_F(E-\mu_n-e\phi_n)A^{\nu}_n({\bf k},E)/(2\pi)$ 
per energy in the state $(n,{\nu},{\bf k})$. 
Summing all states and integrating over energy, 
the electron density in well $n$ (in units 1/cm$^2$) is given by 
\begin{eqnarray} 
n_n&=&\sum_{\nu}\frac{2\mbox{(for spin)}}{2\pi A}\sum_{{\bf k},\nu} 
\int \d E\, 
A^{\nu}_n({\bf k},E)n_F(E-\mu_n-e\phi_n)\\ 
&=&\sum_{\nu}\int \d E\, \rho_n^{\nu}(E) 
n_F(E-\mu_n) 
\label{Eq3STdensity} 
\end{eqnarray} 
with the density of states 
\begin{equation} 
\rho_{n}^{\nu}(E)=\frac{2 \mbox{(for Spin)}}{2\pi A} 
\sum_{\bf k}A_n^{\nu}({\bf k},E)=\frac{\rho_0}{2\pi} 
\int_0^{\infty} \d E_k \, 
A_n^{\nu}({\bf k},E) \label{Eq3DosST} 
\end{equation} 
in well $n$ belonging to the level $\nu$. 
Regarding the transitions from level $\mu$ in well 
$m$ to level $\nu$ in well $n$, 
Eq.~(\ref{Eq3goldenrule}) is modified as follows: 
\begin{itemize} 
\item The energy $E$ is conserved instead of the free particle energy 
$E^{\nu}+E_k+e\phi_n$. 
\item The energy conserving $\delta$-function is replaced 
by $A^{\nu}_n({\bf k}',E)/2\pi$. 
\item At each energy the net particle flow 
from well $m$ to well $n$ is proportional to 
($n_F(E-e\phi_m-\mu_m)\left[1-n_F(E-e\phi_n-\mu_n)\right]$) while 
the net particle flow 
from $n$ to $m$ is proportional to 
($n_F(E-e\phi_n-\mu_n)\left[1-n_F(E-e\phi_m-\mu_m)\right]$). 
Therefore the total current is proportional the difference 
$\left[n_F(E-e\phi_m-\mu_m)-n_F(E-e\phi_n-\mu_n)\right]$ in occupation 
between both wells. 
Defining the electrochemical potential 
$E_F(n)= e\phi_n+\mu_n$ it becomes clear that the current is 
driven by the difference in the electrochemical potential. 
\end{itemize} 
Then the current density from level $\mu$ in well 
$m$ to level $\nu$ in well $n$ is given by 
\begin{equation}\begin{split} 
J_{m\to n}^{\mu \to \nu}=&\frac{2\mbox{(for spin)}e}{2\pi \hbar A } 
\sum_{{\bf k},{\bf k}'}\left|H_{m,{\bf k};n,{\bf k}'}^{\mu,\nu}\right|^2 
\int \d E\, A^{\mu}_m({\bf k},E)A^{\nu}_n({\bf k}',E)\\ 
&\times \left[n_F(E-e\phi_m-\mu_m)-n_F(E-e\phi_n-\mu_n)\right] 
\, . 
\label{Eq3JST}\end{split} 
\end{equation} 
While the derivation given above is heuristic, a microscopic derivation 
is given in Sec.~9.3 of   \cite{MAH90}. In appendix \ref{AppSTderivation} 
it will be shown that Eq.~(\ref{Eq3JST}) is the limiting case of the 
full quantum transport theory based on nonequilibrium Green functions 
in the limit $2T_1\ll \Gamma$. The same approach has been used 
for tunneling between neighboring two-dimensional 
electron gases \cite{MUR95,TUR96}. It should be noted, that  Eq.~(\ref{Eq3JST}) 
only holds if the correlations between scattering events 
in different wells are not significant, otherwise disorder vertex 
corrections must be taken into account \cite{ZHE93}. 
 
An important task is the determination of the matrix elements 
$H_{m,{\bf k};n,{\bf k}'}^{\mu,\nu}$. One possibility is to start 
with eigenfunctions of single quantum wells 
and consider the overlap between those functions obtained for 
different wells $m,n$. This is the procedure suggested by Bardeen \cite{BAR61} 
which essentially has been applied in   \cite{PRE94,AGU97} 
for superlattice transport. 
Another possibility is to start from the Wannier states 
(see section \ref{SecWannier}). If a constant electric 
field is applied to the superlattice structure, the electric potential 
reads $\phi_n=-eFd$ and the 
matrix elements can be obtained directly from $\hat{H}_1$ in 
Eq.~(\ref{Eq2hamW1}) or from $\hat{H}_1^{\rm ren}$ in 
Eq.~(\ref{Eq2hamW1ren}). In subsection \ref{SecSTresults} it will be shown 
that the latter Hamiltonian is more appropriate (as already suggested in 
  \cite{KAZ72}) by a comparison 
with experimental data. 
These matrix elements are calculated for a perfect superlattice structure. 
Thus they conserve the parallel momentum {\bf k}. In addition 
scattering processes at impurities, phonons, or interface fluctuations 
may cause transitions between different wells. The respective 
matrix elements can be obtained from the respective scattering potential 
and the Wannier functions as well. Examples will be given in 
section \ref{SecSTresults}, where it is shown that these processes 
give a background current while the peaks in the current-field 
relation is typically dominated the {\bf k}-conserving 
terms of $\hat{H}_1^{\rm ren}$. 
 
An important feature of Eq.~(\ref{Eq3JST}) is the fact that 
the current is driven by the difference of the  electrochemical potential 
in both wells. This is in contrast to the findings 
of   \cite{KAZ72} (based on density 
matrix theory) 
where the current is driven by the difference 
$f^{\mu}_m({\bf k})-f^{\nu}_n({\bf k}')$ where 
$f^{\nu}_n({\bf k})=n_F(E^{\nu}+E_k-\mu_n)$ is the occupation of the 
state  $(n,{\nu},{\bf k})$. 
In the latter case the current vanishes for equivalent levels ($\mu=\nu$) 
if the matrix element $H_{m,{\bf k};n,{\bf k}'}^{\mu,\nu}$ is diagonal 
in ${\bf k}$. In Appendix \ref{AppDMT} it will be shown how the factor 
$\left[n_F(E-e\phi_m-\mu_m)-n_F(E-e\phi_n-\mu_n)\right]$ 
is recovered from density matrix theory. 
 
\subsubsection{Evaluation for constant broadening\label{SecSTsimple}} 
Here we want to derive simple expressions for the current 
under the assumption of a constant broadening for the electronic states. 
We assume that only the 
lowest level $\mu=a$ is occupied (i.e., $\mu_m,k_BT\ll E^b$). 
 
First we consider nearest neighbor tunneling 
with {\bf k}-conserving matrix elements. 
Then  Eq.~(\ref{Eq3JST}) can be rewritten as follows: 
\begin{equation}\begin{split} 
J_{m\to m+1}^{a\to \nu}= 
\frac{e}{\hbar}|H_{m+1,m}^{\nu,a}|^2&\int \d E\, 
\rho^a_m(E)\langle A_{m+1}^{\nu}\rangle (E,F_m)\\ 
&\left[n_F(E-e\phi_m-\mu_m)-n_F(E-e\phi_{m+1}-\mu_{m+1})\right] \label{Eq3JSTmod} 
\end{split}\end{equation} 
with 
\begin{equation} 
\langle A_{m+1}^{\nu}\rangle (E,F)= 
\frac{\int_0^{\infty} \d E_k \, 
A_m^{a}({\bf k},E) A_{m+1}^{\nu}({\bf k},E)} 
{\int_0^{\infty} \d E_k\, A_m^a({\bf k},E)} \, . 
\label{Eq3hilfST} 
\end{equation} 
Here the effective field $F=(\phi_m-\phi_{m+1})/d$ has been 
introduced and the density of states $\rho_{m}^{a}(E)$ from 
Eq.~(\ref{Eq3DosST}) has been applied. 
Let us assume a constant self-energy 
$\Sigma^{\nu\, {\rm ret}}({\bf k},E)=
-\imai\Gamma^{\nu}/2$ in Eq.~(\ref{Eq3hilfST}) 
for the sake of simplicity. Then the 
spectral functions become Lorentzians 
$A_m^{\nu}({\bf k},E)=\Gamma^{\nu}/[(E-E_k-E^{\nu}-e\phi_{m})^2+(\Gamma^{\nu}/2)^2]$. 
As $\rho^a_m(E)$ is essentially zero for energies below $E^a$ 
we may restrict ourselves to $E\gtrsim E^a$ in the evaluation of 
Eq.~(\ref{Eq3hilfST}). For these energies the function 
$A^{\nu}_{m+1}({\bf k},E)$ takes its maximum at $E_k>eFd+E^a-E^{\nu}$ 
which is larger than zero provided we restrict ourselves 
to $eFd>E^{\nu}-E^a$ (this is always the case for the 
$a\to a$ resonance and $eFd>0$). 
In this case the integrand of Eq.~(\ref{Eq3hilfST}) does not 
take large values for $E_k<0$ and it is justified to 
extend the lower limit of integration to $-\infty$. A straightforward 
evaluation using the calculus of residues yields: 
\begin{equation} 
\langle A_{m+1}^{\nu}\rangle=\frac{\Gamma^{\nu, {\rm eff}}} 
{(E^{\nu}-E^a-eFd)^2+(\Gamma^{\nu,{\rm eff}}/2)^2} 
\quad \mbox{with}\quad  \Gamma^{\nu, {\rm eff}}=\Gamma^a+\Gamma^{\nu} 
\end{equation} 
which only depends on $F$. 
Numerical evaluations indicate that these simplifications 
are quite good for fields above the resonance, i.e. $eFd>E^{\nu}-E^a$. 
In the case of $\nu\neq a$ and $eFd<E^{\nu}-E^a$ the choice 
$\Gamma^{\nu,{\rm eff}}=\Gamma^{\nu}$ may be better. 
Note that this simple model with a constant self-energy cannot be used 
in  the calculation of the density of states (\ref{Eq3DosST}) as 
the integral for the electron density  (\ref{Eq3STdensity}) 
diverges. Therefore we use 
the free-electron density of states $\rho_m(E)=\rho_0\Theta(E-E^a-e\phi_m)$ 
in the following. 
With these simplifications Eq.~(\ref{Eq3JSTmod}) becomes: 
\begin{equation} 
J_{m\to m+1}^{a\to \nu}=e\frac{|H_{m+1,m}^{\nu,a}|^2}{\hbar} 
\frac{\Gamma^{\nu, {\rm eff}}}{(E^{\nu}-E^a-eFd)^2+(\Gamma^{\nu,{\rm eff}}/2)^2} 
n_{\rm eff}(eFd,n_m,n_{m+1}) 
\label{Eq3JSTsimp} 
\end{equation} 
with the effective electron density (here $\tilde{E}=E-e\phi_m$) 
\begin{equation} 
n_{\rm eff}(F,n_m,n_{m+1})= 
\int_{E^a}^{\infty} \d \tilde{E}\, 
\rho_0\left[n_F(\tilde{E}-\mu_m)-n_F(\tilde{E}-\mu_{m+1}+eFd)\right]\label{Eq3neff} 
\end{equation} 
which describes the difference of occupation in both wells. 
The electron density is related to the chemical potential via: 
\begin{equation} 
n_m=\int_{E^a}^{\infty} \d \tilde{E}\, \rho_0 n_F(\tilde{E}-\mu_m)= 
\rho_0k_BT\log\left[ 1+\exp\left(\frac{\mu_m-E^a}{k_BT}\right)\right]\, . 
\end{equation} 
Here we assumed that only the lowest level is occupied. The inclusion of 
higher levels is straightforward, but leads to more complicated expressions. 
Inserting into Eq.~(\ref{Eq3neff}) yields: 
\begin{equation} 
n_{\rm eff}(F,n_m,n_{m+1})=n_m-\rho_0 k_B T 
\log\left[\left(\e^{\frac{n_{m+1}}{\rho_0 k_B T}}-1\right) 
\e^{-\frac{eFd}{k_BT}}+1\right]\, .\label{Eq3neff1} 
\end{equation} 
In  the {\em nondegenerate limit} limit we have $n_{m+1}\ll \rho_0k_B T$ and 
Eq.~(\ref{Eq3neff1}) is further simplified: 
\begin{equation} 
n_{\rm eff}(F,n_m,n_{m+1})= 
n_m-n_{m+1}\exp\left(-\frac{eFd}{k_BT}\right)\label{Eq3neff2}\, . 
\end{equation} 
It is interesting to note that the total current can be written as a 
discrete version of the drift-diffusion model in this case: 
\begin{equation} 
J_{m\to m+1}=e\frac{n_m}{d}v(F)- 
eD(F)\frac{n_{m+1}-n_m}{d^2}\label{Eq3STdriftdiff} 
\end{equation} 
with the velocity 
\begin{equation} 
v(F)=\sum_{\nu} 
\frac{d}{\hbar}|H_{m+1,m}^{\nu,a}|^2 
\frac{\Gamma^{\nu, {\rm eff}}}{(E^{\nu}-E^a-eFd)^2+(\Gamma^{\nu,{\rm eff}}/2)^2} 
\left[1-\exp\left(-\frac{eFd}{k_BT}\right)\right] 
\end{equation} 
and the diffusion coefficient 
\begin{equation} 
D(F)=\sum_{\nu} 
\frac{d^2}{\hbar}|H_{m+1,m}^{\nu,a}|^2 
\frac{\Gamma^{\nu, {\rm eff}}}{(E^{\nu}-E^a-eFd)^2+(\Gamma^{\nu,{\rm eff}}/2)^2} 
\exp\left(-\frac{eFd}{k_BT}\right) 
\end{equation} 
which satisfy the Einstein relation $D(F)=v'(F)k_BT/e$ for $F=0$. 
Remember that these expressions only hold for $eFd\ge 0$. For 
$eFd< 0$ the expressions for the opposite direction can be applied. 
 
In the last part of this subsection  some 
explicit results for the current $a\to a$ are given. 
Here we assume equal densities $n_m=n_{m+1}=n$ in the wells 
and set $H_{m+1,m}^{a,a}=T_1^a$ according to Eq.~(\ref{Eq2hamW1}). 
 
In the  {\em degenerate limit} ($n\gg \rho_0 k_B T$) we find 
$n_{\rm eff}= 
n-(n-\rho_0 eFd)\Theta(n-\rho_0 eFd)$. 
For  $eFd<n/\rho_0$ we obtain 
\begin{equation} 
J_{\rm ST}(F)=e\rho_0\frac{|T_1^a|^2}{\hbar} 
\frac{2\Gamma^a eFd}{(eFd)^2+(\Gamma^a)^2} 
\quad \mbox{for}\quad \left\{ \begin{array}{rl} 
n&\gg \rho_0 k_B T\\ 
n&>\rho_0 eFd 
\end{array}\right.\, . 
\label{Eq3JSTdeg} 
\end{equation} 
In the {\em nondegenerate limit} we obtain for $eFd\ll k_BT$ 
\begin{equation} 
J_{\rm ST}(F)=\frac{en}{k_BT}\frac{|T_1^a|^2}{\hbar} 
\frac{2\Gamma^a eFd}{(eFd)^2+(\Gamma^a)^2} 
\quad \mbox{for}\quad \left\{ \begin{array}{rl} 
n&\ll \rho_0 k_B T\\ 
k_BT&\gg eFd 
\end{array}\right.\, . 
\label{Eq3JSTnondeg} 
\end{equation} 
In both cases the field dependence is given by the Esaki-Tsu result 
(\ref{Eq3EsakiTsu}) with $\Gamma^a=\hbar/\tau$. 
 
\subsubsection{Results\label{SecSTresults}} 
A variety of calculations for different superlattice structures 
have been performed within this model. These calculations 
consist of the following steps: 
\begin{enumerate} 
\item Calculation of the miniband structure $E^{\nu}(q)$ and 
the associated wave functions $\varphi_q(z)$ according to 
Sec.~\ref{SecMiniband}. 
\item Evaluation of the Wannier-functions and respective couplings 
$T_h^{\nu},R_h^{\mu,\nu}$, see Sec.~\ref{SecWannier}. 
\item Evaluation of (intrawell) scattering matrix elements 
$\langle \Psi^{\nu}_{n,{\bf k}'}|\hat{H}^{\rm scatt}| 
\Psi^{\nu}_{n,{\bf k}}\rangle$  for the dominant scattering processes. 
For doped superlattices ionized impurity scattering dominates and 
the respective calculations including screening are presented in 
\cite{WAC97d,WAC98}. Scattering processes at interface roughness 
\cite{WAC98} and phonons have been considered as well. 
\item Calculation of the self-energies $\Sigma^{{\rm ret}\, \nu}_n({\bf k},E)$ 
within different approximation schemes such as the self-consistent 
single-site approximation \cite{WAC97d,SER89,WAC97b} for impurity scattering. 
\item Determination of the chemical potential for given electron 
density provided by the donors from Eq.~(\ref{Eq3STdensity}). 
\item Renormalizing of the matrix elements for each electric field 
according to Eq.~(\ref{Eq2Matren}). 
\item Evaluation of the current density according to Eq.~(\ref{Eq3JST}). 
\end{enumerate} 
 
\begin{figure} 
\epsfig{file=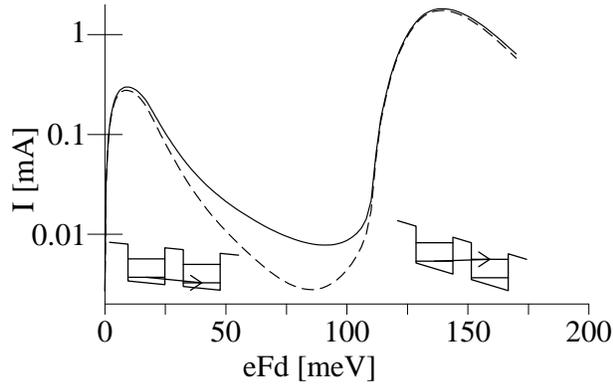,width=8cm} 
\caption[a]{Current-field relation for the superlattice 
studied in   \cite{GRA91} for constant electron density 
$n_m=n_{m+1}=N_D$. The dashed line shows the current from resonant 
transitions with momentum independent matrix elements, while 
both contributions from resonant and non-resonant currents 
contribute to the full line. An electron temperature 
$T_e=20$ K was used in the calculation. 
\label{Fig3STkenngrahn}} 
\end{figure}

Results for the superlattice studied experimentally in   \cite{GRA91,KWO95} 
(9 nm wide GaAs wells, 4 nm AlAs barriers, doping density 
$N_D=1.5\times 10^{11}/{\rm cm}^2$, cross section 
$A=1.13\times 10^{-4}{\rm cm}^2$) 
are shown in Fig.~\ref{Fig3STkenngrahn}. 
Two pronounced peaks can be identified at low and moderate fields. 
The low-field peak is due to tunneling between the lowest levels 
($a\to a$), while the peak around $eFd\approx E^b-E^a=130$ meV 
is due to $a\to b$ tunneling. 
While for the dashed line only  {\bf k}-conserving transitions 
are taken into account, the full line also includes the contributions 
of interwell scattering matrix elements 
$\langle \Psi^{\nu}_{n+1,{\bf k}'}|\hat{H}^{\rm scatt}| 
\Psi^{\nu}_{n,{\bf k}}\rangle$ evaluated for interface roughness, see 
  \cite{WAC98}. (The respective matrix elements for impurity scattering 
are negligible.) These scattering events represent an additional 
current channel in Eq.~(\ref{Eq3JST}) yielding a background current 
which dominates between the resonances, but is negligible compared to 
the resonant currents. 
The height of the $a\to b$ peak 
($I_{\rm max}=1.75$ mA) is in 
good agreement with experimental data exhibiting $I_{\rm max}=1.45$ mA 
(Fig. 6 of   \cite{KWO95}). 
The low-field peak is not resolved experimentally 
due to domain formation yielding a current of about 0.076 mA. 
 
\begin{figure} 
\epsfig{file=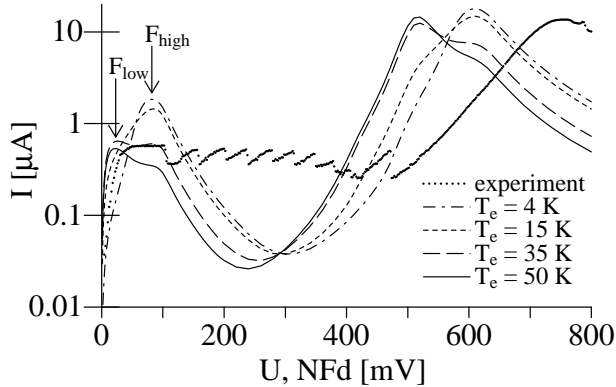,width=8cm} 
\caption[a]{Current-field relation for the superlattice 
studied in   \cite{KEA95b} for constant electron density 
$n_m=n_{m+1}=N_D$. The bias was taken to be NFd for the 
theoretical results, where $N=10$ is the number of quantum wells 
(from   \cite{WAC97d}). 
\label{Fig3STkennzeun}} 
\end{figure}

Results for the superlattice studied  in   \cite{KEA95b} 
(15 nm wide GaAs wells, 5 nm Al$_{0.3}$Ga$_{0.7}$As barriers, 
doping density $N_D=6\times 10^{9}/{\rm cm}^2$, cross section 
$A=8\mu{\rm m}^2$) 
are shown in Fig.~\ref{Fig3STkennzeun}. The general shape is 
like the results from the sample mentioned before. 
While the latter (highly-doped) sample did not exhibit a 
strong temperature dependence, 
the situation is different for the low-doped sample considered here. 
For low electron temperatures $T_e\le 15$ K the electrons are located 
in the impurity band (about 10 meV below the free electron states). 
Therefore a current exhibits a peak at $eF_{\rm high}d\approx 10$ meV 
when these electrons can tunnel into the free electron states 
of the next well. For higher temperatures the electrons 
occupy the free electron states and the maximum occurs at 
$eF_{\rm low}d\approx \Gamma^a\approx 2$ meV as suggested 
by Eq.~(\ref{Eq3JSTnondeg}). Due to the same effect 
the peak at $eFd\approx 50$ meV due to $a\to b$ tunneling 
shifts with temperature. 
The experimental data (taken at a lattice temperature of 4K) 
are shown for comparison. While the low-field conductance is in good agreement 
with the $T_e=4$ K calculations, close to the first maximum the agreement 
becomes better for the $T_e=35$ K curve, which can be caused by electron 
heating. The heights of both maxima are in excellent agreement 
between theory and experiment. The difference in position 
of the second maxima may be caused by an additional voltage drop 
in the contact, which is not taken into account in the theory, where 
the field was just multiplied by the length of the sample. 
Finally, the saw-tooth shape of the experimental current-voltage characteristic 
is due to the formation of electric field domains as discussed 
in section \ref{ChapDomains}. 

\begin{figure} 
\epsfig{file=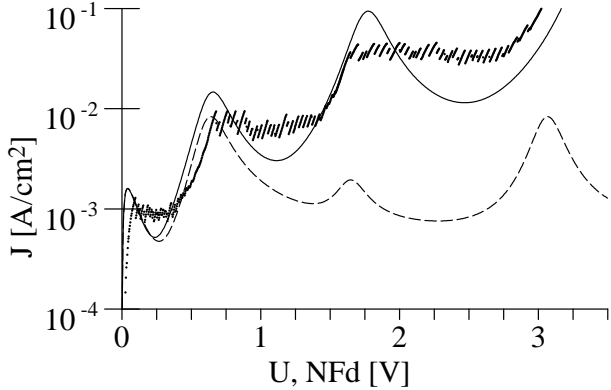,width=8cm} 
\caption[a]{Current-field relation for the superlattice 
studied in   \cite{MUR00}. The bias was taken to be NFd for the 
theoretical results, where $N=30$ is the number of quantum wells.
Full line: Result from sequential tunneling with renormalized matrix elements.
Dashed line: Result from sequential tunneling with bare matrix elements.
Dots: Experimental data (courtesy of Yu.~A.~Mityagin).
\label{Fig3mityagin}} 
\end{figure}

In Fig.~\ref{Fig3mityagin} results are shown for the superlattice
structure from Ref.~\cite{MUR00}
(25 nm wide GaAs wells, 10 nm Al$_{0.3}$Ga$_{0.7}$As barriers, 
doping density $N_D=1.75\times 10^{10}/{\rm cm}^2$).
Due to the large well width the level seperation is small
and several resonances $E^a\to E^{\nu}$ can be observed with increasing
field. 
The calculations (for $T=4$ K and within the
approximation (\ref{Eq3JSTsimp}) applying phenomenological broadenings
$\Gamma^{\nu}=4$ meV for all levels) have been performed
with and without taking into account the renormalization
of the matrix elements according to  Eq.~(\ref{Eq2Matren}) 
including the lowest 6 levels.
Fig.~\ref{Fig3mityagin} shows that the result with
renormalized matrix elements (full line), see Eq.~(\ref{Eq2hamW1ren}),
is in good agreement with the experimental
result but exhibits higher peak currents. This may be due to an overestimation
of the couplings in the calculation.
E.g. assuming barriers of 10.6 nm, the current drops by a factor
of 2. An increase of the Al-content in the barriers would
give a similar trend.
The results with the bare matrix elements (dashed line), 
see Eq.~(\ref{Eq2hamW1}), 
deviates strongly from the experimental result. In particular,
the peak currents do not increase for resonances at higher fields.
This shows that the renormalization procedure
is essential if higher resonances are considered.

 
\subsubsection{Tunneling over several barriers} 
\begin{figure} 
\epsfig{file=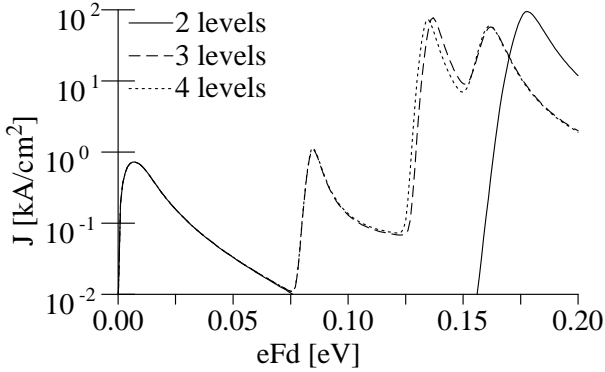,width=8cm} 
\caption[a]{Current-field relation for the superlattice 
studied in   \cite{HEL99} for constant electron density 
$n_m=N_D$. Transitions between nearest and next nearest 
wells have been taken into account  and 
the number of levels was 
$\nu_{\rm max}=2$ (full line), $\nu_{\rm max}=3$ (dashed line), 
and $\nu_{\rm max}=4$ (dotted line). 
$T=20$ K was used in the calculation. 
\label{Fig3STkennhelm}} 
\end{figure}

Up to now the discussion was restricted to next-neighbor coupling, 
which is described by matrix elements 
$\tilde{H}_1^{\mu\nu}$ in Eq.~(\ref{Eq2hamW1ren}). 
The extension to tunneling over $h$ barriers can be 
treated analogously taking into account the matrix element 
$\tilde{H}_h^{\mu\nu}$. The discussion 
of Sec.~\ref{SecSTsimple} can be performed in the 
same way applying the bias drop $heFd$. 
Thus we expect resonances at field strengths 
$eFd=(\tilde{E}^{\nu}-\tilde{E}^{a})/h$. 
The total current is then given by 
\begin{equation} 
J_{\rm ST}=\sum_{h=1}^{h_{\rm max}}\sum_{\nu=1}^{\nu_{\rm max}} 
hJ_{0\to h}^{a \to \nu} 
\end{equation} 
where the individual current densities $J_{0\to h}^{a \to \nu}$ 
are evaluated according to Eq.~(\ref{Eq3JST}). Here it is assumed 
that only the lowest level $a$ is occupied. 
For the samples discussed in the last section, as well as for 
most other samples considered, the respective currents for $h>1$ 
are negligible. 
This is different for the sample discussed in   \cite{HEL99} 
(5 nm wide GaAs wells, 8 nm Al$_{0.29}$Ga$_{0.71}$As barriers, 
doping density $N_D=2.25\times 10^{11}/{\rm cm}^2$), where 
the second miniband is located around the conduction band of the 
barrier and the subsequent minibands $\nu\ge 3$ 
resemble free particle states. Results of the calculation 
with $h_{\rm max}=2$, i.e. taking into account tunneling 
to next-nearest neighbor wells, are shown in Fig.~\ref{Fig3STkennhelm}. 
If the calculation is restricted to the two 
lowest levels ($\nu_{\rm max}=2$) the current-field relation 
resembles the findings of Fig.~\ref{Fig3STkenngrahn}. 
There is a peak at low fields due to $a\to a$ tunneling and a peak 
at $eFd\approx\tilde{E}^{b}-\tilde{E}^{a}=0.177$ meV 
due to tunneling  $a\to b(1)$ into nearest neighbor well. 
The matrix element $\tilde{H}_2^{ab}$ is small, thus no 
transitions $a\to b(2)$ to the next nearest neighbor well can be seen. 
This changes completely if the third level ($\nu_{\rm max}=3$) 
is taken into account for 
the renormalization of the energy levels and couplings. 
First the strong coupling to the third level 
diminishes  the value of $\tilde{E}^{b}$ by 15 meV close to the $a\to b(1)$ 
resonance. Thus the position of this resonance is shifted to 
$eFd=162$ meV where the new resonance condition is 
fulfilled. Secondly a new peak arises at $eFd=85$ meV$\approx 
(\tilde{E}^{b}-\tilde{E}^{a})/2$ due to next nearest well tunneling 
$a\to b(2)$ (remember that the renormalized level energies $\tilde{E}^{b}$ 
are field dependent and thus the local field must be taken 
into account at each comparison). The reason is the strong admixture 
of $R^{bc}_2$ (which is quite large for the superlattice structure considered) 
in the renormalization of the matrix element $\tilde{H}_2^{ab}$ 
due to Eq.~(\ref{Eq2Matren}). 
For the same reason a third peak appears at 
$eFd=136$ meV$\approx(\tilde{E}^{c}-\tilde{E}^{a})/2$ 
due to $a\to c(2)$ tunneling. 
If the fourth level ($\nu_{\rm max}=4$) is taken into account as well, 
the result is hardly changed, thus providing confidence into 
the results. These findings are in agreement 
with the experiments \cite{HEL99} where a strong increase 
of the current was observed at field strengths of $eFd\approx 80$ meV and 
the current density becomes larger than 0.15 kA/cm$^2$. 
Nevertheless, no current peak has been observed so far in this 
field region. 
Tunneling over more than one barrier has also been observed experimentally 
in   \cite{KRI98,SIB98}. 
Current peaks at $eFd=(E^b-E^a)/h$ corresponding to
resonances between $h^{\rm th}$ next neighbors have also been found
in the calculation by Zhao and Hone \cite{ZHA00}. The hight of these
peaks was quite small, probably due to the neglect of
interband couplings $R_h^{ab}$ in their calculation.

These findings show that next-nearest neighbor tunneling is possible 
in superlattice structures. Nevertheless the quantitative description 
is still an open issue. The inclusion of 
results from Zener tunneling \cite{DIC94} may be helpful in  future 
research here.

\subsection{Comparison of the approaches} 
Let us now compare the results from the different approaches 
miniband transport (MBT), sequential 
tunneling (ST), and Wannier-Stark hopping (WSH). 
 
MBT-ST: Comparing Figs.~\ref{Fig3MCrott} and \ref{Fig3STkenngrahn} 
one notices that the global behavior with linear increase 
of the current for low fields and a maximum at moderate fields 
is in qualitative agreement for the MBT and ST approach. 
While the current scales with the square of 
the coupling for ST, the Esaki-Tsu drift velocity is proportional 
to $T_1$. This discrepancy is resolved if either the temperature 
or the electron density is high and a $|T_1|^2$ dependence of 
the current density is recovered for MBT as well, see 
Eqs.~(\ref{Eq3MBT-deg},\ref{Eq3MBT-T}). Comparing 
these results with Eqs.~(\ref{Eq3JSTdeg},\ref{Eq3JSTnondeg}) 
we find that the simplified expression become identical for 
MBT and ST if either $k_BT$ or $n/\rho_0$ are large with respect to 
both $2T_1^a$ and $eFd$. 
This explains  the 
$1/T$ dependence of the current density 
observed experimentally in   \cite{BRO90a,SIB93a} for 
superlattices exhibiting a rather small coupling $T_1^a\approx 1$ meV. 
As the experiments are performed at $T>77$ K, the estimations 
(\ref{Eq3MBT-T},\ref{Eq3JSTnondeg}) hold simultaneously and 
the findings cannot be taken as a manifestation of 
miniband transport. 
 
ST-WSH: Both approaches exhibit negative differential conductivity for 
high electric fields. Let us restrict ourselves to the $a\to a$ resonance 
and consider a superlattice with 
nearest neighbor coupling $T_1^a$. 
For large electric fields 
$eFd\gg 2 {\rm Im}\left\{\Sigma^{a\, {\rm ret}}\right\}=\Gamma$ 
the term $A^{a}_0({\bf k},E)A^{a}_1({\bf k},E)$ 
from Eq.~(\ref{Eq3JST}) 
exhibits a two-peak structure 
\begin{equation}\begin{split} 
A^{a}_0({\bf k},E)A^{a}_1({\bf k},E)\approx & 
2\pi \delta(E-E_k-E^a) 
\frac{2 {\rm Im}\left\{\Sigma^{a\, {\rm ret}}_1({\bf k},E_k+E^a)\right\}} 
{(eFd)^2}\\ 
&+\frac{2 {\rm Im}\left\{\Sigma^{a\, {\rm ret}}_0({\bf k},E_k+E^a-eFd)\right\}} 
{(eFd)^2} 
2\pi \delta(E-E_k-E^a+eFd)\, . 
\end{split}\end{equation} 
Within the Born approximation for the scattering 
\begin{equation} 
{\rm Im}\left\{\Sigma^{a\, {\rm ret}}_m({\bf k},E)\right\}=\sum_{\bf k'} 
\frac{\pi}{\hbar} 
\left|\langle \Psi^a_{m,{\bf k}'}|\hat{H}^{\rm scatt}|\Psi^a_{m,{\bf k}} 
\rangle\right|^2\delta(E-E^a-E_{k'}+meFd)\, . 
\end{equation} 
Applying the  approximation (\ref{Eq3WSHhf}) for $eFd\gg 2T_1^a$ 
the expression (\ref{Eq3JST}) for the current in the ST model 
becomes after several lines of algebra 
\begin{equation} 
J_{0\to 1}^{a \to a}=\frac{2e}{\hbar A } 
\sum_{{\bf k},{\bf k}'} 
R_{0,{\bf k}\to 1,{\bf k}'}n_F(E_k+E_a-\mu) 
\left[1-\exp\left(-\frac{eFd}{k_BT}\right)\right] 
\, . 
\end{equation} 
This is the dominating term of the current 
for Wannier-Stark hopping (\ref{Eq3WSHthermcurrent}). Thus, 
the expressions of ST and WSH become identical in the limit 
of $eFd\gg \Gamma$ and $eFd\gg 2|T_1|$. This is just the overlapping 
region between the ranges of validity of both approaches as depicted in 
Fig.~\ref{Fig3regimes}. 
 
The transition between WSH and MBT is even more difficult. 
MBT typically exhibits a $1/F$ behavior 
for $eFd\gg \Gamma$ as predicted for the Esaki-Tsu relation, while 
$1/F^r$ with various exponents $r\ge 2$ is found 
for $eFd\gg 2|T_1^a|$ from the WSH model, see Sec.~\ref{SecWSH}. 
As mentioned there, a $1/F$ behavior 
can be recovered from the WSH approach 
by summing all contribution $h$ in Eq.~(\ref{Eq3WSHcurrent}) 
for $eFd\ll 2|T_1^a|$. In   \cite{ROT98,ROT97} it is shown 
that in the field region $\Gamma\ll eFd\ll 2|T_1^a|$ 
the results of both approaches agree fairly well. 
Again this agrees with the joint range of validity depicted in 
Fig.~\ref{Fig3regimes}.

\section{Quantum transport\label{ChapNGFT}} 
In semiconductor superlattices 
the miniband width $\Delta$, the potential drop per period 
$eFd$, and the scattering induced broadening $\Gamma$ are often 
of comparable magnitudes. This requires the application of a consistent 
quantum transport theory combining scattering and the quantum mechanical 
temporal evolution. 
Different formulations applying nonequilibrium Green functions \cite{LAK97}, 
density matrix theory \cite{ROS98a}, 
the master-equation approach \cite{FIS99}, or Wigner 
functions \cite{BOR99a} have recently been used to tackle this 
general problem for a variety of different model structures. 
 
Here the formalism of nonequilibrium Green functions 
is applied to study electrical transport in superlattices. 
This approach allows for a systematic study of both 
quantum effects and scattering 
to arbitrary order of perturbation theory. 
Although the calculations involved are quite tedious (as well as 
the acquaintance with this method) such calculations are of importance 
for two purposes: On one hand it is possible to derive simpler 
expressions like those studied in the preceding section from a 
general theory, thus shedding light into the question of 
applicability. On the other hand there are situations 
where no simple theory exists and thus one has to pay the 
price to work with a more elaborate formalism. 
 
A variety of different quantum transport calculations for semiconductor 
superlattices have been 
reported in the literature: 
In \cite{SUR84} an analysis within the density matrix theory has 
been presented, which was simplified to different approaches 
for low, medium, and high field. The same method was applied to study 
transport in a perpendicular magnetic field \cite{SUR90}. 
A similar approach was performed in \cite{BRY97}, where the 
quantum kinetic approach was solved in the limit of Wannier 
Stark hopping and the nature of phonon resonances were analyzed. 
The formation of Landau levels in a longitudinal magnetic field causes 
additional resonances \cite{KLE97a}. 
A transport model based on nonequilibrium Greens 
function \cite{LAI93} has been proposed as well, although explicit 
calculations could only be performed in the high temperature limit 
and within hopping between next neighbor Wannier-Stark states 
there. 
 
This section is organized as follows: 
At first the general formalism of nonequilibrium Green functions 
for stationary transport is briefly 
reviewed in a form  which can be applied to a variety of devices. 
The special notation  to consider transport in 
homogeneous semiconductor superlattices as well as the approximations 
used are described in the second subsection. 
In the third subsection the standard approaches (miniband transport, 
Wannier-Stark hopping, and sequential tunneling as discussed in 
section \ref{ChapStandard}) will be explicitly derived as 
limiting cases of the quantum 
transport model. This proves the regions of validity given in
Fig.~\ref{Fig3regimes}. 
Finally, in the fourth subsection results are presented for different 
samples. The results from the self-consistent quantum transport model 
will be compared with simpler calculations within the standard approaches 
applying identical sample parameters. 
This will demonstrate that the standard approaches work {\em quantitatively 
well} in their respective range of applicability. 
The reader  who is less interested in the theoretical 
concept and underlying equations may skip subsections 
\ref{SecfromNGFT}--\ref{SecNGFconstsig} and continue with the results 
in subsection \ref{SecQtresults}. 
 
\subsection[Nonequilibrium Green functions]{Nonequilibrium Green functions 
applied to stationary transport\label{SecfromNGFT}} 
In this subsection the underlying theory of nonequilibrium Green functions 
is briefly reviewed. 
The notation of \cite{HAU96,MAH90} is followed 
here and the reader is referred to these textbooks for 
a detailed study as well as for proofs 
of several properties addressed here. 
 
We consider a set of one-particle basis states $|\alpha\rangle$ and 
$a_{\alpha}(t),a^{\dag}_{\alpha}(t)$ are the corresponding 
annihilation and creation operators.  The time dependence 
stems from the Heisenberg picture. 
Most  physical one-particle observables can be expressed by the 
one-particle density matrix 
\begin{equation} 
\rho_{\alpha,\beta}(t)=\langle a^{\dag}_{\alpha}(t)a_{\beta}(t)\rangle 
={\rm Tr}\left\{\hat{\rho}a^{\dag}_{\alpha}(t)a_{\beta}(t)\right\} 
\end{equation} 
which is the corresponding quantum mechanical expectation 
value with the density operator $\hat{\rho}$. 
In particular, the occupation of the state $|\alpha\rangle$ is given by 
$\rho_{\alpha,\alpha}(t)$. The task of any 
many-particle quantum theory is the evaluation of $\rho_{\alpha\beta}(t)$ 
in the presence of a Hamiltonian 
\begin{equation} 
\hat{H}=\hat{H}_0+\hat{U}+\hat{H}_{\rm scatt} 
\end{equation} 
where 
\begin{equation} 
\hat{H}_0=\sum_{\alpha}E_{\alpha}a^{\dag}_{\alpha}(t)a_{\alpha}(t) 
\end{equation} 
is diagonal in the basis $|\alpha\rangle$, 
\begin{equation} 
\hat{U}=\sum_{\alpha,\beta}U_{\alpha,\beta}(t) 
a^{\dag}_{\alpha}(t)a_{\beta}(t) 
\end{equation} 
describes an additional potential term,  and 
$\hat{H}_{\rm scatt}$ refers to interactions with phonons, random 
impurity potentials (which are treated within impurity averaging), 
or interactions between the particles. 
 
Within density matrix theory the temporal evolution of 
$\rho_{\alpha,\beta}(t)$ is studied directly by applying 
Heisenberg's equation of motion for the product 
$a^{\dag}_{\alpha}(t)a_{\beta}(t)$. 
E.g., the time dependence of the 
occupation $\rho_{\alpha,\alpha}(t)$ is given by: 
\begin{equation}\begin{split} 
\frac{\d}{\d t} 
\langle a_{\alpha}^{\dag}(t)a_{\alpha}(t)\rangle&= 
\frac{\imai }{\hbar} 
\langle [\hat{H},a_{\alpha}^{\dag}(t)a_{\alpha}(t)]\rangle\\ 
&=\sum_{\beta} 
\frac{\imai }{\hbar}\, 
\left[ 
U_{\beta,\alpha}\langle a_{\beta}^{\dag}(t)a_{\alpha}(t)\rangle- 
U_{\alpha,\beta}\langle a_{\alpha}^{\dag}(t)a_{\beta}(t)\rangle 
+\langle[\hat{H}_{\rm scatt},a_{\alpha}^{\dag}(t)a_{\alpha}(t)]\rangle\right] 
\label{Eq4dmttempevol} 
\end{split}\end{equation} 
where $[a,b]=ab-ba$ denotes the commutator. 
In order to satisfy the equation of continuity, the particle currents  $j(t)$ 
between the basic states have to be identified by 
\begin{equation} 
j^{\beta\to\alpha}(t)=\frac{2}{\hbar} 
\Re\left\{\imai U_{\beta,\alpha} 
\langle a_{\beta}^{\dag}(t)a_{\alpha}(t)\rangle\right\} 
+j^{\beta\to\alpha}_{\rm scatt}(t)\label{Eq4jdmt} 
\end{equation} 
which satisfy $j^{\beta\to\alpha}(t)=-j^{\alpha\to\beta}(t)$ 
as each part of $\hat{H}$ is a Hermitian operator. 
The scattering induced current $j^{\beta\to\alpha}_{\rm scatt}(t)$ 
can be determined once $\hat{H}_{\rm scatt}$ is specified. This term 
typically contains higher order density matrices like 
$\langle a_{\beta'}^{\dag}(t)a_{\beta}^{\dag}(t)a_{\alpha}(t)a_{\alpha'}(t) 
\rangle$ in the case of electron-electron scattering. 
Thus there is no closed set of dynamical equations 
for $\rho_{\alpha,\beta}(t)$ and 
the dynamical evolution generates a hierarchy of 
many-particle density matrices, which has to be closed 
by approximations, see, e.g., \cite{HAU94,KUH98} as well as references cited therein. 
 
A conceptually different approach to many-particle physics 
constitutes the theory of nonequilibrium Green functions which has been 
developed by Kadanoff and Baym \cite{KAD62} and independently 
by Keldysh \cite{KEL65}. In this theory the time dependence of 
$a^{\dag}_{\alpha}(t_1)$ and $a_{\beta}(t_2)$ is considered 
separately, thus two different times appear in the calculation. 
The corresponding generalization of the density matrix is 
the correlation function (or 'lesser' Green function) 
\begin{equation} 
G^<_{\alpha_1,\alpha_2}(t_1,t_2)= 
\imai\langle a^{\dag}_{\alpha_2}(t_2)a_{\alpha_1}(t_1)\rangle 
\end{equation} 
which describes the occupation of the states together 
with the respective correlations both in time and state index. 
Note the unusual order of indices, which will be helpful in later 
stages of the theory. The notation follows 
\cite{HAU96,MAH90} here. Sometimes (e.g., \cite{KAD62,LAN76a}) 
the factor $\imai$ is dropped so that $G^<$ agrees directly with the 
density matrix for equal times. Next to this correlation function 
the retarded and advanced Green functions 
are defined by 
\begin{eqnarray} 
G^{\rm ret}_{\alpha_1,\alpha_2}(t_1,t_2)&=& 
-\imai\Theta(t_1-t_2) 
\langle \left\{a_{\alpha_1}(t_1),a^{\dag}_{\alpha_2}(t_2)\right\}\rangle\\ 
G^{\rm adv}_{\alpha_1,\alpha_2}(t_1,t_2)&=& 
\imai\Theta(t_2-t_1) 
\langle \left\{a_{\alpha_1}(t_1),a^{\dag}_{\alpha_2}(t_2)\right\}\rangle 
=\left[G^{\rm ret}_{\alpha_2,\alpha_1}(t_2,t_1)\right]^* 
\end{eqnarray} 
respectively, where $\{a,b\}=ab+ba$ denotes the anticommutator which is 
appropriate for fermion operators $a_{\alpha}$ considered here. 
These functions describe the response of the system at time 
$t_1$ in state $\alpha_1$ which is excited at time 
$t_2$ in state $\alpha_2$. 
 
\subsubsection{Temporal evolution} 
The time dependence of these Green functions 
is given by the following set equations 
\begin{align} 
\begin{split} 
\left(\imai\hbar \pabl{}{t_1}-E_{\alpha_1}\right)& 
G^<_{\alpha_1,\alpha_2}(t_1,t_2) 
-\sum_{\beta} U_{\alpha_1,\beta}(t_1)G^{<}_{\beta,\alpha_2}(t_1,t_2)\\ 
&=\sum_{\beta}\int \frac{\d t}{\hbar}\left[ 
\Sigma^{\rm ret}_{\alpha_1,\beta}(t_1,t)G^{<}_{\beta,\alpha_2}(t,t_2) 
+\Sigma^{<}_{\alpha_1,\beta}(t_1,t)G^{\rm adv}_{\beta,\alpha_2}(t,t_2) 
\right]\label{Eq4Glesst1}\end{split}\\ 
\begin{split} 
\left(-\imai\hbar \pabl{}{t_2}-E_{\alpha_2}\right)& 
G^<_{\alpha_1,\alpha_2}(t_1,t_2) 
-\sum_{\beta} G^<_{\alpha_1,\beta}(t_1,t_2)U_{\beta,\alpha_2}(t_2)\\ 
&=\sum_{\beta}\int \frac{\d t}{\hbar}\left[ 
G^{\rm ret}_{\alpha_1,\beta}(t_1,t)\Sigma^{<}_{\beta,\alpha_2}(t,t_2) 
+G^{<}_{\alpha_1,\beta}(t_1,t)\Sigma^{\rm adv}_{\beta,\alpha_2}(t,t_2) 
\right]\label{Eq4Glesst2}\end{split}\\ 
\begin{split} 
\left(\imai\hbar \pabl{}{t_1}-E_{\alpha_1}\right)& 
G^{\rm ret/adv}_{\alpha_1,\alpha_2}(t_1,t_2) 
-\sum_{\beta} U_{\alpha_1,\beta}(t_1)G^{\rm ret/adv}_{\beta,\alpha_2}(t_1,t_2)\\ 
&=\hbar\delta(t_1-t_2)\delta_{\alpha_1,\alpha_2} 
+\sum_{\beta} \int \frac{\d t}{\hbar} 
\Sigma^{\rm ret/adv}_{\alpha_1,\beta}(t_1,t) 
G^{\rm ret/adv}_{\beta,\alpha_2}(t,t_2) 
\label{Eq4Grett1}\end{split} 
\end{align} 
\begin{align} 
\begin{split} 
\left(-\imai\hbar \pabl{}{t_2}-E_{\alpha_2}\right)& 
G^{\rm ret/adv}_{\alpha_1,\alpha_2}(t_1,t_2) 
-\sum_{\beta} G^{\rm ret/adv}_{\alpha_1,\beta}(t_1,t_2)U_{\beta,\alpha_2}(t_2)\\ 
&=\hbar\delta(t_1-t_2)\delta_{\alpha_1,\alpha_2} 
+\sum_{\beta} \int \frac{\d t}{\hbar} 
G^{\rm ret/adv}_{\alpha_1,\beta}(t_1,t) 
\Sigma^{\rm ret/adv}_{\beta,\alpha_2}(t,t_2) 
\label{Eq4Grett2}\end{split} 
\end{align} 
which are derived in section 5 of \cite{HAU96}. The same 
result is obtained from the matrix equations 
(3.7.5) and (3.7.6) of \cite{MAH90} if the 
definitions of the retarded and advanced 
Green functions are inserted. 
 
While the equations for $G^{\rm ret}$ and 
$G^{\rm adv}$ exhibit a $\delta(t_1-t_2)$ inhomogeneity typical for 
Green functions, this is not the case for $G^<$ which is, strictly 
speaking, not a Green function (although this term is often used). 
The self-energies $\Sigma$ describe the influence of 
scattering (compare the simplified description in Sec.~\ref{SecSTgeneral}). 
They can be expressed by functionals of the Green functions 
which depend on the approximation used, 
such as the self-consistent Born approximation. 
Here it is crucial to pay attention to the fact that 
$\Sigma^{\rm ret},\Sigma^{\rm adv}$, and $\Sigma^{<}$ 
belong to the same quantity $\Sigma$ (a matrix \cite{MAH90} or a 
self-energy defined on the complex temporal contour \cite{HAU96}) 
and thus the same approximation has to be performed for each quantity. 
In this way one obtains a closed set of integro-differential equations, 
which governs the temporal evolution of the Green functions. 
On the right-hand side of Eqs.~(\ref{Eq4Glesst1},\ref{Eq4Glesst2}) 
the lesser Green functions and lesser self-energies only depend on the time 
arguments $(t_1',t_2')$ with $t_1'\le t_1$ and $t_2'\le t_2$ 
due to the properties of retarded and advanced Green functions. 
Therefore these differential equations for $G^<(t_1,t_2)$ 
can (at least in principle) be solved explicitly by forward integration. 
 
Equations (\ref{Eq4Glesst1},\ref{Eq4Glesst2}) are solved 
by the integral equation (sometimes called Keldysh relation) 
\begin{equation} 
G^<_{\alpha_1,\alpha_2}(t_1,t_2)=\sum_{\beta,\beta'} 
\int \frac{\d t}{\hbar}\int \frac{\d t'}{\hbar} 
G^{\rm ret}_{\alpha_1,\beta}(t_1,t)\Sigma^{<}_{\beta,\beta'}(t,t') 
G^{\rm adv}_{\beta',\alpha_2}(t',t_2)\label{Eq4Keldysht} 
\end{equation} 
which has a nice interpretation: 
$\Sigma^{<}_{\beta,\beta'}(t,t')$ can be considered as an in-scattering 
term, which creates a correlated one-particle excitation 
at times $(t,t')$ as a result of a scattering event. 
The retarded and advanced Green functions provide the action of 
this excitation at the later times  $t_1$ and $t_2$, at which 
the correlation function $G^<$ is observed. 
The relation Eq.~(\ref{Eq4Keldysht}) is a particular solution 
of the differential equations (\ref{Eq4Glesst1},\ref{Eq4Glesst2}). 
The general solution contains a further term (proportional 
to the free evolution of $G^<$ without scattering) 
to satisfy initial conditions, see, e.g., Eq.~(5.11) of \cite{HAU96}. 
Typically, the contribution of these terms decays in time if scattering 
is present, so that 
Eq.~(\ref{Eq4Keldysht}) holds in the long time limit. 
 
If we consider a stationary state without any time dependence of 
the external potential $U$, 
all functions depend only on the time difference $t_1-t_2$ and it is 
convenient to work in Fourier space defined by 
\begin{eqnarray} 
F_{\alpha_1,\alpha_2}(E)&=&\frac{1}{\hbar}\int \d t\, 
\e^{\imai Et/\hbar} F_{\alpha_1,\alpha_2}(t+t_2,t_2)\\ 
F_{\alpha_1,\alpha_2}(t_1,t_2)&=&\frac{1}{2\pi}\int \d E\,\e^{-\imai E(t_1-t_2)/\hbar} F_{\alpha_1,\alpha_2}(E) 
\end{eqnarray} 
both for self-energies and Green functions\footnote{Different 
definitions have been suggested which produce gauge invariant 
equations when  $U$ is due to a combination of an electric and 
magnetic field \cite{BER91,HAU96}. This is important for various approximations 
to treat slowly varying fields.}. 
Then the following relations hold: 
\begin{equation} 
\left\{G^{\rm ret}_{\alpha,\beta}(E)\right\}^*= 
G^{\rm adv}_{\beta,\alpha}(E) 
\quad\mbox{and}\quad 
G^{<}_{\alpha,\beta}(E)=-\left\{G^{<}_{\beta,\alpha}(E)\right\}^* 
\label{Eq4symmetry} 
\end{equation} 
Eqs.~(\ref{Eq4Glesst1},\ref{Eq4Glesst2},\ref{Eq4Grett1},\ref{Eq4Grett2}) 
yield 
\begin{align} 
\begin{split} 
\left(E-E_{\alpha_1}\right)G^<_{\alpha_1,\alpha_2}(E) 
-&\sum_{\beta} U_{\alpha_1,\beta}G^{<}_{\beta,\alpha_2}(E)\\ 
&=\sum_{\beta}\left[ 
\Sigma^{\rm ret}_{\alpha_1,\beta}(E)G^{<}_{\beta,\alpha_2}(E) 
+\Sigma^{<}_{\alpha_1,\beta}(E)G^{\rm adv}_{\beta,\alpha_2}(E) 
\right]\label{Eq4GlessE1}\end{split}\\ 
\begin{split} 
\left(E-E_{\alpha_2}\right)G^<_{\alpha_1,\alpha_2}(E) 
-&\sum_{\beta} G^<_{\alpha_1,\beta}(E)U_{\beta,\alpha_2}(E)\\ 
&=\sum_{\beta}\left[ 
G^{\rm ret}_{\alpha_1,\beta}(E)\Sigma^{<}_{\beta,\alpha_2}(E) 
+G^{<}_{\alpha_1,\beta}(E)\Sigma^{\rm adv}_{\beta,\alpha_2}(E) 
\right]\label{Eq4GlessE2}\end{split}\\ 
\begin{split} 
\left(E-E_{\alpha_1}\right) 
G^{\rm ret/adv}_{\alpha_1,\alpha_2}(E) 
-&\sum_{\beta} U_{\alpha_1,\beta}G^{\rm ret/adv}_{\beta,\alpha_2}(E) 
\\ 
&=\delta_{\alpha_1,\alpha_2}+ 
\sum_{\beta}\Sigma^{\rm ret/adv}_{\alpha_1,\beta}(E) 
G^{\rm ret/adv}_{\beta,\alpha_2}(E) 
\label{Eq4GretE} 
\end{split}\\ 
\begin{split} 
\left(E-E_{\alpha_2}\right) 
G^{\rm ret/adv}_{\alpha_1,\alpha_2}(E) 
-&\sum_{\beta} G^{\rm ret/adv}_{\alpha_1,\beta}(E)U_{\beta,\alpha_2} 
\\ 
&=\delta_{\alpha_1,\alpha_2}+ 
\sum_{\beta}G^{\rm ret/adv}_{\alpha_1,\beta}(E) 
\Sigma^{\rm ret/adv}_{\beta,\alpha_2}(E) 
\label{Eq4GretE2}\end{split}\, . 
\end{align} 
If $U_{\beta,\alpha}$ and $\Sigma^{\rm ret/adv}_{\beta,\alpha}(E)$ 
are symmetric matrices, then  $G^{\rm ret/adv}_{\alpha_1,\alpha_2}(E) 
=G^{\rm ret/adv}_{\alpha_2,\alpha_1}(E)$ holds as well. 
This can be shown by subtracting Eq.~(\ref{Eq4GretE2}) 
from Eq.~(\ref{Eq4GretE}), where $\alpha_1$ and $\alpha_2$ are exchanged. 
 
In the same way the Keldysh relation becomes 
\begin{equation} 
G^<_{\alpha_1,\alpha_2}(E)=\sum_{\beta,\beta'} 
G^{\rm ret}_{\alpha_1,\beta}(E)\Sigma^{<}_{\beta,\beta'}(E) 
G^{\rm adv}_{\beta',\alpha_2}(E)\label{Eq4KeldyshE}\, . 
\end{equation} 
A quite elementary derivation of Eqs.~(\ref{Eq4GretE},\ref{Eq4KeldyshE}) 
is given in section 8 of \cite{DAT95}. 
From $G^<(E)$ the density matrix 
can be evaluated directly via 
\begin{equation} 
\rho_{\alpha,\beta}=-\imai G^<_{\beta,\alpha}(t,t)= 
-\imai\int\frac{\d E}{2\pi} 
G^{<}_{\beta,\alpha}(E)\label{Eq4rhoG} 
\end{equation} 
which provides us with the one-particle expectation values 
for most quantities of interest. 
 
\subsubsection{Self-energies\label{Secselfenergy}} 
The self-energies can be obtained from the usual 
diagrammatic rules which are derived in most textbooks 
on many-particle theory such as \cite{MAH90}. 
In the following the results are given within the self-consistent 
Born-approximation. 
 
For {\em impurity scattering} one considers an impurity potential 
$V(\pol{r};\{\pol{r}_i\})$ which depends on the locations 
$\pol{r}_i$ of the impurities. The respective Hamiltonian 
is given by 
\begin{equation} 
\hat{H}_{\rm imp}= 
\sum_{\beta_1,\beta_2} 
V_{\beta_1,\beta_2}(\{\pol{r}_i\})a^{\dag}_{\beta_1} 
a_{\beta_2}\label{Eq4Vimp} 
\end{equation} 
with $V_{\beta_1,\beta_2}(\{\pol{r}_i\})= 
\langle \beta_1|V(\pol{r};\{\pol{r}_i\})|\beta_2\rangle $. 
For large systems 
the average $\langle V\ldots V \rangle_{\rm imp}$ over all possible 
impurity configurations 
$\{\pol{r}_i\}$ has to be taken 
(interface roughness can be treated in a similar 
way). Within the self-consistent 
Born-approximation only correlations between two scattering matrix 
elements are taken into account. Thus one finds 
\begin{equation} 
\Sigma^{</{\rm ret/adv}}_{\alpha_1,\alpha_2}(E)= 
\sum_{\beta_1,\beta_2} 
\langle V_{\alpha_1,\beta_1}(\{\pol{r}_i\}) 
V_{\beta_2,\alpha_2}(\{\pol{r}_i\})\rangle_{\rm imp} 
G^{</{\rm ret/adv}}_{\beta_1,\beta_2}(E)\label{Eq4Sigimp} 
\end{equation} 
 
For {\em phonon scattering} the respective Hamiltonian reads 
\begin{equation} 
\hat{H}_{\rm phonon}= 
\sum_{\pol{p},l}\left[ \hbar\omega_l(\pol{p})b_{\pol{p},l}^{\dag}b_{\pol{p},l} 
+\sum_{\beta_1,\beta_2} 
M_{\beta_1,\beta_2}(\pol{p},l)(b_{\pol{p},l}+b_{-\pol{p},l}^{\dag}) 
a^{\dag}_{\beta_1}a_{\beta_2}\right]\label{Eq4Vphonon} 
\end{equation} 
where $b_{\pol{p},l},b_{\pol{p},l}^{\dag}$ are the (bosonic) 
annihilation and creation operators of the phonon mode $l$ 
(such as acoustic/optical or longitudinal/transverse) 
with wave vector $\pol{p}$. 
Using Langreth rules (section 4.3 of \cite{HAU96}), 
one obtains the retarded self-energy 
within the self-consistent Born approximation: 
\begin{equation}\begin{split} 
\Sigma^{\rm ret}_{\alpha_1,\alpha_2}(E)=& 
\imai\sum_{\pol{p},l}\sum_{\beta_1,\beta_2} 
M_{\alpha_1,\beta_1}(\pol{p},l)M_{\beta_2,\alpha_2}(\pol{p},l) 
\int\frac{\d E_1}{2\pi} 
\Bigg[ 
G^{\rm ret}_{\beta_1,\beta_2}(E-E_1) 
D^{\rm ret}_l(\pol{p},E_1)\\ 
&+G^{\rm ret}_{\beta_1,\beta_2}(E-E_1) 
D^{<}_l(\pol{p},E_1)+ 
G^{<}_{\beta_1,\beta_2}(E-E_1) 
D^{\rm ret}_l(\pol{p},E_1)\Bigg] 
\label{Eq4Sigretphonon1} 
\end{split}\end{equation} 
where $D_l(\pol{p},E_1)$ refers to the phonon Green function. 
Now we replace the phonon Green function by its unperturbed 
equilibrium values (section 4.3 of \cite{HAU96} and 
Eq.~(2.9.9)\footnote{Note the sign error for $D^{\rm adv}_0$ in Eq.~(2.9.9) 
of \cite{MAH90}} 
of \cite{MAH90}). 
\begin{align} 
D^{\rm ret\, 0}_l(\pol{p},E_1) 
=&\frac{1}{E_1-\hbar\omega_l(\pol{p})+\imai 0^+} 
-\frac{1}{E_1+\hbar\omega_l(\pol{p})+\imai 0^+}\\ 
D^{\rm adv\, 0}_l(\pol{p},E_1)=&\frac{1}{E_1-\hbar\omega_l(\pol{p})-\imai 0^+} 
-\frac{1}{E_1+\hbar\omega_l(\pol{p})-\imai 0^+}\\ 
D^{<\, 0}_l(\pol{p},E_1)=& 
-2\pi \imai
\left\{n_B(\hbar\omega_l(\pol{p}))\delta(E_1-\hbar\omega_l(\pol{p})) 
+\left[n_B(\hbar\omega_l(\pol{p}))+1\right] 
\delta(E_1+\hbar\omega_l(\pol{p}))\right\}\, , 
\end{align} 
i.e., $D^{<\, 0}_l(\pol{p},E_1)=n_B(E_1)[D^{\rm ret\, 0}_l(\pol{p},E_1) 
-D^{\rm adv\, 0}_l(\pol{p},E_1)]$ 
where $n_B(E)=[\exp(E/k_BT)-1]^{-1}$ is the Bose distribution. 
As $G^{\rm ret}_{\beta_1,\beta_2}(E-E_1)$ 
has only poles for $\Im\{E_1\}>0$, we find 
\[ 
\int_{-\infty}^{\infty}\frac{\d E_1}{2\pi} 
G^{\rm ret}_{\beta_1,\beta_2}(E-E_1) 
D^{{\rm ret}\, 0}_l(\pol{p},E_1)= 
-\imai\left[G^{\rm ret}_{\beta_1,\beta_2}(E-\hbar\omega_l(\pol{p}))- 
G^{\rm ret}_{\beta_1,\beta_2}(E+\hbar\omega_l(\pol{p}))\right] 
\] 
from the residua of the contour over the complex plane with $\Im \{E_1\}<0$. 
Putting things together we obtain: 
\begin{equation}\begin{split} 
\Sigma^{\rm ret}_{\alpha_1,\alpha_2}(E)&= 
\sum_{\pol{p},l} \sum_{\beta_1,\beta_2}
M_{\alpha_1,\beta_1}(\pol{p},l)M_{\beta_2,\alpha_2}(\pol{p},l) 
\Bigg[\left[n_B(\hbar\omega_l(\pol{p}))+1\right] 
G^{\rm ret}_{\beta_1,\beta_2}(E-\hbar\omega_l(\pol{p}))\\ 
&+n_B(\hbar\omega_l(\pol{p}))G^{\rm ret}_{\beta_1,\beta_2}(E+\hbar\omega_l(\pol{p}))+\frac{1}{2}G^{<}_{\beta_1,\beta_2}(E-\hbar\omega_l(\pol{p})) 
-\frac{1}{2}G^{<}_{\beta_1,\beta_2}(E+\hbar\omega_l(\pol{p}))\\ 
&+\imai\int\frac{\d E_1}{2\pi} 
G^{<}_{\beta_1,\beta_2}(E-E_1) 
\left(\mathcal{P}\left\{\frac{1}{E_1-\hbar\omega_l(\pol{p})}\right\}- 
\mathcal{P}\left\{\frac{1}{E_1+\hbar\omega_l(\pol{p})}\right\}\right)\Bigg] 
\label{Eq4Sigretphon} 
\end{split}\end{equation} 
The lesser self-energy reads 
\begin{equation} 
\begin{split} 
\Sigma^{<}_{\alpha_1,\alpha_2}(E)=& 
\imai\sum_{\pol{p},l}\sum_{\beta_1,\beta_2} 
M_{\alpha_1,\beta_1}(\pol{p},l)M_{\beta_2,\alpha_2}(\pol{p},l) 
\int\frac{\d E_1}{2\pi} 
G^{<}_{\beta_1,\beta_2}(E-E_1) 
D^{<}_l(\pol{p},E_1)\\ 
=&\sum_{\pol{p},l}\sum_{\beta_1,\beta_2} 
M_{\alpha_1,\beta_1}(\pol{p},l)M_{\beta_2,\alpha_2}(\pol{p},l) 
\Big[n_B(\hbar\omega_l(\pol{p}))G^{<}_{\beta_1,\beta_2} 
(E-\hbar\omega_l(\pol{p}))\\ 
&\phantom{\sum_{\pol{p},l}\sum_{\beta_1,\beta_2} 
M_{\alpha_1,\beta_1}(\pol{p},l)M} 
+\left[n_B(\hbar\omega_l(\pol{p}))+1\right]G^{<}_{\beta_1,\beta_2} 
(E+\hbar\omega_l(\pol{p}))\Big]\, . 
\label{Eq4Siglessphon} 
\end{split} 
\end{equation} 
which describes the in-scattering from correlated 
states $\beta_1,\beta_2$ by 
phonon absorption as well as stimulated and spontaneous emission 
of phonons. 

The Coulomb interaction can be easily included within the
Hartree-Fock approximation, which provides an additional
potential (depending on $ \int \d E\, G^{<}(E)$), see, e.g., chapter 8 
of Ref.~\cite{DAT95}. Higher order approximations 
(describing electron--electron scattering) 
are difficult to implement. All these effects have been neglected
in this work.

The combination of these functionals for the self-energies 
with  Eqs.~(\ref{Eq4GretE},\ref{Eq4KeldyshE}) for the Green functions 
provides a coupled set of equations which has to be solved 
self-consistently to obtain the functions $G^<(E)$ for the 
stationary state. Afterwards the physical quantities of interest 
can be evaluated by Eq.~(\ref{Eq4rhoG}). 
 
\subsubsection{Thermal equilibrium} 
In thermal equilibrium the electron distribution is 
governed by the Fermi distribution. 
In the language of Green functions this can be written as 
\begin{equation} 
G^{<}_{\alpha_1,\alpha_2}(E)= 
\imai n_F(E) A_{\alpha_1,\alpha_2}(E) 
\label{Eq4Glesstherm} 
\end{equation} 
with the spectral function 
\begin{equation} 
A_{\alpha_1,\alpha_2}(E)=\imai\left[G^{\rm ret}_{\alpha_1,\alpha_2}(E) 
-G^{\rm adv}_{\alpha_1,\alpha_2}(E)\right]\label{Eq4A} 
\end{equation} 
which is derived in section 3.7 of \cite{MAH90}, 
e.g.. 
As discussed before, the occupation of the state $\alpha$ is given by 
\begin{equation} 
\rho_{\alpha,\alpha}=\frac{1}{2\pi \imai} 
\int \d E\, G^{<}_{\alpha,\alpha}(E) 
=\int \d E\, \frac{1}{2\pi}A_{\alpha,\alpha}(E)n_F(E)\, . 
\end{equation} 
Thus $A_{\alpha,\alpha}(E)/2\pi$ represents the 
energy resolved density of state $\alpha$. 
Here it is crucial to note the difference to the 
classical value of the occupation $n_F(E_{\alpha})$, which is 
only recovered in the free-particle case $U=\Sigma=0$ when 
$A_{\alpha,\alpha}(E)=2\pi\delta(E-E_{\alpha})$ holds. 
In contrast $\rho_{\alpha,\alpha}$ and $n_F(E_{\alpha})$ 
will in general differ, if scattering induced broadening 
leads to a finite width of the spectral function. 
 
These effects can be estimated assuming $U=0$ and 
$\Sigma_{\alpha_1,\alpha_2}^{\rm ret}(E)\approx 
-\imai\Theta(E)\delta_{\alpha_1,\alpha_2}\Gamma/2$ 
which mimics the fact that there are no scattering states 
below a band edge at $E=0$. Then one finds 
from Eqs.~(\ref{Eq4GretE},\ref{Eq4A}): 
\begin{equation} 
A_{\alpha_1,\alpha_2}(E)\approx\delta_{\alpha_1,\alpha_2} 
\frac{\Gamma}{(E-E_{\alpha})^2+\Gamma^2/4}\Theta(E) 
\end{equation} 
and in the limit $E_{\alpha}\gg k_BT$ one obtains 
\begin{equation} 
\rho_{\alpha,\alpha}\sim\frac{C k_BT}{2\pi} 
\frac{\Gamma}{(E_{\alpha})^2+\Gamma^2/4} \label{Eq4occupation} 
\end{equation} 
for a non-degenerate distribution $n_F(E)\approx C\e^{-E/k_BT}$. 
Thus, the occupation of the high energy states 
is larger than one would estimate from a semiclassical 
distribution $n_F(E_{\alpha})\approx C\e^{-E_{\alpha}/k_BT}$. 
 
Finally, it should be pointed out, that the different 
Green functions $G^<$, $G^{\rm ret}$, and $G^{\rm adv}$ are related 
to each other in thermal equilibrium which allows for 
a description in terms of a single Green function. 
Thus, the theory of equilibrium Green functions is significantly 
simpler than its nonequilibrium counterpart discussed here.

\subsubsection{Spatial boundary conditions and contacts} 
 
Although we are concerned with homogeneous 
infinite systems  in this section, 
some remarks concerning boundary conditions in real 
structures are appropriate. They are needed  in the discussion 
of transmission through superlattices \cite{WAC99}. 
 
In order to solve the system of equations discussed above 
in a finite system, boundary conditions have to be specified. 
Here two types can be distinguished. On the one hand there are regions where 
the device is terminated by an insulating layer. 
Here it is appropriate to neglect states in these regions, as their 
energy is significantly larger than 
the relevant energies in the device. A far more interesting 
point is the treatment of contacts, which act as a source or 
drain for the electric current. 
 
We separate the system into a central region with index $C$ and 
lead region with index $L$. The matrix $G_{\alpha,\beta}$ 
can than be divided in submatrices of the type 
${\bf G}_{CL}$, where the index $\alpha$ belongs to the central region 
and $\beta$ to one of the leads. 
Then the matrix equation (\ref{Eq4GretE}) 
can be written in the form 
\begin{equation} 
\begin{split} 
\begin{pmatrix} E-{\bf E}_C +\imai 0^+& 0\\ 
0& E-{\bf E}_{L}+\imai 0^+\end{pmatrix} 
&\cdot 
\begin{pmatrix} {\bf G}_{CC}^{\rm ret}(E) &{\bf G}_{CL}^{\rm ret}(E)\\ 
{\bf G}_{LC}^{\rm ret}(E) &{\bf G}_{LL}^{\rm ret}(E)\end{pmatrix} 
= 
\begin{pmatrix} {\bf 1} & 0\\ 
0& {\bf 1} \end{pmatrix} 
\\ 
+&\begin{pmatrix} {\bf U}_{CC} &{\bf U}_{CL}\\ 
{\bf U}_{LC} &{\bf U}_{LL}\end{pmatrix} 
\cdot 
\begin{pmatrix} {\bf G}_{CC}^{\rm ret}(E) &{\bf G}_{CL}^{\rm ret}(E)\\ 
{\bf G}_{LC}^{\rm ret}(E) &{\bf G}_{LL}^{\rm ret}(E)\end{pmatrix}\\ 
+& 
\begin{pmatrix} \boldsymbol{\Sigma}_{CC}^{\rm ret}(E) 
&\boldsymbol{\Sigma}_{CL}^{\rm ret}(E)\\ 
\boldsymbol{\Sigma}_{LC}^{\rm ret}(E) 
&\boldsymbol{\Sigma}_{LL}^{\rm ret}(E)\end{pmatrix} 
\cdot 
\begin{pmatrix} {\bf G}_{CC}^{\rm ret}(E) &{\bf G}_{CL}^{\rm ret}(E)\\ 
{\bf G}_{LC}^{\rm ret}(E) &{\bf G}_{LL}^{\rm ret}(E)\end{pmatrix} 
\label{Eq4Gretmatrix}\end{split} 
\end{equation} 
Now let us assume that $\boldsymbol{\Sigma}_{CL}=\boldsymbol{\Sigma}_{LC}=0$, 
i.e., there is no scattering between the lead regions and the central region. 
Then we can write: 
\begin{equation} 
(E-{\bf E}_{L}+\imai 0^+){\bf G}_{LC}^{\rm ret}(E)= 
{\bf U}_{LC}\cdot {\bf G}_{CC}^{\rm ret}(E) 
+{\bf U}_{LL}\cdot {\bf G}_{LC}^{\rm ret}(E) 
+\boldsymbol{\Sigma}_{LL}^{\rm ret}(E)\cdot {\bf G}_{LC}^{\rm ret}(E) 
\end{equation} 
This equation is solved by 
\begin{equation} 
{\bf G}_{LC}^{\rm ret}(E)={\bf G}_{L0}^{\rm ret}(E) 
\cdot {\bf U}_{LC}\cdot {\bf G}_{CC}^{\rm ret}(E) 
\end{equation} 
where ${\bf G}_{L0}^{\rm ret}(E)$ is 
the Green function of the lead satisfying the 
equation 
\begin{equation} 
(E-{\bf E}_{L}+\imai 0^+) 
\cdot{\bf G}_{L0}^{\rm ret}(E) 
={\bf 1}+{\bf U}_{LL}\cdot{\bf G}_{L0}^{\rm ret}(E) 
+\boldsymbol{\Sigma}_{LL}^{\rm ret}(E)\cdot{\bf G}_{L0}^{\rm ret}(E)\, . 
\end{equation} 
It is important to note, that 
${\bf G}_{L0}^{\rm ret}(E)$ is {\em not} exactly the Green function of 
the pure lead. In fact  $\boldsymbol{\Sigma}_{LL}^{\rm ret}(E)$ 
has to be evaluated from the full Green function ${\bf G}^{\rm ret}(E)$ 
and not only from  ${\bf G}_{L0}^{\rm ret}(E)$. This difference vanishes 
under the usual  assumption  that scattering is negligible in the leads. 
Now the part of equation (\ref{Eq4Gretmatrix}) for 
${\bf G}_{CC}^{\rm ret}(E)$ can be written as 
\begin{equation}\begin{split} 
(E-{\bf E}_{C}+\imai 0^+) 
\cdot {\bf G}_{CC}^{\rm ret}(E) 
=&{\bf 1}+{\bf U}_{CC}\cdot{\bf G}_{CC}^{\rm ret}(E)\\ 
&+\left[\boldsymbol{\Sigma}_{CC}^{\rm ret}(E) 
+{\bf U}_{CL}\cdot{\bf G}_{L0}^{\rm ret}(E) 
\cdot {\bf U}_{LC}\right]\cdot {\bf G}_{CC}^{\rm ret}(E)\, . 
\label{Eq4GretCC} 
\end{split}\end{equation} 
This is a closed equation for the matrix 
${\bf G}_{CC}^{\rm ret}(E)$ concerning the states 
inside the structure. The term 
${\bf U}_{CL}\cdot {\bf G}_{L0}^{\rm ret}(E) \cdot {\bf U}_{LC}$ 
can be viewed as an additional self-energy, 
due to the transitions between the central region and the lead. 
 
In a similar way Eq.~(\ref{Eq4Gretmatrix}) yields: 
\begin{equation} 
{\bf G}_{LL}^{\rm ret}(E)={\bf G}_{L0}^{\rm ret}(E) 
\cdot \left[{\bf 1}+{\bf U}_{LC}\cdot {\bf G}_{CL}^{\rm 
ret}(E)\right] \label{Eq4GretLL} 
\end{equation} 
From Eq.~(\ref{Eq4GretE2}) a similar structure as 
Eq.~(\ref{Eq4Gretmatrix}) can be obtained which provides: 
\begin{equation} 
{\bf G}_{CL}^{\rm ret}(E)={\bf G}_{CC}^{\rm ret}(E) 
\cdot {\bf U}_{CL}\cdot {\bf G}_{L0}^{\rm ret}(E) 
\label{Eq4GretCL} 
\end{equation} 
Furthermore all relations hold for the advanced Green functions 
in the same  way. 
 
With these ingredients the Keldysh relation(\ref{Eq4KeldyshE}) 
can be rewritten as 
\begin{equation}\begin{split} 
{\bf G}_{CC}^{<}(E) 
=&{\bf G}_{CC}^{\rm ret}(E)\cdot\boldsymbol{\Sigma}_{CC}^{<}(E) 
\cdot{\bf G}_{CC}^{\rm adv}(E) 
+{\bf G}_{CL}^{\rm ret}(E)\cdot\boldsymbol{\Sigma}_{LL}^{<}(E) 
\cdot{\bf G}_{LC}^{\rm adv}(E)\\ 
=&{\bf G}_{CC}^{\rm ret}(E)\cdot 
\left[\boldsymbol{\Sigma}_{CC}^{<}(E) 
+{\bf U}_{CL}\cdot {\bf G}_{L0}^{<}(E)\cdot{\bf U}_{LC}\right] 
\cdot{\bf G}_{CC}^{\rm adv}(E) 
\label{Eq4GlessCC} 
\end{split}\end{equation} 
where ${\bf G}_{L0}^{<}(E)$ defined by 
\begin{equation} 
{\bf G}_{L0}^{<}(E)= {\bf G}_{L0}^{\rm ret}(E)\cdot 
\boldsymbol{\Sigma}_{LL}^{<}(E)\cdot {\bf G}_{L0}^{\rm adv}(E) 
\end{equation} 
is {\em not} exactly the lesser Green function of 
the pure lead as $\boldsymbol{\Sigma}_{LL}^{<}(E)$ 
depends on the full Green function ${\bf G}^{\rm ret}(E)$. 
E.g., this reflects the fact, that the presence of a current through the 
central region will in principle effect the electron distribution 
in the lead. Nevertheless this reaction is typically negligible. 
Again the complications vanish under the assumption, that 
scattering is neglected in the leads. 
In the same way one obtains 
\begin{equation}\begin{split} 
{\bf G}_{CL}^{<}(E) 
=&{\bf G}_{CC}^{\rm ret}(E)\cdot\boldsymbol{\Sigma}_{CC}^{<}(E) 
\cdot{\bf G}_{CL}^{\rm adv}(E) 
+{\bf G}_{CL}^{\rm ret}(E)\cdot\boldsymbol{\Sigma}_{LL}^{<}(E) 
\cdot{\bf G}_{LL}^{\rm adv}(E)\\ 
=&{\bf G}_{CC}^{\rm ret}(E)\cdot\boldsymbol{\Sigma}_{CC}^{<}(E) 
\cdot{\bf G}_{CC}^{\rm adv}(E)\cdot {\bf U}_{CL}\cdot{\bf G}_{L0}^{\rm adv}(E)\\ 
&+{\bf G}_{CC}^{\rm ret}(E)\cdot {\bf U}_{CL}\cdot{\bf G}_{L0}^{\rm ret}(E) 
\cdot\boldsymbol{\Sigma}_{LL}^{<}(E) 
\cdot{\bf G}_{L0}^{\rm adv}(E)\left[{\bf 1}+{\bf U}_{LC}\cdot 
{\bf G}_{CL}^{\rm adv}(E)\right] \\ 
=&{\bf G}_{CC}^{<}(E)\cdot {\bf U}_{CL}\cdot{\bf G}_{L0}^{\rm adv}(E) 
+{\bf G}_{CC}^{\rm ret}(E)\cdot {\bf U}_{CL}\cdot{\bf G}_{L0}^{<}(E) 
\label{Eq4GlessCL} 
\end{split}\end{equation} 
where Eqs~(\ref{Eq4GretLL},\ref{Eq4GretCL},\ref{Eq4GlessCC}) 
have been subsequently applied (partially in the form for advanced functions). 
 
Let us consider a typical structure where the central region is 
connected to several independent leads $\ell$, which are 
translational invariant in their current direction. 
It is assumed that each lead $\ell$ is disorder-free so that the eigenstates 
can be separated into transverse and longitudinal parts, 
$\phi_{{\ell}\lambda q}(\pol{r})=\chi_{{\ell}\lambda}({\bf r}) 
\varphi^{\ell}_{q}(z)$, 
where $z$ is the spatial coordinate in the direction towards the central 
structure and ${\bf r}$ is a two-dimensional vector perpendicular to $z$. 
(A different coordinate system is applied for each lead.) 
The index $\lambda$ numbers the transverse modes within a given lead and 
$q$ denotes the behavior far away from the central region where 
$\varphi^{\ell}_q(z) \sim \e^{\imai qz}$ is assumed. 
The corresponding matrices ${\bf G}_{L0}$ are diagonal with matrix 
elements $G_{\ell \lambda q}(E)$ in this basis. 
 
The electric current from lead $\ell$ and mode $\lambda$ 
into the central region can be obtained from 
Eqs.~(\ref{Eq4jdmt},\ref{Eq4rhoG}). 
\begin{equation} 
I_{\ell\lambda}=2\mbox{(for spin)}e\frac{2}{\hbar}\int\frac{\d E}{2\pi} 
\sum_q\sum_{\alpha} 
\Re\left\{U_{\ell\lambda q,\alpha} 
G^{<}_{\alpha,\ell\lambda q}(E)\right\} 
\end{equation} 
where then sum $\sum_{\alpha}$ runs over all states belonging to 
the central region. Eq.~(\ref{Eq4GlessCL}) provides: 
\begin{equation}\begin{split} 
I_{\ell\lambda}=&\frac{4e}{\hbar}\int\frac{\d E}{2\pi} \sum_q 
\sum_{\alpha} 
\Re\left\{U_{\ell\lambda q,\alpha}\sum_{\beta}\left[ 
G^{<}_{\alpha\beta}(E)U_{\beta,\ell\lambda q}G_{\ell\lambda q}^{\rm adv}(E) 
+G^{\rm ret}_{\alpha\beta}(E)U_{\beta, \ell\lambda q} 
G_{\ell\lambda q}^{<}(E)\right]\right\}\\ 
=&\frac{2e}{\hbar}\int\frac{\d E}{2\pi} \sum_q 
\sum_{\alpha\beta} 
U_{\beta,\ell\lambda q}U_{\ell\lambda q,\alpha}\Big[ 
G^{<}_{\alpha\beta}(E)G_{\ell\lambda q}^{\rm adv}(E) 
-G^{<}_{\alpha\beta}(E)G_{\ell\lambda q}^{\rm ret}(E)\\ 
&\phantom{\frac{2e}{\hbar}\int\frac{\d E}{2\pi} \sum_q 
\sum_{\alpha\beta} 
U_{\beta,\ell\lambda q}U_{\ell\lambda q,\alpha}\Big[} 
+G^{\rm ret}_{\alpha\beta}(E)G_{\ell\lambda q}^{<}(E) 
-G^{\rm adv}_{\alpha\beta}(E)G_{\ell\lambda q}^{<}(E) 
\Big]\\ 
=&\frac{2e}{\hbar}\int\frac{\d E}{2\pi} 
{\rm Tr}\left\{\boldsymbol{\Gamma}_{CC}(\ell\lambda,E)\cdot 
\left[\imai{\bf G}^{<}_{CC}(E)+f_{\ell\lambda}(E) 
\imai\left({\bf G}^{\rm ret}_{CC}(E) 
-{\bf G}^{\rm adv}_{CC}(E)\right) 
\right]\right\} 
\label{Eq4Jcontact} 
\end{split} 
\end{equation} 
In the second line, the real part was taken by adding the complex 
conjugated expression (with $\alpha$ and $\beta$ exchanged). 
In the last line the definition 
\begin{equation} 
\Gamma_{\beta\alpha}(\ell\lambda,E) 
= \imai\sum_q U_{\beta,\ell\lambda q}U_{\ell\lambda q,\alpha} 
\left[G_{\ell\lambda q}^{\rm ret}(E)-G_{\ell\lambda q}^{\rm adv}(E)\right] 
\end{equation} 
is introduced and it is assumed that the occupation of the modes 
$\lambda$ in the lead $\ell$ 
can be treated by a  distribution function 
$f_{\ell\lambda}(E)$ with 
\begin{equation} 
G_{\ell\lambda q}^{<}(E)=f_{\ell\lambda}(E) 
\left[G_{\ell\lambda q}^{\rm adv}(q,E)-G_{\ell\lambda q}^{\rm ret}(q,E)\right] 
\, . 
\end{equation} 
Equation~(\ref{Eq4Jcontact}) has been derived in 
\cite{MEI92,DAT95,HAU96}. Together with 
Eqs.~(\ref{Eq4GretCC},\ref{Eq4GlessCC}) one obtains a closed 
set of equations for the Green functions in the central 
regions which is 
a convenient starting point for the simulation of quantum devices, 
see. 
Equation~(\ref{Eq4Jcontact}) can also be used as a starting point 
to derive the Landauer-B{\"u}ttiker formalism \cite{BUE86} which can be easily 
applied if $\Sigma=0$ inside the central region \cite{DAT90}. 
Otherwise some complications arise as discussed in \cite{WAC99}. 
A generalization to take into account time-dependent phenomena 
is straightforward \cite{JAU94}.

\subsection{Application to the superlattice structure\label{SecApplSL}} 
While the discussion of nonequilibrium Green functions was quite general 
in the preceding subsection, 
the general formalism will now be applied to a superlattice 
structure, which is assumed to be infinitely long. 
Therefore we restrict ourselves to the central region here. 
We use the basis given by the products of Wannier states multiplied by 
plane waves in the direction parallel to the layers, 
$\Psi_n^{\nu}(z)\e^{\imai {\bf k}\cdot {\bf r}}/\sqrt{A}$. Then 
the general states are given by 
$|\alpha\rangle=|n,\nu,{\bf k}\rangle$. The Hamiltonian $\hat{H}_0$ 
is given by Eq.~(\ref{Eq2hamW0}) and $\hat{U}$ is given by 
Eqs.~(\ref{Eq2hamW1},\ref{Eq2hamW2}). 
For simplicity we restrict ourselves 
to the lowest level $\mu=a$ (and omit the respective indices) 
and nearest neighbor coupling $T_1$ in the following. 
Furthermore we set $E^a=0$. (The inclusion of higher levels
is straightforward but tedious.)
We assume that the superlattice is 
(after impurity averaging) spatially homogeneous in the 
$x,y$ plane. Then the expectation values 
$\langle [a_m({\bf k}_1)]^{\dag} 
a_n({\bf k}_2)\rangle$ must vanish for ${\bf k}_1\neq{\bf k}_2$.
Thus the Green functions are diagonal in 
the wave vector {\bf k} and can be written as 
$G_{n,m}({\bf k};t_1,t_2)$ in the following. 
 
\subsubsection{Basic equations} 
The Wannier functions are essentially localized within a single 
quantum well. Therefore the scattering matrix elements connecting 
states of different quantum wells are small 
compared to those describing intrawell scattering. 
Thus, we restrict ourselves to scattering matrix elements 
$V_{m{\bf k},n{\bf k}'}(\{\pol{r}_i\})$, 
$M_{m{\bf k},n{\bf k}'}(\pol{p},l)$ 
which are diagonal in the well indices $m,n$ in the following 
(no interwell scattering). 
 
In this case the scattering-induced currents between different 
wells vanish in Eq.~(\ref{Eq4jdmt}) and the total electric 
current density  well $m$ 
to well $m+1$ is given by 
\begin{equation} 
J_{m\to m+1}=\frac{2\mbox{(for Spin)}e}{A} 
\sum_{{\bf k}}\frac{2}{\hbar}\, \int \frac{\d E}{2\pi} 
\Re\left\{ T_1
G_{m+1,m}^{<}({\bf k},E)\right\} 
\label{Eq4JNGFT} 
\end{equation} 
where Eq.~(\ref{Eq4rhoG}) has been applied. 
Similarly the electron density in well $m$ is given by 
\begin{equation} 
n_m= 
\frac{2\mbox{(for Spin)}}{A} 
\sum_{{\bf k}} \int\frac{\d E}{2\pi} \Im 
\left\{G_{m,m}^{<}({\bf k},E)\right\} 
\label{Eq4nNGFT} 
\end{equation} 
 
We consider impurity and phonon scattering within the 
self-consistent Born approximation. 
Furthermore we neglect correlations between the 
scattering matrix elements in different wells. 
This means that 
$\langle V_{m{\bf k},m{\bf k}'}(\{\pol{r}_i\})
V_{n{\bf k}',n{\bf k}}(\{\pol{r}_i\})\rangle_{\rm imp}$ 
vanishes in Eq.~(\ref{Eq4Sigimp}) for $n\neq m$.
This assumption is realistic for short range potentials 
of random impurities but becomes problematic if  interface 
roughness scattering is considered where significant 
correlations between neighboring wells may occur 
\cite{HOL94,MEN95}. 
For phonon scattering this approximation makes sense 
if localized phonons such as 
interface phonons are considered which are different in each well. 
Under this assumption the self-energies become diagonal 
in the well index and can be written as 
\begin{equation} 
\Sigma^{</{\rm ret, imp}}_{n}({\bf k},E) 
=\sum_{{\bf k}'} 
\langle V_{n{\bf k},n{\bf k}'}(\{\pol{r}_i\})
V_{n{\bf k}',n{\bf k}}(\{\pol{r}_i\})\rangle_{\rm imp}
G^{</{\rm ret}}_{n,n}({\bf k}',E) 
\label{Eq4sigimpSL} 
\end{equation} 
and in the same way for phonon scattering with phonon modes 
$l,\pol{p}$ we find from 
Eqs.~(\ref{Eq4Sigretphon},\ref{Eq4Siglessphon}): 
\begin{equation}\begin{split} 
\Sigma^{\rm ret, phonon}_{n}&({\bf k},E)= 
\sum_{\pol{p},l,{\bf k}'} 
|M_{n{\bf k},n{\bf k}'}(\pol{p},l)|^2
\Bigg[ 
\left[n_B(\hbar\omega_l(\pol{p}))+1\right]G^{\rm ret}_{n,n} 
({\bf k}',E-\hbar\omega_{\ell}(\pol{p}))\\ 
+&n_B(\hbar\omega_l(\pol{p}))G^{\rm ret}_{n,n}({\bf k}',E+\hbar\omega_l(\pol{p}))\\ 
+&\frac{1}{2}G^{<}_{n,n}({\bf k}',E-\hbar\omega_l(\pol{p})) 
-\frac{1}{2}G^{<}_{n,n}({\bf k}',E+\hbar\omega_l(\pol{p}))\\ 
+&\imai\int\frac{\d  E_1}{2\pi} 
G^{<}_{n,n}({\bf k}',E-E_1) 
\left( \mathcal{P}\left\{\frac{1}{E_1-\hbar\omega_l(\pol{p})}\right\} - 
\mathcal{P}\left\{\frac{1}{E_1+\hbar\omega_l(\pol{p})}\right\}\right)\Bigg] 
\label{Eq4sigretphononSL} 
\end{split}\end{equation} 
and 
\begin{equation}\begin{split} 
\Sigma^{<,{\rm phonon}}_{n}({\bf k},E) 
=&\sum_{\pol{p},l,{\bf k}'} |M_{n{\bf k},n{\bf k}'}(\pol{p},l)|^2 
\big[n_B(\hbar\omega_l(\pol{p}))G^{<}_{n,n} 
({\bf k}',E-\hbar\omega_l(\pol{p}))\\ 
&+\left[n_B(\hbar\omega_l(\pol{p}))+1\right]G^{<}_{n,n} 
({\bf k}',E+\hbar\omega_l(\pol{p})\big] 
\label{Eq4siglessphononSL} 
\end{split}\end{equation}
Then Eq.~(\ref{Eq4GretE}) becomes 
\begin{equation}\begin{split} 
\left(E-E_{k}+eFdm_1 -\Sigma^{\rm ret}_{m_1}({\bf k},E) 
\right) 
G^{\rm ret}_{m_1,m_2}({\bf k},E)& 
-T_1G^{\rm ret}_{m_1-1,m_2}({\bf k},E) 
-T_1G^{\rm ret}_{m_1+1,m_2}({\bf k},E)\\ 
&=\delta_{m_1,m_2} 
\label{Eq4GretSL} 
\end{split}\end{equation} 
and the Keldysh relation becomes 
\begin{equation} 
G^{<}_{m_1,m_2}({\bf k},E)=\sum_n 
G^{\rm ret}_{m_1,n}({\bf k},E)\Sigma^{<}_{n}({\bf k},E) 
G^{\rm adv}_{n,m_2}({\bf k},E) 
\label{Eq4KeldyshSL}\, . 
\end{equation} 
Together with the functionals for the self-energies, 
Eqs.~(\ref{Eq4GretSL},\ref{Eq4KeldyshSL}) form a closed 
set of equations which will be solved in subsection 
\ref{SecQtresults}, where explicit results are presented. 
 
In the same way the difference 
between Eq.~(\ref{Eq4GlessE1}) and Eq.~(\ref{Eq4GlessE2}) 
gives 
\begin{equation} 
\begin{split} 
(m-n)eFd G^<_{m,n}({\bf k},E)=&T_1\left[G^<_{m-1,n}({\bf k},E) 
+G^<_{m+1,n}({\bf k},E)-G^<_{m,n-1}({\bf k},E) 
-G^<_{m,n+1}({\bf k},E) 
\right]\\ 
&+\Sigma^{\rm ret}_{m}({\bf k},E) 
G^<_{m,n}({\bf k},E) 
+\Sigma^{<}_{m}({\bf k},E)G^{\rm adv}_{m,n}({\bf k},E)\\ 
&-G^{\rm ret}_{m,n}({\bf k},E) 
\Sigma^{<}_{n}({\bf k},E) 
-G^{<}_{m,n}({\bf k},E) 
\Sigma^{\rm adv}_{n}({\bf k},E) 
\label{Eq4GlessEdiff} 
\end{split}\end{equation} 
which will be the starting point for the derivation of 
the miniband conduction and Wannier-Stark hopping model 
in appendix \ref{AppDerivation}.

\subsubsection{Constant scattering matrix elements\label{Secconstmat}} 
 
For numerical calculations the {\bf k}-dependence of the self-energy 
is difficult to  handle, as a two dimensional array 
of $\Sigma({\bf k},E)$ has to be evaluated and stored in each 
calculation cycle. Thus the problem becomes much simpler if 
constant (i.e. momentum independent) scattering matrix elements 
are assumed. 
Then the self-energy for impurity scattering can be written as 
\begin{equation}\begin{split} 
\Sigma^{</{\rm ret}}_{n}({\bf k},E) 
=&\sum_{{\bf k}'} 
\langle V_{nn}(\{\pol{r}_j\}) 
V_{nn}(\{\pol{r}_j\})\rangle_{\rm imp} 
G^{</{\rm ret}}_{n,n}({\bf k}',E)\\ 
=&A\langle V_{nn}(\{\pol{r}_j\}) 
V_{nn}(\{\pol{r}_j\})\rangle_{\rm imp}\frac{\rho_0}{2}\int_0^{\infty} \d E_{k'} 
G^{</{\rm ret}}_{n,n}({\bf k}',E) 
\end{split} 
\end{equation} 
which does not depend on {\bf k}. (Note, that 
$G^{</{\rm ret}}_{n,n}$ only depends on $|{\bf k}|=\sqrt{2E_km}/\hbar$ 
due to the rotational symmetry in the $(x,y)$-plane.) 
As $G^{{\rm ret}}_{n,n}({\bf k}',E)\sim -1/E_{k'}$ for large 
$E_{k'}$, the respective integral exhibits a logarithmic divergence. 
This can be either cured by applying a finite cut-off $E_k^{\rm max}$ or 
by adding $1/(E_{k'}+\Gamma)$ in the integrand with an 
arbitrary but constant value $\Gamma>0$. Throughout this work the total 
scattering rate of free-particle states is used here. 
This procedure adds a constant real term to the retarded 
and advanced self-energy, which effectively renormalizes the energy scale 
to $E+\frac{1}{2}A\langle V_{nn}(\{\pol{r}_j\})V_{nn}(\{\pol{r}_j\}) 
\rangle_{\rm imp}\rho_0 \log(E_k^{\rm max}/\Gamma +1)$ 
but does not change the physics, which only depends on energy 
differences. 
 
For free-particle Green functions 
$G^{{\rm ret}}_{n,n}({\bf k}',E)=1/(E-E_{k'}+neFd+\imai 0^+)$ 
the integral yields 
\begin{equation} 
\Sigma^{{\rm ret}}_{n}({\bf k},E) 
=A\langle V_{nn}(\{\pol{r}_j\}) 
V_{nn}(\{\pol{r}_j\})\rangle_{\rm imp}\frac{\rho_0}{2} 
\left[ 
\log\left|\frac{\Gamma}{E+neFd}\right| 
-\imai\pi \Theta(E+neFd) 
\right] 
\end{equation} 
implying a scattering rate (for $E>-neFd$) 
\begin{equation} 
\frac{1}{\tau_{\rm imp}}=-\frac{2}{\hbar}\Im\{\Sigma^{\rm ret}\} 
=\frac{\pi\rho_0 A\langle V_{nn}(\{\pol{r}_j\}) 
V_{nn}(\{\pol{r}_j\})\rangle_{\rm imp}}{\hbar} 
\end{equation} 
which will be used as the parameter for the impurity 
scattering strength in the subsequent calculations.\footnote{Note
that the product $A\langle V_{nn}(\{\pol{r}_j\}) 
V_{nn}(\{\pol{r}_j\})\rangle_{\rm imp}$ typically  does not
depend on the sample area $A$.}
 
Phonon scattering is treated in the same way with a constant 
matrix element $M_n(\pol{p},l)=M_l$. 
Here the scattering rate is defined via 
\begin{equation} 
\frac{1}{\tau_{{\rm phonon},l}} 
=\frac{\pi\rho_0 A M_l^2}{\hbar } 
\end{equation} 
which is the free-particle spontaneous phonon emission rate if the 
final state is energetically available and Pauli blocking 
is negligible at low densities.

\subsection{Solution for constant self-energy\label{SecNGFconstsig}} 
The nature of the retarded Green function determined by 
Eq.~(\ref{Eq4GretSL}) can be analyzed by an analytical 
evaluation for the constant self-energy 
$\Sigma^{\rm ret}_{n}({\bf k},E)=-\imai\Gamma/2$.
Then one finds with Eq.~(9.1.27) of \cite{ABR66}: 
\begin{equation} 
G^{\rm ret}_{m,n}({\bf k},E)=\sum_{j} 
\frac{J_{m-j}\left(\frac{2T_1}{eFd}\right) 
J_{n-j}\left(\frac{2T_1}{eFd}\right)} 
{E+jeFd-E_k+\imai\frac{\Gamma}{2}} 
\label{Eq4GretGammaE} 
\end{equation} 
which is a superposition of broadened Wannier-Stark states, 
compare Eq.~(\ref{Eq2WSsimp}). 
The time-dependent Green function can be evaluated with help 
of Eq.~(9.1.79) of \cite{ABR66}: 
\begin{equation}\begin{split} 
G^{\rm ret}_{m,n}({\bf k};t_1,t_2)=&-\imai \Theta(t_1-t_2)\imai^{n-m} 
\e^{-\imai \left(E_k-\frac{m+n}{2}eFd\right)(t_1-t_2)/\hbar} 
\e^{-\frac{\Gamma}{2\hbar} (t_1-t_2)} \\ 
&\times J_{m-n}\left[\frac{4T_1}{eFd} 
\sin \left(\frac{eFd}{2\hbar}(t_1-t_2)\right) 
\right] 
\label{Eq4GretGammat} 
\end{split}\end{equation} 
This formulation is quite helpful in order to study the different 
limits. 
At first one notes that the off-diagonal elements of $G^{\rm ret}_{m,n}$ 
start to contribute significantly if 
\begin{equation} 
\left|\frac{4T_1}{eFd}\sin \left(\frac{eFd}{2\hbar}(t_1-t_2)\right) 
\right|\gtrsim 1 
\label{Eq4limit} 
\end{equation} 
Now only the times $0\le (t_1-t_2)\lesssim 2\hbar /\Gamma$ contribute 
to $G^{\rm ret}$ due to the exponential decay. 
If $eFd>\Gamma$ the sine  may oscillate over many periods 
with the average  absolute value of $2/\pi \approx 1/2$. 
Then Eq.~(\ref{Eq4limit}) gives $2|T_1|\gtrsim eFd$. 
If otherwise $eFd<\Gamma$ we may replace  $\sin(x)\approx x$ and 
Eq.~(\ref{Eq4limit}) gives $2|T_1|\gtrsim \Gamma$ at 
$t_1-t_2=\hbar/\Gamma$. 
Thus we conclude that the states are essentially 
delocalized if 
\begin{equation} 
2|T_1|\gg \Gamma \quad \mbox{and} \quad 2|T_1|\gg eFd 
\label{Eq4condMBT} 
\end{equation} 
On the other hand for 
\begin{equation} 
2|T_1|\ll \Gamma \quad \mbox{or} \quad  2|T_1|\ll eFd 
\label{Eq4condST} 
\end{equation} 
$G^{\rm ret}_{m,n}$ becomes small for $m\neq n$ 
and the states are essentially localized. 
Furthermore for 
\begin{equation} 
eFd\gg \Gamma 
\label{Eq4condWSH} 
\end{equation} 
the poles at $E=-\imai\Gamma/2+jeFd$ are clearly resolved 
in the energy dependence of Eq.~(\ref{Eq4GretGammaE}) 
which indicates the persistence of the Wannier-Stark ladder under scattering. 
The ranges (\ref{Eq4condMBT},\ref{Eq4condST},\ref{Eq4condWSH}) 
correspond to the regimes of validity for miniband transport, 
sequential tunneling and Wannier-Stark hopping, respectively, 
as given in Fig.~\ref{Fig3regimes}. In Appendix \ref{AppDerivation}
it will be shown explicitly that the respective transport equations 
can be recovered from the model of nonequilibrium Green functions 
in these ranges. 
 
\subsection{Results\label{SecQtresults}} 
Now the results from explicit calculations of 
the full quantum transport model 
are presented. 
Let us first summarize the underlying assumptions as 
applied successively in subsection \ref{SecApplSL}. 
\begin{itemize} 
\item Stationary transport in a homogeneous electric field 
is considered where only the 
lowest subband of the superlattice is taken into account. 
\item The scattering matrix elements are assumed to be diagonal 
in the well index (no interwell scattering) and no 
correlations between scattering matrix elements belonging 
to different wells exist. 
\item The scattering is treated within the self-consistent 
Born approximation (thus neglecting effects due to weak localization 
\cite{SZO92,SZO93}) using momen\-tum-independent matrix 
elements. 
(The latter assumption is equivalent to a localized scattering 
potential, which may be problematic for phonon scattering) 
\end{itemize} 
Within these assumptions it is possible to solve the system 
of equations 
Eqs.~(\ref{Eq4sigimpSL},\ref{Eq4sigretphononSL},\ref{Eq4siglessphononSL}, 
\ref{Eq4GretSL},\ref{Eq4KeldyshSL}) self-consistently. 
Note, that in all these calculations temperature only enters the occupation
of the phonon modes, which are assumed to be in thermal equilibrium.
The electronic distribution is calculated self-consistently
without any assumption of (heated) equilibrium.
 
Throughout this subsection we use the scattering time 
$\tau_{\rm opt}=0.125$ ps and the phonon energy 
$\hbar\omega_{\rm opt}=36$ meV for optical phonon scattering, 
which is the dominating energy relaxation process in III-V materials. 
In order to guaranty energy relaxation for particle 
energies below 36 meV, acoustic phonon scattering has to be 
taken into account. We mimic these phonons by a second 
phonon with constant energy 
$\hbar\omega_{\rm ac}=\hbar\omega_{\rm opt}(\sqrt{5}-1)/10\approx 4.4498$ 
meV. Here $\hbar \omega_{\rm ac}$ should be less than $k_BT$ 
($\approx6.4$ meV at 77 K), 
so that this mechanism can be efficient close to thermal 
equilibrium. Furthermore the ratio $\omega_{\rm ac}/\omega_{\rm opt}$ 
was chosen irrational in order to avoid spurious resonances. 
The respective scattering time is chosen $\tau_{\rm ac}=5$ ps. 
 
\begin{figure} 
\noindent\epsfig{file=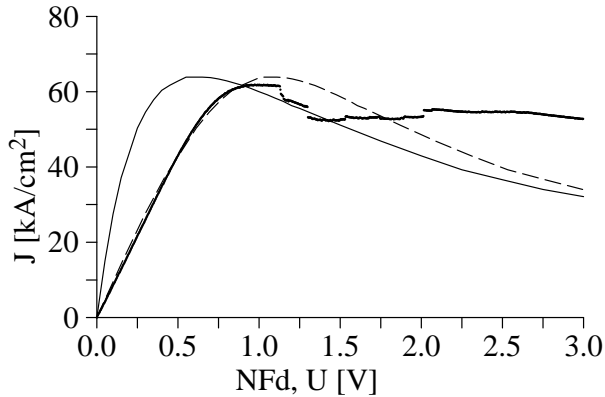,width=8cm} 
\caption[a]{Current-voltage characteristic for the superlattice 
studied in \cite{SCH98f}. Full line: Result from quantum 
transport 
for $N=100$ wells. Dots: 
Experimental data (courtesy of E. Schomburg). Dashed line: 
Theoretical result with a serial resistance of $7.2\times 
10^{-6}\Omega {\rm cm}^2/A$ 
($T=300$ K). 
\label{Fig4kenn2836}} 
\end{figure}

At first consider the strongly coupled superlattice 
studied experimentally in \cite{SCH98f}. 
It consists of 100 periods with 3.45 nm GaAs wells and 
0.96 nm AlAs barriers, yielding a coupling $T_1=-20.5$ meV. 
The doping density provides $N_D= 3.6\times 10^{10}/{\rm cm}^2$ 
and an impurity  scattering time $\tau_{\rm imp}=0.12$ ps 
is applied, which can be estimated from the impurity scattering rate 
for this doping range \cite{WAC97d}. 
Results for the lattice temperature $T=300$ K (which only enters 
the phonon occupation number $n_B$) are shown in Fig.~\ref{Fig4kenn2836} 
(full line). 
One encounters the typical shape for superlattice transport. 
While the current peak is in good quantitative 
agreement with the experimental observation the peak position 
is shifted significantly. The agreement becomes 
excellent if a serial resistance of the order of $10\Omega$ (for the 
experimental sample area) is 
included which may result from contacts, leads, or 
the substrate. 
 
\begin{figure} 
\noindent\epsfig{file=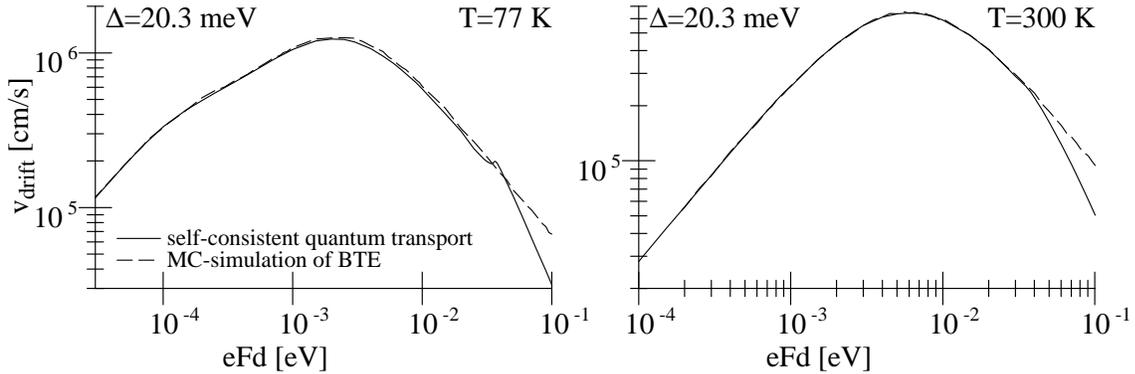,width=15cm} 
\caption[a]{Drift velocity versus field for a 
superlattice  with 
$\Delta=20.3$ meV, $d=5.1$ nm, $N_D=5.1\times 10^{9}/{\rm cm}^2$, 
and $\tau_{\rm imp}=0.333$ ps. 
Full line: Quantum transport. Dashed line: Miniband transport. 
(From \cite{WAC99b}). 
\label{Fig4kennrott}} 
\end{figure}

\begin{figure} 
\noindent\epsfig{file=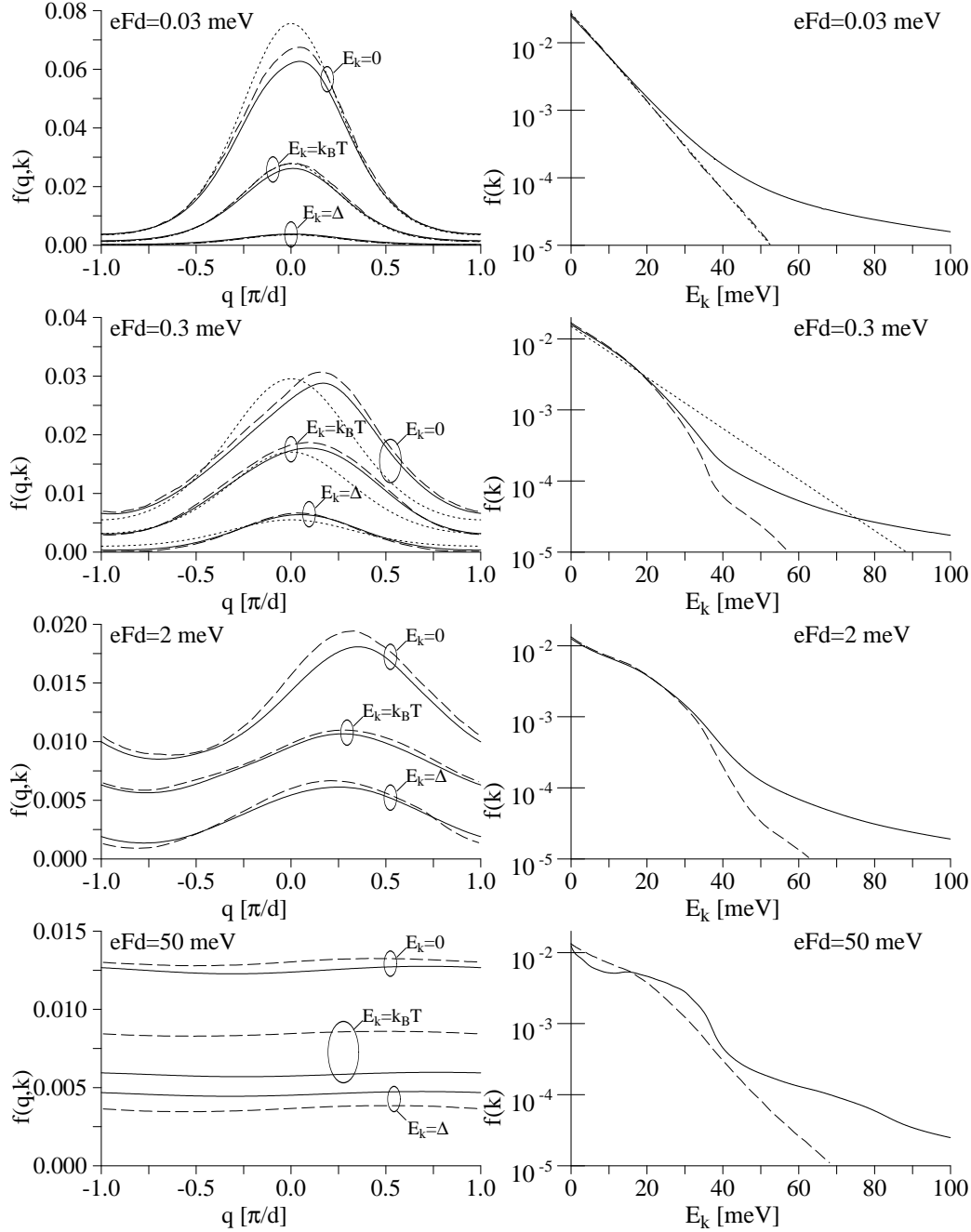,width=13.8cm} 
\caption[a]{Distribution functions $f(q,{\bf k})$ and 
$f(E_k)=d/(2\pi)\int \d q\,f(q,{\bf k})$ 
of the carriers for different field strengths for the superlattice 
studied in Fig.~\ref{Fig4kennrott} at $T=77$ K. 
Full line: Diagonal elements of the density matrix $\rho(q,{\bf k})$ 
as calculated from the quantum transport 
model. Dashed line: semiclassical distribution function 
as calculated from the Boltzmann equation for miniband transport. 
Dotted line: Thermal distribution $n_F(E_q+E_k-\mu)$ applying 
an electron temperature 
$T_e=77$ K for $eFd=0.03$ meV  and 
$T_e=140$ K for $eFd=0.3$ meV. 
\label{Fig4vertrott77}} 
\end{figure}

Several simulations have been performed for a model superlattice 
with $T_1=-5.075$ meV, $d=5.1$ nm, and  $N_D=5.1\times 10^{9}/{\rm cm}^2$, 
which has been extensively studied by S.~Rott \cite{ROT99a} 
(see also Figs.~\ref{Fig3MCrott} and \ref{Fig3WSHrott} in section 3). 
The rather low doping is taken into account by 
an impurity scattering time $\tau_{\rm imp}=0.333$ ps in agreement 
with semiclassical Monte-Carlo simulations. 
The results for the drift velocity $v_{\rm drift}=J/(eN_D/d)$ for 
$T=77$ K and $T=300$ K are shown in Fig.~\ref{Fig4kennrott} (full line). 
For comparison,  Monte-Carlo simulations within the miniband transport 
model (Sec.~\ref{SecMBT}) 
have been performed by S.~Rott and A.~Markus (Institut f{\"u}r Technische 
Physik der Universit{\"a}t Erlangen) applying the same scattering matrix 
elements. 
For low and moderate field strengths up to $eFd\approx 20$ meV 
the relation obtained from miniband transport agrees extremely 
well with the full quantum transport result. At larger 
fields, $eFd$ exceeds the miniband width 
and miniband transport is no longer valid as discussed 
in subsection \ref{SecNGFconstsig}. 
In Fig.~\ref{Fig4vertrott77} the electron distribution functions 
(for $T=77$ K) are 
depicted for various electric field strengths. 
Comparison of the dashed and full lines for miniband conduction 
and quantum transport shows that the distribution functions 
agree very well for $eFd\lesssim 20$ meV in the energy range 
$E_k<30$ meV. For higher values of $E_k$ the quantum mechanical 
distribution function is larger than its semiclassical counterpart. 
Here the occupation is quite small, so that the tail from the 
broadened spectral function $\sim \Gamma/E_k^2$ becomes 
visible \cite{REG87,REG97}, in good agreement with the 
estimate of Eq.~(\ref{Eq4occupation}). In compensation 
the quantum result is slightly smaller than its semiclassical 
counterpart for low values of $E_k$, as the total density 
has to be the same. 
Nevertheless, these effects are small compared to the typical 
occupation numbers and one can conclude that the Boltzmann equation 
for miniband transport gives reliable results both 
for the current density and the electron distribution. 
Actually, the quality of agreement is stunning regarding 
the crude assumptions necessary for the derivation of 
the Boltzmann equation in Appendix~\ref{AppBTEderivation}. 
 
The velocity-field relations from Fig.~\ref{Fig4kennrott} 
can be directly compared with those in Fig.~\ref{Fig3MCrott}, 
where the correct matrix elements have been applied within the semiclassical 
miniband transport model. There are no qualitative 
differences, so that one may conclude that the assumption 
of constant matrix 
elements as well as the artificial acoustic phonons do not cause 
unphysical results. 
 
Let us now study the details in the velocity-field relation 
and through a glance at the respective distribution functions. 
At very low fields one encounters a linear increase of 
the drift velocity with the electric field. 
Here the distribution function 
resembles a thermal distribution with the lattice temperature 
(dotted lines for $eFd=0.03$ meV). 
The slight shift can be easily treated within linear response, 
yielding a field-independent mobility. 
 
For higher fields ($eFd=0.3$ meV) one encounters a sublinear 
increase of the drift velocity. Here significant heating 
occurs, which can be seen from the respective distribution 
function in Fig.~\ref{Fig4vertrott77}. The distribution 
functions resembles a shifted Fermi distribution with an electron 
temperature $T_e\approx 140$ K  for $E_k\lesssim \hbar \omega_{\rm opt}$. 
For higher energies optical phonon scattering is still efficient 
and one encounters a steeper decrease of $f(E_k)$ with energy. 
This shows that the concept of an electron temperature 
makes sense in this range both for quantum and semiclassical 
transport even without electron-electron scattering. 
For electric fields close to the current maximum 
($eFd=2$ meV) the distribution functions 
strongly deviates form any thermal or heated distribution function. 
In the semiclassical picture this can be viewed as a result 
from frequent Bragg-scattering where particle reach the states 
with $q=\pi/d$. If the electric field becomes even stronger, 
the electrons may traverse the $q$-Brillouin zone several times without 
scattering for $eFd\gg \hbar/\tau$ in  the semiclassical miniband picture. 
This leads to a flat electron distribution in $q$ 
as can be observed for $eFd=50$ meV in Fig.~\ref{Fig4vertrott77}. 
The same holds for the quantum distribution function although 
the reason is different. 
In this field range ($2|T_1|\ll eFd$) the Wannier-Stark states are essentially 
localized to a single well. Thus a semiclassical occupation of 
the Wannier-State creates a flat distribution in $q$-space as 
well. 
 
While both approaches explain the flat distribution in $q$-space, 
significant differences in scattering arise. 
The electron running through the  $q$ states exhibits different 
energies $E(q)$ during passage. This provides different 
selection rules for scattering then the presence of a 
Wannier-Stark states with fixed energy. 
Therefore the $f(E_k)$-distribution calculated by the quantum 
transport model exhibits pronounced features on the energy 
scales $eFd$ and $\hbar\omega_{\rm opt}$ as well as the difference 
$eFd-\hbar\omega_{\rm opt}$ because at these energies new 
scattering channels appear. 
If these scales match, the phonon resonance at $eFd=\hbar\omega_{\rm opt}$ 
appears in the velocity-field characteristics, see Fig.~\ref{Fig4kennrott}. 
In contrast, the semiclassical result exhibits a rather flat $f(E_k)$ 
distribution and no phonon resonance can be observed. 
 
\begin{figure} 
\noindent\epsfig{file=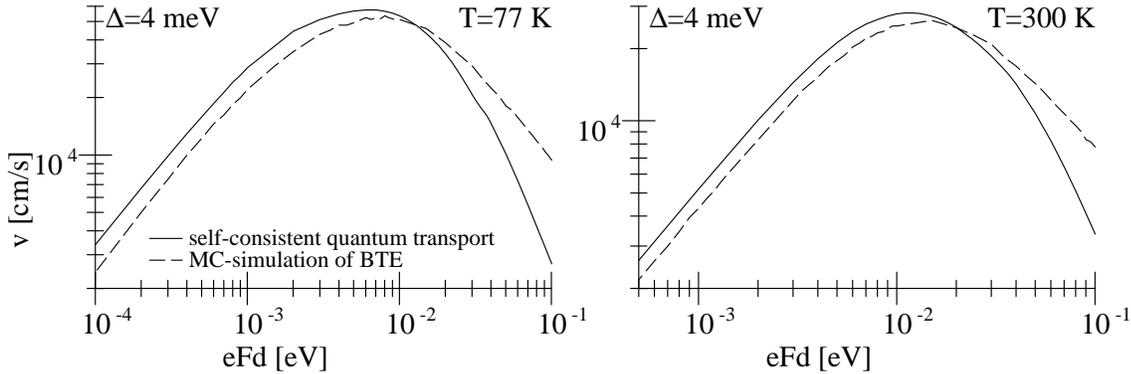,width=15cm} 
\caption[a]{Drift velocity versus field for a 
superlattice  with 
$T_1=1$ meV, $d=5.1$ nm,  $N_D=5.1\times 10^{9}/{\rm cm}^2$, 
and $\tau_{\rm imp}=0.0666$ ps. 
Full line: Quantum transport. Dashed line: Miniband transport. 
\label{Fig4kennwac}} 
\end{figure}

The situation changes for small coupling and strong scattering, 
when $\Gamma>2|T_1|$ holds. 
Fig.~\ref{Fig4kennwac} shows the velocity field 
relations for $T_1=-1$ meV and $\tau_{\rm imp}=0.0666$ ps. 
Here the semiclassical miniband transport calculation 
strongly deviates from the full quantum result both in the 
low-field and in the high-field region. Nevertheless, the 
agreement gets better for higher temperatures. 
 
\begin{figure} 
\noindent\epsfig{file=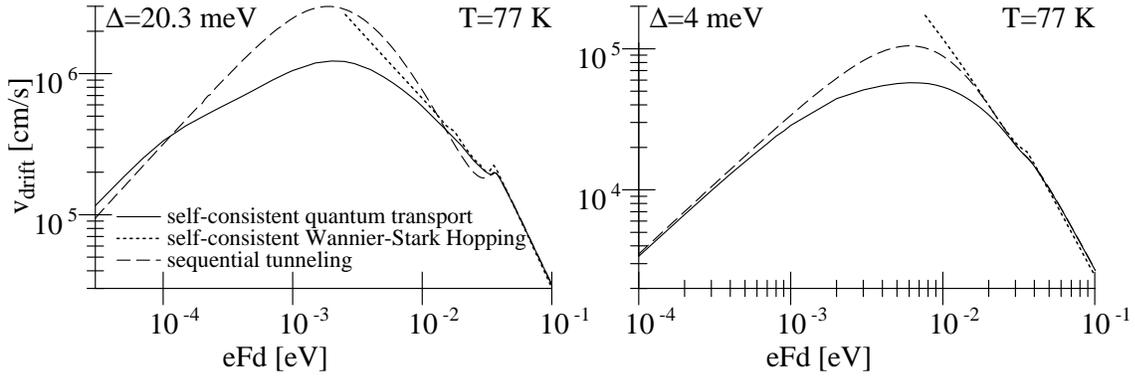,width=15cm} 
\caption[a]{Drift velocity versus field for a 
superlattice  with 
$d=5.1$ nm,  $N_D=5.1\times 10^{9}/{\rm cm}^2$, 
$T_1=5.075$ meV, $\tau_{\rm imp}=0.33$ ps (left) as well as 
$T_1=1$ meV, $\tau_{\rm imp}=0.0666$ ps (right). 
Full line: Quantum transport. Dashed line: sequential tunneling. 
Dotted line: self-consistent Wannier-Stark hopping. 
\label{Fig4kennrottwac77}} 
\end{figure}

In Fig.~\ref{Fig4kennrottwac77} the quantum transport calculations 
are compared with results from the models of 
Wannier-Stark hopping and sequential tunneling. 
In all calculations identical scattering matrix elements are used. 
The calculations for Wannier-Stark hopping have been performed by 
S.~Rott and A.~Markus (Institut f{\"u}r Technische 
Physik der Universit{\"a}t Erlangen) using 
self-consistent distribution functions, see section~\ref{SecWSH} 
for details. 
One finds that the Wannier-Stark hopping model provides 
good results (including the phonon resonance) in the high-field 
region where $eFd\gg \Gamma$ ($\Gamma\approx 7$ meV and 15 meV 
for the left and right superlattice, respectively, for energies 
at which phonon emission is possible). 
The agreement deteriorates significantly if the 
simple version of Wannier-Stark hopping without 
self-consistency is applied. 
 
The sequential 
tunneling model clearly fails at low and moderate field strengths 
for the strongly coupled superlattice (left part of 
Fig.~\ref{Fig4kennrottwac77}) as $2|T_1|>\Gamma$. 
In the high field region $eFd\gg 2|T_1|$ the results becomes valid 
and the phonon resonance can be observed within the 
sequential tunneling model. 
For weakly coupled superlattice (right part of 
Fig.~\ref{Fig4kennrottwac77}), the sequential tunneling model 
should be applicable 
and the low-field conductance is in good agreement with the full quantum 
transport result. Significant deviations occur at 
intermediate fields because electron heating is not included in the 
sequential tunneling model. As the current density drops with electron 
temperature (see section \ref{SecSTsimple}), the drift velocities are 
too high, if heating is neglected. These heating effects have been 
recently taken into account within the sequential tunneling 
model \cite{STE99,STE99a}. 
 
\begin{figure} 
\noindent\epsfig{file=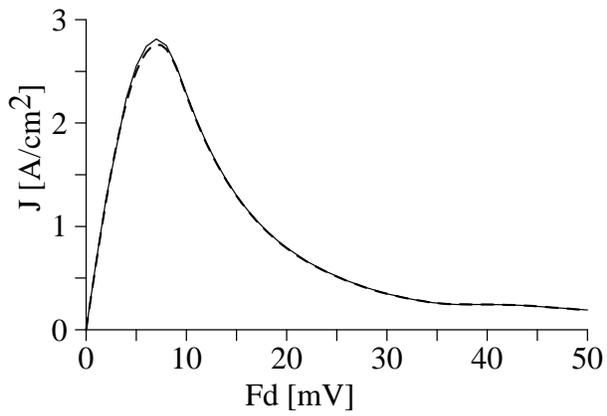,width=8cm} 
\caption[a]{Current-voltage characteristic for 
the superlattice studied in \cite{GRA91,KWO95} 
with parameters $T_1=-0.02$ meV, $N_{D}=1.5\times 10^{11}/{\rm cm}^2$, 
$d=13$ nm, $\tau_{\rm imp}=0.0666$ ps, and $T=4$ K. 
Full line: Quantum transport. Dashed line: sequential tunneling. 
\label{Fig4kennNGFgrahn}} 
\end{figure}

In Fig.~\ref{Fig4kennNGFgrahn}, a comparison between quantum 
transport and sequential tunneling is shown for the 
weakly coupled superlattice studied in \cite{GRA91,KWO95}, see 
also section \ref{SecSTresults}. Excellent 
agreement is found between both approaches and the results 
agree well with those presented in Fig.~\ref{Fig3STkenngrahn} applying 
realistic impurity scattering matrix elements and neglecting phonon scattering. 
 
The  observations presented here correspond to the 
boundaries of validity given in Fig.~\ref{Fig3regimes}, 
which can be considered as a reliable guide. 
Furthermore the results indicate 
that the models of miniband transport and sequential tunneling give 
qualitatively reasonable results even outside their range of 
applicability. Thus they can be used as a rough guide to obtain 
a correct  order of magnitude of the current.

\section{Formation of field domains\label{ChapDomains}}
\begin{figure}
\noindent\epsfig{file=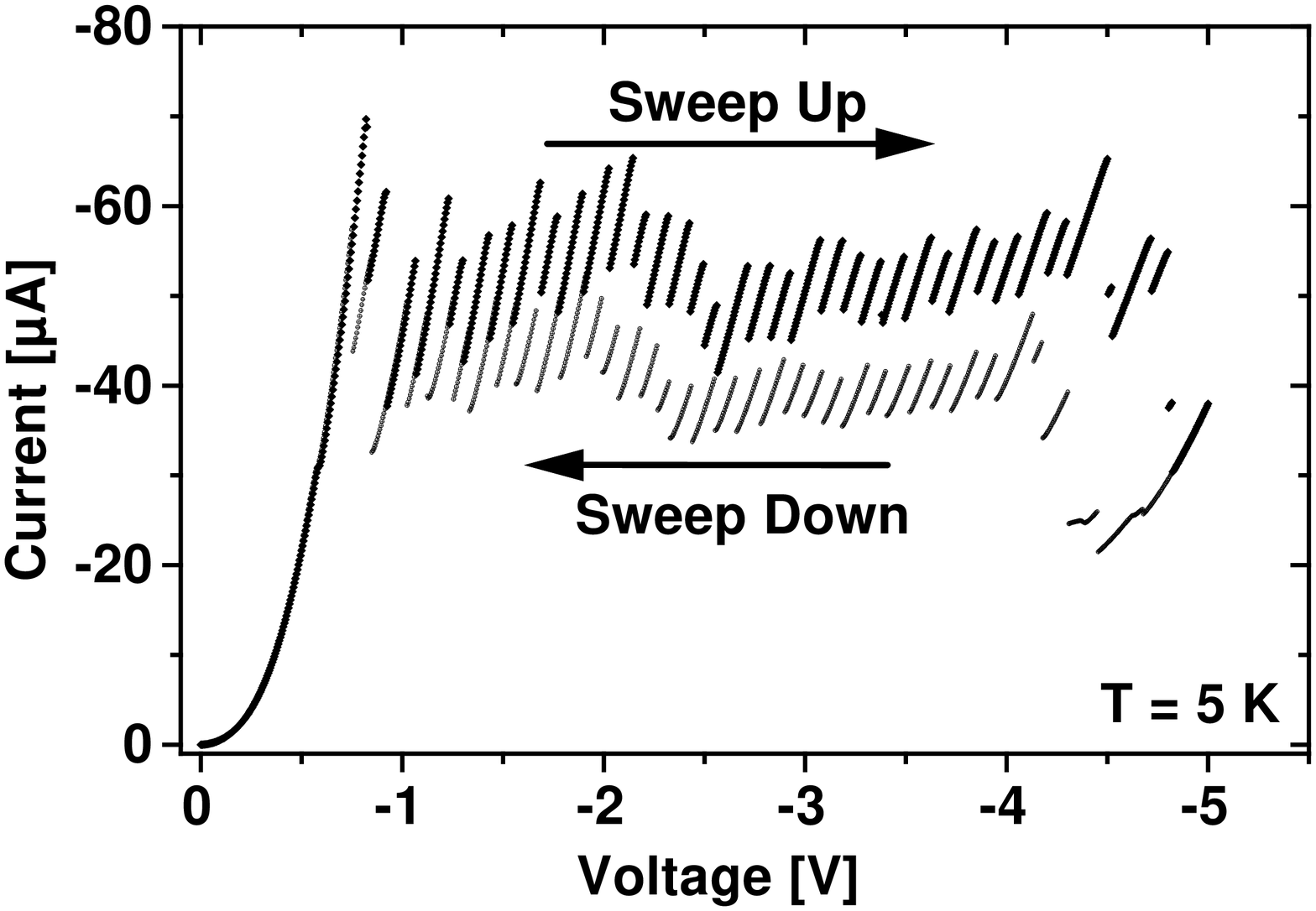,width=7.0cm}
\hspace{0.5cm}
\epsfig{file=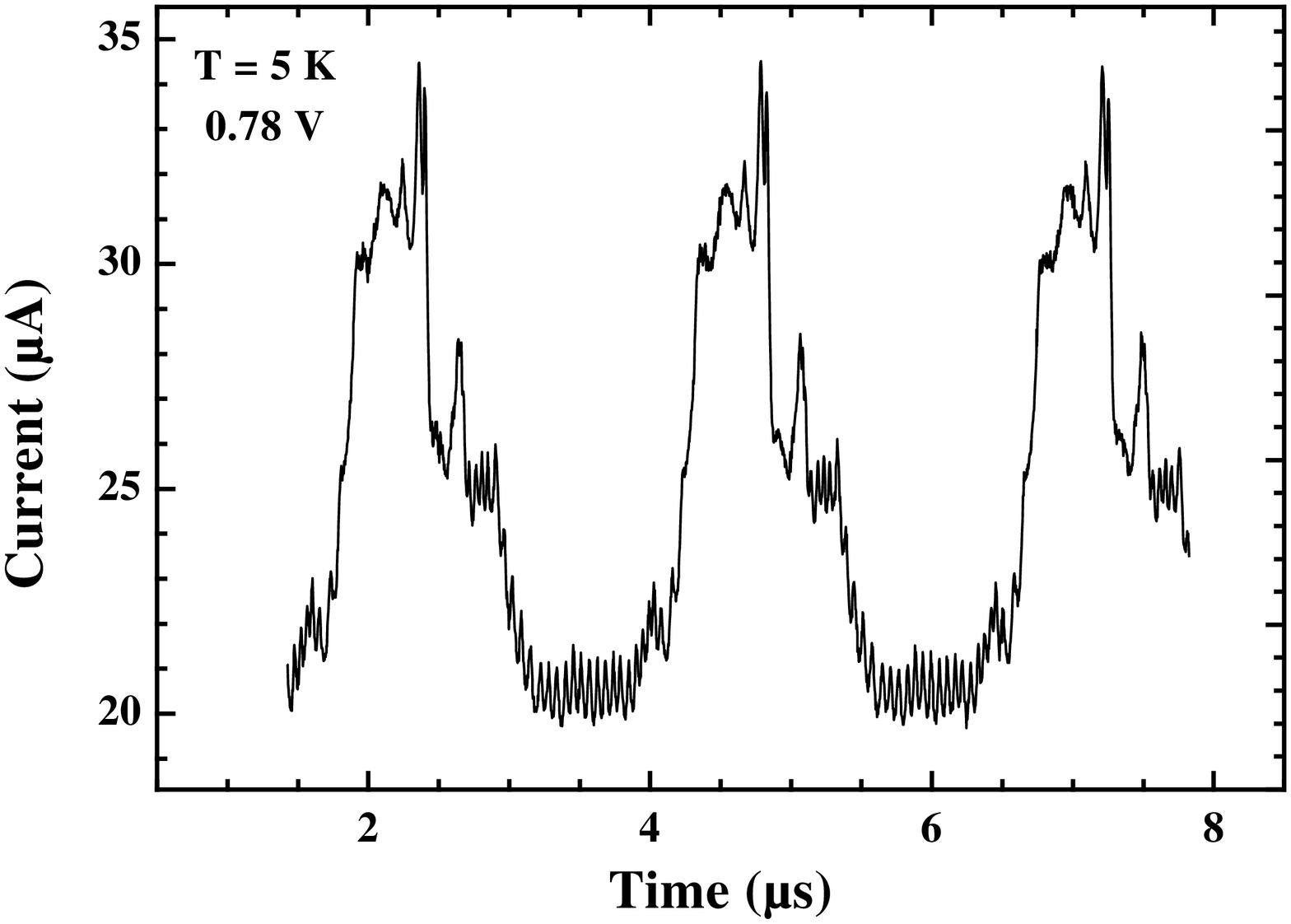,width=7.6cm}\\[-0.6cm]
(a)\hspace{7.4cm}(b)
\caption[a]{Experimental results for
a semiconductor superlattice with 40 periods consisting of
4 nm AlAs barriers and 9 nm GaAs wells. The doping density is
$N_D=1.5\times 10^{11}/{\rm cm}^2$ and the sample area is
$A=1.13\times 10^{-8} {\rm m}^2$.
(a) Current-voltage characteristic for negative bias applied to the
top contact, where a stationary current is observed
(from  \cite{KAS94}).
(b) Current-time signal for a constant positive bias exhibiting
self-sustained current oscillations
(from  \cite{KAS97}).}
\label{Fig5Expkastrup}
\end{figure}

As shown in sections \ref{ChapStandard} and \ref{ChapNGFT},
semiconductor superlattices typically exhibit
ranges of negative differential conductivity.
This occurs after the first
current peak for miniband transport (or sequential tunneling)
as well as after the subsequent resonances at higher fields
when different levels align.
The shape of the current-field relation is thus typically
of N-type and domain formation effects are likely to occur
(see \cite{SHA92} for a general overview).
The prototype of an extended device with N-shaped
current-field relation is the Gunn diode which exhibits
self-sustained current oscillation due to traveling field
domains \cite{GUN64,KRO64,MCC66,KNI67,BOE69,HIG92,BON97b}.
A similar behavior has been suggested for semiconductor
superlattices \cite{BUE77,BUE79}, and oscillatory behavior has indeed
been found experimentally in the last years \cite{KAS95,HOF96} with
frequencies over 100 GHz \cite{SCH99h,SCH99d}.
In contrast to the Gunn diode, semiconductor superlattices
frequently exhibit the formation of stable stationary
domains which lead to a characteristic saw-tooth 
pattern in the
current-voltage characteristic (see Fig.~\ref{Fig5Expkastrup}a)
as observed by many different groups
\cite{ESA74,KAW86,GRA91,KWO95,CHO87,HEL89,HEL90,HAN95a,MIT97}. (See also
\cite{SHI97,SCH98j} for domain formation in a parallel magnetic field.)
The measurements are typically performed
by applying a continuous sweep of the bias. As the branches 
overlap, one observes different parts of the branches for sweep up
and sweep down of the bias\footnote{Ofcourse the variation of the bias 
must be slow, so that the field domains
can follow adiabatically. Otherwise, the sawtooth structure 
disappears \cite{SHI97a,AMA01a}.}
Sometimes a quite complicated behavior is observed as well.
In particular for the
superlattice structure of Fig.~\ref{Fig5Expkastrup}
stationary field domains \cite{KAS94},
self-sustained oscillations \cite{KAS95}, as well as bistability
between stationary and oscillatory behavior \cite{ZHA97} has been
observed under fixed bias conditions within the first plateau of
the current-voltage characteristic.

In this section such complex behavior will be analyzed
by a comparison between numerical results based on the sequential 
tunneling model from section \ref{SecST} with analytical studies.
It will be shown
that most of the observed effects   can be understood as the
result of a competition between two mechanisms: (i) The
motion of fronts connecting  low- and high-field domains and (ii)
the dynamical evolution of the field close to the injecting
contact, i.e., the cathode.

This section is organized as follows: First a model is described
which is able to reproduce most of the experimental findings
for weakly coupled superlattices. In subsection~\ref{SecNumerics} 
numerical results
are presented reproducing the behavior shown in Fig.~\ref{Fig5Expkastrup}.
In the subsequent subsections the key elements, namely the
traveling fronts (subsection~\ref{SecFronts}) and the cathode  behavior 
(subsection~\ref{SecContact}) are
discussed separately. In subsection~\ref{SecGlobal} 
it will be demonstrated how their
combination explains the behavior observed both in experiment and simulation.
Finally, the findings are summarized and an instruction for the
analysis of nonlinear superlattice transport behavior is presented.

\subsection{The model\label{SecModelDomain}}
\begin{figure}
\noindent\epsfig{file=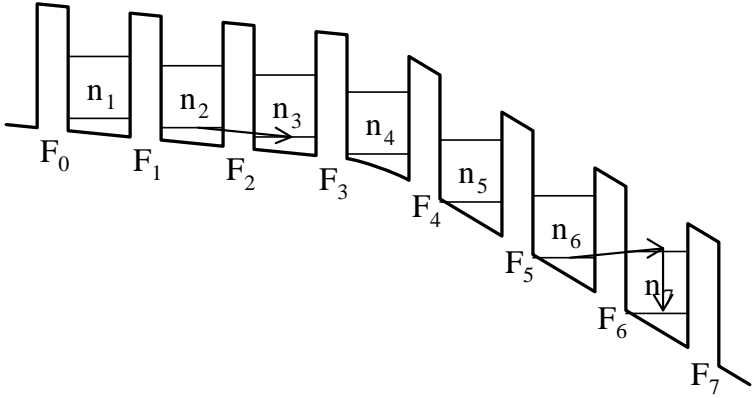,width=10cm}
\caption[a]{Sketch of a superlattice structure with an inhomogeneous
field distribution.}
\label{Fig5modelskizze}
\end{figure}

In weakly-coupled multiple quantum wells the electronic states
are essentially localized in single wells forming energy
levels $E^{\nu}$. Transport then occurs by sequential tunneling
between neighboring wells. The current from well $m$ to well $m+1$
is modeled by a function $J_{m\to (m+1)}=J(F_m,n_m,n_{m+1})$,
where $F_m$ is the average field drop between the respective wells and
$n_m$ denotes the electron density (per unit area) in well $m$
as depicted in Fig.~\ref{Fig5modelskizze}.
Considering a superlattice with $N$ wells embedded between $N+1$
barriers, the dynamics is determined by the continuity equation
\begin{equation}
e\frac{\d  n_m}{\d  t}=J_{(m-1)\to m}-J_{m\to (m+1)}\quad \mbox{for }
m=1,\ldots N\, .
\label{Eq5continuity}
\end{equation}
The electric field satisfies Poisson's equation
\begin{equation}
\epsilon_r\epsilon_0 (F_m-F_{m-1})=e(n_m-N_D)
\quad \mbox{for } m=1,\ldots N
\label{Eq5poisson}
\end{equation}
where $N_D$ is the doping density per period (per unit area),
and $\epsilon_r$ and $\epsilon_0$ are the relative and absolute
permittivities. Finally, the total voltage $U$ is determined by
\begin{equation}
U(t)=\sum_{m=0}^N  F_m d
\label{Eq5voltage}
\end{equation}
where we have neglected a possible voltage drop at the contacts for
simplicity.
Equations (\ref{Eq5continuity}-\ref{Eq5voltage}) can be transformed
into an equivalent set of equations:
\begin{alignat}{2}
\epsilon_r\epsilon_0\frac{\d  F_m}{\d  t}&=
\left(J(t)-J_{m\to (m+1)}\right) &
\quad &\mbox{for } m=0,\ldots N
\label{Eq5Ampere}\\
n_m&=N_D+\frac{\epsilon_r\epsilon_0}{e} (F_m-F_{m-1})&\quad &\mbox{for }
m=1,\ldots N
\label{Eq5Poisson2}\\
(N+1)J(t)&=\sum_{j=0}^N J_{m\to (m+1)}
+\frac{\epsilon_r\epsilon_0}{d}\frac{\d  U(t)}{\d  t} &&\label{Eq5global}
\end{alignat}
This shows that the dynamical evolution  of the local fields
is driven by the total current density $J(t)$, which itself is determined
by the global behavior of the sample. Thus, $J(t)$ represents a
global coupling.
In order to obtain a closed set of equations, the currents across
the first (cathode) and the last (anode) barrier, $J_{0\to 1}$ and
$J_{N\to (N+1)}$, respectively, have to be specified.
Here the following approaches have been taken previously:

In \cite{PRE94,BON94,SCH96b,WAC97a} these contact currents
were calculated within the assumption of two fictitious additional
wells, one  before the first and one after the last barrier. Then the current
$J_{0\to 1}$ is given by $J(F_0,n_0,n_{1})$ for tunneling between
two wells, where the fictitious density $n_0$ has to be specified,
usually assuming $n_0=(1+c)N_D$, with $c>-1$.
The current across the last barrier is treated analogously by introducing
a fictitious density $n_{N+1}$.
This approach will be referred to as
{\em constant density boundary condition} in the following.

Alternatively, one may assume that the current is proportional
to the local field, i.e., $AJ_{0\to 1}=\sigma F_0d$ with an Ohmic
conductance $\sigma$.
For the anode condition $J_{N\to (N+1)}$ one has to take into
account,  that the current must vanish if $n_N$ tends to zero,
as otherwise the density $n_N$ can become negative. This
can be ensured by an additional factor $n_N/N_D$
and $AJ_{N\to (N+1)}=\sigma F_Nd\, n_N/N_D$ is a reasonable choice.
This will be referred to
as {\em Ohmic boundary condition}.

The actual potential distribution at the boundary could be
taken into account within a transmission-type formalism \cite{AGU97}.
Here I will restrict myself to the two approaches sketched above, which
are simpler and provide an understanding of  most of the
underlying physics. They contain the main
ingredients to understand more complicated and physically better
motivated versions.

\subsection{Numerical results\label{SecNumerics}}
\begin{figure}
\noindent\epsfig{file=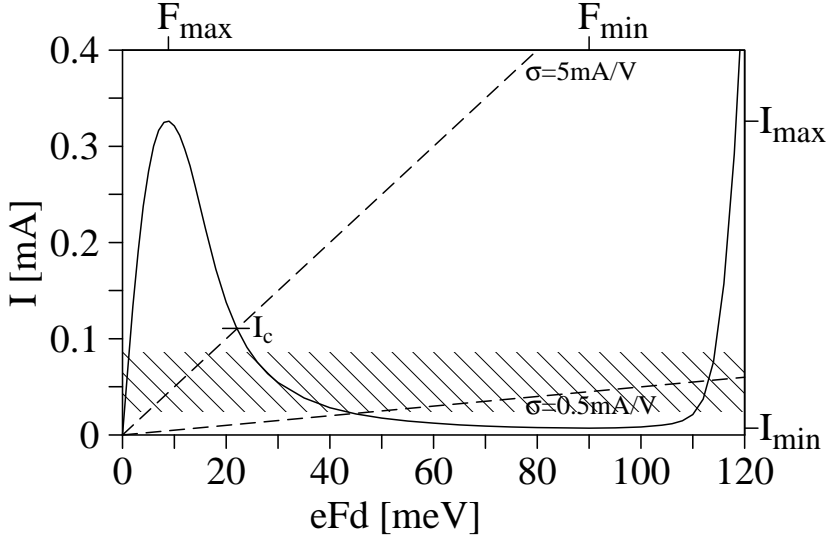,width=11cm}
\caption[a]{Current-field relation $I(eFd,n_m=N_D,n_{m+1}=N_D)$
calculated from the microscopic model for the superlattice
of Fig.~\ref{Fig5Expkastrup}.
The relation for the cathode current
$I_{0\to 1}={\rm sgn}(e)\sigma Fd$ is also shown
for two different values of $\sigma$ (broken lines).
The hatched area indicates the current range
$I_{\rm min}^{\rm dom}<I
<I_{\rm max}^{\rm dom}$ where stable
domains are found.}
\label{Fig5homGrahn}
\end{figure}

Let us consider the GaAs/AlAs superlattice structure
from Fig.~\ref{Fig5Expkastrup} (which exhibits
a rather small miniband width of $0.08$ meV)
as a model system for the subsequent calculations.
In the following the function $J(F_m,n_m,n_{m+1})$ is calculated from the
nominal sample parameters by the model described in section \ref{SecST}.
(See also  Fig.~\ref{Fig3STkenngrahn}.)
The result for a homogeneous electron density 
$n_m=n_{m+1}=N_D$ is shown in Fig.~\ref{Fig5homGrahn}.
Here, the current $I={\rm sgn}(e)AJ$ is shown
applying the experimental sample area $A=1.13\times 10^{-8}
{\rm m}^2$ in order to facilitate comparison with experimental
data. In this section the factor ${\rm sgn}(e)$ is included,
so that the sign of the current equals the particle current in
transport direction which simplifies the following discussion.
Fig.~\ref{Fig5homGrahn} shows that the current-field characteristics
exhibits a linear conductivity for low fields, a
maximum at $eFd=eF_{\rm max}d$,
a range with negative differential conductivity for $eFd>eF_{\rm max}d$
and a second sharp rise of the current for higher fields $eFd>eF_{\rm
min}d$ due to  resonant tunneling from the lowest level to the second
level of adjacent wells.

Different approaches to obtain the local current density 
$J(F_m,n_m,n_{m+1})$ can be applied as well and the overall
results do not depend on this choice, provided the general
shape resembles Fig.~\ref{Fig5homGrahn}. 
E.g., in Refs.~\cite{KAS95,PRE94,AGU97,BON94,MIT97,SAN99}
one can find similar results like those presented in this section,
where varios types of local current density functions are applied.

\begin{figure}
\noindent\epsfig{file=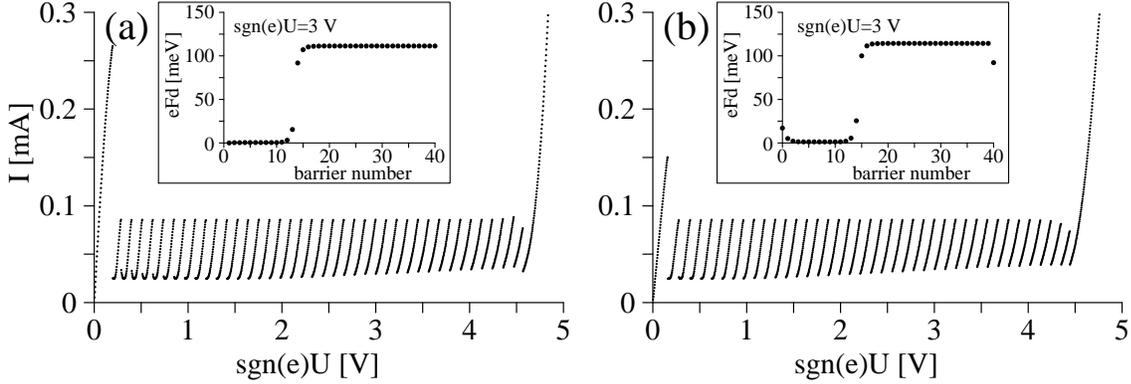,width=15cm}
\caption[a]{Current-voltage characteristics exhibiting
stationary field domains as shown in the insets
for different boundary conditions:
(a) $I_{0\to 1}=I(F_0,1.5N_D,n_{1})$ and $I_{N\to
N+1}=I(F_0,n_{N},1.5N_D)$,
(b) $I_{0\to 1}=\sigma F_0d$ and $I_{N\to N+1}=\sigma F_Nd (n_{N}/N_D$)
with $\sigma=5 mA/V$. (Parameters as in Fig.~\ref{Fig5Expkastrup}).}
\label{Fig5kennGrahn}
\end{figure}

By simulating  equations (\ref{Eq5continuity}-\ref{Eq5voltage}) for a
fixed bias $U$ until a stationary state is reached
the stationary current-voltage characteristic shown in
Fig.~\ref{Fig5kennGrahn} are obtained. Here the initial condition for the
calculation is taken from the result of the previous
voltage point, simulating a  sweep-up of the bias $U$. In
Fig.~\ref{Fig5kennGrahn}a we use the constant density boundary
condition with $c=0.5$, while the Ohmic boundary condition with
$\sigma=5$ mA/V was applied in Fig.~\ref{Fig5kennGrahn}b.
Both characteristics are almost identical and in reasonable quantitative
agreement with the experimental data presented in Fig.~\ref{Fig5Expkastrup}a.
As shown in the insets, the branches are due to the formation of
electric field domains inside the sample, where a low-field domain
is located at the cathode and a high-field domain is located at the anode.
Close to the contacts we find a small transition layer, which
depends on the contact boundary conditions.
The domain branches span a fixed current range
$I_{\rm min}^{\rm dom}<I<I_{\rm max}^{\rm dom}$ with
$I_{\rm min}^{\rm dom}=24.7\mu$A and
$I_{\rm max}^{\rm dom}=85.5\mu$A,  indicated by the hatched
area in Fig.~\ref{Fig5homGrahn}.

\begin{figure}
\noindent\epsfig{file=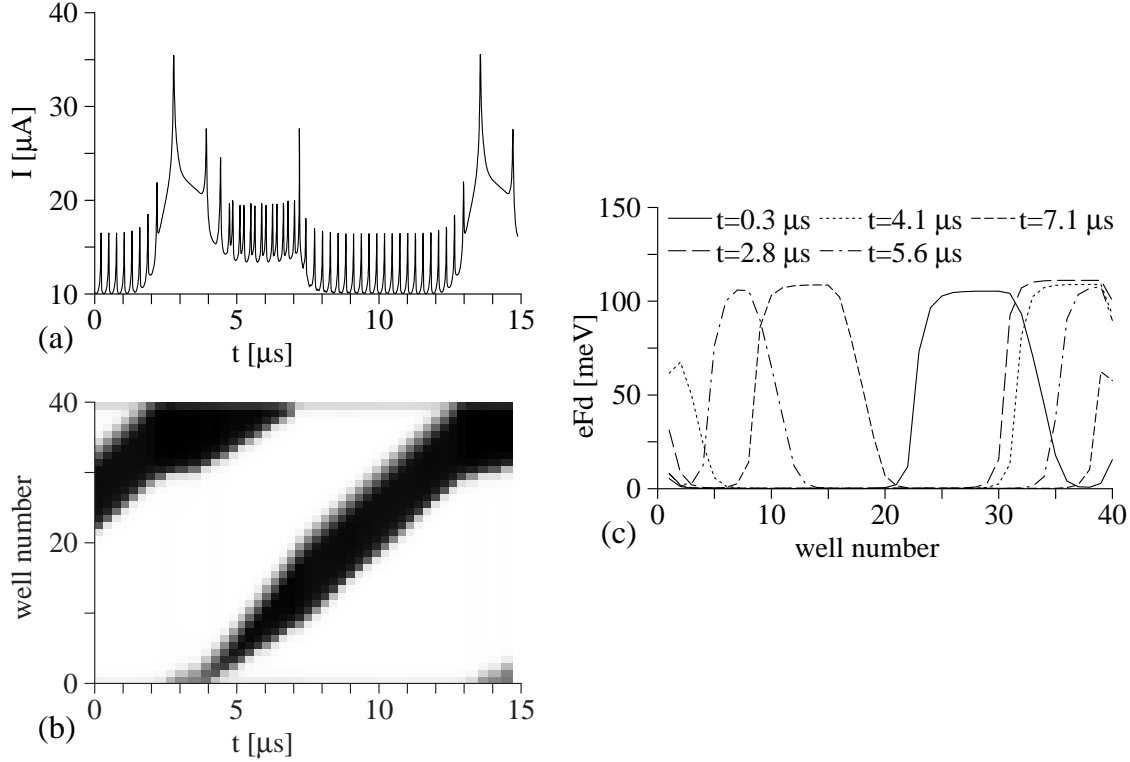,width=15cm}
\caption[a]{Self-sustained current oscillations
for fixed bias conditions sgn$(e)U=1.2$ V and an Ohmic boundary condition
$\sigma=0.5 mA/V$. (a) Current $I$ versus time $t$. (b) Density plot of the
electric field distribution. The high field region is black. (c) Electric
field profile at different times. (Parameters as Fig.~\ref{Fig5Expkastrup}).}
\label{Fig5oscGrahn}
\end{figure}

For different boundary conditions oscillatory behavior can be
obtained for the same sample parameters.
This is shown in Fig.~\ref{Fig5oscGrahn} for the
Ohmic boundary condition with $\sigma=0.5$ mA/V.
The current signal resembles
the measured signal displayed in Fig.~\ref{Fig5Expkastrup}b. Nevertheless
different current signals have been observed as well, both experimentally
and in numerical simulations (see, e.g., Fig.~10.12 in  \cite{WAC98} for
results obtained from the same model with different boundary conditions).
The oscillations reported here  are due
to traveling high-field domains which is different from the
calculations presented in \cite{KAS97,BON95,PAT98} where
traveling monopoles have been reported. This is due to the
difference in the boundary conditions and in the transport model
used in both calculations, see also Refs.~\cite{SAN99,BON00}, where
this point is analyzed.

\subsection{Traveling fronts\label{SecFronts}}
In this subsection traveling fronts will be examined which will form
one of the building blocks to understand the global dynamics.
Eq.~(\ref{Eq5Ampere}) shows that the dynamical evolution
of the electric field is determined by the total current
$I(t)$. The corresponding dynamics can
be studied most easily for a constant total current
corresponding to current controlled conditions.
For $I_{\rm min}=7.1\mu$A $ < I
< I_{\rm max}=326 \mu$A we
have three intersections of $I$ with the homogeneous current-field
relation shown in Fig.~\ref{Fig5homGrahn}. They correspond
to three stationary homogeneous field distributions
$eF^{I}<eF^{II}<eF^{III}$. Linearization of  Eq.~(\ref{Eq5Ampere})
shows that the field $F^{II}$ is unstable under 
current-controlled conditions as $\d J/\d F<0$. Thus a homogeneous initial
field distribution will either tend to $F^{I}$ or to $F^{III}$
in its temporal evolution.

\begin{figure}
\noindent\epsfig{file=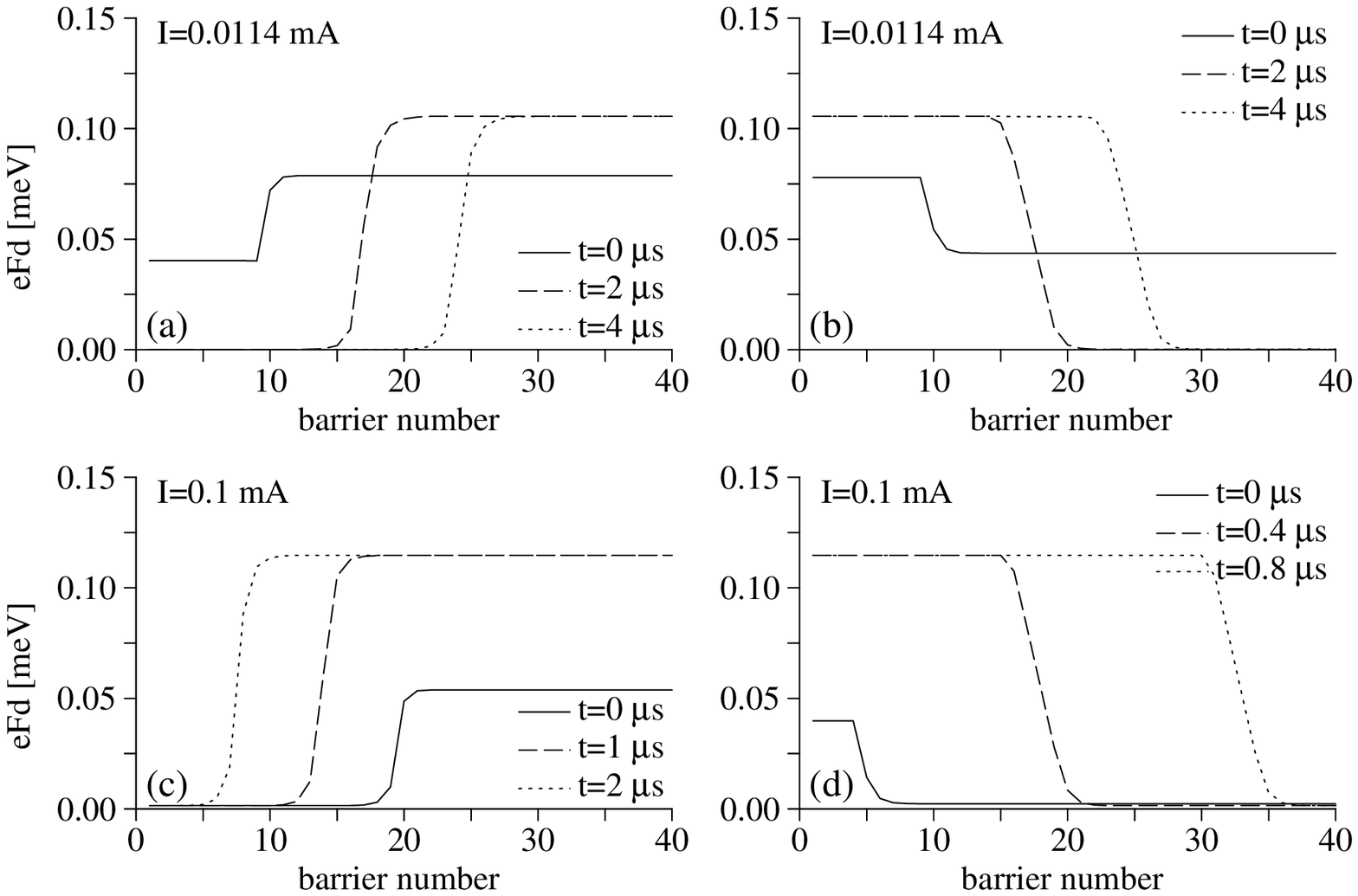,width=15cm}
\caption[a]{Temporal evolution of different initial field profiles
under constant current conditions. (Parameters as in
Fig.~\ref{Fig5Expkastrup}).
\label{Fig5fronts}}
\end{figure}

For an appropriate inhomogeneous initial field distribution
a part develops to $F^{I}$ while another part reaches
$F^{III}$. Thus a front between these two spatial regimes appears.
Calculations on the basis of Eqs.~(\ref{Eq5Ampere}) - (\ref{Eq5Poisson2})
show that the front develops on a typical
time scale of less than  $1\, \mu$s and afterwards travels
through the sample with unaltered shape, as shown in
Fig.~\ref{Fig5fronts}.
Here two types of fronts appear. {\em Accumulation fronts}
connect a low-field region on the left side to a high-field region
on the right side as shown in Figs.~\ref{Fig5fronts}a,\ref{Fig5fronts}c.
Eq.~(\ref{Eq5Poisson2}) shows that the carriers accumulate ($n_m>N_D$)
in the transition region. If, on the other hand, the
high-field region is located on the left side, see
Figs.~\ref{Fig5fronts}b,\ref{Fig5fronts}d, a
{\em depletion front} is present.

\begin{figure}
\noindent\epsfig{file=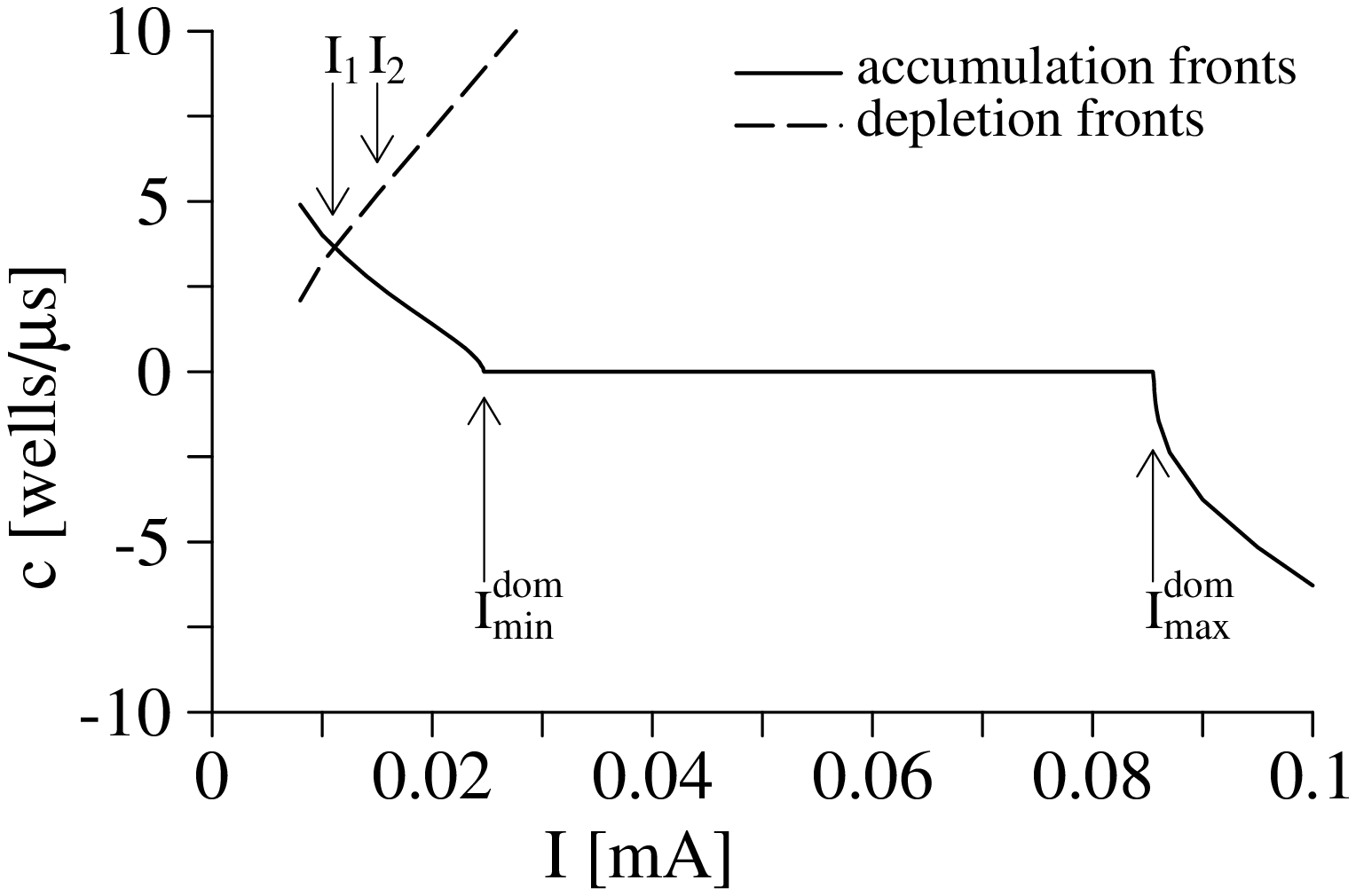,width=11cm}
\caption[a]{Front velocities for accumulation fronts
($c_{\rm acc}$, full line) and depletion fronts ($c_{\rm dep}$, dashed lines).
The velocities are approximately linear functions
of the current outside the range shown here. $I_1$ and $I_2$ denote
the current where
$c_{\rm acc}=c_{\rm dep}$ and $2c_{\rm acc}=c_{\rm dep}$, respectively.
(Parameters as in Fig.~\ref{Fig5Expkastrup}).}
\label{Fig5velocities}
\end{figure}

As can be seen from Fig.~\ref{Fig5fronts}
the front velocities depend on the external current and differ
for  accumulation fronts  with velocity $c_{\rm acc}$
and depletion fronts  with velocity $c_{\rm dep}$.
These velocities have been determined from a series of simulations
as a function of $I$
and are given in Fig.~\ref{Fig5velocities}. Note that $c_{\rm acc}(I)$ becomes
zero in a finite range of currents $24.7\mu A<I<85.5\mu A$
corresponding
to the range $I_{\rm min}^{\rm dom}<I<I_{\rm max}^{\rm dom}$ of the
stationary domains discussed above. (Such a stationary
front is shown in the inset of Fig.~\ref{Fig5kennGrahn}.)
For higher currents $c_{\rm acc}(I)$
is negative, i.e., the front travels upstream against the direction of the
average drift velocity of the electrons as shown in Fig.~\ref{Fig5fronts}c.
In contrast  $c_{\rm dep}(I)$ is positive for all
currents for the sample parameters used. These functions
$c_{\rm acc}(I),c_{\rm dep}(I)$ have been shown to be very helpful  to
understand and analyze
Gunn oscillations \cite{BON97b,BON97a}. Here they are applied
in the context of semiconductor superlattices where some peculiarities can
be found.
In the following two subsections the special shape of $c_{\rm acc}(I)$ and
$c_{\rm dep}(I)$ will be explained.

\subsubsection{Continuum limit and fronts traveling backwards}
In order to understand the simulations, let us first discuss
the continuum limit of vanishing superlattice period.
We can approximate
\begin{equation}
J_{m\to (m+1)}\approx
e\frac{n_m}{d}v(F_m)-eD(F_m)\frac{n_{m+1}-n_{m}}{d^2}\, ,
\label{Eq5STdriftdiff}
\end{equation}
with the average drift velocity $v(F)$ and an effective diffusion constant
$D(F)$ given by
\begin{equation}
v(F)=\frac{d}{eN_D}J(F,N_D,N_D) \quad \mbox{and} \quad
D(F)=-\frac{d^2}{e}\pabl{J(F,N_D,n_2)}{n_2}\, .
\end{equation}
This approximation
becomes exact in the nondegenerate limit, see Eq.~(\ref{Eq3STdriftdiff}),
which is not applicable for the sample discussed here.
Replacing finite differences by derivatives,
Eqs.~(\ref{Eq5Ampere},\ref{Eq5Poisson2}) are transformed to
\begin{equation}
\frac{\partial F(z,t)}{\partial t}=
\frac{J}{\epsilon_r\epsilon_0}-\frac{eN_D}{d\epsilon_r\epsilon_0}v(F)
-v(F)\frac{\partial F(z,t)}{\partial z}+D(F)\frac{\partial^2
F(x,t)}{\partial z^2}\,.
\label{Eq5Gunn}
\end{equation}
This is the standard equation for the Gunn effect,
for which
$\d  c_{\rm acc}(I)/\d  I<0$ and
$\d  c_{\rm dep}(I)/\d  I>0$ holds \cite{KNI67}.
Neglecting diffusion, the fronts can only travel in the direction of
the drift velocity, i.e. $c(I)>0$, which can be shown by the method of
characteristics \cite{BON97}. We conclude
\[\mbox{without diffusion term:}\qquad
c_{\rm dep}>0 \mbox{ and } c_{\rm acc}>0
\quad \mbox{for }I_{\rm min}<I<I_{\rm max}
\]
In contrast, if
the drift term $v(F)\,\partial F/\partial z$ is neglected, one obtains a
nonlinear reaction-diffusion system. Here solutions of traveling
accumulation fronts $F(z,t)=f(z-c_{\rm acc}t)$ exist with a monotonic
function f(z) and
$f(-\infty)=F^I$ and $f(\infty)=F^{III}$. Furthermore one finds
$c_{\rm acc}>0$ if $A_{\rm area}(J)<0$ where
\begin{equation}
A_{\rm area}(J)=\int_{F^I}^{F^{III}}\d  F\,
\frac{1}{D(F)}\left[\frac{J}{\epsilon_r\epsilon_0}-
\frac{eN_D}{d\epsilon_r\epsilon_0}v(F)\right]\, ,
\end{equation}
see section 2 of  \cite{MIK94}.
The depletion fronts can be obtained applying the symmetry operation
$z\to - z$, thus $c_{\rm dep}(J)=-c_{\rm acc}(J)$.
Typically, the integral $A_{\rm area}$ is positive for
$I_{\rm co}<I<I_{\rm max}$
and negative for $I_{\rm min}<I<I_{\rm co}$,
where $I_{\rm co}$ is the current, where both field domains coexist.
(This value satisfies the equal area rule \cite{SCH88c} $A_{\rm area}(I_{\rm
co})=0$.)
Then one finds the generic scenario:
\[\mbox{without drift term:}\qquad
\begin{array}{ccl}
c_{\rm dep}<0 &\mbox{ and } c_{\rm acc}>0 & \mbox{ for } I_{\rm min}<I<I_{\rm co} \\
c_{\rm dep}=0 &\mbox{ and } c_{\rm acc}=0 & \mbox{ for } I=I_{\rm co} \\
c_{\rm dep}>0 &\mbox{ and } c_{\rm acc}<0 & \mbox{ for } I_{\rm co}<I<I_{\rm max}
\end{array}
\]
Including the drift term, it is obvious that both velocities will
increase and the intersection point $c_{\rm dep}=c_{\rm acc}$
is shifted to a positive velocity
in agreement with Fig.~\ref{Fig5velocities}.
If the diffusion constant is not too small, a range of negative
velocity is likely to remain.  This can be estimated in the following way:
Let $F_0$ satisfy $J=eN_Dv(F_0)$. Linearization of
Eq.~(\ref{Eq5Gunn}) for $F(z,t)=F_0+F(k)\exp[\imai (kz-\omega t)]$
gives the dispersion
\begin{equation}
\omega=-\imai\gamma+v_0k-\imai D_0k^2
\end{equation}
with $v_0=v(F_0)$, $D_0=D(F_0)$, and the dielectric relaxation rate $\gamma=
eN_Dv'(F_0)/(d\epsilon_r\epsilon_0)$. Thus an initial
condition $F(z,0)=F_0+\varepsilon\delta(z-z_0)$ has the solution
\begin{equation}
F(z,t)=\frac{\varepsilon}{\sqrt{4\pi D_0t}}
\exp\left[-\gamma t-\frac{(z-z_0-v_0t)^2}{4D_0t}\right]
\label{Eq5perturb}
\end{equation}
which essentially travels in the direction of the drift velocity.
Nevertheless $F(z,t)$ also grows exponentially in the range
$z<z_0$ provided $-\gamma>v_0^2/4D_0$.  This can be interpreted as
the occurrence of an absolute instability  \cite{GUE71} for
\begin{equation}
-e\frac{\d  v(F)}{\d  F}N_D>\frac{v(F)^2\epsilon_r\epsilon_0d}{4D(F)}
\label{Eq5Gueret}
\end{equation}
when  a local perturbation from a homogeneous field spreads in
both directions. If $D$ is larger than  the bound given by (\ref{Eq5Gueret}),
spatial variations not necessarily travel through the structure and
either stationary domain structures or backward traveling field domains
are likely to occur. This condition has been applied to superlattice
transport in  \cite{IGN85} and seems
to be in good agreement with data from a variety of strongly coupled
superlattices \cite{GRE98}. Nevertheless the condition for a
stationary front ($c=0$) is only valid for a specific current, as
the functions $c(I)$ are strictly monotonic for the continuous drift
diffusion model (\ref{Eq5Gunn}). This is in
contrast to the results shown  in Fig.~\ref{Fig5velocities} where
$c_{\rm acc}(I)$ is zero in a finite interval.
Thus we can draw the following conclusions from
the continuum limit:
\begin{itemize}
\item Backward traveling fronts can occur for large values of
$DN_D$ both for accumulation and depletion fronts.
\item The front velocities $c(I)$ are  decreasing with $I$ for
accumulation fronts and increasing for depletion fronts.
\end{itemize}
Finally note that
the simple model for the current used in   \cite{KAS97,BON94,WAC97a}
(where the $n_{m+1}$-dependence of $J(F,n_m,n_{m+1})$ is neglected)
implies $D=0$ in Eq.~(\ref{Eq5STdriftdiff}). These models
are appropriate for the second plateau but differences occur for the
first plateau discussed here. In particular these models cannot
reveal fronts traveling backwards.

\subsubsection{Discreteness of the superlattice and stationary fronts}
While the considerations of the previous subsection
apply to a continuous system, weakly-coupled
superlattices form a system where the discretization due to the
finite superlattice period is essential. As explained in
\cite{WAC98,WAC97a,MIT96} this leads to stationary domain
states, provided the transition region between the two domains
becomes of the order of the  superlattice period. In this case the
accumulation or depletion front gets trapped within one well.
As this pinning can occur within a certain range of currents
$I_{\rm min}^{\rm dom}<I<I_{\rm max}^{\rm dom}$ one observes extended
branches in the current-voltage characteristic
(Figs.~\ref{Fig5Expkastrup},\ref{Fig5kennGrahn}).
As shown in  \cite{WAC98,WAC97a} the sufficient
condition for a stationary  accumulation front  reads
\begin{equation}
N_D\gtrsim N_D^{\rm acc}\equiv \frac{v_{\rm min}}{v_{\rm max}-v_{\rm min}}
\frac{\epsilon_r\epsilon_0}{e}(F_{\rm min}-F_{\rm max})\, .
\label{Eq5stabdiskret}
\end{equation}
(Similar results have been given in \cite{BON94,SCH96b,MIT96} as well.)
For the superlattice structure considered, one obtains
$N_D^{\rm acc}=1.2\times 10^{10}/$cm$^2$, which is smaller
than the actual doping density $N_D=1.5\times 10^{11}/$cm$^2$.
Therefore stationary accumulation fronts can
exist in a certain range of currents for this sample.
Let us remark that the condition (\ref{Eq5stabdiskret}) strongly
resembles (\ref{Eq5Gueret}) if  the
diffusion constant $D(F)=v(F)d/2$ for shot noise \cite{IGN85} is used.
Again the discreteness of the structure is responsible for the
occurrence of stable domains.

For more strongly doped superlattices  the velocity of
the depletion fronts may become zero as well. The corresponding
condition is given by \cite{WAC98}
\begin{equation}
N_D\gtrsim N_D^{\rm dep}\equiv
\frac{v_{\rm max}}{v_{\rm max}-v_{\rm min}}
\frac{\epsilon_r\epsilon_0}{e}(F_{\rm min}-F_{\rm max})\, .
\label{Eq5stabdep}
\end{equation}
For the superlattice structure considered this relation give
$N_D^{\rm dep}=4.4\times 10^{11}/$cm$^2$, which is larger  than
the actual doping. Therefore no stationary depletion fronts occur
in this sample.
Together with the monotony arguments depicted in the previous subsection
these estimations explain the shape of the $c(I)$ functions shown
in Fig.~\ref{Fig5velocities}.
Rigorous proofs of  some of the features discussed here
are  given in  \cite{CAR00}.

In general one can distinguish three types of superlattice:
For $N_D\lesssim N_D^{\rm acc}$ there are no stationary
fronts. For $N_D^{\rm acc} \lesssim N_D\lesssim
N_D^{\rm dep}$ accumulation fronts are stationary within a
certain current range. Finally for highly doped superlattices with
$N_D\gtrsim N_D^{\rm dep}$
both accumulation fronts and depletion fronts can be stationary, i.e., pinned
at a certain well.

\subsection{The injecting contact\label{SecContact}}
Now the influence of the contact boundary
condition at the cathode $J_{0\to 1}(F_0)$ shall be investigated
which is essential for the
dynamical behavior. For a given current $J$, the evolution
of the field at the cathode, $F_0$, is given by:
\begin{eqnarray}
\epsilon_r\epsilon_0\frac{\d F_0}{\d t}&=&
J-J_{0\to 1}(F_0)\label{Eq5contact}
\end{eqnarray}
Let  $F_c(J)$ be the solution of $J_{0\to 1}(F)=J$, which
forms an attracting point of Eq.~(\ref{Eq5contact}), i.e.,
$\d  J_{0\to 1}/\d  F>0$. Then
$F_0(t)$ will tend towards $F_c(J)$.
Provided the relaxation time at
the contact is much smaller than the corresponding time scale
on which $J$ changes, one may use the boundary condition $F_0 = F_c(J(t))$
to describe superlattice dynamics. If $F_c(J)$ is close to the value of
$F^{I}(J)$ or $F^{III}(J)$, a low-field domain
or a high-field domain will be injected into the sample,
respectively \cite{BON97}.
For a pure drift system the
condition for the injection of a low-field domain
is given by $eF_c(J)<eF^{II}(J)$ \cite{SHA92}.
This gives a qualitative bound but this does not  hold strictly
in the superlattice system due to discreteness and diffusion.
For the Ohmic boundary condition depicted in Fig.~\ref{Fig5homGrahn}
we thus find that a low-field domain forms close to the
injecting contact for $I\lesssim I_c$ and a high-field domain forms
there for  $I\gtrsim I_c$, where $I_c$ is the current at the intersection
of $I_{0\to 1}(F_0)$ with the homogeneous current-field characteristic.

In contrast to the cathode where electrons are injected into the
sample, the anode contact conditions do not play a major role.
A boundary layer exists there, which is typical stable
(see e.g.  \cite{BON97} for a discussion within a drift model).
This can be understood from the fact, that the perturbations mainly
travel through the sample in the direction of the current flow,
see Eq.~(\ref{Eq5perturb}), even if there might be some response
in the opposite direction as well.

\subsection{Global behavior\label{SecGlobal}}
The behavior found numerically in Sec.~\ref{SecNumerics} will now be explained
within the  interplay between the dynamics of fronts and contacts.
Let us restrict ourselves to rather long superlattices where the
$N_{3D}L$ criterion
\begin{equation}
N_DN>N_{D}^{nL}\equiv 2.09\frac{v(F)\epsilon_r\epsilon_0}{-e \frac{\d  v}{\d F}}
\label{Eq5NLcrit}
\end{equation}
is satisfied so that the homogeneous field
distribution is unstable \cite{KRO64,MCC66}.
In this case either stationary domain states,
self-sustained periodic current oscillations, or aperiodic behavior occur.
Calculations yield $N_{D}^{nL}\approx 10^{11}/$cm$^2$ for the superlattice
structure under consideration, which is much
smaller than $40N_{D}$. This criterion has been successfully applied
to a variety of superlattice structures \cite{GRE98}.

\subsubsection{Formation of stable stationary field domains}
For the sample considered, accumulation fronts  become
stationary in the current range
$I_{\rm min}^{\rm dom}<I<I_{\rm max}^{\rm dom}$.
Obviously, these domain states are only possible if the
low-field domain is maintained at the cathode, i.e.
$eF_c(I)\lesssim eF^{II}(I)$, as discussed in the previous subsection.
For $\sigma=5$ mA/V, we  find $I_c=110\mu$A$\, >I_{\rm max}^{\rm dom}$.
Thus these domain states are allowed in agreement with
the numerical findings from Fig.~\ref{Fig5kennGrahn}.
The same behavior can be found for the
constant density boundary condition with $c>0$. In this case
the field $F_0$ remains in the low field region as long
as $I\lesssim (1+c)I_{\rm max}$, which is hardly exceeded
unless the superlattice is operated in the second tunneling resonance.
Therefore similar behavior is observed for both boundary conditions.

\begin{figure}
\noindent\epsfig{file=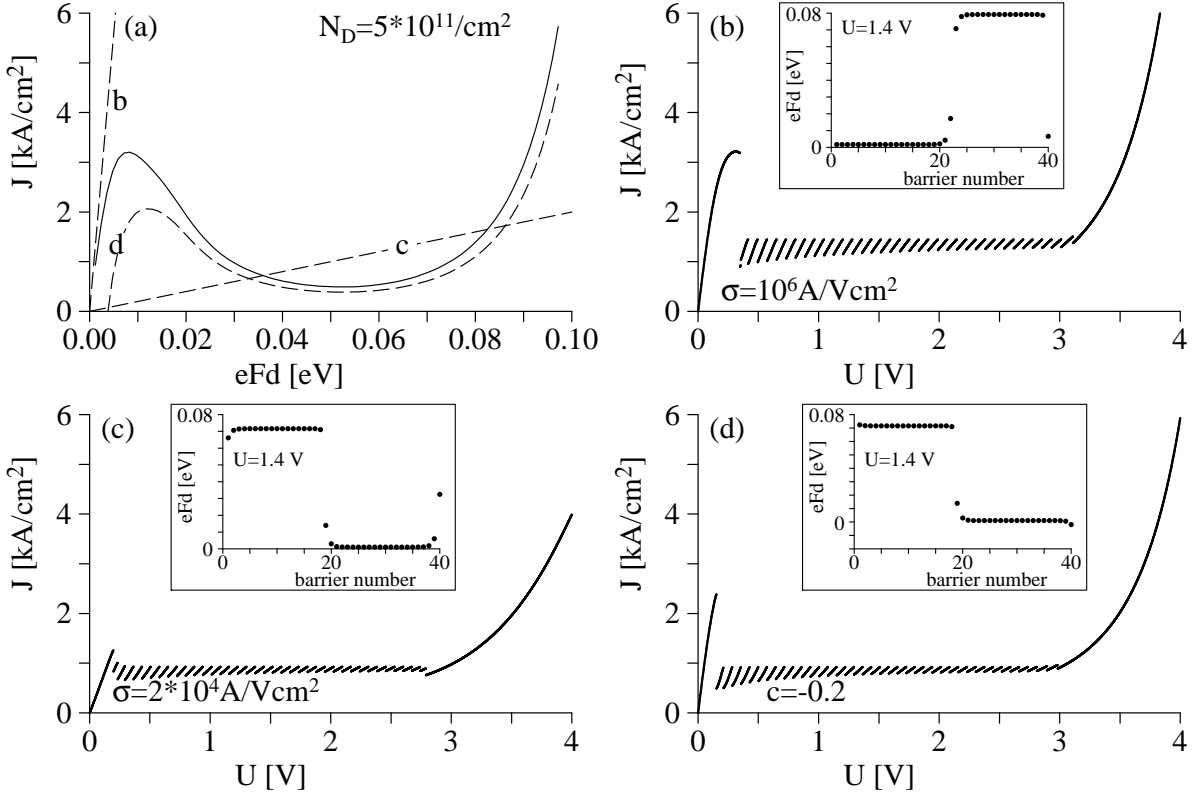,width=16cm}
\caption[a]{Results for a superlattice with 5 nm Al$_{0.3}$Ga$_{0.7}$As
barriers and 8 nm GaAs quantum wells with  a high 
doping density $N_D=5\times 10^{11}/{\rm cm}^2$. 
(a) Current-field relation (full line) evaluated by 
Eq.~(\ref{Eq3JSTsimp}) with $\Gamma^a=\Gamma^b=8$ meV. 
The dashed lines depict the cathode current densities used in parts (b), (c), 
and (d), respectively.
(b) Current-bias relation for an Ohmic boundary condition with
large $\sigma$. (c) Current-bias relation for an Ohmic boundary condition with
small $\sigma$. (d) Current-bias relation for a constant-density
boundary condition with $c=-0.2$. The insets display examples of the
respective field distributions.}
\label{Fig5HighDope}
\end{figure}

For the superlattice structure considered here, stationary
depletion fronts do not occur. Therefore one expects that the high-field
domain is always located at the anode for stationary behavior.
This agrees  with experimental findings for this superlattice \cite{KWO95b}.
For superlattices with a higher doping the situation is different
and field distributions have been observed where
the high-field domain is located at the cathode \cite{HEL90}.
These experimental findings are in good agreement with
the criterion (\ref{Eq5stabdep}). Simulations \cite{WAC97b} give
such a field distribution for appropriate contact conditions.
A further example is shown in Figs.~\ref{Fig5HighDope}c,d.

These stationary fronts appear within a finite current
range $I_{\rm min}^{\rm dom}<I<I_{\rm max}^{\rm dom}$.
If the maximum accumulation occurs in well $m_{\rm dom}$,
the total bias for accumulation fronts is given by
\begin{equation}
U(I,m_{\rm dom})=m_{\rm dom}F^{I}(I)d+(N+1-m_{\rm dom})F^{III}(I)d
+U_c(I)+U_{\rm front}(I)\, .\label{Eq5branch}
\end{equation}
Here $U_c$ and $U_{\rm front}$ are correction terms due to
the inhomogeneous field profiles at the contacts and in the front
region, which  do not depend on the front position $m_{\rm dom}$,
provided this position is not to close to either of the contacts.
Due to the periodicity of the superlattice, a shift of the front by one
period does not change the current. Thus, one obtains
up to $N$ branches $U(I,m_{\rm dom})$
of Eq.~(\ref{Eq5branch}) with $m_{\rm dom}=1,\ldots N$.
This results in the characteristic sawtooth
pattern in the current-voltage characteristics for
Figs.~\ref{Fig5Expkastrup}a,\ref{Fig5kennGrahn}.
In addition, two field distributions exist,
which exhibit a homogeneous (except for boundary effects)
field distribution $F_m=F^{I}(I)$ and $F_m=F^{III}(I)$ for all $m$.
The actual number of branches can be smaller due
to the inhomogeneous field distributions at the boundaries
as shown in the insets in Fig.~\ref{Fig5kennGrahn}.
The voltages corresponding to neighboring branches differ
by $[F^{III}(I)-F^{I}(I)]d$. As this difference depends
on the current, these  branches are not exactly reproduced for different
$m_{\rm dom}$, but
change their shape. Neglecting boundary effects, their slopes are given by
\begin{equation}
\frac{\d  U}{\d  I}=
m_{\rm dom}d\frac{\d F^{I}(I)}{\d  I}
+(N-m_{\rm dom}+1)d\frac{\d F^{III}(I)}{\d  I}
\end{equation}
which varies between
$L\, \d F^{I}(I)/\d  I$ for small biases (accumulation front close
to the right contact) and $L\, \d F^{III}(I)/\d  I$ for large bias
(accumulation front close to the left contact).
The behavior for depletion fronts is identical except that they are located
close to the left/right contact for small/large bias and the
fields $F^{III}$ and $F^{I}$ have to be exchanged in  Eq.~(\ref{Eq5branch}).

\subsubsection{Self-sustained current oscillations}
For $\sigma=0.5$ mA/V we  find $I_c=22\mu$A, which is smaller
than $I_{\rm min}^{\rm dom}$. Therefore no stationary
domain states are possible. In contrast, self-sustained
oscillations occur, as shown in Fig.~\ref{Fig5oscGrahn}. The mechanism of these
oscillations can be understood by using ideas reminiscent
of the Gunn effect asymptotic \cite{BON97b}:
for $t=0.3 \mu$s the total bias is distributed between a
high-field domain and a low-field domain. As the total bias is
constant, there is a certain part of each oscillation period
during which both boundaries must travel at the same velocity
\cite{BON97b,BON97a}. Then $c_{\rm acc}(I_1)=c_{\rm dep}(I_1)$ giving
$I_{1}=0.0114$ mA (see Fig.~\ref{Fig5fronts}). As
$I_1<I_c$ the
cathode remains in the the low-field domain. We observe a current
signal which is constant in average and exhibits fast oscillations
with the period $c/d$ due to well-to-well hopping of the
accumulation front. This feature has been discussed in the analysis
of switching behavior \cite{PRE95,KAS96,AMA01} and is explained in detail
in  \cite{SAN99}.

After the leading edge of the high-field domain
has reached the anode, the size of the high-field domain shrinks
as the trailing edge travels further with unaltered velocity.
In order to maintain the total bias, the fields increase in both
domains, leading to an increase of $I(t)$ via the global condition
(\ref{Eq5global}), as can be seen at $t=2.8\mu$s. When $I(t)$ becomes larger than
$I_c$ the field at the cathode injects a high-field domain into the
superlattice ($t=4.1\mu$s). As $c_{\rm acc}$ is quite large in this
range of currents
the newly formed domain expands relatively fast and the fields
drop in order to maintain the bias. Therefore the current
shrinks again below $I_c$ and the  cathode injects
a low-field domain into the superlattice ($t=5.6\mu$s).

In this situation three boundaries are present
in the superlattice. The old accumulation front (around well 35) and the
depletion and accumulation front limiting the newly formed high-field
domain (from well 5 to 10). In this situation the sum of the
extensions of both high-field regions is kept constant if the depletion
front travels with twice the velocity of the accumulation fronts, i.e.
$2c_{\rm acc}(I_2)=c_{\rm dep}(I_2)$
yielding $I_{2}=0.015$ mA (see Fig.~\ref{Fig5fronts}). This is
just the average current observed in the simulation observed in the
plateau around $t=5.6\mu$s.
The oscillatory part of the current exhibits a lower fundamental
frequency compared to the frequency observed around
$t= 0.3\mu$s as $c_{\rm acc}(I_2)<c_{\rm acc}(I_1)$.
In addition one observes two peaks per period resulting from the
presence of two accumulation fronts.
This behavior is maintained until $t=7.1\mu$s when the old
accumulation front reaches the anode and only the newly formed high-field
domain remains. Afterwards the cycle is repeated.

\begin{figure}
\noindent\epsfig{file=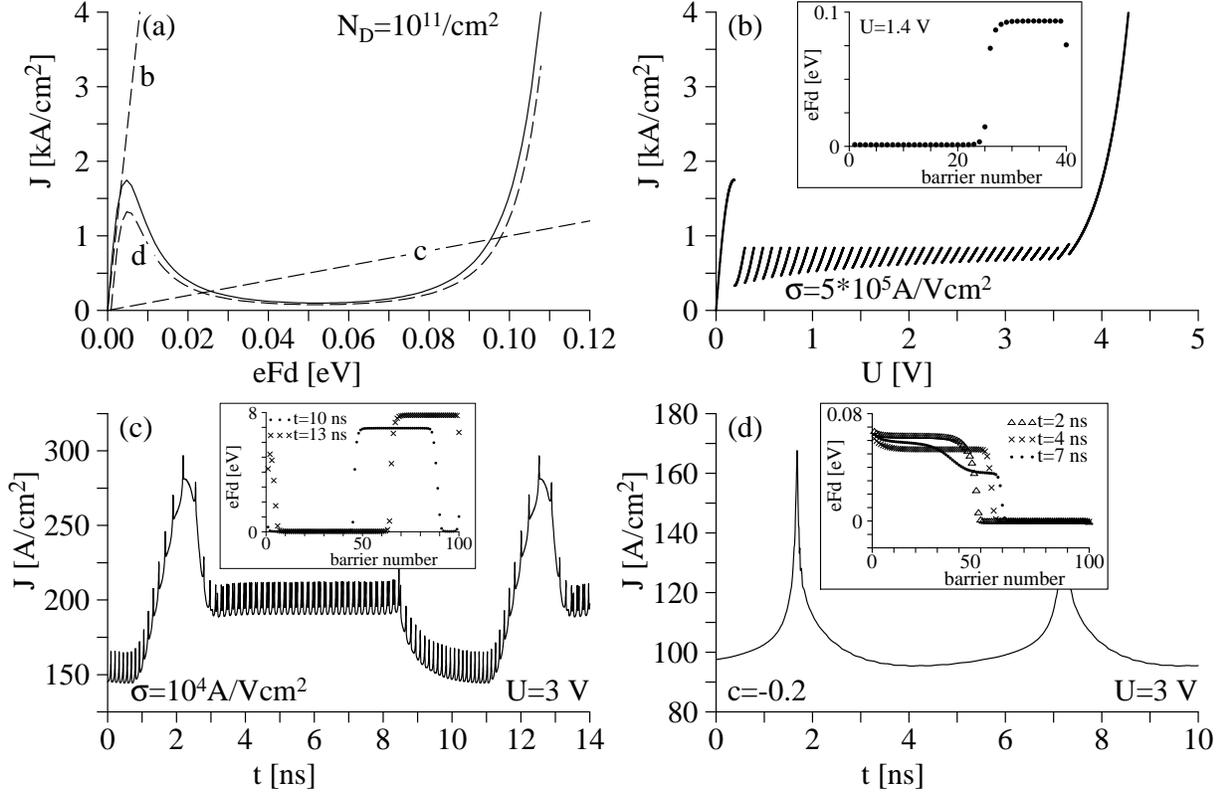,width=16cm}
\caption[a]{Same as Fig.~\ref{Fig5HighDope} for a medium
doping density $N_D= 10^{11}/{\rm cm}^2$. 
(c) and (d) show current oscillations at a fixed bias.}
\label{Fig5MediumDope}
\end{figure}

\begin{figure}
\noindent\epsfig{file=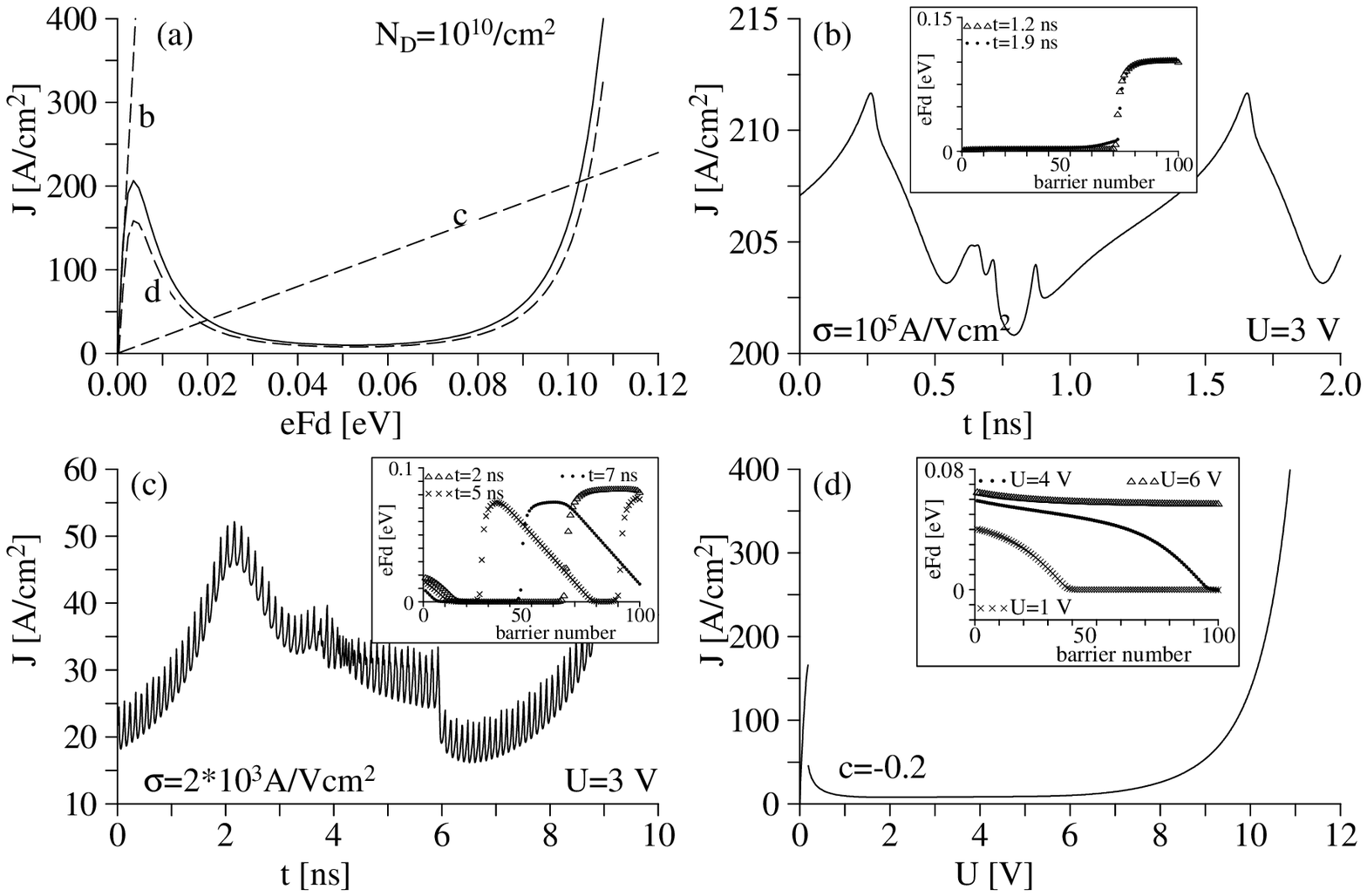,width=16cm}
\caption[a]{Same as Fig.~\ref{Fig5HighDope} for a low
doping density $N_D= 10^{10}/{\rm cm}^2$. 
(b) and (c) show current oscillations at a fixed bias.}
\label{Fig5LowDope}
\end{figure}

Different scenarios for oscillatory behavior are also possible.
For the constant density boundary condition with $c<0$ oscillations due to
traveling depletion fronts have been found for the same model,
see  Refs.~\cite{KAS97,WAC98} and Fig.~\ref{Fig5MediumDope}d for
a further example. In this case the field at the cathode
remains above $eF_{\rm min}$, once this range of fields has been
reached. For $eU<NeF_{\rm min}$ the high-field domain cannot
extend over the whole superlattice and thus a range of fields with
$eFd<eF_{\rm min}d$ must be present. This requires the existence
of a depletion front, which cannot be stationary for the
superlattice considered here as
$N_D < N_D^{\rm dep}$. Thus the front travels towards the anode
increasing the width of the high-field domain. In order to keep the
bias constant, the field inside the domain shrinks during this
process. If $F$ becomes lower than $F_{min}$ the domain becomes
unstable and the field tends to the low field value $F^{I}$. As the
field $F_0$ remains high due to the boundary condition a new
depletion front is formed and the cycle repeats itself. An equivalent
oscillation mechanism is discussed in detail in
\cite{KAS97,BON97} for accumulation fronts. (This 
mechanism is also active in the oscillation shown in  
Fig.~\ref{Fig5LowDope}b.)
It becomes  relevant for lower doped samples, i.e.,
for $N_D < N_D^{\rm acc}$, when these fronts cannot become
stationary.

\subsection{Summary}
Let us summarize the findings from the previous subsections
in order to obtain a general outline for the analysis of
semiconductor superlattices. To visualize the general trends,
various types of behavior are shown for a  superlattice test structure
in Figs.~\ref{Fig5HighDope}--\ref{Fig5LowDope} for different doping 
densities.

\begin{enumerate}
\item Determine the local current-field relation $J(F_m,n_m,n_{m+1})$
from some transport model, see, e.g., section \ref{ChapStandard}.
The corresponding results for $n_m=n_{m+1}=N_D$ are displayed
in Figs.~\ref{Fig5HighDope}a,\ref{Fig5MediumDope}a,\ref{Fig5LowDope}a.
\item Determine $N_D^{\rm acc}$ and $N_D^{\rm dep}$ from
Eqs.~(\ref{Eq5stabdiskret},\ref{Eq5stabdep}).
According to their definition $N_D^{\rm acc}<N_D^{\rm dep}$ holds.
From Fig.~\ref{Fig5MediumDope}a we find $N_D^{\rm acc}\approx
1.7\times 10^{10}/{\rm cm}^2$
and $N_D^{\rm dep}\approx 3\times 10^{11}/{\rm cm}^2$.
\item Compare theses quantities with the actual doping density $N_D$

\begin{description}
\item[$N_D^{\rm dep}<N_D$:] Stable domains form
both for good and bad contacts at the cathode. 
The domain boundaries can be formed by accumulation
layers, see Fig.~\ref{Fig5HighDope}b,  or
depletions layers, see Figs.~\ref{Fig5HighDope}c,d.
Both may even coexist if the respective ranges
$[I^{\rm dom}_{\rm min},I^{\rm dom}_{\rm max}]$ overlap.
In all cases the current-voltage characteristic exhibits the 
typical saw-tooth behavior, where the number of jumps
roughly equals the number of periods.

\item[$N_D^{\rm acc}<N_D<N_D^{\rm dep}$:] The behavior
depends crucially on the boundary condition at the injecting
contact: If the current can be injected at
a fairly low electric field in the cathode (good contact),
stable domains are found which exhibit accumulation layers, i.e.,
the high-field  region is located at the anode, 
see Fig.~\ref{Fig5MediumDope}b.
Otherwise, for bad contacts,  one observes self-sustained current
oscillations, see Figs.~\ref{Fig5MediumDope}c,d.
The shape of these oscillations as well as their frequency
depends on the type of contacts. 

\item[$N_D<N_D^{\rm acc}$:] 
Stable domains associated with a  saw-tooth
current-voltage characteristic do not occur.
The behavior of the sample is either
dominated by current oscillations, 
see Figs.~\ref{Fig5LowDope}b,c or a rather smooth
stationary current-voltage relation,
see Fig.~\ref{Fig5LowDope}d. According to the
$N_{3D}L$-criterion (\ref{Eq5NLcrit}), the latter case
dominates for short samples and lower doping. 
Here it seems to be crucial that the inhomogeneous
field distribution essentially takes values
from a range where $\d v/\d F$ is small, suggesting
large values of $N_{D}^{nL}$. (Similar behavior is found
for $\sigma=10^5$A/Vcm$^2$ if the number of periods is reduced.)
\end{description}

\end{enumerate}

Experimentally, this scenario  can be
verified by varying the electron density by optical irradiation 
\cite{OHT98,OHT00a}. Transtions between oscillating and stationary domains
can also be provoked by changing the lattice temperature \cite{WAN00a}.
This effectively alters the shape of the $j(F)$ relation and thereby via 
Eqs.~(\ref{Eq5stabdiskret},\ref{Eq5stabdep}) the critical doping
densities.
There are several further aspects complicating the picture sketched above,
which are not addressed in the discussion given here:

Real superlattices are not perfect structures, where the properties
of each well are repeated exactly. In contrast there will be fluctuations
in the doping density, the barrier and well width as well as the
material composition from period to period. Some information
can be obtained by X-ray analysis and indeed it has been possible
to relate some global properties to the extend of disorder \cite{GRE98,GRE98a}
in the  respective samples.
By extensive simulations it has been shown that
the presence of such disorder inside the superlattice affects
the behavior significantly \cite{PAT98,SCH96a}. The
nature of the current oscillations, the actual shape
of the domain branches,  as well as critical doping
densities can be affected by the amount of disorder.
Frequently, one observes a direct correlation between some
global current signals and the
actual realization in a particular sample \cite{STE99a,SCH96d}.

Recently, an additional S-type current-voltage characteristics has been 
found in strongly coupled superlattices due to electron heating 
\cite{STE00a}.
The combination of N-type and S-type negative differential conductivity
may provide additional interesting effcets.

In all calculations performed here the transport model for sequential
tunneling has been applied, which provides rate equations
between the quantum wells.
For strongly coupled superlattices it is questionable if the electrons
can be confined to accumulation layers extending over a few wells.
Therefore it is not clear, in how far these stationary domain structures
persist. The alternative is to start from the miniband model.
Such calculations have been performed in  \cite{PER92,BLO97} using
the drift velocity from the relaxation time model.
A more microscopic approach can be performed within
the hydrodynamic model \cite{BUE79,CAO99}.
Nevertheless one has to be aware that
the miniband transport model becomes questionable for large field
strengths (compare Fig.~\ref{Fig3regimes}), which
are typically reached within a high-field domain.
Thus a quantum transport calculation would be desirable to
clarify the situation.
Except for the stability analysis of the
homogeneous state with respect to spatial fluctuations in
\cite{LAI93} I am not aware of any quantum transport simulations
concerning inhomogeneous field distributions in 
superlattices\footnote{Very recently  stationary inhomogeneous field 
distributions could be modeled for short superlattices
within the quantum transport model discussed in Section \ref{ChapNGFT}.
First results are in qualitative agreement with 
those from the sequential tunneling model \cite{DAM01}.}.
Thus it remains an open question in how far quantum effects
modify the behavior discussed in this section.

\section{Transport under irradiation\label{ChapIrr}} 
In this chapter we consider superlattice transport 
under irradiation by an external 
microwave field with frequency $\Omega/2\pi$. 
In this case a  further energy scale, 
the photon energy $\hbar\Omega$ of the radiation field, 
comes into play. 
For frequencies in the THz range (1 THz$\triangleq 4.14$ meV) 
this energy is of the same order of magnitude as typical 
miniband widths, scattering induced broadening, and the potential 
drop per period. This provides an interesting field to study 
various types of quantum effects. 
 
Transport under irradiation has first been studied 
theoretically within the simple Esaki-Tsu model. 
In this context it was shown that negative differential 
dynamical  conductance \cite{KTI72,PAV76} occurs. 
In the rectified response 
replica of the current peaks appear at field strengths obeying 
$eFd=eF_{\rm peak}d+\ell\hbar\Omega$ (with $\ell\in \mathbb{Z}$) 
indicating the quantum nature 
of the radiation field \cite{IGN91,IGN93}. Furthermore 
absolute negative conductance (i.e. a negative current for positive bias) 
is possible under certain conditions 
\cite{IGN76a,IGN78}. 
 
With the development of the free-electron laser 
as a high power THz source it became possible to study 
these effects experimentally and indeed the photon-assisted 
replica of current peaks \cite{GUI93,KEA95a} as well 
as absolute negative conductance \cite{KEA95b} 
were observed. The superlattices used in these experiments 
exhibited rather small miniband widths, so that the 
application of miniband transport (as done in the theories 
mentioned above)  is questionable as 
discussed  in chapters \ref{ChapStandard} and \ref{ChapNGFT}. 
Nevertheless, these findings could be explained 
qualitatively \cite{KEA95b,ZEU96,INA96,PLA97} 
within the standard theory of photon-assisted tunneling \cite{TIE63,TUC79,TUC85} 
as well, which is applicable for sequential tunneling. A 
quantitative description of these experiments \cite{WAC97d} was possible 
within the sequential tunneling approach described 
in Sec.~\ref{SecST}. 
 
Photon-assisted peaks in the current-voltage characteristic could also 
be observed in strongly coupled superlattices \cite{UNT96} although 
the results are less clear in this case. 
Furthermore the reduction of current due to the 
irradiation could be nicely demonstrated in  experiments 
for superlattices with large miniband widths \cite{WIN97}, which 
gave an excellent agreement with the simple miniband models 
mentioned above. 
 
In this section the basic ingredients of the transport theory 
under irradiation are reviewed. 
The first subsection deals with the simple quasi-static response, 
which holds for low frequencies. This will be the basis 
for the discussion of quantum effects in the subsequent subsections. 
The main results within the miniband transport will be 
reviewed in the second subsection. Here a form will be chosen 
which simplifies the comparison with sequential tunneling 
which is discussed in the third subsection. 
 
Throughout this chapter 
we consider a homogeneous electric field 
$F(t)$ along the superlattice structure which can be separated 
into a  dc-part $F_{\rm dc}$ and a cosine-shaped time dependence 
with amplitude $F_{\rm ac}\ge 0$, i.e., 
\begin{equation} 
F(t)=F_{\rm dc}+F_{\rm ac}\cos(\Omega t)\, . 
\label{Eq6Fvont} 
\end{equation} 
The transport problem has been considered for $F_{\rm ac}=0$ in the preceding 
chapters where the relation $I_{\rm dc}(F_{\rm dc})$ was obtained. 
Now we are looking for periodic solutions with period $2\pi/\Omega$, neglecting 
transient effects as well as the possibility of aperiodic behavior. 
Then the general current response can be written as 
\begin{equation} 
I(t)=I_0+\sum_{h=1}^{\infty} 
\left[I_{h}^{\rm cos}\cos(h\Omega t) 
+I_{h}^{\rm sin}\sin(h\Omega t)   \right]\, . 
\label{Eq6Ivont} 
\end{equation} 
In general, four different aspects of transport under irradiation can be 
identified: 
\begin{enumerate} 
\item 
The {\em rectified response} $I_0$ is considered in many 
different experiments. It is easily accessible and can be used 
for the detection 
of high frequency signals. 
\item 
The {\em active current} $I_1^{\rm cos}$ 
provides the direct interaction with the 
irradiation field. If $\d I_1^{\rm cos}/\d F_{\rm ac}<0$ 
we observe gain at the given frequency $\Omega$. 
\item The {\em reactive current} 
$I_1^{\rm sin}$ describes the response out of phase, which can be 
described by an {\em inductance} ($I_1^{\rm sin}=U_{\rm ac}/L\Omega$) 
or a {\em capacitance} ($I_1^{\rm sin}=-C\Omega U_{\rm ac}$) 
in standard circuit theory. 
There are two possibilities to define inductive and capacitive 
effects: 
(i) One assumes that $C$ and $L$ are always positive. 
Then positive/negative $I_1^{\rm sin}$ is referred 
to as  an  inductive/capacitive effect, respectively \cite{FU93}. 
(ii) One regards the low frequency limit. It will be shown later 
that typically  $I_1^{\rm sin}\propto \Omega $ for low frequencies. 
This resembles the behavior of a capacitor which can either be 
positive or negative \cite{ERS98}. 
\item 
{\em Harmonic generation} $I_h$ for $h\ge 2$: These terms describe 
the occurrence of higher harmonics and can be used to generate 
higher frequencies \cite{GRE95,WAN96a,GHO99}. If $F_{\rm dc}=0$ only the odd 
multiplies $h$ are present for symmetric structures with $I_{\rm 
dc}(F)=-I_{\rm dc}(-F)$. 
\end{enumerate} 
 
In this chapter we restrict ourselves to homogeneous field 
distributions with the time dependence (\ref{Eq6Fvont}). 
In this case the current is  homogeneous over the superlattice 
direction and no charge accumulation inside the structure 
occurs. Such complications can be treated 
within the general formalism discussed in \cite{BUE98,PED98}. Domain 
formation effects in superlattices under 
irradiation have been studied in \cite{AGU98}. If higher harmonics 
are present in the time dependence of the field, the superlattice 
may act as a rectifier \cite{GOY98}. Furthermore the response of an 
external circuit is neglected here. A detailed discussion of the 
latter issue can be found in \cite{IGN99}. 
 
\subsection{Low frequency limit\label{Secrflimit}} 
In the range of radio frequencies (say $\Omega \ll 1$ THz) the 
frequency $\Omega$ is slow with respect to the internal degrees of 
freedom, such as carrier heating or $\Delta E/\hbar$ (where $\Delta 
E$ describes typical energy scales of the transport problem). Than 
one can assume, that the current follows the field instantaneously: 
\begin{equation} 
\begin{split} 
I_{\rm rf}(t)=&I_{\rm dc}(F(t))\\ 
=&\sum_{n=0}^{\infty} \frac{1}{n!} 
\frac{\d^n I_{\rm dc}(F_{\rm dc})}{\d F^n} 
\left(F_{\rm ac}\frac{\e^{\imai \Omega t}+ \e^{-\imai \Omega t}}{2}\right)^n\\ 
=&\sum_{j=0}^{\infty}\frac{1}{(2j)!} 
\left(\frac{F_{\rm ac}}{2}\right)^{2j} 
\frac{\d^{2j} I_{\rm dc}(F_{\rm dc})}{\d F^{2j}} 
\left[\binom{2j}{j} 
+\sum_{k=1}^{j} 
\binom{2j}{j+k} 
\left(\e^{\imai 2k \Omega t}+ \e^{-\imai  2k \Omega t}\right)\right]\\ 
&+\sum_{j=0}^{\infty}\frac{1}{(2j+1)!} 
\left(\frac{F_{\rm ac}}{2}\right)^{2j+1} 
\frac{\d^{2j+1} I_{\rm dc}(F_{\rm dc})}{\d F^{2j+1}} 
\sum_{k=0}^{j} 
\binom{2j+1}{j+1+k} 
\left(\e^{\imai (2k+1) \Omega t}+ \e^{-\imai  (2k+1) \Omega t}\right) 
\end{split} 
\end{equation} 
This shows that all terms $I_h^{\rm sin}$ vanish in the radio-frequency 
limit. Furthermore, we obtain the following expressions 
in lowest order of the irradiation field: 
\begin{eqnarray} 
I_{0,{\rm rf}}&=&I_{\rm dc}+\frac{F_{\rm ac}^2}{4} 
\frac{\d^{2} I_{\rm dc}(F_{\rm dc})}{\d F^2} \\ 
I_{1,{\rm rf}}^{\rm cos}&=& 
F_{\rm ac}\frac{\d  I_{\rm dc}(F_{\rm dc})}{\d F}\label{Eq6dynresprf}\\ 
I_{h,{\rm rf}}^{\rm cos}&=& 
\frac{2}{h!} \left(\frac{F_{\rm ac}}{2}\right)^h 
\frac{\d^h I_{\rm dc}(F_{\rm dc})}{\d F^h}\label{Eq6hrf} 
\end{eqnarray} 
They will be compared with the results discussed in the next subsections. 
 
\subsection{Results for miniband transport} 
In the miniband transport model the time dependence of 
the electric field enters the Boltzmann equation 
(\ref{Eq3Boltzmann}) which complicates the problem tremendously. 
The easiest way to deal with this situation is the 
relaxation-time approximation. Assuming 
$\tau_m=\tau_e=\tau$, Eqs.~(\ref{Eq3dynJ},\ref{Eq3dync}) can be solved 
for an arbitrary time dependence of the electric field. 
One obtains \cite{IGN76a,IGN95} 
\begin{eqnarray} 
J(t)&=& \frac{2e|T_1|c_{\rm eq}(\mu,T)}{\hbar\tau} 
\int_{-\infty}^{t}\d t_1 \e^{-(t-t_1)/\tau} 
\sin\left[\int_{t_1}^t\d t_2 \frac{eF(t_2)d}{\hbar}\right]\\ 
c(t)&=& \frac{c_{\rm eq}(\mu,T)}{\tau} 
\int_{-\infty}^{t}\d t_1 \e^{-(t-t_1)/\tau} 
\cos\left[\int_{t_1}^t\d t_2 \frac{eF(t_2)d}{\hbar}\right]\, . 
\end{eqnarray} 
Let us now consider the field dependence (\ref{Eq6Fvont}). 
The crucial parameter in the following will be the 
ratio between the ac-field strength and the photon energy 
\begin{equation} 
\alpha=\frac{eF_{\rm ac}d}{\hbar\Omega} 
\end{equation} 
which will appear in most of the following results as the argument 
of the integer Bessel functions $J_n$. In order to avoid 
confusion with the  symbol $J$ for the 
current density, the results are given in terms of the current 
$I=AJ$ in the following. 
Then one finds: 
\begin{equation} 
\begin{split} 
I(t)=& \frac{2eA|T_1|c_{\rm eq}(\mu,T)}{\hbar\tau} 
\int_{-\infty}^{t}\d t_1 \e^{-(t-t_1)/\tau} 
\Im \left\{ \exp\left[\imai \frac{eF_{\rm dc}d}{\hbar}(t-t_1) 
+\imai\alpha (\sin(\Omega t)-\sin(\Omega t_1)\right]\right\}\\ 
=& \frac{2eA|T_1|c_{\rm eq}(\mu,T)}{\hbar\tau}\Im \left\{ 
\int_{-\infty}^{t}\d t_1 \e^{-(t-t_1)/\tau} 
\e^{\imai \frac{eF_{\rm dc}d}{\hbar}(t-t_1)} 
\sum_{\ell'}J_{\ell'}(\alpha) \e^{\imai \ell'\Omega t} 
\sum_{\ell}J_{\ell}(\alpha) \e^{-\imai \ell\Omega t_1}\right\}\\ 
=& \frac{2eA|T_1|c_{\rm eq}(\mu,T)}{\hbar\tau} 
\sum_{\ell'}\sum_{\ell}J_{\ell'}(\alpha) J_{\ell}(\alpha) 
\Im \left\{ 
\frac{1}{\frac{1}{\tau}-\imai\frac{eF_{\rm dc}d}{\hbar}-\imai\ell\Omega} 
\e^{\imai (\ell'-\ell)\Omega t}\right\}\\ 
=& \frac{2eA|T_1|c_{\rm eq}(\mu,T)}{\hbar\tau} 
\sum_{h}\sum_{\ell}J_{h+\ell}(\alpha) J_{\ell}(\alpha) 
\left[\Im \left\{ 
\frac{1}{\frac{1}{\tau}-\imai\frac{eF_{\rm dc}d}{\hbar}-\imai\ell\Omega} 
\right\}\cos(h\Omega t)\right.\\ 
&\phantom{ \frac{2eA|T_1|c_{\rm eq}(\mu,T)}{\hbar\tau} 
\sum_{h}\sum_{\ell}J_{h+\ell}(\alpha) J_{\ell}(\alpha) } 
\quad+\left. 
\Re \left\{ 
\frac{1}{\frac{1}{\tau}-\imai\frac{eF_{\rm dc}d}{\hbar}-\imai\ell\Omega} 
\right\}\sin(h\Omega t)\right] 
\end{split} 
\end{equation} 
From Eq.~(\ref{Eq3J2relax}) one can identify 
\begin{equation} 
I_{\rm dc}(eFd)=\frac{2eA|T_1|c_{\rm eq}(\mu,T)}{\hbar\tau} 
\, \Im \left\{ 
\frac{1}{\frac{1}{\tau}-\imai\frac{eFd}{\hbar}} 
\right\} 
=\frac{2eA|T_1|c_{\rm eq}(\mu,T)}{\hbar} 
\frac{\Gamma\, eFd}{(eFd)^2+\Gamma^2} 
\label{Eq6Iminiband} 
\end{equation} 
with $\Gamma=\hbar/\tau$. Furthermore we define 
\begin{equation}\begin{split} 
K(eFd)=&\frac{2eA|T_1|c_{\rm eq}(\mu,T)}{\hbar\tau} 
\, \Re \left\{ 
\frac{1}{\frac{1}{\tau}-\imai\frac{eFd}{\hbar}}\right\}= 
\frac{2eA|T_1|c_{\rm eq}(\mu,T)}{\hbar} 
\frac{\Gamma^2}{(eFd)^2+\Gamma^2}   \\ 
=&\int \frac{\d \mathcal{E}}{\pi} 
\mathcal{P}\left\{\frac{I_{\rm dc}(\mathcal{E})}{\mathcal{E}-eFd}\right\} 
\label{Eq6Kminiband} 
\end{split}\end{equation} 
where the Kramers-Kronig relation, connecting the imaginary and real 
part of $\hbar/(\Gamma-\imai\mathcal{E})$, 
has been applied. 
Then we can evaluate the components of Eq.~(\ref{Eq6Ivont}) in the form 
\begin{eqnarray} 
I_0&=&\sum_{\ell} \left(J_{\ell}(\alpha) \right)^2 
I_{\rm dc}\left(eF_{\rm dc}d+\ell\hbar \Omega\right)\label{Eq6I0}\\ 
I_h^{\rm cos}&=&\sum_{\ell} J_{\ell}(\alpha) 
\left( J_{\ell+h}(\alpha)+J_{\ell-h}(\alpha)\right) 
I_{\rm dc}\left(eF_{\rm dc}d+\ell\hbar \Omega\right)\label{Eq6Ihcos}\\ 
I_h^{\rm sin}&=&\sum_{\ell} J_\ell(\alpha) 
\left( J_{\ell+h}(\alpha)-J_{\ell-h}(\alpha)\right) 
K\left(eF_{\rm dc}d+\ell\hbar \Omega\right) 
\, .\label{Eq6Ihsin} 
\end{eqnarray} 
 
\begin{figure} 
\epsfig{file=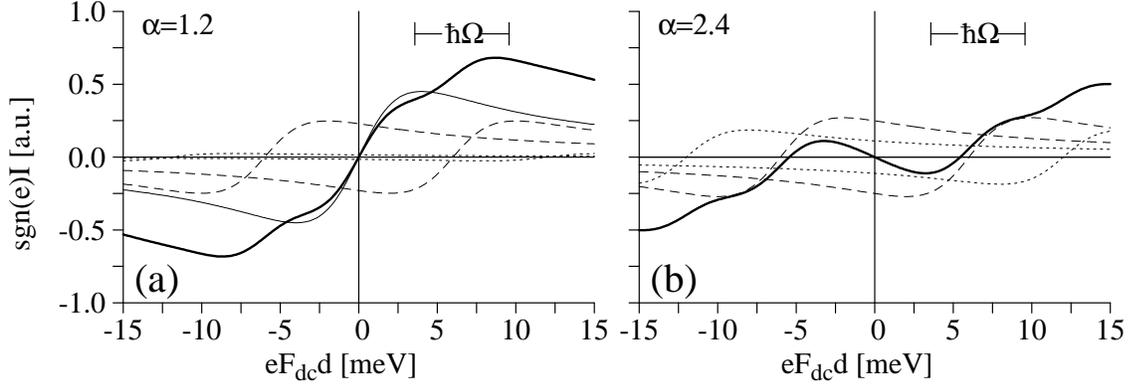,width=15cm} 
\caption[a]{Rectified current-field relation from Eq.~(\ref{Eq6I0}) 
within the Esaki-Tsu model for 
$\alpha=1.2$ (a)  and $\alpha=2.4$ (b). 
The thick line gives the total current $I_0(F_{\rm dc})$. 
The thin lines show the contributions 
$(J_{0}(\alpha))^2I_{\rm dc}(eF_{\rm dc}d)$ (full line), 
$(J_{\pm 1}(\alpha))^2 I_{\rm dc}(eF_{\rm dc}d\pm\hbar \Omega)$ (dashed line), 
and $(J_{\pm 2}(\alpha))^2 I_{\rm dc}(eF_{\rm dc}d\pm 2\hbar \Omega)$ 
(dotted line). Parameters $\Gamma=4$ meV, $\hbar \Omega=6$ meV. 
\label{Fig6MBTeinstr}} 
\end{figure}

Let us first consider the rectified response $I_0$, which is 
a sum of several dc-curves 
(as shown in Fig.~\ref{Fig3esakitsu}b shifted by integer 
multiples of the photon energy.
This explains the occurrence of photon-assisted peaks 
at biases $eF_{\rm dc}d\approx \Gamma+\ell \hbar\Omega$
as shown in Fig.~\ref{Fig6MBTeinstr}a. 
If the coefficient for $\ell=0$ becomes small, i.e., 
close to a zero of $J_0(\alpha)$ (the first zero occurs 
at $\alpha= 2.4048\ldots$) the terms with $\ell=\pm 1$ dominate, which 
can provide absolute negative conductance 
for $\hbar \Omega \gtrsim \Gamma$ as shown in 
Fig.~\ref{Fig6MBTeinstr}b \cite{IGN78}. 
 
For small ac-fields $eF_{\rm ac}d \ll \hbar \Omega$ we may 
set $J_0(\alpha)\approx 1, 
J_{\pm 1}(\alpha)\approx \pm eF_{\rm ac}d/(2\hbar \Omega)$, while 
the higher order Bessel functions are approximately  zero. 
Thus $I_h\approx 0$ for $h\ge 2$ and 
\begin{alignat}{2} 
I_1^{\rm cos}\approx &\frac{I_{\rm dc}(eF_{\rm dc}d+\hbar\Omega)- 
I_{\rm dc}(eF_{\rm dc}d-\hbar\Omega)}{2\hbar \Omega} 
eF_{\rm ac}d&&=:g_{\rm dyn}(\Omega)F_{\rm ac}d\label{Eq6I1coslin}\\ 
I_1^{\rm sin}\approx &-\frac{\hbar \Omega}{2} 
\frac{K(eF_{\rm dc}d+\hbar\Omega)-2K(eF_{\rm dc}d) 
+K(eF_{\rm dc}d-\hbar\Omega)}{(\hbar \Omega)^2} 
eF_{\rm ac}d&&=:-\Omega c_t(\Omega) F_{\rm ac}d\label{Eq6I1sinlin} 
\end{alignat} 
which can be viewed as a resistor $1/g_{\rm dyn}$ and a 
capacitor $c_t$ in parallel 
yielding a complex admittance $z^{-1}=g_{\rm dyn}+\imai\Omega c_t$ 
(small letters indicate quantities per period and the 
engineering convention $I(t)\propto \e^{\imai \Omega t}$ is used here). 
Similar expressions were derived in Ref.~\cite{KTI72} for 
the model with two scattering times.
Both $g_{\rm dyn}$ and $c_t$ may take positive or negative values. 
Furthermore note that $c_t$ is {\em not} the sample capacitance 
which has to be added in parallel as well, but originates from 
a quantum effect. 
(For small frequencies $c_t\approx \frac{e\hbar}{2} 
\frac{\d^2 K(\mathcal{E})}{\d \mathcal{E}^2}$ vanishes 
in the limit $\hbar \to 0$, if the functions $I_{\rm dc}(\mathcal{E})$ 
and $K(\mathcal{E})$ are kept constant.)
 
Eq.~(\ref{Eq6I1coslin}) shows that 
the derivative in the low-frequency response 
(\ref{Eq6dynresprf}) is replaced by a finite difference on 
the quantum scale. A straightforward calculation for the Esaki-Tsu 
model gives $g_{\rm dyn}(\Omega)<0$ for 
$|eF_{\rm dc}d|>\sqrt{\Gamma^2+(\hbar\Omega)^2}$. 
In this range the superlattice structure can provide gain. 
Nevertheless one has to note, that this occurs in the range 
of negative differential conductivity, where the homogeneous 
field distribution is typically unstable as discussed in 
section \ref{ChapDomains}. 
 
In the limit of small scattering $\Gamma\to 0$, the functions 
$I_{\rm dc}(eFd)$ and $K(eFd)$ vanish unless $F\approx 0$. From 
Eqs.~(\ref{Eq6I0},\ref{Eq6Ihcos},\ref{Eq6Ihsin}) one finds 
for $F_{\rm dc}\approx 0$ all components $I_0,I_h^{\rm cos},I_h^{\rm sin}$ 
of the current vanish for $J_0(\alpha)=0$. 
This can be interpreted as a dynamical 
localization \cite{DUN86} or the collapse of the miniband 
\cite{HOL92}: for a certain strength of the irradiation field the 
periodic structure does not conduct any current. In addition, a 
finite conductivity appears at $eF_{\rm dc}d\approx 
\ell \hbar\Omega$ opening up new transport channels, which are 
not present  for $\alpha=0$ in the limit of $\Gamma\to 0$. 
 
All these results have been obtained within the simple Esaki-Tsu model. 
Calculations within the energy balance model \cite{LEI98} provide 
similar results. 
If the dielectric relaxation is included, chaotic behavior \cite{ALE96} 
as well as a spontaneous generation of dc current \cite{ALE98b} 
can be found. Recently, Monte-Carlo simulations of the Boltzmann 
equation under THz-irradiation have been performed, too \cite{MAR00}. 
 
\subsection{Sequential tunneling\label{SecSTeinstr}} 
For sequential tunneling between two neighboring wells 
($m$ and $m+1$) 
Eqs.~(\ref{Eq6I0},\ref{Eq6Ihcos},\ref{Eq6Ihsin}) hold again 
with the dc-expression from Eq.~(\ref{Eq3JST}): 
\begin{equation} 
\begin{split} 
I_{\rm dc}(eFd)=2e\sum_{{\bf k},\nu,\mu} 
|H_{1}^{\nu,\mu}|^2 
\int_{-\infty}^{\infty}& 
\frac{dE}{2\pi \hbar}\tilde{A}^{\mu}_{m}({\bf k},E)\tilde{A}^{\nu}_{m+1}({\bf k},E+eFd)\\ 
&\times\left[n_F(E-\mu_m)-n_F(E+eFd-\mu_{m+1})\right] 
\label{Eq6IST} 
\end{split} 
\end{equation} 
and  the quantity 
\begin{equation} 
\begin{split} 
K(eFd)=&-4e\sum_{{\bf k},\nu,\mu} 
|H_{1}^{\nu,\mu}|^2 
\int_{-\infty}^{\infty} 
\frac{dE}{2\pi \hbar}\left[ 
n_F(E-\mu_m)\tilde{A}^{\mu}_{m}({\bf k},E) 
\Re\{\tilde{G}^{\nu {\rm adv}}_{m+1}({\bf k},E+eFd)\} \right.\\ 
&\left.+ \Re\{\tilde{G}^{\mu {\rm ret}}_{m}({\bf k},E)\}n_F(E+eFd-\mu_{j+1}) 
\tilde{A}^{\nu}_{m+1}({\bf k},E+eFd) \right] \label{Eq6KST}\\ 
=&\int \frac{\d  \mathcal{E}}{\pi} 
\mathcal{P}\left\{\frac{I_{\rm dc}(\mathcal{E})}{\mathcal{E}-eFd}\right\} 
\end{split} 
\end{equation} 
where $\tilde{G}_m({\bf k},E)$ equals $G_m({\bf k},E-meFd)$ in the limit 
of decoupled wells. (The definition used here differs by an $m$-dependent shift 
of the energy scale from the one used 
in Sections \ref{SecST} and \ref{AppSTderivation}.) 
These expressions have been derived 
in  \cite{TUC79} for a constant matrix elements $H_{1}^{\nu,\mu}$. 
A similar derivation is provided in Appendix \ref{AppEinstr}. 
Furthermore it is shown there that Eq.~(\ref{Eq6I0}) also applies 
to the case of a field-dependent matrix element 
$H_{1}^{\nu,\mu}=eFdR_{1}^{\nu,\mu}$, which is relevant for tunneling 
between nonequivalent levels. Note that the THz field also 
couples the different subbands via the term $R_0^{\mu\nu}$, 
which may cause further effects \cite{WAG96}. 
 
For sequential tunneling Eq.~(\ref{Eq6I0}) has a simple interpretation. 
In the evaluation of the dc-current, $eFd$ gives the energy 
mismatch between the levels in well $m$ and $m+1$. Under irradiation 
photons of energy $\hbar \Omega$ 
can be absorbed or emitted during the tunneling process, which 
provides the energy mismatch $eF_{\rm dc}d\pm \ell\hbar\Omega$ 
for the absorption/emission of $\ell$ photons during the tunneling 
process. The Bessel functions represent the 
probability that an $\ell$-photon process occurs assuming a classical 
radiation field (i.e. neglecting spontaneous emission). Eq.~(\ref{Eq6I0}) 
then consists of the weighted sum of all possible photon-assisted 
tunneling processes. Similar results have been obtained in 
 \cite{HAR97b}, where a strictly one-dimensional tight-binding 
lattice coupled to a heat bath has been 
considered (the approximation of incoherent tunneling dynamics applied there 
corresponds to sequential tunneling).  An extended discussion of the methods 
used there can be found in  \cite{GRI98}. 
 
As discussed in section \ref{ChapStandard} 
the structure of the  first peak at low electric fields 
$eFd\approx \Gamma$ is similar for miniband transport and sequential 
tunneling. Therefore the discussion for miniband transport 
given in the preceding subsection holds for sequential tunneling as well. 
In addition,  Eqs.~(\ref{Eq6IST},\ref{Eq6KST}) 
also describe the current peaks at resonances between different 
levels (a,b) in neighboring quantum wells. There one typically finds, 
see Eq.~(\ref{Eq3JSTsimp}) 
\begin{equation} 
I_{\rm dc}(eFd)=eN_{D}A\frac{|H^{ab}|^2}{\hbar} 
\frac{\Gamma^{b,{\rm eff}}}{(eFd+E^a-E^b)^2+(\Gamma^{b,{\rm eff}}/2)^2} 
\label{Eq6Iab}\, . 
\end{equation} 
With the Kramers-Kronig relation  one obtains 
\begin{equation} 
K(eFd)=-eN_{D}A\frac{|H^{ab}|^2}{\hbar} 
\frac{2(eFd+E^a-E^b)}{(eFd+E^a-E^b)^2+(\Gamma^{b,{\rm eff}}/2)^2} 
\label{Eq6Kab} 
\end{equation} 
where the field dependence of $H^{ab}$ has been neglected. 
 
In the following some results are presented for the superlattice 
structure studied experimentally in  \cite{KEA95b,WAC97d,ZEU96} 
(15 nm wide GaAs wells, 5 nm Al$_{0.3}$Ga$_{0.7}$As barriers, 
doping density $N_D=6\times 10^{9}/{\rm cm}^2$, cross section 
$A=8\mu{\rm m}^2$). 
Results for this structure have been presented already in 
Fig.~\ref{Fig3STkennzeun}, where the complicated temperature dependence 
was discussed. Here we will focus to the behavior under irradiation. 
For these calculations a constant electron temperature $T_e=35$ K 
is assumed, which provides best agreement around the first 
maximum. Due to the presence of the irradiation field, it seems 
realistic that the electron gas does not  even 
reach thermal equilibrium for vanishing dc-bias.

\subsubsection{Rectified THz response} 
\begin{figure} 
\epsfig{file=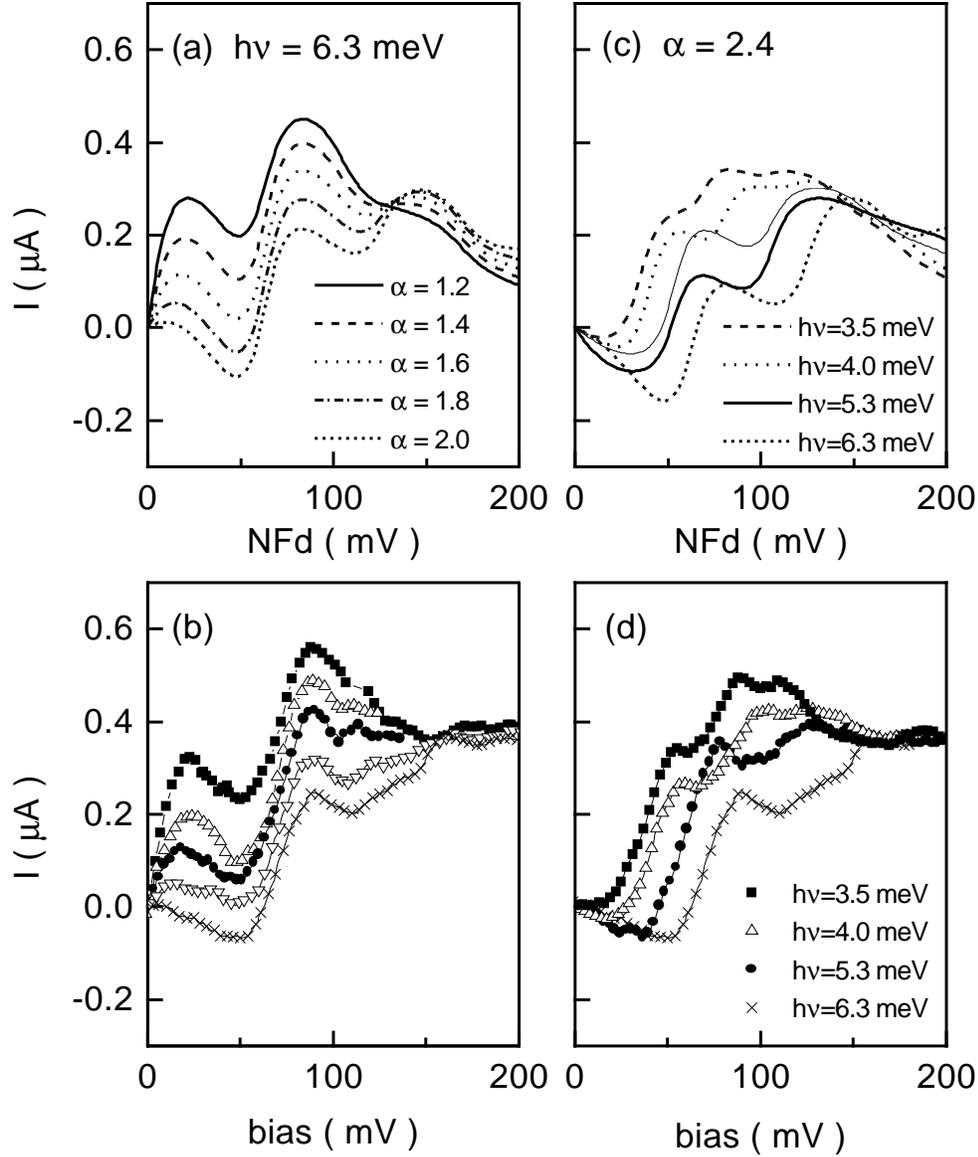,width=13cm} 
\caption[a]{Rectified current response for the superlattice 
of  \cite{KEA95b} displaying absolute negative conductance. 
(a) Theoretical results for $\hbar \Omega =6.3$ meV and different 
field strength $eF_{\rm ac}d=\alpha \hbar \Omega $ of the irradiation. 
(b) Experimental results for  $ \hbar \Omega =6.3$ meV and different 
laser intensities increasing from the top to the bottom. 
The actual  values $F_{\rm ac}$ inside the sample are not accessible. 
(c) Theoretical results  for $\alpha=2.4$ and different 
photon energies. The thin line depicts $\hbar \Omega =5.3$ meV and 
$\alpha=2.1$. (d) Experimental results for different photon energies. 
The laser intensity  was tuned to give maximum  negative conductance. 
(From  \cite{WAC97d}). 
\label{Fig6negativleit}} 
\end{figure}

In Fig.~\ref{Fig6negativleit} results are shown 
for the rectified current response $I_0(eF_{\rm dc}d)$ under different 
strengths and frequencies of the irradiation field. 
Both from experiment and theory one observes a range 
of absolute negative conductance (i.e., $I_0<0$ for $F_{\rm dc}>0$) 
for low biases. Furthermore photon-assisted peaks are visible at 
field strengths $eF_{\rm dc}d\approx eF_{\rm max}d+\ell\hbar \Omega$. 
These findings are in qualitative agreement with the discussion 
of Fig.~\ref{Fig6MBTeinstr}. 
Quantitative agreement between theory (Fig.~\ref{Fig6negativleit}a) 
and experiment (Fig.~\ref{Fig6negativleit}b) is found for 
$\hbar\Omega=6.3$ meV (1.5 THz) for different strengths of the laser field. 
The low-field peak occurs at 
$U_{\rm dir}=N F_{\rm max}d\approx 20$ mV corresponding to direct 
tunneling.  Photon replicas can be observed 
at $U\approx U_{\rm dir}+N\hbar\Omega/e=83$ mV  and 
$U\approx U_{\rm dir}+2N \hbar\Omega/e=146$ mV. 
For low bias and high 
intensities ($\alpha=2.0$) there is a region of absolute negative 
conductance \cite{KEA95b}, which will be discussed in the following. 
 
In Fig.~\ref{Fig6negativleit}d the laser intensity has been tuned 
such that maximal absolute negative conductance occurred for 
each of the  different laser frequencies. 
Then one  observes a minimum in the current at 
$U\approx -U_{\rm dir}+ N\hbar\Omega/e$ which is just the first 
photon replica of the direct tunneling peak on the negative bias side. 
This replica dominates the current if the direct tunneling channel is 
suppressed close to the zero of $J_0(\alpha)$ in Eq.~(\ref{Eq6I0}), 
i.e., $\alpha\approx 2.4$, as used in the calculation of 
Fig.~\ref{Fig6negativleit}c. 
Both the theoretical and experimental results show that 
absolute negative conductance persists in a wide 
range of frequencies but becomes less pronounced 
with decreasing photon energy. 
In the calculation absolute negative conductance vanishes for 
$\hbar\Omega <1.8$ meV which is approximately equal to $\hbar\Omega \lesssim 
eF_{\rm max}d$. 
(The latter relation has been verified by calculations 
for different samples as well.) 
For $\hbar\Omega=5.3$ meV  a smaller value 
of $\alpha=2.1$ (thin line) agrees better with the experimental data 
(in the same sense the value $\alpha=2.0$ agrees better for 
$\hbar\Omega=6.3$ meV, compare Fig.~\ref{Fig6negativleit}a. 
This may be explained as follows: 
If strong NDC is present in doped superlattices, the homogeneous 
field distribution becomes unstable and either self-sustained 
oscillations or stable field domains form as discussed in section 
\ref{ChapDomains}. 
Then the current-voltage characteristic deviates from the 
relation for homogeneous field distribution, where $U=NFd$, and typically 
shows less pronounced NDC. Therefore maximal negative conductance 
is observed at a laser field corresponding to a value of $\alpha<2.4$, 
where the NDC is weaker and the  field distribution is still homogeneous. 
The presence of an  inhomogeneous field distribution  could also explain the 
deviations between theory and experiment for $U>150$ mV. 
Quantitative agreement between theory and experiment 
regarding the rectified response was also obtained for 
a different superlattice structure, where up to seven 
photon replica of the first current peak could be observed \cite{WAC97c}.

\subsubsection{Negative dynamical conductance\label{SecNDynCon}} 
A semiconductor element is able to give gain at the given 
frequency if  $\d I_1^{\rm cos}/\d F_{\rm ac}<0$. 
In the low frequency limit (subsection~\ref{Secrflimit}) one obtains 
$I_{1,{\rm rf}}=\d I_{\rm dc}(F)/\d F\times F_{\rm ac}$ 
if $F_{\rm ac}$ is not too large. This means that gain is related 
to negative differential conductance in the static current-field 
relation $I_{\rm dc}(F)$. 
As such a situation is typically unstable with respect to 
the formation of inhomogeneous field distributions, it is difficult 
to apply this concept for a real device. 
Thus it would be 
of interest to have a system exhibiting 
$\d I_1^{\rm cos}/\d F_{\rm ac}<0$ in the THz 
frequency range considered but a positive differential 
conductance at $\Omega \to 0$. 
Inspection of Eq.~(\ref{Eq6I1coslin}) for 
the standard expressions (\ref{Eq6Iminiband},\ref{Eq6Iab}) 
shows that $I_1^{\rm cos}$ is always 
positive as long as $\d I_{\rm dc}/\d F>0$ holds. 
Thus, it is a nontrivial task to find the opposite case in a 
semiconductor superlattice. 
 
\begin{figure} 
\noindent\epsfig{file=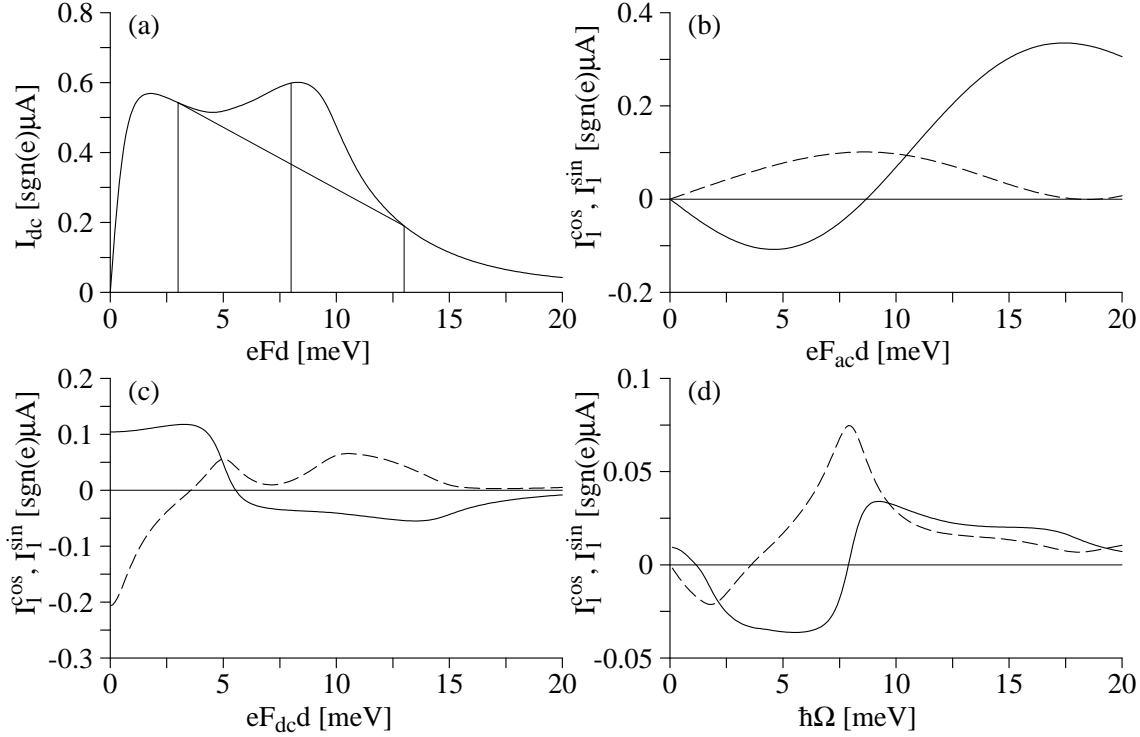,width=15cm} 
\caption[a]{Dynamical response for the 
superlattice of \cite{KEA95b}. (a) Current-field relation without 
irradiation. (b) Current response, full line $I_1^{\rm cos}$, 
dashed line $I_1^{\rm sin}$ for fixed $eF_{\rm dc}d=8$ meV and 
$\hbar \Omega=5$ meV. (c) dito for fixed $eF_{\rm ac}d=1$ meV and 
$\hbar \Omega=5$ meV. (d) dito for fixed $eF_{\rm ac}d=1$ meV and 
$eF_{\rm dc}d=8$ meV.} 
\label{Fig6zeungain} 
\end{figure}

Fig.~\ref{Fig6zeungain}a shows the calculated 
current-field relation $I_{\rm dc}(eFd)$ for the superlattice structure 
discussed here. 
It exhibits a kind of plateau in the range 2 meV$<eFd<$ 9 meV, due to the 
presence of impurity bands which is observed experimentally 
as well (see Fig.~\ref{Fig3STkennzeun}). Experiments\footnote{private 
communication from Stefan Zeuner} show that this 
plateau  is almost unchanged for lattice temperatures between 
4 K and 35 K. 
At $eFd=8$ meV we have positive differential 
conductance, but the finite distance derivative for $\hbar \Omega=5$ meV 
is clearly negative as indicated in the figure (see also  \cite{WAC97e}). 
Thus $I_1^{\rm cos}$ 
will be negative as long as $eF_{\rm ac}d$ is not too large and the terms 
with $\ell>1$ become important in Eq.~(\ref{Eq6Ihcos}). From 
Fig.~\ref{Fig6zeungain}b one obtains 
a negative dynamical conductance  $g_{\rm dyn}=-33\mu$A/V 
at $\hbar \Omega=5$ meV. Due to the higher order 
terms the dynamical conductance becomes positive for 
ac-field strengths larger  than 8.5 mV per period. 
Fig.~\ref{Fig6zeungain}c,d show that negative dynamical conductance 
persists over a wide range of dc-bias and frequency. 
Nevertheless one must note that for $eF_{\rm dc}d>9$ meV the 
static conductance $\d I_{\rm dc}/\d  (Fd)$ 
becomes negative yielding domain formation as 
observed experimentally (see Fig.~\ref{Fig3STkennzeun}). 
 
Unfortunately, the negative  dynamical conductance of the device 
considered here is compensated by the contact 
resistance. Measurements of the temperature dependent 
conductance of the sample yield values up to $20\mu$S 
(around $T=30$ K) at zero bias \cite{WAC97d}. As a part of the resistance 
is from the superlattice itself one may conclude 
that the contact resistance $R_c$ is definitely smaller 
than 50 k$\Omega$ in the 
sample. In comparison the dynamical resistance for 10 wells 
is $R_{\rm dyn}=10/g_{\rm dyn}=-300$ k$\Omega$, which seems to dominate. 
But capacitive effects have to be taken into account: 
The sample capacitance per period is given by 
$c_s=A\epsilon\epsilon_0/d\approx 46$ fF. 
This gives a total impedance 
\begin{equation} 
Z=R_c+N\frac{1}{g_{\rm dyn}+\imai\Omega (c_s+c_t)}=R_c+ 
R_{\rm dyn} 
\frac{1-\imai\Omega (c_s+c_t)/g_{\rm dyn}} 
{1+(\Omega (c_s+c_t)/g_{\rm dyn})^2} 
\end{equation} 
As $c_s/g_{\rm dyn}=-1.4$ ns, the negative dynamical resistance will 
be compensated  even by a small contact resistance at 
THz frequencies. Thus, significant larger values 
of $g_{\rm dyn}$ (i.e. higher current densities) are necessary 
for the observation of gain. 
Nevertheless, the effect discussed here is quite general and 
therefore  gain should be observable  in superlattice 
structures with special shapes for the dc-characteristic.

\subsubsection{Tunneling capacitance} 
\begin{figure} 
\noindent\epsfig{file=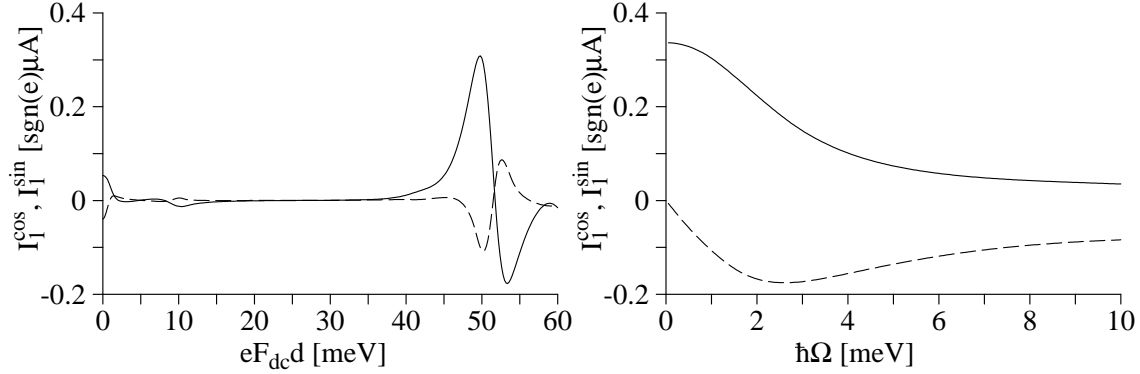,width=15cm} 
\caption[a]{Dynamical response for the superlattice of  \cite{KEA95b}. 
(a) Current response (full line $I_1^{\rm cos}$, 
dashed line $I_1^{\rm sin}$) for fixed $\hbar \Omega=1$ meV, 
(b) for fixed $eF_{\rm dc}d=50$ meV. 
In both cases $eF_{\rm ac}d=0.1$ meV.} 
\label{Fig6quantenc} 
\end{figure}

Now we want to investigate the reactive current $I_1^{\rm sin}$ 
from Eq.~(\ref{Eq6Ihsin}). As can be seen in Fig.~\ref{Fig6zeungain}d 
as well as in Fig.~\ref{Fig6quantenc}b) $I_1^{\rm sin}\propto \Omega$ for 
low frequencies in the linear response region (i.e., small 
irradiation fields). This can be interpreted 
as a tunneling capacitance 
\begin{equation} 
I_1^{\rm sin}\approx -c_t\Omega F_{\rm ac}d\label{Eq6capaz} 
\end{equation} 
as shown in  Eq.~(\ref{Eq6I1sinlin}). 
In Fig.~\ref{Fig6quantenc}a the reactive current 
$I_1^{\rm sin}$  is displayed as  a function of $F_{\rm dc}$ 
using a small irradiation 
field and a small frequency, so that Eq.~(\ref{Eq6capaz}) holds. 
The quantity $I_1^{\rm sin}$ shows a very characteristic 
behavior around the $a\to b$ resonance with a 
minimum at $eF_{\rm dc}d=50$meV, where the differential 
conductance is positive. This behavior can be understood within 
the approximation (\ref{Eq6Kab}), yielding 
\begin{equation} 
c_t=\frac{e\hbar}{2}\frac{\d^2K(eFd)}{\d (eFd)^2} 
=\e^2N_{D}A|H^{ab}|^2 
\frac{2(eFd+E^a-E^b)\left[(eFd+E^a-E^b)^2-3(\Gamma^{b,{\rm eff}}/2)^2\right]} 
{\left[(eFd+E^a-E^b)^2+(\Gamma^{b,{\rm eff}}/2)^2\right]^3}\label{Eq6ctest} 
\end{equation} 
which provides just the structure observed around $eF_{\rm dc}e\approx 50$ meV. 
From Fig.~\ref{Fig6quantenc}b one obtains 
$c_t=0.67$ fF at $eF_{\rm dc}d=50$meV, which is about 1.7\% of the sample 
capacitance $c_s$. 
It would be interesting if this tunneling capacitance can be measured. 
Here it may be useful to study superlattice structures 
with higher doping and 
larger coupling, which enhances the ratio between $c_t/c_s$, 
as can be seen from Eq.~(\ref{Eq6ctest}).

\subsubsection{Harmonic generation} 
\begin{figure} 
\noindent\epsfig{file=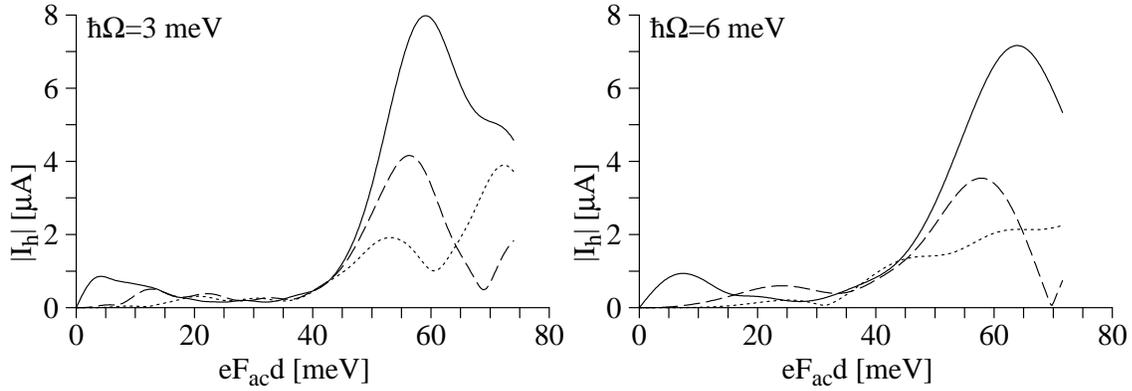,width=15cm} 
\caption[a]{Generation of harmonics $|I_1|$ (full line), 
$|I_3|$ (dashed line), and $|I_5|$ (dotted line) 
for two different frequencies for the superlattice of 
\cite{KEA95b}. As $F_{\rm dc}=0$, the even harmonics vanish.} 
\label{Fig6zeungeneration} 
\end{figure}

In order to investigate harmonic generation the 
quantities $|I_h|=\sqrt{(I_h^{\rm cos})^2+(I_h^{\rm sin})^2}$ have 
been plotted for $F_{\rm dc}=0$ in Fig.~\ref{Fig6zeungeneration}. 
The basic frequency $|I_1|$ dominates for 
low irradiation fields, while the other quantities vanish like 
$|I_3|\sim (eF_{\rm ac}d)^3$ and $|I_5|\sim (eF_{\rm ac}d)^5$ as 
expected from the 
low-frequency limit (\ref{Eq6hrf}). 
The current of the third harmonic is in the range of the dc-current 
at the first peak indicating strong harmonic generation. 
In the range $eF_{\rm ac}d>50$ meV the $a\to b$ resonance becomes 
of importance which can be understood from the classical behavior 
$I_{\rm dc}(eF_{\rm ac}d\cos(\Omega t))$. This provides 
a possibility to probe the second resonance even if it is not 
directly accessible due to sample heating. A similar idea 
to probe a peak by its response to an applied frequency 
has been performed in  \cite{SCH96c}. 
 
\subsection{Discussion} 
In this chapter it has been shown that transport under irradiation 
is  essentially governed by Eqs.~(\ref{Eq6I0},\ref{Eq6Ihcos},\ref{Eq6Ihsin}). 
This scheme holds both for miniband transport within the 
relaxation time approximation and sequential tunneling 
for constant coupling matrix elements, albeit with different 
functions $I_{\rm dc}(eFd)$ and $K(eFd)$. Therefore the question 
arises, if this structure might be general. Although this issue has 
not been settled finally, I believe that this is not the case. If 
one considers next-nearest neighbor tunneling processes within the 
model of sequential tunneling, it becomes obvious, that 
photon-assisted peaks occur at field strengths $2eF_{\rm 
dc}d=2eF_{\rm peak}d+\ell\hbar\Omega$ in contrast to the structure 
of Eq.~(\ref{Eq6I0}). (This might provide an interesting tool to 
investigate tunneling processes where tunneling occurs between 
wells separated by more than one barrier \cite{KRI98,SIB98,HEL99}.) 
The same behavior is found in a miniband model containing higher 
Fourier components in the band structure \cite{ZHA97c}. Therefore 
it seems that the structure of 
Eqs.~(\ref{Eq6I0},\ref{Eq6Ihcos},\ref{Eq6Ihsin}) is limited to 
next-neighbor tunneling, or a cosine-shaped band structure $E(q)$, 
respectively. Further deviations may occur if the THz-field causes 
additional heating of the electron gas\footnote{This was pointed 
out by A.~A. Ignatov (private communication, 1998)}, which had been 
neglected in the models applied here. 
 
The generic structure of the functions $I(eFd)$ and $K(eFd)$ is given by 
Eqs.~(\ref{Eq6Iminiband},\ref{Eq6Kminiband},\ref{Eq6Iab},\ref{Eq6Kab}). 
These imply the following typical effects under irradiation: 
\begin{itemize} 
\item Photon-assisted peaks in the rectified current response at 
characteristic field strengths 
$eFd\approx eF_{\rm peak}d+\ell \hbar \Omega$ with integer values 
$\ell$ for next neighbor coupling. 
\item Absolute negative conductance if the normalized ac-field 
strength $\alpha$ is close the zeros of $J_0(\alpha)$ and 
$\hbar\Omega\gtrsim \Gamma$. 
\item Gain in most of the region of negative differential 
conductivity $\d I_{\rm dc}/\d F<0$. 
\item A quantum capacitance with the 
characteristic dependence (\ref{Eq6ctest}) on 
the dc-field close to the resonances. 
\item Generation of higher harmonics. 
\end{itemize} 
These effects rely on the structure of 
Eqs.~(\ref{Eq6I0},\ref{Eq6Ihcos},\ref{Eq6Ihsin}) and should 
therefore (at least approximately) hold both for wide and 
narrow minibands. In addition, it has been shown that 
gain in the THz range is also possible in the region of 
positive differential conductance for appropriate shapes of 
the $I_{\rm dc}(eFd)$--relation. 

\section{Summary}
In this review the transport properties of semiconductor 
superlattices have been analyzed. Strong emphasis has been given 
to the microscopic modeling of the stationary transport for a homogeneous 
electrical field  as well as the formation of inhomogeneous field 
distributions leading to stationary field domains and self-sustained 
current oscillations. 
 
The three different standard approaches, miniband transport, 
Wannier-Stark hopping, and sequential tunneling, have been reviewed 
in detail. Although the concepts applied are quite different, each 
approach provides negative differential conductivity for 
sufficiently high electric fields. While miniband transport and 
sequential tunneling provide an Ohmic behavior for low electric 
fields together with a maximum of the current around $eFd\approx 
\Gamma$, Wannier-Stark hopping fails in the low-field region. In 
particular, good quantitative agreement with various transport 
measurements in weakly coupled superlattice structures has been 
obtained within the sequential tunneling model, both with and 
without irradiation.

The relation between the standard transport approaches could be 
identified by considering quantum transport 
based on nonequilibrium Green functions. It has been explicitly 
shown that the equations used for these simplified models can be 
obtained from the quantum transport model by 
applying various approximations. This justifies each of these 
approaches and sheds light on the respective ranges of applicability 
sketched in Fig.~\ref{Fig3regimes}. 
Good quantitative agreement was found between self-consistent 
solutions of the quantum transport model with each of the 
simplified approaches in their respective ranges of applicability. 
 
We have shown how different aspects of nonlinear pattern 
formation in semiconductor superlattices, such as the formation 
of stationary field domains as well as self-sustained 
oscillations, 
can be understood by the properties of traveling fronts. 
Fig.~\ref{Fig5velocities} shows that depletion and 
accumulation fronts travel with different velocities, which 
depend on the total current acting as a global coupling. 
This situation is similar to the Gunn diode albeit the 
discreteness of the superlattice structure allows for stationary 
fronts in a finite interval of current, when the front becomes 
trapped. As this trapping may occur in each of the wells, one 
obtains several branches (of the order of superlattice periods) 
in the global current voltage characteristics, which exhibits a typical 
sawtooth shape. The calculated 
current-voltage characteristics as well as the self-sustained 
oscillations are in reasonable quantitative agreement with experimental 
findings. Furthermore, the conditions for the occurrence of self-sustained 
oscillations have been discussed. 

Under strong irradiation by a THz field, photon-assisted
peaks appear, and absolute negative conductivity
is possible. Both effects are predicted in a similar way
by the sequential tunneling model and miniband transport.
It has been demonstrated that 
there is a possibility for gain even in the region, where the 
low-frequency conductivity is positive (i.e. no field-domain 
formation effects are expected). In addition, the tunneling 
processes are connected with characteristic variations in the 
capacitance of the structure, which have not been observed so far. 

\section{Outlook}

Even 30 years after the proposal by Esaki and Tsu, semiconductor
superlattices continue to be  a hot topic of ongoing research.
Presently the following directions stand out:
\begin{description}
\item[Superlattices with lower dimension:]
In this work superlattice structures have been considered,
where a free electron behavior is present in the two directions
parallel to the layers. Presently, the first experiments
have been performed where these lateral directions are confined.
In Ref.~\cite{MIL99} the measurement of  
the conduction through a stack of 50 InAs
quantum dots was reported (see also \cite{PRY98}). 
Such structures can be regarded
as a superlattice structure consisting of
zero-dimensional boxes.
Negative differential resistance was observed recently 
in a superlattice formed by quantum wires, fabricated by
the method of cleaved edge overgrowth \cite{DEU00}.
Due to the restricted phase space perpendicular
to the transport direction, scattering should be strongly reduced
in these structures, and strong effects related to the
miniband structure are likely to be observed.
Theoretical approaches to transport in such structures
can be found in Refs.~\cite{LYA95,BRY00} 
for quantum box superlattices and in
Ref.~\cite{KLE97b} for a superlattice formed by quantum wires.

\item[Self-sustained current oscillations:]
As discussed in section \ref{ChapDomains}, traveling field
domains cause self-generated  current oscillations
in superlattices. While the first experiments reported
frequencies in the MHz range \cite{KAS95} at 5 K 
for a weakly coupled superlattice, 
frequencies up to 150 GHz at could be observed even at room
temperature  in specially designed  
superlattices with a large miniband width \cite{SCH99h}. 
While it may be difficult
to increase the fundamental frequency, the use of
higher harmonics can be helpful 
for possible devices in the THz-range, although the 
prospects of this approach are under debate \cite{KRO00}.

\item[Field domains in strongly coupled superlattices:]
The formation of field domains is usually considered
within the model of rate equation presented in Section \ref{SecModelDomain}.
This model is justified for weakly coupled superlattices,
where the electrons can be considered to be localized in
single quantum wells. This localization becomes questionable
for strongly coupled superlattices and it would be desireable
to develop a quantum transport model, which takes into
account both the inhomogeneity and the time dependence
of the  field distributions. (See also the discussion at the end of
Section \ref{ChapDomains}.)

\item[Chaos and nonlinear dynamics:]
If the superlattice  is driven by an ac-bias, chaotic behavior is 
likely to occur due to the presence of incommensurable frequencies, which
is well known for the Gunn diode \cite{MOS90}. Simulations
for semiconductor superlattices \cite{BUL95,CAO99a} yield similar
results. These chaotic oscillations could be observed 
experimentally \cite{ZHA96,LUO98b} and provide a nice example, where 
many aspects of chaotic systems \cite{BUL99} can be observed.
There are also reports indicating undriven chaos under dc bias conditions
in superlattices \cite{ZHA96}, which is not
understood yet. In addition, frequency locking associated with
the the occurrence of a Devil's staircase has been
observed \cite{SCH99j}. These examples provide an interesting
field to study general phenomena of nonlinear physics.

\item[Generation of THz signals:]
Since the original proposal by Esaki and Tsu, it has been tempting
to use the Bloch oscillator as a source for THz radiation.
Unfortunately, only transient signals have been observed so far.
Another possibility lies in  the occurrence of a negative dynamical 
conductance. Here the problem arises that 
domain formation is a competing process
in this region of parameter space. Two different scenarios have
been proposed as a solution to this problem:
In Ref.~\cite{WAC97e} it was proposed that for special shapes
of the local current-field characteristic gain is possible 
even in the region where the low-frequency conductivity is 
positive, i.e., no field-domain 
formation effects are expected (see also section \ref{SecNDynCon}).
A different possibility could be related to the
fact that the low-frequency conductivity 
can become positive for large amplitudes of the ac fields \cite{KRO00},
thus stabilizing the oscillation.
It is not clear by now, if one of these effects may be
useful to establish THz devices on the basis of semiconductor 
superlattices.
\end{description}

\begin{ack}
 
This work  resulted from  a long collaboration with many colleagues 
from several different groups. 
First I want to thank E.~Sch{\"o}ll for initiating the project 
as well as for many stimulating discussions. In particular several of the 
concepts to combine semiconductor transport and nonlinear physics have 
their origin in his suggestions. 
During my stay at the Mikroelektronik Centret in Lyngby (DK) 
as well as in the time afterwards I profited 
substantially from A.-P.~Jauho, who introduced me to the 
application of nonequilibrium Green functions in semiconductor 
transport. From my visits at the groups of 
L.L.~Bonilla (Madrid) and S.J.~Allen (Santa Barbara, USA) 
I gained much insight into the mathematical 
modeling of instabilities and the problems related to transport 
under strong THz irradiation, respectively. 
I appreciated very much to work together with A.~Amann, S.~Bose, J.~Damzog, 
G. D{\"o}hler, F.~Elsholz, E. Gornik, H.T.~Grahn, J.~Grenzer, M. Helm, 
B.~Yu-Kuang Hu, A.~Ignatov, K.~Johnson, 
J.~Kastrup, G.~Kie{\ss}lich, A.~Kristensen, A.~Markus, Y.A. Mityagin, 
M.~Moscoso, 
P.E.~Lindelof, M.~Patra, G.~Platero, F.~Prengel, C.~Rauch, K.F.~Renk, S.~Rott, 
E.~Schomburg, G.~Schwarz, J.S.~Scott, H.~Steuer, 
M.C.~Wanke, S.~Winnerl, and S.~Zeuner, 
who contributed substantially to the development of the 
basic ideas reported here.  Last not least I would like to thank 
the Deutsche Forschungsgemeinschaft for providing me with 
a grant for my stay in Lyngby as well as for 
financial support within the Sfb296. 

\end{ack}

\appendix
\section{Sequential tunneling with density matrices\label{AppDMT}}
As discussed in Section~\ref{SecST} the electrical
transport in weakly coupled superlattices can be described
by sequential tunneling. The idea of this concept is
that the sates are essentially localized in single wells
and the residual coupling causes transitions between
neighboring wells. As this coupling ($T_1$ for equivalent levels)
is small, one can restrict the theory to the lowest order
$T_1^2$, which provides essentially
Fermi's golden rule. Already in Sec.~\ref{SecST}
it was mentioned that scattering induced broadening is essential
to recover the correct behavior for finite fields.
This complication was treated within the theory of
nonequilibrium Green functions which provided
Eq.~(\ref{Eq3JST}) as derived in Appendix~\ref{AppSTderivation}.

An alternative way to treat quantum transport is density
matrix theory which also gives an exact treatment of
both quantum effects and scattering.
A recent overview can be found in  \cite{KUH98}.
Density matrix theory has been applied to superlattice transport
in  \cite{KAZ72} for the evaluation of currents
between nonequivalent levels.
To my knowledge no such calculations exist regarding transport
between equivalent levels with identical particle densities.
In this case the calculation of   \cite{KAZ72} provides
zero current independent of the electric field as shown below.
A possible resolution of this problem will be presented in
this appendix  and a form similar to  Eq.~(\ref{Eq3JSTsimp})
will be derived. This demonstrates
the equivalence of both approaches and highlights the
differences in performance of both approaches.

\subsection{The model}

We use the Hamiltonian (\ref{Eq2hamW0},\ref{Eq2hamW1}) in the basis
of Wannier states and restrict ourselves to the lowest
band for simplicity. Furthermore impurity and phonon scattering
is taken into account within the restriction of intrawell scattering.
Then the Hamiltonin reads $\hat{H}=\hat{H}_0+\hat{U}+\hat{V}_{\rm imp}
+\hat{V}_{\rm phon}$ with:
\begin{eqnarray}
\hat{H}_0&=&\sum_{n,{\bf k}}
\left(E^a+E_k-eFdn\right)a_n^{\dag}({\bf k})a_n({\bf k})\\
\hat{U}&=&\sum_{n,{\bf k}}T_1 \left[a_{n+1}^{\dag}({\bf k})a_n({\bf k})
+a_{n}^{\dag}({\bf k})a_{n+1}({\bf k})\right]\\
\hat{V}_{\rm imp}&=&\sum_{n,{\bf k},{\bf k}'}
V_{k'k}^n a_n^{\dag}({\bf k}')a_n({\bf k})\\
\hat{V}_{\rm phon}&=&\sum_{n,{\bf p}}\hbar\omega_p
b_n^{\dag}({\bf p})b_n({\bf p})
+\sum_{n,{\bf k},{\bf p}}
M_{p}^n a_n^{\dag}({\bf k}+{\bf p})\left[ b_n({\bf p})+b_n^{\dag}(-{\bf p})
\right]a_n({\bf k})\, .
\end{eqnarray}
In the following we assume that phonon scattering is strong enough
to establish thermal equilibrium. Thus a thermal distribution
function with chemical distribution $\mu_n$ can be assumed for each well.
Furthermore, let us assume
that impurity scattering is stronger than
phonon scattering and dominates the broadening of the states,
which simplifies the following calculations essentially.
Correlations between the scattering matrix elements $V^n$ in different
wells are neglected.
(These drastic approximations make sense as we are mainly interested
in the structure of the theory, not in quantitative results.
The inclusion of different scattering mechanism should be
possible in an analogous way.)
In this case the self-energy within the Born approximation for impurity
averaging is given by
\begin{equation}
\begin{split}
\Sigma^{\rm ret}_n({\bf k},E)=&\sum_{\bf k'}
\left|V_{kk'}^n\right|^2\frac{1}{E-E_{k'}+neFd+\imai 0^+}\\
\approx& -\imai\pi\sum_{\bf k'}\left|V_{kk'}^n\right|^2 \delta(E-E_{k'}+neFd)
\approx -\imai\Gamma/2
\label{EqaGamma}
\end{split}
\end{equation}
where we neglected the real part and assumed that the scattering rate
is energy independent for simplicity.
Using Eq.~(\ref{Eq4jdmt}) one obtains the current density
from well $n$ to well $n+1$
\begin{eqnarray}
J_{n\to n+1}=  \frac{2\mbox{(for spin)}e}{A}\sum_{{\bf k}}\frac{2}{\hbar}\,
\Im \left\{T_1
\langle a_{n+1}^{\dag}({\bf k})a_{n}({\bf k})\rangle\right\}\, .
\label{EqaJdmt}
\end{eqnarray}
The task is the evaluation of the current in lowest order
$T_1^2$ with respect to the coupling (sequential tunneling).
Under these conditions the theory of nonequilibrium Green functions
provides Eq.~(\ref{Eq3JST}) which can be simplified to
Eq.~(\ref{Eq3JSTsimp}) for the self energy (\ref{EqaGamma}).

\subsection{Density matrix theory}
The key point is the temporal evaluation of
$\langle a_{n+1}^{\dag}({\bf k},t)a_{n}({\bf k},t)\rangle$.
Similar to Eq.~(\ref{Eq4dmttempevol}) the dynamics is given by
\begin{equation}
\begin{split}
\frac{\hbar}{\imai }\frac{\d}{\d t}
\langle a_{n+1}^{\dag}({\bf k},t)a_{n}({\bf k},t)\rangle=&
\langle [\hat{H},a_{n+1}^{\dag}({\bf k})a_{n}({\bf k})\rangle\\
=&-eFd\langle a_{n+1}^{\dag}({\bf k})a_{n}({\bf k})\rangle
+T_1\left(\langle a_{n}^{\dag}({\bf k})a_{n}({\bf k})\rangle
-\langle a_{n+1}^{\dag}({\bf k})a_{n+1}({\bf k})\rangle\right) \\
& +\sum_{\bf k'} \left[
V_{k'k}^{n+1} \langle a_{n+1}^{\dag}({\bf k}')a_{n}({\bf k})\rangle-
V_{kk'}^{n} \langle a_{n+1}^{\dag}({\bf k})a_{n}({\bf k}')\rangle  \right]\, .
\label{EqaCurrdyn}
\end{split}
\end{equation}
Here (and in the following) terms containing density matrices
$\langle a_{n+2}^{\dag}({\bf k}')a_{n}({\bf k})\rangle$
will be neglected as they provide terms of order $T_1^4$ in the current.
As stationary states will be considered,
the time dependence is dropped in most of the density matrices.
On the right side new expressions of the type
$\langle a_{n+1}^{\dag}({\bf k}')a_{n}({\bf k})\rangle$ appear.
Their temporal evolution is given by
\begin{equation}
\begin{split}
\frac{\hbar}{\imai }\frac{\d}{\d t}&
V_{k'k}^{n+1}\langle a_{n+1}^{\dag}({\bf k}',t)a_{n}({\bf k},t)\rangle
=(E_{k'}-E_{k}-eFd)V_{k'k}^{n+1}
\langle a_{n+1}^{\dag}({\bf k}')a_{n}({\bf k})\rangle\\
&+\sum_{\bf k''} V_{k''k'}^{n+1}V_{k'k}^{n+1}
\langle a_{n+1}^{\dag}({\bf k}'')a_{n}({\bf k})\rangle
-\sum_{\bf k''} V_{kk''}^{n}V_{k'k}^{n+1}
\langle a_{n+1}^{\dag}({\bf k}')a_{n}({\bf k}'')\rangle\\
&+T_1\left[V_{k'k}^{n+1}\langle a_{n}^{\dag}({\bf k}')a_{n}({\bf k})\rangle-
V_{k'k}^{n+1}\langle a_{n+1}^{\dag}({\bf k}')a_{n+1}({\bf k})\rangle\right]
\label{EqaDynhelp}
\end{split}
\end{equation}
and
\begin{equation}
\begin{split}
\frac{\hbar}{\imai }\frac{\d}{\d t}&
V_{kk'}^{n}\langle a_{n+1}^{\dag}({\bf k},t)a_{n}({\bf k}',t)\rangle
=(E_{k}-E_{k'}-eFd)V_{kk'}^{n}
\langle a_{n+1}^{\dag}({\bf k})a_{n}({\bf k}')\rangle\\
&+\sum_{\bf k''} V_{k''k}^{n+1}V_{kk'}^{n}
\langle a_{n+1}^{\dag}({\bf k}'')a_{n}({\bf k}')\rangle
-\sum_{\bf k''} V_{k'k''}^{n}V_{kk'}^{n}
\langle a_{n+1}^{\dag}({\bf k})a_{n}({\bf k}'')\rangle\\
&+T_1\left[V_{kk'}^{n}\langle a_{n}^{\dag}({\bf k})a_{n}({\bf k}')\rangle-
V_{kk'}^{n}\langle a_{n+1}^{\dag}({\bf k})a_{n+1}({\bf k}')\rangle\right]
\label{EqaDynhelp1}\, .
\end{split}
\end{equation}
These equations have again to be solved in the stationary state.
In the following the key quantities of interest are
the occupation $f_n({\bf k})$ of the mode ${\bf k}$ in well $n$
and the corresponding polarizations $P_n({\bf k})$  which provide the
current. They are given by
\begin{equation}
f_n({\bf k})
=\langle a_{n}^{\dag}({\bf k})a_{n}({\bf k})\rangle\quad
\mbox{and}\quad
P_n({\bf k})=\langle a_{n+1}^{\dag}({\bf k})a_{n}({\bf k})\rangle\, .
\end{equation}
In the following, the more complicated
density matrices appearing in Eqs.~(\ref{EqaDynhelp},\ref{EqaDynhelp1})
have to be related to these quantities.

\subsubsection{Lowest order calculation}
In  \cite{KAZ72}  only  the lowest order terms within
the impurity averaging process and the coupling have been considered
for the evaluation of Eq.~(\ref{EqaCurrdyn}).
As the current is already of order
$T_1\langle a_{n+1}^{\dag}({\bf k})a_{n}({\bf k})\rangle$
only terms up to $T_1$ or $V^2$ will be taken into account.
Due to impurity averaging
only the  term $V_{kk'}^{n+1}V_{k'k}^{n+1}=|V_{k'k}^{n+1}|^2$
remains then in Eq.~(\ref{EqaDynhelp}) and we
find in the stationary state
\begin{equation}
V_{k'k}^{n+1}\langle a_{n+1}^{\dag}({\bf k}')a_{n}({\bf k})\rangle=
\frac{-1}{E_{k'}-E_{k}-eFd+\imai 0^+}|V_{k'k}^{n+1}|^2
P_{n}({\bf k})
\end{equation}
where the term $\imai 0^+$ ensures that the correlations vanish
for $t\to -\infty$. (An alternative way to obtain the factor
$\imai 0^+$ is the application of the
Markov limit, see  \cite{KUH98}.)
Using the same argument we obtain  from Eq.~(\ref{EqaDynhelp1}):
\begin{equation}
V_{kk'}^{n} \langle a_{n+1}^{\dag}({\bf k})a_{n}({\bf k}')\rangle=
\frac{1}{E_{k}-E_{k'}-eFd+\imai 0^+}|V_{k'k}^{n}|^2
P_{n}({\bf k})
\end{equation}
Inserting into Eq.~(\ref{EqaCurrdyn}) gives
\begin{equation}
\frac{\hbar}{\imai }\frac{\d}{\d t}
P_{n}({\bf k},t)
=-eFd P_{n}({\bf k})
+T_1\left[f_{n}({\bf k})-f_{n+1}({\bf k})\right]
+i\Gamma P_{n}({\bf k})
\end{equation}
where the approximation (\ref{EqaGamma}) has been  used.
This equation has also been applied in  \cite{FER96a} to study time
dependent phenomena.
The stationary solution yields
\begin{equation}
P_{n}({\bf k})=
\frac{-1}{-eFd+\imai\Gamma}T_1\left[f_{n}({\bf k})-f_{n+1}({\bf k})\right]
\end{equation}
and we obtain the current density via Eq.~(\ref{EqaJdmt})
\begin{equation}
J_{n\to n+1}=  2e\frac{T_1^2}{\hbar A}\sum_{{\bf k}}
\frac{2\Gamma}{(eFd)^2+\Gamma^2}
\left[f_{n}({\bf k})-f_{n+1}({\bf k})\right] \, .
\label{EqaJdmtsimp}
\end{equation}
Taking into account that
$\rho(E)=2\mbox{(for spin)}/A\sum_{{\bf k}}\delta(E-E_k)$
is the density of states, this expression is almost identical with
Eq.~(\ref{Eq3JSTsimp}). Nevertheless there is a significant difference:
While in Eq.~(\ref{Eq3JSTsimp})
the transport is driven by the difference
of the {\em occupation at the same energy}, now the transport is driven by
the difference of the  {\em occupation of the state ${\bf k}$ in both wells}.
The latter difference becomes zero, if both wells have the
same electron density. This has lead to the conclusion that
no resonant tunneling peak occurs in weakly coupled superlattices
for tunneling between equivalent levels \cite{SHI75},
in contrast to the findings of section~\ref{SecST}.
Regarding tunneling between the ground state and excited states
(which are typically empty)
the respective formula provides a finite current which essentially
agrees with the corresponding result of Eq.~(\ref{Eq3JSTsimp}).
The corresponding calculations have been presented in  \cite{KAZ72},
where interwell correlations  between scattering
matrix elements were also taken into account.

\subsubsection{Improved treatment}
This problem can be circumvented by taking into account
the last term in Eqs.~(\ref{EqaDynhelp},\ref{EqaDynhelp1}) as well.
Neglecting terms containing the coupling
$T_1$ (which provide terms $\sim T_1^4$ in the current)
the dynamics of
$V_{k'k}^{n'}\langle a_{n}^{\dag}({\bf k}')a_{n}({\bf k})\rangle$ is given
by:
\begin{equation}\begin{split}
\frac{\hbar}{\imai }\frac{\d}{\d t}&
V_{k'k}^{n'}\langle a_{n}^{\dag}({\bf k}',t)a_{n}({\bf k},t)\rangle
=(E_{k'}-E_{k})V_{k'k}^{n'}
\langle a_{n}^{\dag}({\bf k}')a_{n}({\bf k})\rangle\\
&+\sum_{\bf k''} V_{k''k'}^{n}V_{k'k}^{n'}
\langle a_{n}^{\dag}({\bf k}'')a_{n}({\bf k})\rangle
-\sum_{\bf k''} V_{kk''}^{n}V_{k'k}^{n'}
\langle a_{n}^{\dag}({\bf k}')a_{n}({\bf k}'')\rangle
\end{split}
\end{equation}
By impurity averaging all terms $V^2$
vanish unless  $n'=n$ and  $k''=k$ or $k''=k'$.
Thus, we find in the stationary state:
\begin{equation}
V_{k'k}^{n'}\langle a_{n}^{\dag}({\bf k}')a_{n}({\bf k})\rangle
=\frac{-1}{E_{k'}-E_{k}+\imai 0^+}|V_{k'k}^{n}|^2\delta_{n,n'}
\left[f_n({\bf k})-f_n({\bf k}')\right]
\end{equation}
Then the evaluation of Eq.~(\ref{EqaDynhelp}) yields
\begin{equation}
\begin{split}
&V_{k'k}^{n+1}\langle a_{n+1}^{\dag}({\bf k}')a_{n}({\bf k})\rangle\\
&=\frac{-1}{E_{k'}-E_{k}-eFd+\imai 0^+}
\left\{|V_{k'k}^{n+1}|^2P_{n}({\bf k})
+\frac{T_1}{E_{k'}-E_{k}+\imai 0^+}|V_{k'k}^{n+1}|^2
\left[f_{n+1}({\bf k})-f_{n+1}({\bf k}')\right]\right\}\\
&= \frac{-1}{E_{k'}-E_{k}-eFd+\imai 0^+}|V_{k'k}^{n+1}|^2
P_{n}({\bf k})\\
&\phantom{=}+\frac{T_1}{eFd}|V_{k'k}^{n+1}|^2\left(
\frac{1}{E_{k'}-E_{k}+\imai 0^+}-\frac{1}{E_{k'}-E_{k}-eFd+\imai 0^+}\right)
\left[f_{n+1}({\bf k})-f_{n+1}({\bf k}')\right]
\end{split}
\end{equation}
and Eq.~(\ref{EqaDynhelp1}) gives
\begin{equation}
\begin{split}
V_{kk'}^{n}&\langle a_{n+1}^{\dag}({\bf k})a_{n}({\bf k}')\rangle=
\frac{1}{E_{k}-E_{k'}-eFd+\imai 0^+}|V_{k'k}^{n}|^2
P_{n}({\bf k})\\
&-\frac{T_1}{eFd}|V_{k'k}^{n}|^2\left(
\frac{1}{E_{k}-E_{k'}+\imai 0^+}-\frac{1}{E_{k}-E_{k'}-eFd+\imai 0^+}\right)
\left[f_{n}({\bf k}')-f_{n}({\bf k})\right]
\end{split}
\end{equation}
This provides further terms
for Eq.~(\ref{EqaCurrdyn}), which becomes
\begin{equation}
\begin{split}
\frac{\hbar}{\imai }\frac{\d}{\d t}&
P_{n}({\bf k},t)
=-eFd P_{n}({\bf k})
+T_1\left[f_{n}({\bf k})-f_{n+1}({\bf k})\right]
+\imai\Gamma P_{n}({\bf k})\\
&-\frac{T_1}{eFd}\sum_{k'}
\left(\frac{1}{E_{k'}-E_{k}+\imai 0^+}-\frac{1}{E_{k'}-E_{k}-eFd+\imai 0^+}\right)
|V_{k'k}^{n+1}|^2\left[f_{n+1}({\bf k}')-f_{n+1}({\bf k})\right]
\\
&+\frac{T_1}{eFd}\sum_{k'}
\left(\frac{1}{E_{k}-E_{k'}+\imai 0^+}-\frac{1}{E_{k}-E_{k'}-eFd+\imai 0^+}\right)
|V_{k'k}^{n}|^2\left[f_{n}({\bf k}')-f_{n}({\bf k})\right]
\end{split}
\end{equation}
Now we apply the relation
$1/(x+\imai 0^+)=\mathcal{P}\{1/x\}-\imai\pi\delta(x)$ and restrict ourselves
to the imaginary parts.
The isotropy in ${\bf k}$-space gives
$f_n({\bf k})=f_n(E_k)$ and we find with
Eq.~(\ref{EqaGamma}) the stationary solution
\begin{equation}
\begin{split}
P_{n}({\bf k})=&
\frac{1}{eFd-\imai\Gamma}\Bigg\{
T_1\left[f_{n}(E_k)-f_{n+1}(E_k)\right] \\
&+\frac{T_1}{eFd}\frac{\imai \Gamma}{2}
\left[f_{n+1}(E_k)-f_{n+1}(E_k+eFd)-f_{n}(E_k)+f_{n}(E_k-eFd)
\right]\Bigg\}
\end{split}
\end{equation}
and the current (\ref{EqaJdmt}) becomes
\begin{equation}\begin{split}
J_{n\to n+1}=&  2e\frac{T_1^2}{\hbar A}\sum_{{\bf k}}
\frac{2\Gamma}{(eFd)^2+\Gamma^2}\\
&\times\left[\frac{f_{n}(E_k)+f_{n}(E_k-eFd)}{2}
-\frac{f_{n+1}(E_k+eFd)+f_{n+1}(E_k)}{2}\right] \, .
\end{split}\end{equation}
In contrast to Eq.~(\ref{EqaJdmtsimp}) the current is now driven by
the occupation difference  taken at different values of $E_k$
so that the total energy $E_k-neFd$ is equal in both wells.
This structure agrees with Eq.~(\ref{Eq3JSTsimp})
and a finite current is found  for identical occupation
function $f_n(E_k)=f_{n+1}(E_k)$.
This shows that the terms
$V_{k'k}^{n'}\langle a_{n}^{\dag}({\bf k}')a_{n}({\bf k})\rangle$,
which  had been neglected in the derivation  of Eq.~(\ref{EqaJdmtsimp}),
are of crucial importance. They describe the internal correlations
in the single quantum wells.
These correlations correspond to the broadening of states
in the Green function formalism, where  they are taken into
account by treating the probe energy $E$ and the energy of the
bare state $E_k$ separately. As the broadening of the states
is of crucial importance for the tunneling current, it becomes
clear, that density matrix theory gives a wrong result
if the corresponding matrices are neglected.

\section{Derivation of the standard approaches\label{AppDerivation}} 
In this appendix the relations between the quantum transport 
equations of section \ref{ChapNGFT}
and the standard approaches (miniband transport, 
Wannier-Stark hopping, and sequential tunneling as discussed in section
\ref{ChapStandard}) are 
examined. 
It will be shown, that the transport equations for the different 
standard approaches can
be  derived explicitly from the quantum transport model 
using various types of approximations. In each case
the respective approximations can be justified within
the range of  validity of the given standard approach 
sketched in Fig.~\ref{Fig3regimes}
and motivated in subsection \ref{SecNGFconstsig}.
 
\subsection{Sequential tunneling\label{AppSTderivation}} 
In the parameter ranges $2T_1\ll \Gamma$ or $2T_1\ll eFd$ 
the diagonal elements of $G_{n,m}$ dominate 
(see subsection \ref{SecNGFconstsig}) and an expansion 
of Eqs.~(\ref{Eq4GretSL},\ref{Eq4KeldyshSL}) in $T_1$ 
is appropriate. In this way the formula for sequential 
tunneling (\ref{Eq3JST}) will be recovered as the 
leading order in $T_1$ of the general Eq.~(\ref{Eq4JNGFT}). 
 
For $T_1=0$ we obtain 
$G^{</{\rm ret}}_{m_1,m_2}({\bf k},E)=\delta_{m_1,m_2} 
\tilde{G}^{</{\rm ret}}_{m_1}({\bf k},E)$ 
which are determined by 
\begin{equation} 
\left(E-E_{k}+eFdm -\tilde{\Sigma}^{\rm ret}_{m}({\bf k},E) 
\right) 
\tilde{G}^{\rm ret}_{m}({\bf k},E)=1 
\label{Eq4Grettilde} 
\end{equation} 
and 
\begin{equation} 
\tilde{G}^{<}_{m}({\bf k},E)= 
\tilde{G}^{\rm ret}_{m}({\bf k},E)\tilde{\Sigma}^{<}_{m}({\bf k},E) 
\tilde{G}^{\rm adv}_{m}({\bf k},E) 
\end{equation} 
where the self-energies $\tilde{\Sigma}_{m}$ are 
evaluated applying the Green-functions $\tilde{G}_{m}$. 
These equations decouple in the well 
index. As no current flows in this case, 
one obtains the equilibrium solution 
\begin{eqnarray} 
\tilde{G}^{<}_{m}({\bf k},E)&=&\imai\tilde{A}_m({\bf k},E)n_F(E-\mu_m+meFd)\\ 
\tilde{\Sigma}^{<}_{m}({\bf k},E)&=& 
\left[\tilde{\Sigma}^{\rm adv}_m({\bf k},E) 
-\tilde{\Sigma}^{\rm ret}_m({\bf k},E)\right] 
n_F(E-\mu_m+meFd) 
\end{eqnarray} 
with the spectral function 
\begin{equation} 
\tilde{A}_{m}({\bf k},E)=\imai\left[ 
\tilde{G}^{\rm ret}_{m}({\bf k},E)- 
\tilde{G}^{\rm adv}_{m}({\bf k},E)\right] 
=-2\Im\left\{\tilde{G}^{\rm ret}_{m}({\bf k},E)\right\} 
=2\Im\left\{\tilde{G}^{\rm adv}_{m}({\bf k},E)\right\} 
\end{equation} 
To first order in $T_1$,  Eq.~(\ref{Eq4GretSL}) 
gives together with Eq.~(\ref{Eq4Grettilde}): 
\begin{eqnarray} 
G^{\rm ret}_{m\pm 1,m}({\bf k},E)&=& 
\tilde{G}^{\rm ret}_{m\pm 1}({\bf k},E) 
T_1\tilde{G}^{\rm ret}_{m}({\bf k},E)+\mathcal{O}(T_1^2)\\ 
G^{\rm ret}_{m,m}({\bf k},E)&=& 
\tilde{G}^{\rm ret}_{m}({\bf k},E)+\mathcal{O}(T_1^2)\\ 
G^{\rm ret}_{n,m}({\bf k},E)&=&\mathcal{O}(T_1^2)\quad \mbox{for} 
\quad |m-n|\ge 2 
\end{eqnarray} 
and Eq.~(\ref{Eq4KeldyshSL}) gives 
\begin{equation}\begin{split} 
G^{<}_{m+1,m}({\bf k},E)=& 
\tilde{G}^{\rm ret}_{m+1}({\bf k},E) 
T_1\tilde{G}^{\rm ret}_{m}({\bf k},E) 
\tilde{\Sigma}^{<}_{m}({\bf k},E) 
\tilde{G}^{\rm adv}_{m}({\bf k},E)\\ 
&+\tilde{G}^{\rm ret}_{m+1}({\bf k},E) 
\tilde{\Sigma}^{<}_{m+1}({\bf k},E) 
\tilde{G}^{\rm adv}_{m+1}({\bf k},E) 
T_1\tilde{G}^{\rm adv}_{m}({\bf k},E) 
+\mathcal{O}(T_1^2)\\ 
=&T_1\Big[\tilde{G}^{\rm ret}_{m+1}({\bf k},E) 
\tilde{G}^{<}_{m}({\bf k},E) 
+\tilde{G}^{<}_{m+1}({\bf k},E)\tilde{G}^{\rm adv}_{m}({\bf k},E)\Big] 
+\mathcal{O}(T_1^2)\\ 
=&\imai T_1\Big[\tilde{G}^{\rm ret}_{m+1}({\bf k},E) 
\tilde{A}_{m}({\bf k},E)n_F(E-\mu_m+meFd)\\ 
&+\tilde{A}_{m+1}({\bf k},E)n_F(E-\mu_{m+1}+(m+1)eFd) 
\tilde{G}^{\rm adv}_{m}({\bf k},E)\Big] 
+\mathcal{O}(T_1^2) 
\end{split}\end{equation} 
Then the current is evaluated by Eq.~(\ref{Eq4JNGFT}) 
\begin{equation} 
\begin{split} 
J_{m\to m+1}^{a\to a}=& 
\frac{2e}{A} 
\sum_{{\bf k}}\frac{2}{\hbar}\, \int \frac{\d E}{2\pi} 
\Re\left\{T_1G_{m+1,m}^{<}({\bf k},E)\right\}\\ 
=&\frac{2e}{A} 
\sum_{{\bf k}}\frac{|T_1|^2}{\hbar}\, 
\int \frac{\d E}{2\pi} 
\tilde{A}_{m+1}({\bf k},E)\tilde{A}_{m}({\bf k},E)\\ 
&\times \left[n_F(E-\mu_m+meFd)-n_F(E-\mu_{m+1}+(m+1)eFd)\right] 
+\mathcal{O}(T_1^3) 
\end{split} 
\end{equation} 
which is just the expression (\ref{Eq3JST}) used for sequential 
tunneling. 
 
\subsection{Miniband conduction\label{AppBTEderivation}} 
If $2|T_1|\gg \Gamma,eFd$, the states in a semiconductor 
superlattice are essentially delocalized, as shown in
subsection \ref{SecNGFconstsig}.
In this case it makes sense to work in an extended basis 
like the Bloch states $q$. The key point is the idea, that 
the occupation of these states can be treated as a semiclassical 
distribution function $f(q,{\bf k})$ neglecting quantum mechanical 
correlations. We will show that 
the Boltzmann equation for $f(q,{\bf k})$ as well 
as the formula for the current density 
(see Sec.~\ref{SecMBT}) can be derived 
from the full quantum transport model under this assumption. 
 
We consider a 
superlattice with a homogeneous electric field $F$ and 
a homogeneous carrier distribution. In this case it makes sense 
to use a local energy scale $\mathcal{E}=E-e\phi_n$ which refers to the 
bottom of the respective quantum well (here  $e\phi_n=-neFd$ holds). 
We define 
\begin{equation} 
\bar{G}_{n,m}\left({\bf k},\mathcal{E}\right)= 
G_{n,m}\left({\bf k},\mathcal{E}-eFd\frac{n+m}{2}\right) 
\end{equation} 
and $\bar{\Sigma}_{n,m}({\bf k},\mathcal{E})$ is the same way. 
Now, $\bar{G}_{m,n}({\bf k},\mathcal{E})$ only depends 
on the difference $m-n$ and $\bar{\Sigma}_m({\bf k},\mathcal{E})= 
\bar{\Sigma}({\bf k},\mathcal{E})$ 
due to the homogeneity of the system 
under stationary transport. Therefore 
it is helpful to define the spatial Fourier transform 
\begin{equation} 
\bar{G}(q,{\bf k},\mathcal{E})=\sum_h 
\e^{-\imai qhd} 
\bar{G}_{n+h,n}({\bf k},\mathcal{E}) 
\end{equation} 
This corresponds to the Fourier representation 
\begin{equation} 
\bar{G}(q,{\bf k},\mathcal{E}) 
=\sum_h\frac{1}{\hbar}\int \d t\, 
\e^{\imai \left(\mathcal{E}+e\frac{\phi_n+\phi_m}{2}\right)\frac{t}{\hbar}} 
\e^{-\imai qhd}G_{n+h,n}({\bf k};t+t_2,t_2) 
\end{equation} 
which is the special case for a homogeneous electric field 
of the general gauge invariant version used in section 7 of \cite{HAU96} 
to obtain gauge invariant quantities \cite{BER91}. 
The respective diagonal element of the density matrix 
is defined by 
\begin{equation} 
f(q,{\bf k})=\frac{1}{2\pi \imai}\int \d \mathcal{E}\, 
\bar{G}^{<}(q,{\bf k},\mathcal{E})= 
\frac{1}{2\pi \imai}\int \d \mathcal{E}\, 
\sum_h \e^{-\imai qhd} 
\bar{G}^{<}_{n+h,n}({\bf k},\mathcal{E})\, 
\end{equation} 
Applying Eq.~(\ref{Eq4GlessEdiff}) we obtain with these definitions 
\begin{equation} 
\begin{split} 
\imai eF\pabl{}{q}&\bar{G}^<(q,{\bf k},\mathcal{E})= 
2\imai\sin(qd)T_1\left[\bar{G}^<
\left(q,{\bf k},\mathcal{E}+\frac{eFd}{2}\right) 
-\bar{G}^<\left(q,{\bf k},\mathcal{E}-\frac{eFd}{2}\right)\right]\\ 
&+\sum_h\e^{-\imai qhd}\Bigg[ 
\bar{\Sigma}^{<}\left({\bf k},\mathcal{E}+\frac{heFd}{2}\right) 
\bar{G}^{\rm adv}_{h,0}({\bf k},\mathcal{E}) 
-\bar{G}^{\rm ret}_{h,0}({\bf k},\mathcal{E}) 
\bar{\Sigma}^{<}\left({\bf k},\mathcal{E}-\frac{heFd}{2}\right)\\ 
&\phantom{+\sum_h\e^{-\imai qhd}\Bigg[} 
-\left( 
\bar{\Sigma}^{\rm adv}\left({\bf k},\mathcal{E}-\frac{heFd}{2}\right) 
-\bar{\Sigma}^{\rm ret}\left({\bf k},\mathcal{E}+\frac{heFd}{2}\right)\right) 
\bar{G}^<_{h,0}({\bf k},\mathcal{E})\Bigg] 
\label{Eq4KadBaym} 
\end{split}\end{equation} 
which is still exact. 
In the same way Eq.~(\ref{Eq4GretSL}) yields: 
\begin{equation} 
\begin{split} 
\left(\mathcal{E}-E_k+\frac{\imai eF}{2}\pabl{}{q}\right) 
&\bar{G}^{\rm ret}(q,{\bf k},\mathcal{E})- 
\e^{\imai qd}T_1\bar{G}^{\rm ret}\left(q,{\bf k},\mathcal{E}+\frac{eFd}{2}\right) 
-\e^{-\imai qd}T_1\bar{G}^{\rm ret}\left(q,{\bf k},\mathcal{E}-\frac{eFd}{2}\right)\\ 
&=1+\sum_h\e^{-\imai qhd} 
\bar{\Sigma}^{\rm ret}\left({\bf k},\mathcal{E}+\frac{heFd}{2}\right) 
\bar{G}^{\rm ret}_{h,0}({\bf k},\mathcal{E}) 
\end{split}\end{equation}
 
Assuming that the self-energy does not depend strongly 
on $\mathcal{E}$\footnote{This implies that the density of final states
for scattering processes does not vary strongly with energy. This variation
occurs typically on the energy scale of the miniband width. Therefore
this assumption can be justified for $2T_1\gg eFd$.}
within the energy scale $eFd$, i.e. 
$\bar{\Sigma}^{\rm ret}\left({\bf k},\mathcal{E}+heFd/2\right)\approx
 \bar{\Sigma}^{\rm ret}\left({\bf k},\mathcal{E}\right)$,
this equation is solved by 
\begin{equation} 
\bar{G}^{\rm ret}(q,{\bf k},\mathcal{E})= 
\frac{1}{\mathcal{E}-E_k-2T_1\cos(qd)
-\bar{\Sigma}^{\rm ret}\left({\bf k},\mathcal{E}\right)} 
+\mathcal{O}\left\{(eFd)^2 
\pabl{^2\bar{G}^{\rm ret}(q,{\bf k},\mathcal{E})}{\mathcal{E}^2} 
\right\}\, . 
\end{equation} 
On the energy scale $2|T_1|$ the Green function 
$\bar{G}^{\rm ret}(q,{\bf k},\mathcal{E})$ essentially 
resembles the free particle Green function 
$1/(E-E_k-2T_1\cos(qd)+\imai 0^+)$ if 
$\bar{\Sigma}^{\rm ret}\sim \Gamma\ll 2|T_1|$ 
and $eFd \ll 2|T_1|$ (so that the last term is 
negligible), which is just the 
condition (\ref{Eq4condMBT}). 
 
Now we integrate both sides of Eq.~(\ref{Eq4KadBaym}) over energy 
$\mathcal{E}$, 
so that the $T_1$ terms on the right-hand side cancel each other. 
In the terms containing the self-energies the following 
approximations are performed: 
\begin{itemize} 
\item The $heFd/2$ terms in energy dependence of the self-energies 
are neglected. 
\item The expression 
$[\bar{G}^{\rm adv}(q,{\bf k},\mathcal{E})- 
\bar{G}^{\rm ret}(q,{\bf k},\mathcal{E})]$ is approximated by 
$2\pi \imai \delta(\mathcal{E}-2T_1\cos(qd)-E_k)$ 
which holds exactly for the free particle Green functions. 
\item We use 
$\bar{G}^{<}(q,{\bf k},\mathcal{E})\approx f(q,{\bf k}) 
2\pi \imai \delta(\mathcal{E}-2T_1\cos(qd)-E_k)$. 
This means that the energetical width of the 
respective states is neglected. As information about 
quantum mechanical correlations  is stored in the 
energy dependence (the Fourier transform 
of the time difference $t_1-t_2$), this approximation 
provides quasiclassical particles with specific momenta 
$q,{\bf k}$. 
\end{itemize} 
Then one finds 
\begin{equation} 
\imai eF\pabl{}{q}f(q,{\bf k})= 
\bar{\Sigma}^{<}\left({\bf k},E_q+E_k\right)- 
\left(\bar{\Sigma}^{\rm adv}\left({\bf k},E_q+E_k\right) 
-\bar{\Sigma}^{\rm ret}\left({\bf k},E_q+E_k\right)\right) 
f(q,{\bf k}) 
\end{equation} 
Now the same approximations are used in the evaluations 
of the self-energy for impurity scattering 
(\ref{Eq4sigimpSL}): 
\begin{equation}\begin{split} 
\Sigma^{<}({\bf k},E_q+E_k) 
=&\sum_{{\bf k}'} 
\langle V_{n{\bf k},n{\bf k}'}(\{\pol{r}_i\})
V_{n{\bf k}',n{\bf k}}(\{\pol{r}_i\})\rangle_{\rm imp}
\frac{d}{2\pi} 
\int_{-\pi/d}^{\pi/d}  \d q'\, G^{<}(q',{\bf k}',E_q+E_k)\\ 
=&\sum_{{\bf k}'}\frac{d}{2\pi}\int_{-\pi/d}^{\pi/d}  \d q'\, 
2\pi 
\langle V_{n{\bf k},n{\bf k}'}(\{\pol{r}_i\})
V_{n{\bf k}',n{\bf k}}(\{\pol{r}_i\})\rangle_{\rm imp}\\
&\times f(q',{\bf k}')\imai \delta(E_q+E_k-E_{q'}-E_{k'}) 
\end{split}\end{equation} 
which is just (up to the factor $\imai$) the in-scattering term 
from Fermi's golden rule. 
In the same way the  term $\Sigma^{\rm adv}-\Sigma^{\rm ret}$ 
provides the out-scattering rate. It is instructive to note, that 
the insertion of Eqs.~(\ref{Eq4sigretphononSL},\ref{Eq4siglessphononSL}) 
gives the respective phonon scattering terms 
including the Pauli blocking factors $(1-f(q,{\bf k}))$. 
Thus, the Boltzmann equation can be derived as a limiting case 
of the full Kadanoff Baym equation 
if the scattering induced broadening as well as the field 
dependence is neglected in the scattering term. 
The same results essentially holds if 
the full space and time dependence is maintained, as shown 
in many textbooks, such as \cite{HAU96}. Some of these 
approximations can be relaxed. In particular, a significant improvement 
can be made by the generalized Kadanoff Baym ansatz 
\cite{LIP86,SPI94,SPI95,KRA97}. 
 
Finally the current density is evaluated via Eq.~(\ref{Eq4JNGFT}) 
which takes the form: 
\begin{equation}\begin{split} 
J_{n\to n+1}=&\frac{2e}{A} 
\sum_{{\bf k}}\frac{2}{\hbar}\, \int \frac{\d \mathcal{E}}{2\pi} 
\Re\left\{T_1 
\frac{d}{2\pi}\int_{-\pi/d}^{\pi/d} \d q\, \e^{\imai qd} 
G^{<}(q,{\bf k},\mathcal{E})\right\}\\ 
=&\frac{2e}{A} 
\sum_{{\bf k}}\frac{2}{\hbar}\, 
\frac{d}{2\pi}\int_{-\pi/d}^{\pi/d} \d q\, 
\Re\left\{T_1 \e^{\imai qd} 
\imai f(q,{\bf k})\right\}\\ 
=&\frac{2e}{A} 
\sum_{{\bf k}} 
\frac{1}{2\pi}\int_{-\pi/d}^{\pi/d} \d q\, 
\frac{-2T_1d \sin(qd)}{\hbar} 
f(q,{\bf k}) 
\end{split} 
\end{equation} 
which is just Eq.~(\ref{Eq3Boltzmann-J}). 
 
\subsection{Wannier-Stark hopping\label{AppWSHderivation}} 
In subsection \ref{SecNGFconstsig} it was shown that the
Wannier-Stark states become resolved for $eFd\gg \Gamma$.
Here it will be  shown that the equations 
(\ref{Eq3WSHcurrent},\ref{Eq3WSHselfconsist}) for the self-consistent 
Wannier-Stark hopping model  can be derived from the 
of the quantum transport in this limit.
Similar to the derivation of miniband transport, the key point is 
the idea, that the occupation of the Wannier-Stark states $j$ can be 
treated as a semiclassical 
distribution function $f_j({\bf k})$ neglecting quantum mechanical 
correlations. 
Similar derivations have been presented 
in \cite{CAL84,LYA95}, where essentially 
the same approximations have been applied like in the quantum transport 
model discussed in this work.
 
According to Eq.~(\ref{Eq2WSsimp}) 
the Wannier-Stark states are 
$|\Phi_j\rangle=\sum_n 
J_{n-j}(\beta)|\Psi_{n}\rangle$ 
with $\beta=2T_1/eFd$. 
The respective Green functions are given by 
\begin{equation} 
G_{{\rm WS }j_1,j_2}({\bf k},E)=\sum_{m_1,m_2} 
J_{m_1-j_1}(\beta)J_{m_2-j_2}(\beta)G_{m_1,m_2}({\bf k},E)\, . 
\end{equation} 
Within this new basis Eq.~(\ref{Eq4GretSL}) becomes 
\begin{equation} 
\left(E-E_{k}+j_1eFd\right) 
G^{\rm ret}_{{\rm WS }j_1,j_2}({\bf k},E) 
=\delta_{j_1,j_2}+\sum_j\Sigma^{\rm ret}_{{\rm WS }j_1,j}({\bf k},E) 
G^{\rm ret}_{{\rm WS }j,j_2}({\bf k},E) 
\label{Eq4DysonWSH} 
\end{equation} 
with the definition 
\begin{equation}\begin{split} 
\Sigma_{{\rm WS }j_1,j_2}({\bf k},E) 
=&\sum_{m}J_{m-j_1}(\beta)J_{m-j_2}(\beta) 
\Sigma_{m}({\bf k},E)\\ 
=&\sum_{m,j_3,j_4}\sum_{{\bf k}'}J_{m-j_1}(\beta)J_{m-j_2}(\beta) 
\left|V_{{\bf k}'{\bf k}}\right|^2 
J_{m-j_3}(\beta)J_{m-j_4}(\beta) G_{{\rm WS }j_3,j_4}({\bf k}',E) 
\label{Eq4SigmaWS} 
\end{split} 
\end{equation} 
where the Born approximation was applied 
under the assumption, that scattering is diagonal 
in the Wannier basis, independent on the well number $m$, 
and that no correlations exist between 
different wells $m$. Furthermore we restrict ourselves to 
impurity scattering with matrix element 
$\langle \Psi_{m,{\bf k}'}|\hat{H}^{\rm scatt} 
|\Psi_{m,{\bf k}}\rangle=V_{{\bf k}'{\bf k}}$ for simplicity. 
In the same way Eq.~(\ref{Eq4GlessEdiff}) gives 
\begin{equation} 
\begin{split} 
(j_1-j_2)eFd  G^<_{{\rm WS }j_1,j_2}({\bf k},E)=& 
\sum_j\Bigg[ 
\Sigma^{\rm ret}_{{\rm WS }j_1,j}({\bf k},E) 
G^<_{{\rm WS }j,j_2}({\bf k},E) 
+\Sigma^{<}_{{\rm WS }j_1,j}({\bf k},E) 
G^{\rm adv}_{{\rm WS }j,j_2}({\bf k},E)\\ 
&-G^{\rm ret}_{{\rm WS }j_1,j}({\bf k},E) 
\Sigma^{<}_{{\rm WS }j,j_2}({\bf k},E) 
-G^{<}_{{\rm WS }j_1,j}({\bf k},E) 
\Sigma^{\rm adv}_{{\rm WS }j,j_2}({\bf k},E)\Bigg]\, . 
\label{Eq4GlessWSH} 
\end{split}\end{equation} 
 
The typical energy scale is given by $eFd$. If the self-energies 
(of the order of $\Gamma$) are small in comparison to $eFd$, 
Eq.~(\ref{Eq4DysonWSH}) gives 
\begin{equation} 
G^{\rm ret}_{{\rm WS }j_1,j_2}({\bf k},E)\approx 
\delta_{j_1,j_2}\frac{1}{E-E_{k}+j_1 eFd+\imai 0^+} 
\label{Eq4GretWSfree} 
\end{equation} 
describing  free-particle Wannier-Stark states. 
The occupation of these states is governed by 
a semiclassical distribution $f_{j_1}({\bf k})$ yielding 
\begin{equation} 
G^{<}_{{\rm WS }j_1,j_2}({\bf k},E)\approx 
2\pi \imai\delta_{j_1,j_2} f_{j_1}({\bf k}) \delta(E-E_{k}+j_1eFd) 
\label{Eq4GlessWSfree} 
\end{equation} 
 
Let us first consider the case $j_1=j_2$, when the left-hand side of 
Eq.~(\ref{Eq4GlessWSH}) vanishes. 
Applying  the approximations (\ref{Eq4GretWSfree},\ref{Eq4GlessWSfree}) 
in the scattering term 
on the right-hand side and performing the 
integration $1/(2\pi)\int \d E$ we obtain 
\begin{equation} 
0=\imai\left[\Sigma^{\rm ret}_{{\rm WS }j_1,j_1}({\bf k},E_k-j_1eFd) 
-\Sigma^{\rm adv}_{{\rm WS }j_1,j_1}({\bf k},E_k-j_1eFd)\right] 
f_{j_1}({\bf k}) 
+\imai\Sigma^{<}_{{\rm WS }j_1,j_1}({\bf k},E_k-j_1eFd) 
\end{equation} 
Inserting Eq.~(\ref{Eq4SigmaWS}) one obtains together with 
the approximations (\ref{Eq4GretWSfree},\ref{Eq4GlessWSfree}): 
\begin{equation}\begin{split} 
0=&\sum_{j_2,m,{\bf k}'} 
\left[J_{m-j_1}(\beta)\right]^2 
\left|V_{{\bf k}'{\bf k}}\right|^2 
\left[J_{m-j_2}(\beta)\right]^2 2\pi\delta(E_k-j_1eFd-(E_{k'}-j_2eFd)) 
\left[f_{j_1}({\bf k})-f_{j_2}({\bf k}')\right]\\ 
=&\hbar\sum_{{\bf k}'j_2} 
\left[R_{j_1,{\bf k}\to j_2,{\bf k'}}f_{j_1}({\bf k})- 
R_{j_2,{\bf k'}\to j_1,{\bf k}}f_{j_2}({\bf k}')\right] 
\end{split} 
\end{equation} 
where Eq.~(\ref{Eq3R-WSH}) has been inserted. 
This is just the  condition for self-consistency 
(\ref{Eq3WSHselfconsist}) in the stationary case 
for the Wannier-Stark hopping approach. 
 
The current is determined from Eq.~(\ref{Eq4JNGFT}) which can be 
rewritten as
\begin{equation} 
\begin{split}
J_{0\to 1}=&\frac{2e}{A \hbar  } 
\sum_{{\bf k}}\, \int \frac{\d E}{2\pi} 
T_1 \Re\left\{ 
G_{1,0}^{<}({\bf k},E)-G_{0,1}^{<}({\bf k},E)
\right\}\\
=&\frac{2e}{A\hbar} 
\sum_{j_1,{\bf k}}
\sum_{m} m J_{m}(\beta)J_{m-j_1}(\beta)  
\int\frac{\d  E}{2\pi} 
\Re\left\{j_1 eFd   G^<_{{\rm WS}j_1,0}({\bf k},E)\right\} 
\end{split}
\end{equation} 
(To verify this idendity insert all definitions 
into the second line and use $\int \d E\, G_{m,n}^{<}({\bf k},E)=
\int \d E\, G_{m-n,0}^{<}({\bf k},E)$ for the homogeneous system)
Now Eq.~(\ref{Eq4GlessWSH}) can be inserted, where 
the  approximations (\ref{Eq4GretWSfree},\ref{Eq4GlessWSfree}) 
are applied to the scattering terms. This yields: 
\begin{equation}\begin{split} 
J=& \frac{2e}{A\hbar}\sum_{j_1,{\bf k}} 
\sum_{m} m J_{m}(\beta)J_{m-j_1}(\beta)  
\Bigg[
-f_0({\bf k})\Im\left\{\Sigma^{\rm ret}_{{\rm WS }j_1,0}({\bf k},E_k)\right\} 
-\frac{1}{2}\Im\left\{\Sigma^{<}_{{\rm WS }j_1,0}({\bf k},E_k)\right\}\\ 
&-\frac{1}{2}\Im\left\{\Sigma^{<}_{{\rm WS }j_1,0}({\bf k},E_k-j_1eFd)\right\} 
+f_{j_1}({\bf k})\Im\left\{\Sigma^{\rm adv}_{{\rm WS }j_1,0}({\bf k},E_k-j_1eFd)\right\} 
\Bigg]\, . 
\label{Eq4WSHJint} 
\end{split}\end{equation} 
where it has been used that $\Sigma^{<}$ is purely imaginary. 
Now we define the auxiliary function 
\begin{equation} 
f_{\rm aux}(j'-j)= 
\sum_{{\bf k},{\bf k}'} 
|V_{{\bf k}'{\bf k}}|^2 
\pi\delta(E_k-jeFd-E_{k'}+j'eFd) 
\left[f_j({\bf k})-f_{j'}({\bf k}')\right] 
\end{equation} 
which is an odd function of the difference $j'-j$ 
for a homogeneous situation, when the occupation functions 
are independent from the index $j$. 
Inserting Eq.~(\ref{Eq4SigmaWS}) 
the {\em first two summands} of Eq.~(\ref{Eq4WSHJint}) yield: 
\begin{equation}\begin{split} 
\frac{2e}{A\hbar}&\sum_{j_1,m,n,j_2} 
m J_{m}(\beta) J_{m-j_1}(\beta) 
J_{n-j_1}(\beta) J_{n}(\beta) 
\left[J_{n-j_2}(\beta)\right]^2f_{\rm aux}(j_2)\\ 
=&\frac{2e}{A\hbar}\sum_{n,j_2} 
n \left[J_{n}(\beta)\right]^2\left[J_{n-j_2}(\beta)\right]^2 
f_{\rm aux}(j_2)\\ 
=&\frac{2e}{A\hbar}\sum_{n,h\ge 1} 
h \left[J_{n}(\beta)\right]^2\left[J_{n-h}(\beta)\right]^2 
f_{\rm aux}(h) 
\label{Eq4JWSHpart} 
\end{split}\end{equation} 
where in the last line the relation 
$\sum_n n \left[J_{n}(\beta)\right]^2 
\left[J_{n-j_2}(\beta)\right]^2=\frac{j_2}{2} \sum_n 
\left[J_{n}(\beta)\right]^2 
\left[J_{n-j_2}(\beta)\right]^2$ 
and the symmetry for $j_2=\pm h$ was applied. 
Comparing with the relation 
(\ref{Eq3R-WSH}) one obtains: 
\begin{equation} 
\frac{e}{A}\sum_{h\ge 1,{\bf k},{\bf k}'} 
\left[R_{0,{\bf k}\to h,{\bf k'}}f_0({\bf k})- 
R_{h,{\bf k'}\to 0,{\bf k}}f_{h}({\bf k}')\right] 
\end{equation} 
which is just half 
the current density for Wannier-Stark hopping (\ref{Eq3WSHcurrent}). 
Similarly the 
{\em last two summands} of Eq.~(\ref{Eq4WSHJint}) yield: 
\begin{equation}\begin{split} 
\frac{2e}{A\hbar}&\sum_{j_1,m,n,j_2} 
m J_{m}(\beta) J_{m-j_1}(\beta) 
J_{n-j_1}(\beta) J_{n}(\beta) 
\left[J_{n-j_2}(\beta)\right]^2f_{\rm aux}(j_2-j_1)\\ 
&=\frac{2e}{A\hbar}\sum_{j_1,m',n',h} 
(m'+j_1) J_{m'+j_1}(\beta) J_{m'}(\beta) 
J_{n'}(\beta) J_{n'+j_1}(\beta) 
\left[J_{n'-h}(\beta)\right]^2f_{\rm aux}(h)\\ 
&=\frac{2e}{A\hbar}\sum_{n',h} 
n' \left[J_{n'}(\beta)\right]^2 
\left[J_{n'-h}(\beta)\right]^2 
f_{\rm aux}(h) 
\end{split}\end{equation} 
where in the second line $n'=n-j_1$, $m'=m-j_1$, and $h=j_2-j_1$ 
have been introduced. In the derivation of the last line, 
the term with prefactor $j_1$ vanishes by performing the $m'$ sum. 
The final expression is identical to the second line of 
Eq.~(\ref{Eq4JWSHpart}) and thus the full current density 
$J_{\rm WSH}$ for Wannier-Stark hopping (\ref{Eq3WSHcurrent}) 
is recovered from Eq.~(\ref{Eq4WSHJint}).

\section{Quantum transport under irradiation \label{AppEinstr}}
In this appendix the quantum transport in superlattices
under irradiation with a THz field $F_{\rm ac}$ is investigated.
Together with a static field $F_{\rm dc}$ one obtains
a potential with diagonal elements
\begin{equation}
U_n(t)=-eF_{\rm dc}dn-\alpha\hbar\Omega\cos(\Omega t)n\label{EqCpot}
\end{equation}
in the Wannier basis, where $\alpha=eF_{\rm ac}d/\hbar\Omega$ is
the ratio between the radiation field strength and its energy quantum.
Similar to section \ref{ChapNGFT} the theory of nonequilibrium Green
functions is applied. The respective equations are derived in subsection
\ref{SecCformulation} for transport in the lowest miniband.
In subsection \ref{SecCST} it is shown that
Eqs.~(\ref{Eq6I0}-\ref{Eq6Ihsin}) together with
Eqs.~(\ref{Eq6IST},\ref{Eq6KST}) hold for sequential tunneling
between equivalent levels $a\to a$, where the tunneling
matrix element $T_1^a$ does not depend on the field.
In contrast, the matrix element
$H^{ba}=eF(t)dR_1^{ba}=[eF_{\rm dc}d+\hbar\Omega\alpha\cos(\Omega t_1)]R^{ba}_1$
for the transition between different levels [see Eq.~(\ref{Eq2hamW1})]
depends on time which  provides further complications.
This will be analyzed in subsection \ref{SecCab}, where it will
be shown that Eq.~(\ref{Eq6I0}) for the rectified response
still holds in this case.

\subsection{General formulation\label{SecCformulation}}
In this subsection essentially the same approximation
as in section \ref{SecApplSL} are applied.
In particular the self-energies $\Sigma^{</{\rm ret}}_{m}(t_3,t_4) $
are assumed to be diagonal in the well index.
For simplicity we set $E^a=0$ here. Neglecting scattering and coupling
between the wells,
the bare Green function reads
\begin{equation}
g^{\rm ret}_{n}(t_1,t_2)=-\imai\Theta(t_1-t_2)\e^{-\imai (E_k-eF_{\rm dc}dn)(t_1-t_2)}
\e^{\imai \alpha n[\sin(\Omega t_1)-\sin(\Omega t_2)]}
\end{equation}
for the potential (\ref{EqCpot}).
Similar to \cite{BRA97} we define the on-site evolution
\begin{equation}
S_m(t)=\e^{-\imai \alpha m \sin(\Omega t)}
\end{equation}
of the states and apply the following Fourier
expansion for {\em retarded functions}
\begin{equation}
C^{\rm ret}_{m,n}(t_1,t_2)=
S^{\dag}_m(t_1)
\frac{1}{2\pi} \int \d E
\sum_r \e^{-\imai r\Omega t_1}\e^{-\imai E(t_1-t_2)}
C^{\rm ret}_{m,n;r}(E)S_n(t_2)
\end{equation}
In particular this definition gives:
\begin{equation}
g^{\rm ret}_{n;r}(E)=\frac{1}
{E-E_k+neF_{\rm dc}d+\imai 0^+}\delta_{r,0}
\end{equation}
Now investigate
\begin{equation}\begin{split}
\frac{1}{\hbar}\int \d t_3&\,g^{\rm ret}_{m}(t_1,t_3)T_1
G^{\rm ret}_{m\pm 1,n}(t_3,t_2)\\
=&\frac{1}{\hbar}\int \d t_3 S^{\dag}_m(t_1)
\frac{1}{2\pi} \int \d E_1
\e^{-\imai E_1(t_1-t_3)/\hbar}
\frac{1}{E_1-E_k+meF_{\rm dc}d+\imai 0^+}S_m(t_3)\\
&\times T_1 S^{\dag}_{m\pm 1}(t_3)
\frac{1}{2\pi} \int \d E_2
\sum_{r_2} \e^{-\imai r_2\Omega t_3}\e^{-\imai E_2(t_3-t_2)/\hbar}
G^{\rm ret}_{m\pm 1,n;r_2}(E_2)S_n(t_2)\\
=&S^{\dag}_m(t_1)
\frac{1}{2\pi} \int \d E_1
\frac{1}{\hbar}\int \d t_3 \e^{-\imai E_1(t_1-t_3)/\hbar}
\frac{1}{E_1-E_k+meF_{\rm dc}d+\imai 0^+}\\
&\times T_1 \sum_{s} J_s(\alpha) \e^{\pm is\Omega t_3}
\frac{1}{2\pi} \int \d E_2
\sum_{r_2} \e^{-\imai r_2\Omega t_3} \e^{-\imai E_2(t_3-t_2)/\hbar}
G^{\rm ret}_{m\pm 1,n;r_2}(E_2)S_n(t_2)\\
=&S^{\dag}_m(t_1)
\frac{1}{2\pi} \int \d E_2\sum_{r_2,s}
\e^{-\imai (E_2/\hbar+r_2\Omega\mp s\Omega)t_1}
\frac{1}{E_2+r_2\hbar\Omega\mp s\hbar\Omega -E_k+meFd+\imai 0^+}\\
&\times T_1  J_s(\alpha)
  \e^{\imai E_2t_2/\hbar}
G^{\rm ret}_{m\pm 1,n;r_2}(E_2)S_n(t_2)\\
=&S^{\dag}_m(t_1)
\frac{1}{2\pi} \int \d E\sum_{r}
\e^{-\imai r\Omega t_1}\e^{-\imai E(t_1-t_2)/\hbar}  S_n(t_2)\\
&\times\frac{1}{E+r\hbar\Omega-E_k+meF_{\rm dc}d+\imai 0^+}
T_1 \sum_{r_2} J_{\pm(r_2-r)}(\alpha)
G^{\rm ret}_{m\pm 1,n;r_2}(E) \label{EqCgretpart1}\, .
\end{split}\end{equation}
Similarly
\begin{equation}\begin{split}
\frac{1}{\hbar^2}\int \d t_3&\int \d t_4\,g^{\rm ret}_{m}(t_1,t_3)
\Sigma^{\rm ret}_{m}(t_3,t_4)
G^{\rm ret}_{m,n}(t_4,t_2)\\
=&S^{\dag}_m(t_1)
\frac{1}{2\pi} \int \d E\sum_{r}
\e^{-\imai r\Omega t_1}\e^{-\imai E(t_1-t_2)/\hbar}S_n(t_2)\\
&\times
\sum_{r_3}\frac{1}{E+r\hbar\Omega-E_k+meF_{\rm dc}d+\imai 0^+}
\Sigma^{\rm ret}_{m;r-r_3}(E+r_3\hbar\Omega)
G^{\rm ret}_{m,n;r_3}(E)\, .
\end{split}\end{equation}
With these relations Eq.~(\ref{Eq4Grett1}) gives
the recursion for the retarded Green function:
\begin{equation}\begin{split}
(E+r\hbar\Omega&-E_k+meF_{\rm dc}d+\imai 0^+)G^{\rm ret}_{m,n;r}(E)=\delta_{m,n}\delta_{r,0}\\
&+T_1 \sum_{r_2}
\left[J_{(r_2-r)}(\alpha) G^{\rm ret}_{m+1,n;r_2}(E)
+J_{(r-r_2)}(\alpha) G^{\rm ret}_{m-1,n;r_2}(E)\right]\\
&+\sum_{r_3}\Sigma^{\rm ret}_{m;r-r_3}(E+r_3\hbar\Omega)
G^{\rm ret}_{m,n;r_3}(E)\label{EqCGret}
\end{split}\end{equation}
For the {\em advanced functions} we define
\begin{eqnarray}
C^{\rm adv}_{m,n}(t_1,t_2)&=&
S^{\dag}_m(t_1)
\frac{1}{2\pi} \int \d E
\sum_r \e^{\imai r\Omega t_2}\e^{-\imai E(t_1-t_2)/\hbar}
C^{\rm adv}_{m,n;r}(E)S_n(t_2)
\end{eqnarray}
so that $G^{\rm adv}_{m,n;r}(E)=\left\{G^{\rm ret}_{n,m;r}(E)\right\}^*$
holds.  The Fourier expansion of the
{\em lesser functions} is defined by
\begin{equation}
C^<_{m,n}(t_1,t_2)=
S^{\dag}_m(t_1)
\frac{1}{2\pi} \int \d E
\sum_r \e^{-\imai r\Omega (t_1+t_2)/2}\e^{-\imai E(t_1-t_2)/\hbar}
C^<_{m,n;r}(E)S_n(t_2)\, .
\end{equation}
Then we find the product
\begin{equation}\begin{split}
\frac{1}{\hbar^2}&\int \d t_3\int \d t_4\,G^{\rm ret}_{m,m_1}(t_1,t_3)
\Sigma^{<}_{m_1}(t_3,t_4)
G^{\rm adv}_{m_1,n}(t_4,t_2)\\
=&
S^{\dag}_m(t_1)\sum_{r}
\frac{1}{2\pi} \int \d E
\e^{-\imai r\Omega(t_1+t_2)/2} \e^{-\imai E(t_1-t_2)/\hbar}S_n(t_2)
\sum_{r_1,r_3}G^{\rm ret}_{m,m_1;r_1}\left(E+\left(\frac{r}{2}-r_1\right)
\hbar\Omega\right)\\
&\times\Sigma^{<}_{m_1;r-r_1+r_3}\left(E-\frac{r_1+r_3}{2}\hbar\Omega\right)
G^{\rm adv}_{m_1,n;r_3}\left(E-\left(\frac{r}{2}+r_3\right)\hbar\Omega\right)
\end{split}\end{equation}
so that the Keldysh relation becomes
\begin{equation}\begin{split}
G^{<}_{n,m;r}(E)=&
\sum_{r_1,r_3}G^{\rm ret}_{n,m_1;r_1}\left(E+\left(\frac{r}{2}-r_1\right)
\hbar\Omega\right)\\
&\times\Sigma^{<}_{m_1;r-r_1+r_3}\left(E-\frac{r_1+r_3}{2}\hbar\Omega\right)
G^{\rm adv}_{m_1,m;r_3}\left(E-\left(\frac{r}{2}+r_3\right)\hbar\Omega\right)
\, . \label{EqCKeldysh}
\end{split}\end{equation}
Eqs.~(\ref{EqCGret},\ref{EqCKeldysh}) allow for a self-consistent solution
provided the functionals for the self-energy are known.
Within the self-consistent Born approximation they are given
by  the same  functionals as in section
\ref{Secselfenergy} where the same index $r$ is added
in the self-energies as well as the Green-functions.
This diagonal structure in $r$ is due to the fact that
the scattering matrix element is
either not time dependent (for impurity scattering) or only
depends on the time difference $t_1-t_2$ (for phonon scattering).

Finally, the current is given by
\begin{equation}\begin{split}
I_{n\to n+1}=&
\frac{4e}{\hbar}\sum_{{\bf k}}\,{\rm Re}
\left\{T_1 G^<_{n+1,n}(t,t,{\bf k})\right\}\\
=&\frac{4e}{\hbar}
\sum_{{\bf k}}\sum_r\int \frac{\d E}{2\pi}
{\rm Re}\left\{\sum_s J_{s+r}(\alpha)
T_1 G^<_{n+1,n;s}(E,{\bf k})\e^{\imai r\Omega
t}\right\}\label{EqCcurrent}
\end{split}\end{equation}
where $S_{n+1}^{\dag}S_n(t)=\sum_{r'}J_{r'}(\alpha)\e^{\imai r'\Omega t}$
has been used.
Note that for homogeneous systems
\begin{equation}
G_{m,n;r}(E)=G_{m-n,0;r}(E+neF_{\rm dc}d)\quad \mbox{and}\quad
\Sigma_{n;r}(E)=\Sigma_{0;r}(E+neF_{\rm dc}d)
\end{equation}
holds which can simplify the calculation significantly.

\subsection{Sequential tunneling\label{SecCST}}
Now we want to derive the expression for sequential
tunneling used in section \ref{SecSTeinstr}.
Like in Appendix \ref{AppSTderivation}
the lowest order in the coupling
yields  a current $\sim T_1^2$.
For vanishing coupling Eqs.~(\ref{EqCGret},\ref{EqCKeldysh}) give
$G^{</{\rm ret}}_{m_1,m_2;r}({\bf k},E)=\delta_{m_1,m_2}\delta_{r,0}
\tilde{G}^{</{\rm ret}}_{m_1}({\bf k},E+m_1eF_{\rm dc}d)$ and
$\Sigma^{</{\rm ret}}_{m}({\bf k},E)=
\tilde{\Sigma}^{</{\rm ret}}_{m;r}({\bf k},E+meF_{\rm dc}d)\delta_{r,0}$
which are [up to the shift in the energy argument which
eliminates the $F_{\rm dc}$-dependence of
$\tilde{G}_m({\bf k},E)$]
the same functions as applied in Appendix \ref{AppSTderivation}
without irradiation. Thus
\begin{equation}
\tilde{G}^{<}_{m}({\bf k},E)=\tilde{G}^{{\rm ret}}_{m}({\bf k},E)
\tilde{\Sigma}^{<}_{m}({\bf k},E)\tilde{G}^{\rm adv}_{m}({\bf k},E)
=\imai\tilde{A}({\bf k},E)n_F(E-\mu_m)\, .
\end{equation}
In lowest order of the coupling one finds from Eq.~(\ref{EqCGret}):
\begin{align}
\begin{split}
G^{\rm ret}_{m\pm 1,m;r}({\bf k},E)=&
\tilde{G}^{\rm ret}_{m\pm 1}({\bf k},E+(m\pm 1)eF_{\rm dc}d+r\hbar\Omega)
T_1J_{\pm r}(\alpha)\tilde{G}^{\rm ret}_{m}({\bf k},E+meF_{\rm dc}d) \\
&+\mathcal{O}(T_1^2)
\end{split}\\
G^{\rm ret}_{m,m;r}({\bf k},E)=&
\delta_{r,0}\tilde{G}^{\rm ret}_{m}({\bf k},E+meF_{\rm dc}d)+\mathcal{O}(T_1^2)\\
G^{\rm ret}_{n,m;r}({\bf k},E)=&\mathcal{O}(T_1^2)\quad \mbox{for}
\quad |m-n|\ge 2
\end{align}
and in a similar way:
\begin{equation}\begin{split}
G^{\rm adv}_{m\pm 1,m;r}({\bf k},E)=&
\tilde{G}^{\rm adv}_{m\pm 1}({\bf k},E+(m\pm 1)eF_{\rm dc}d)
T_1J_{\mp r}(\alpha)
\tilde{G}^{\rm adv}_{m}({\bf k},E+r\hbar\Omega+meF_{\rm dc}d)\\
&+\mathcal{O}(T_1^2)
\end{split}
\end{equation}
With these relations Eq.~(\ref{EqCKeldysh}) gives
\begin{equation}\begin{split}
G^{<}_{m+1,m;s}&({\bf k},E)=
\tilde{G}^{\rm ret}_{m+ 1}({\bf k},E+(m+1)eF_{\rm dc}d+s/2\hbar\Omega)
T_1J_{s}(\alpha)\tilde{G}^{\rm ret}_{m}({\bf k},E+meF_{\rm
dc}d-s/2\hbar\Omega)\\
&\phantom{+}\times \tilde{\Sigma}^{<}_{m}({\bf k},E+meF_{\rm dc}d-s/2\hbar\Omega)
\tilde{G}^{\rm adv}_{m}({\bf k},E+m_eF_{\rm dc}d-s/2\hbar\Omega)\\
&+\tilde{G}^{\rm ret}_{m+1}({\bf k},E+(m+1)eF_{\rm dc}d+s/2\hbar\Omega)
\tilde{\Sigma}^{<}_{m+1}({\bf k},E+(m+1)eF_{\rm dc}d+s/2\hbar\Omega)\\
&\phantom{+}\times \tilde{G}^{\rm adv}_{m+1}({\bf k},E+(m+1)eF_{\rm dc}d+s/2\hbar\Omega)
T_1J_{s}\tilde{G}^{\rm adv}_{m}({\bf k},E+meF_{\rm
dc}d-s/2\hbar\Omega)\\
=&\imai T_1J_{s}\Big[
\tilde{G}^{\rm ret}_{m+1}({\bf k},\tilde{E}+eF_{\rm dc}d+s\hbar\Omega)
\tilde{A}_{m}({\bf k},E)n_F(\tilde{E}-\mu_m)\\
&+\tilde{A}_{m+1}({\bf k},\tilde{E}+eF_{\rm dc}d+s\hbar\Omega)
n_F(\tilde{E}+eF_{\rm dc}d+s\hbar\Omega-\mu_{m+1})
\tilde{G}^{\rm adv}_{m}({\bf k},\tilde{E})\Big]\\
&+\mathcal{O}(T_1^2)
\end{split}\end{equation}
where $\tilde{E}=E+meF_{\rm dc}d-s/2\hbar\Omega$ was inserted in the last line.
Inserting into Eq.~(\ref{EqCcurrent}) and sorting with respect to
$\cos(r\Omega t)$ and $\sin(r\Omega t)$ we obtain
Eqs.~(\ref{Eq6I0}-\ref{Eq6Ihsin}) with
\begin{equation}
\begin{split}
I_{\rm dc}(eFd)=
\frac{2e}{\hbar}T_1^2
\sum_{{\bf k}}\int \frac{\d E}{2\pi}
&\tilde{A}^{\rm ret}_{m+1}({\bf k},\tilde{E}+eFd)
\tilde{A}_{m}({\bf k},E)\\
&[n_F(\tilde{E}-\mu_m)-n_F(\tilde{E}+eFd-\mu_{m+1})]
\end{split}
\end{equation}
and
\begin{equation}
\begin{split}
K(eFd)=-\frac{4e}{\hbar}T_1^2
&\sum_{{\bf k}}\int \frac{\d E}{2\pi}
\Big[\Re\{\tilde{G}^{\rm ret}_{m+1}({\bf k},\tilde{E}+eFd)\}
\tilde{A}_{m}({\bf k},E)n_F(\tilde{E}-\mu_m)\\
&
+\tilde{A}_{m+1}({\bf k},\tilde{E}+eFd)
n_F(\tilde{E}+eFd-\mu_{m+1})\Re\{\tilde{G}^{\rm adv}_{m}({\bf k},\tilde{E})\}\Big]
\end{split}
\end{equation}
This proves the formulas applied in section \ref{SecSTeinstr}
for $a\to a$ tunneling.

\subsection{Tunneling between different levels\label{SecCab}}
Now we want to consider tunneling between different levels.
We will focus on the transitions from level  $a$ in well $m=0$ to
the level $b$ in well $m=1$ and omit the well indices in the
following.
According to Eq.~(\ref{Eq2hamW1}), the matrix element
for this transition
$[eF_{\rm dc}d+\hbar\Omega\alpha\cos(\Omega t_1)]R^{ba}_1$
becomes time dependent.
Neglecting the coupling one obtains again the functions
$\tilde{G}_{a/b}({\bf k},E)$ as in the preceding subsection.
In the lowest order
Eq.~(\ref{Eq4Grett1}) yields
\begin{equation}
\begin{split}
\left(\imai\hbar \pabl{}{t_1}-E_k-E^b+eF_{\rm dc}d\right)
G^{\rm ret/adv}_{b,a}({\bf k};t_1,t_2)=&
[eF_{\rm dc}d+\hbar\Omega\alpha\cos(\Omega t_1)]R^{ba}_1
\tilde{G}^{\rm ret/adv}_{a,a}({\bf k};t_1,t_2)\\
&+\int \frac{\d t}{\hbar}
\tilde{\Sigma}^{\rm ret/adv}_{b}({\bf k};t_1,t)
G^{\rm ret/adv}_{b,a}({\bf k};t,t_2)
\end{split}
\end{equation}
In the same way as  Eq.~(\ref{EqCgretpart1}) we obtain
\begin{equation}\begin{split}
\frac{1}{\hbar}\int \d t_3&\,g^{\rm ret}_{b}(t_1,t_3)
[eF_{\rm dc}d+\hbar\Omega\alpha\cos(\Omega t_3)]R^{ba}_1
\tilde{G}^{\rm ret}_{a}(t_3,t_2)\\
=&\frac{1}{\hbar}\int \d t_3 S^{\dag}_1(t_1)
\frac{1}{2\pi} \int \d E_1
\e^{-\imai E_1(t_1-t_3)/\hbar}
\frac{1}{E_1-E^b-E_k+eF_{\rm dc}d+\imai 0^+}S_1(t_3)\\
&\times  [eF_{\rm dc}d+\hbar\Omega\alpha\cos(\Omega t_3)]R^{ba}_1
S^{\dag}_{0}(t_3)
\frac{1}{2\pi} \int \d E_2 \e^{-\imai E_2(t_3-t_2)/\hbar}
\tilde{G}^{\rm ret}_{a}(E_2)S_0(t_2)\\
=&\frac{1}{\hbar}\sum_{s}\int \d t_3 S^{\dag}_1(t_1)
\frac{1}{2\pi} \int \d E_1
\e^{-\imai E_1(t_1-t_3)/\hbar} J_s(\alpha)
\frac{1}{E_1-E^b-E_k+eF_{\rm dc}d+\imai 0^+}\\
&\times  \e^{-\imai s\Omega t_3}
\left[eF_{\rm dc}d+\hbar\Omega\alpha\frac{\e^{\imai \Omega t_3}+\e^{-\imai \Omega t_3}}{2}\right]
R^{ba}_1 \frac{1}{2\pi} \int \d E_2 \e^{-\imai E_2(t_3-t_2)/\hbar}
\tilde{G}^{\rm ret}_{a}(E_2)S_0(t_2)\\
=&\frac{1}{2\pi} S^{\dag}_1(t_1) S_0(t_2)
 \int \d E_1
\e^{-\imai E_1t_1/\hbar}\sum_{s} J_s(\alpha)
\frac{1}{E_1-E^b-E_k+eF_{\rm dc}d+\imai 0^+}\\
&\times  R^{ba}_1\Big[eF_{\rm dc}d \e^{\imai (E_1/\hbar-s\Omega)t_2/\hbar}
\tilde{G}^{\rm ret}_{a}(E_1-s\hbar\Omega)\\
&\phantom{\times  R^{ba}_1\Big[}
+\frac{\hbar\Omega\alpha}{2} \e^{\imai (E_1/\hbar-(s+1)\Omega)t_2/\hbar}
\tilde{G}^{\rm ret}_{a}(E_1-(s+1)\hbar\Omega)\\
&\phantom{\times  R^{ba}_1\Big[}
+\frac{\hbar\Omega\alpha}{2} \e^{\imai (E_1/\hbar-(s-1)\Omega)t_2/\hbar}
\tilde{G}^{\rm ret}_{a}(E_1-(s-1)\hbar\Omega)\Big]
\\
=&\frac{1}{2\pi} S^{\dag}_1(t_1)
 \int \d E\sum_{r}\e^{-\imai r\Omega t_1}
\e^{-\imai E(t_1-t_2)/\hbar} S_0(t_2)
\frac{1}{E+r\hbar \Omega-E^b-E_k+eF_{\rm dc}d+\imai 0^+} \\
&\times R^{ba}_1\Big[
J_r(\alpha)eF_{\rm dc}d+
J_{r+1}(\alpha)\hbar\Omega\alpha/2+J_{r-1}(\alpha)\hbar\Omega\alpha/2\Big]
\tilde{G}^{\rm ret}_{a}(E)
\end{split}\end{equation}
Therefore we find
\begin{equation}
G^{\rm ret}_{b,a;r}({\bf k},E)=
\tilde{G}^{\rm ret}_{b}({\bf k},E+eF_{\rm dc}d+r\hbar\Omega)
 J_r(\alpha)R^{ba}_1[eF_{\rm dc}d +r\hbar\Omega]
\tilde{G}^{\rm ret}_{a}({\bf k},E)
+\mathcal{O}(T_1^2)
\end{equation}
and the Keldysh relation becomes in analogy to the preceding
subsection:
\begin{equation}\begin{split}
G^{<}_{b,a;s}({\bf k},E)=&
\imai J_s(\alpha)R^{ba}_1[eF_{\rm dc}d +s\hbar\Omega]
\Big[\tilde{G}^{\rm ret}_{b}({\bf k},\tilde{E}+eF_{\rm dc}d+s\hbar\Omega)
\tilde{A}_{a}({\bf k},E)n_F(\tilde{E}-\mu_0)\\
&+\tilde{A}_{b}({\bf k},\tilde{E}+eF_{\rm dc}d+s\hbar\Omega)
n_F(\tilde{E}+eF_{\rm dc}d+s\hbar\Omega-\mu_{1})
\tilde{G}^{\rm adv}_{a}({\bf k},\tilde{E})\Big]+\mathcal{O}(T_1^2)
\end{split}\end{equation}
Finally, the current density can be obtained
by inserting the time dependent matrix element
into Eq.~(\ref{EqCcurrent})
\begin{equation}\begin{split}
I_{a\to b}
=&\frac{4e}{\hbar}
\sum_{{\bf k}}\sum_{r'}\int \frac{\d E}{2\pi}
\Re\left\{\sum_s J_{s+r'}(\alpha)
[eF_{\rm dc}d+\hbar\Omega\alpha(\e^{\imai \Omega t}+\e^{-\imai \Omega t})/2]
R^{ba}_1 G^<_{b,a;s}(E,{\bf k})\e^{\imai r'\Omega t}\right\}\\
=&
\frac{4e}{\hbar}
\sum_{{\bf k}}\sum_{r}\int \frac{\d E}{2\pi}
\Re\left\{\sum_s J_{s+r}(\alpha)
[eF_{\rm dc}d+(s+r)\hbar\Omega]
R^{ba}_1 G^<_{b,a;s}(E,{\bf k})\e^{\imai r\Omega t}\right\} \, .
\end{split}\end{equation}
For $r=0$,  Eq.~(\ref{Eq6I0}) can be recovered even in the
case of a linear field dependence of the matrix element.
In contrast, for $r\ge 1$  Eqs.~(\ref{Eq6Ihcos},\ref{Eq6Ihsin})
only hold if the field dependence of the matrix
element is negligible, which is appropriate if
$eF_{\rm ac}d\ll eF_{\rm dc}d$ and $\hbar\Omega \ll eF_{\rm dc}d$ hold.


\end{document}